\documentclass[aps,prd, final,showpacs]{revtex4}
\usepackage[latin1]{inputenc}
\usepackage{amsmath}
\usepackage{amsfonts}
\usepackage{amssymb}
\usepackage{amsthm}
\usepackage[dvips]{graphics}
\usepackage[pdftex]{graphicx}
\newcommand{\be}{\begin{equation}}
\newcommand{\ee}{\end{equation}}

\begin{document}
%\title{New Perspectives in Chameleon Field Theories}
\title[Scalar Field Theories with a Strong Coupling to Matter]{Evading Equivalence Principle Violations, Cosmological and other Experimental Constraints in Scalar Field Theories with a Strong Coupling to Matter}

\author{David F. Mota}
\affiliation{Institute for Theoretical Physics, University of Heidelberg, 69120 Heidelberg,Germany} \affiliation{Institute of Theoretical Astrophysics,
University of Oslo N-0315, Oslo, Norway.} \affiliation{Perimeter Institute for
Theoretical Physics, Waterloo, Ontario N2L 2Y5, Canada.}
\email{D.Mota@thphys.uni-heidelberg.de}

\author{Douglas J. Shaw}
\affiliation{DAMTP, Centre for Mathematical Sciences,
University of Cambridge, Wilberforce Road, Cambridge CB3 0WA, UK}
\email{D.Shaw@damtp.cam.ac.uk}

\date{\today}

\begin{abstract}
We show that, as a result of non-linear self-interactions, it is feasible, at least in light of the bounds coming from terrestrial tests of
gravity, measurements of the Casimir force and those constraints imposed by the physics of compact objects, big-bang nucleosynthesis and measurements of the cosmic microwave background, for there to exist, in our Universe, one or more scalar fields that couple to matter much \emph{more strongly} than gravity does.   
These scalar fields behave like chameleons: changing their properties to fit their surroundings. As a result these scalar fields can be not only very strongly coupled to matter, but also remain relatively light over solar system scales. These fields could also be detected by a number of future experiments provided they are properly designed to do so.  
These results open up an altogether new window, which might lead to a
completely different view of the r\^{o}le played by light scalar
fields in particle physics and cosmology.
\end{abstract}

\maketitle

\tableofcontents

\section{Introduction \label{Intro}}
There is wide-spread interest in the possibility that, in addition
to the matter described by the standard model of particle physics,
our Universe may be populated by one or more scalar fields.
%{\footnote{in addition to the Higgs scalar of the standard model}}
These are a general feature in high energy physics beyond
the standard model and are often related to the presence of
extra-dimensions, \cite{morecham3}.  The existence of scalar fields has also been postulated as means to explain the early and late time
acceleration of the Universe \cite{sn1a,inflation,quint,quint1,quint2,easson}.
%As well as any possible variation in the
%fundamental constants of nature.
%
%When scalar fields do exist,
It is almost always the case that such fields interact with
matter: either due to a direct Lagrangian coupling or indirectly
through a coupling to the Ricci scalar or as the result of quantum
loop corrections, \cite{damour,carroll1,carroll2,anupam1,anupam2}.  If the scalar field self-interactions are
negligible, then the experimental bounds on such a field are very
strong: requiring it to either couple to matter much more weakly
than gravity does, or to be very heavy \cite{uzan,bounds,damour1,damour2}.
%In either case, the scalar field
%would have had little or no detectable effect on the history of
%Universe, and would generally be indistinguishable from a cosmological
%constant
%
Recently, a novel scenario was presented by Khoury and Weltman
\cite{cham1} that employed self-interactions of the scalar-field to
avoid the most restrictive of the current bounds.  In the models that they
proposed, a scalar field couples to matter with gravitational
strength, in harmony with general expectations from string theory,
whilst, at the same time, remaining very light on cosmological
scales.   In this paper we will go much further and show, contrary to most expectations, that the scenario presented in \cite{cham1} allows scalar fields, which are very light on cosmological scales, to couple to matter much \emph{more} strongly than gravity does, and yet still satisfy \emph{all} of the current experimental and observational constraints.

The cosmological value of such a field evolves over Hubble
time-scales and could potentially cause the late-time acceleration of our Universe \cite{chameleoncosmology}.  The crucial feature that these models
possess is that the mass of the scalar field depends on the local
background matter density. On Earth, where the density is some $10^{30}$ times higher than the cosmological background, the
Compton wavelength of the field is sufficiently small as to satisfy
all existing tests of gravity.  In the solar system, where the density is
several orders of magnitude smaller, the Compton wavelength of the field can be much larger.  This means that, in those models, it is possible for the scalar field to have a mass in the solar system that is much smaller than was previously thought allowed.   In the
cosmos, the field is lighter still and its energy density evolves
slowly over cosmological time-scales and it could function as an
effective cosmological constant. While the idea of a
density-dependent mass term is not new~\cite{added,pol,others,others1,others2,others3}, the
work presented in \cite{cham1,chameleoncosmology} is novel in that the
scalar field can couple directly to matter with gravitational
strength. If a scalar field theory contains a mechanism by which the
scalar field can obtain a mass that is greater in high-density regions
than in sparse ones, we deem it to possess a \emph{chameleon
mechanism} and be a \emph{chameleon field theory}.   When referring
to chameleon theories, it is common to refer to the scalar field as
the \emph{chameleon}.

%\subsubsection*{Outline of the paper}
We start this article by reviewing the main features of scalar field theories with a chameleon mechanism. 
Afterwards, this paper is divided into roughly two parts: in sections
\ref{sing}, \ref{effmacr} and \ref{forcetwo} we study the
behaviour of chameleon theories as field theories, and derive some
important results. From sections \ref{exper} onwards, we combine
these results with a number of experimental and astrophysical
limits to constrain the unknown parameters of these chameleon
theories $\{\beta, M, \lambda\}$. 
We shall show how the non-linear
effects, identified in sections \ref{sing}-\ref{forcetwo}, allow
for a very large matter coupling, $\beta$, to be compatible with
all the available data.  We also note that some laboratory based
tests of gravity need to be redesigned, if they are to be able to
detect the chameleon. If the design of these experiments can be
adjusted in the required way, and their current precision
maintained of its current level, then a large range of
sub-Planckian chameleon theories could be detected, or ruled out,
in the near future.

In section \ref{sing}, we study how $\phi$ behaves both inside and
outside an isolated body and derive the conditions that must hold
for such a body to have a thin-shell.  In this section, we show
how non-linear effects ensure that the value that the chameleon
takes far away from a body with a thin-shell is \emph{independent}
of the matter-coupling, $\beta$. Whilst such $\beta$-independence
as been noted before for $\phi^4$-theory in \cite{nelson}, this is
the first time that it has been shown to be a generic prediction
of a large class of chameleon theories.  In section \ref{effmacr}
we show the internal, \emph{microscopic}, structure of macroscopic bodies
can unexpectedly alter the \emph{macroscopic} behaviour of the chameleon.
Using the results of section \ref{sing} and \ref{effmacr} we are
then able to calculate the $\phi$-force between two bodies; this
is done in section \ref{forcetwo}.  In each of these sections we
take care to note precisely when linearisation of the chameleon
field equation is invalid.

Laboratory bounds on chameleon field theories are analysed in
section \ref{exper}.  We focus mainly on the E\"{o}t-Wash
experiment reported in \cite{EotWash, Eotnew}, which tests for corrections
to the $1/r^2$ behaviour of gravity, and experimental programmes that measurement the Casimir force \cite{Lamcas,Decca,othercas}. We also look at the variety
of laboratory and solar-system based tests for violations of the
weak equivalence principle (WEP) \cite{LLR, WEP,WEP1,WEP2}.  The extent to which proposed
satellite-based searches for WEP violation will aide in the search
for scalar fields with a chameleon-like behaviour is considered in
this section.  We shall see that for a large range of values of $M$ and $\lambda$, laboratory tests of gravity at unable to place \emph{any} upper-bound on $\beta$.

In sections \ref{compact} and \ref{cosmo} we show how the stability of
white-dwarfs and neutron stars, as well as requirements coming from
big bang nucleosynthesis and the Cosmic Microwave Background, can be
used to bound the parameters of chameleon field theories.  We shall
see that such considerations do result in upper-bounds on
$\beta$.

Finally, in sections \ref{allbounds} and \ref{conclude} we collate
all of the different experimental and astrophysical restrictions
on chameleon theories, use them to plot the allowed values of
$\beta$, $M$ and $\lambda$, and discuss our results and their
implications.

We include a summary of the main results at the end of sections \ref{sing}-\ref{forcetwo} for easy reference. This allows the reader, less interested on the detailed derivation of the formulae, 
to follow the whole article. Throughout this work we take the signature of spacetime to be $(+\,-\,-\,-)$ and set $\hbar = c =1$; $G= 1/M_{pl}^2$ where $M_{pl}$ is the Planck mass.

\section{Chameleon Field Theories}
In the theories proposed in \cite{cham1}, the chameleon mechanism
was realised by giving the scalar field both a potential,
$V(\phi)$, and a coupling to matter, $B(\beta \phi /M_{pl})\rho$;
where $\rho$ is the local density of matter. We shall say more
about how the functions $V$ and $B$ are defined, and the meaning
of $\beta$, below. The potential and the coupling-to-matter
combine to create an effective potential for the chameleon field:
$V^{eff}(\phi) = V(\phi) + B(\beta \phi /M_{pl})\rho$.  The values
$\phi$ takes at the minima of this effective potential will
generally depend on the local density of matter. If at a minima of
$V^{eff}$ we have $\phi = \phi_c$, i.e.
$V^{eff}_{,\phi}(\phi_c)=0$, then the effective `mass' ($m_c$) of
small perturbations about $\phi_{c}$, will be given by the second
derivative of $V^{eff}$, i.e. $m_c^2 = V^{eff}_{,\phi
\phi}(\phi_c)$.  It is usually the case that $\vert V_{,\phi \phi}
\vert \gg \vert B_{,\phi \phi}\rho\vert$ and so $m_{c}$ will be
determined almost entirely by the form of $V(\phi)$ and the value
of $\phi_c$. If $V(\phi)$ is neither constant, linear nor
quadratic in $\phi$ then $V_{,\phi \phi}(\phi_c)$, and hence the
mass $m_{c}$, will depend on $\phi_c$.  Since $\phi_c$ depends on
the background density of matter, the effective mass will also be
density-dependent.  Such a form for $V(\phi)$ inevitably results
in non-linear field equations for $\phi$.

\begin{figure}[tbh]
\begin{center}
\includegraphics[width=7.4cm]{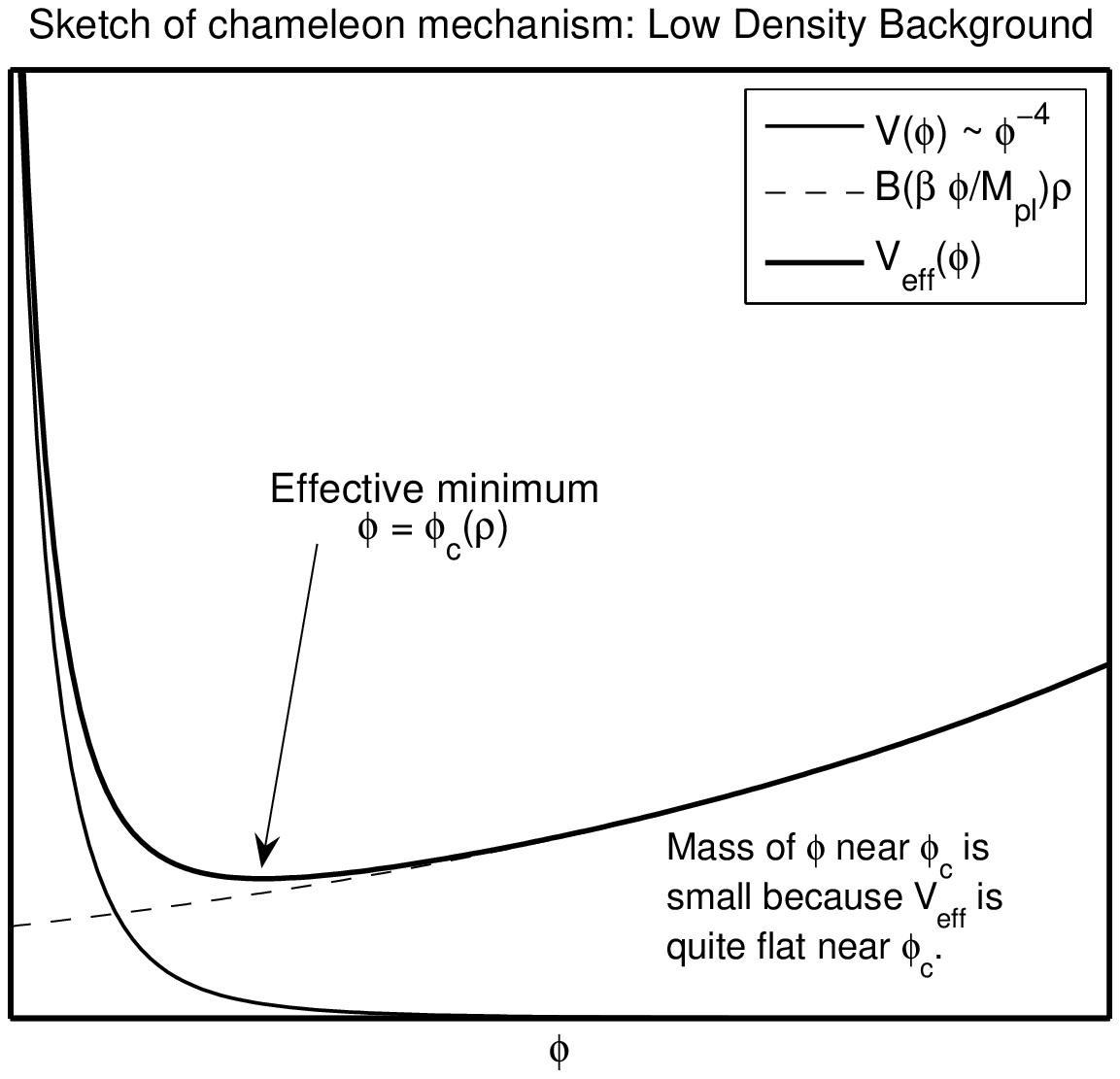}
\includegraphics [width=7.4cm]{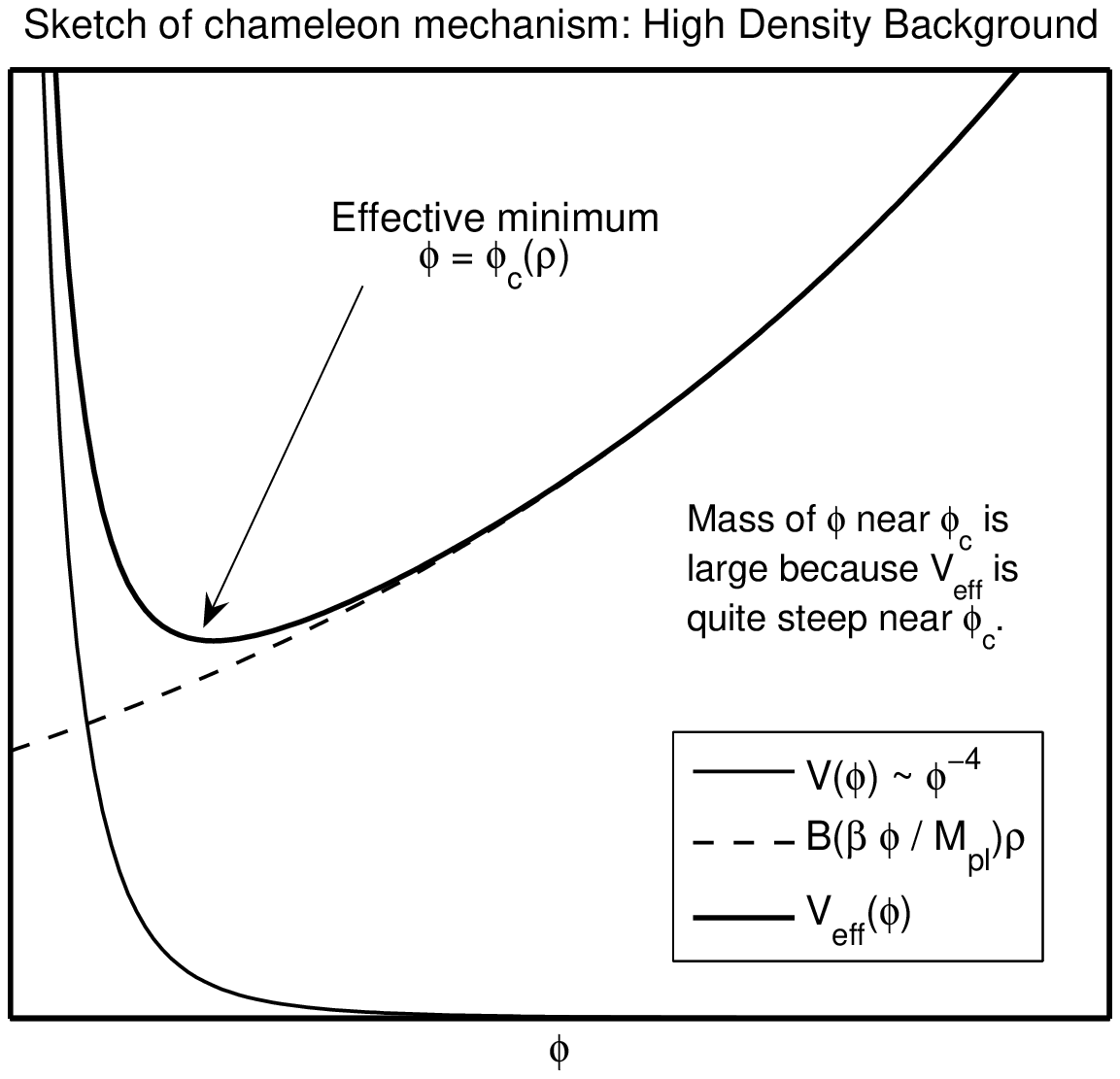}
\end{center}
\caption[Sketch of the chameleon mechanism for a runaway potential]{Sketch of the chameleon mechanism for a runaway potential:
$V \sim \phi^{-4}$.  The sketch on the left is for a low density
background, whereas the drawing of the right shows what occurs when
there is a high density of matter in the surroundings.  We can clearly
see that the position of the effective minimum, $\phi_c$, and the
steepness of the effective potential near that minimum, depends on the
density.  A shallow minimum corresponds to a low chameleon mass. The
mass of the chameleon can be clearly seen to grow with the background
density of matter.}
\label{FIGphirun}
\end{figure}

\begin{figure}[tbh]
\begin{center}
\includegraphics[width=7.4cm]{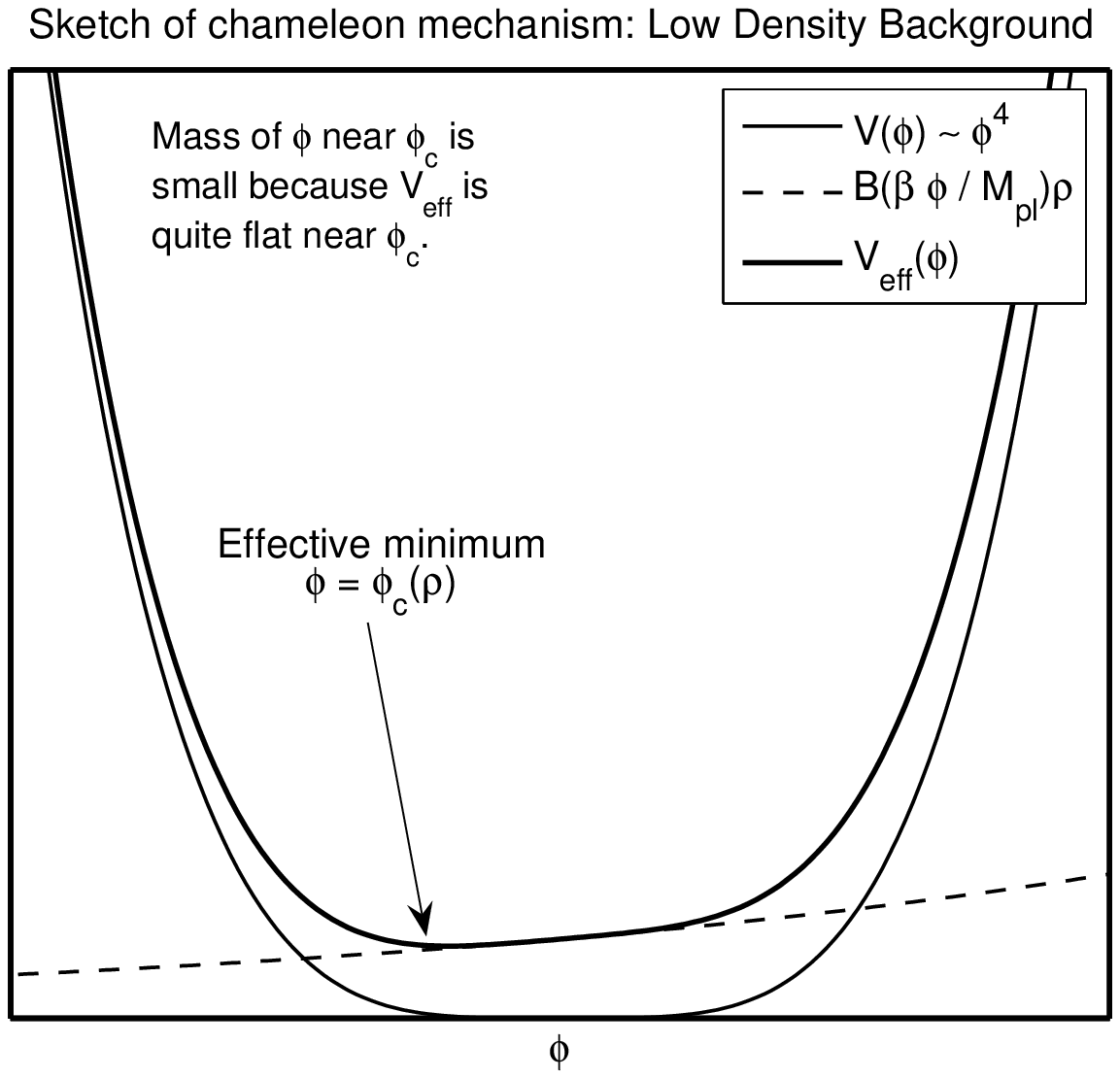}
\includegraphics[width=7.4cm]{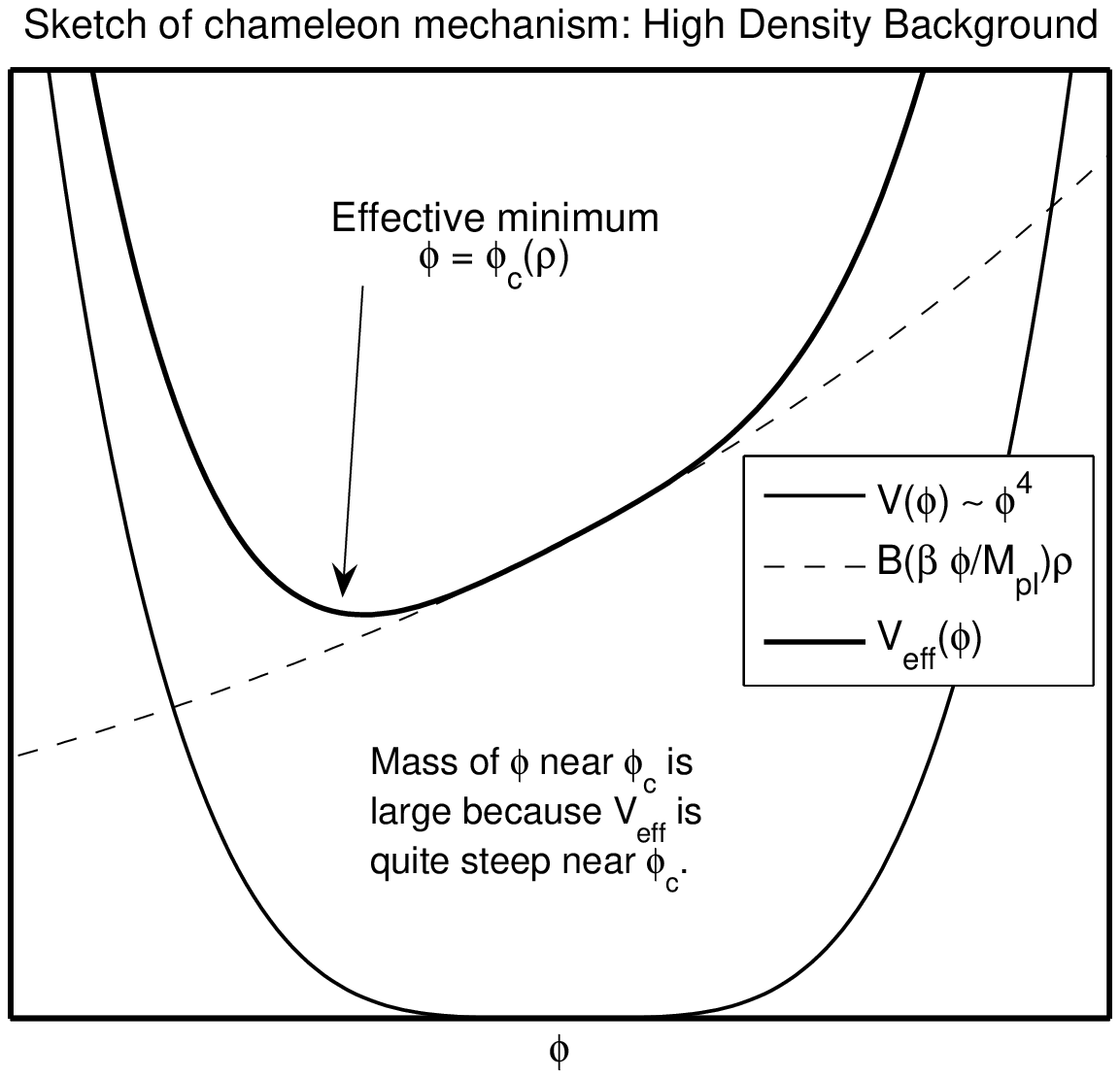}
\end{center}
\caption[Sketch of the chameleon mechanism for a potential with a minimum]{Sketch of the chameleon mechanism for a potential with a
minimum at $\phi=0$: $V \sim \phi^{4}$.  The sketch on the left is for
a low density background, whereas the drawing of the right shows what
occurs when there is a high density of matter in the surroundings.  We
can clearly see that the position of the effective minimum, $\phi_c$,
and the steepness of the effective potential near that minimum, depends
on the density.  A shallow minimum corresponds to a low chameleon
mass. The mass of the chameleon near $\phi_c$ can be clearly seen to grow with the
background density of matter.}
\label{FIGphimin}
\end{figure}

For a scalar field theory to be a chameleon theory, the effective
mass of the scalar must increase as the background density
increases. This implies
$V_{,\phi\phi\phi}(\phi_c)/V_{,\phi}(\phi_c) > 0$.  It is
important to note that it is \emph{not} necessary for either
$V(\phi)$, or $B(\beta \phi /M_{pl})$, to have any minima
themselves for the effective potential, $V^{eff}$, to have
minimum. A sketch of the chameleon mechanism, as described above,
is shown in FIGS. \ref{FIGphirun} and \ref{FIGphimin}.
In FIG \ref{FIGphirun} the potential is taken to be of runaway
form and has no minimum itself. However, It is clear from the sketches that $V^{eff}$ \emph{does} have an minimum, and  that the value
$\phi$ takes at that minimum is density dependent.  In FIG
\ref{FIGphimin} the potential is taken to behave like $\phi^4$ and
so \emph{does} have a minimum at $\phi=0$.  However, the minimum
of the effective potential, $V^{eff}$, does \emph{not} coincide
with that of $V$. Once again, the minimum of $V^{eff}$ is seen to
be density dependent.

\subsection{The Thin-Shell Effect}
This \emph{chameleon mechanism} often results in macroscopic
bodies developing what is called a ``thin-shell".  A body is said
to have a \emph{thin-shell} if $\phi$ is approximately constant
everywhere inside the body apart from in a small region near the
surface of the body. Large ($\mathcal{O}(1)$) changes in the value of
$\phi$ can and do occur in this surface layer or
\emph{thin-shell}. Inside a body with a thin-shell $\vec{\nabla}\phi$
vanishes everywhere apart from in a thin surface layer. Since the force mediated by $\phi$ is proportional to $\vec{\nabla}\phi$, it is only that surface layer, or \emph{thin-shell}, that both
feels and contributes to the `fifth force' mediated by $\phi$.

It was noted in \cite{cham1, chameleoncosmology} that the
existence of such a thin-shell effect allows scalar field theories
with a chameleon mechanism to evade the most stringent
experimental constraints on the strength of the field's coupling
to matter. For example: in the solar system, the chameleon can be
very light thus mediate a long-range force. The limits on such
forces are very tight, \cite{LLR,LLR1}.  However, since the chameleon
only couples to a small fraction of the matter in large bodies
i.e. that fraction in the thin-shell, the chameleon force between
the Sun and the planets is very weak. As a result the chameleon
has no great effect on planetary orbits, and the otherwise tight
limits on such a long-range force are evaded, \cite{chama}.  In
section \ref{sing}, we will show that the presence of a thin-shell
effect is intimately linked to non-linear nature of chameleon
field theories.
\subsection{Chameleon to Matter Coupling}
When a scalar field, $\phi$, couples to a species of matter, the
effect of that coupling is to make the mass, $m$, of that species
of particles $\phi$-dependent. This can happen either at the
classical level (i.e. in the Lagrangian) or a result of quantum
corrections. We parameterise the dependence of $m$ on $\phi$ by
\be
m(\phi) = m_0 C\left(\frac{\beta \phi}{M_{pl}}\right)
\ee where
$M_{pl}$ is the Planck mass and $m_0$ is just some constant with
units of mass whose definition will depend on one's choice of the
function $C\left(\frac{\beta \phi}{M_{pl}}\right)$. $\beta$
defines the strength of the coupling of $\phi$ to matter. We shall
say more about the definition of $\beta$ below. A $\phi$-dependent
mass will cause the rest-mass density of this particle species to
be $\phi$ dependent, specifically \be \rho(\phi)=\rho_0
C\left(\frac{\beta \phi}{M_{pl}}\right). \ee The coupling of $\phi$
to the local energy density of this particle species is given by:
$\partial \rho(\phi) / \partial \phi$ which is: \be
\frac{\partial \rho(\phi)}{\partial \phi} =
B'\left(\frac{\beta \phi}{M_{pl}}\right)\frac{\beta
\rho(\phi)}{M_{pl}} , \ee where $B(x) = \ln C(x)$ and
$B'(x) = dB(x)/dx$.  Throughout this work we will, for simplicity,
assume that our chameleon field, $\phi$, couples to all species of
matter in the same way, however we will keep in mind the fact
that, generically, different species of matter will interact with
the chameleon in different ways. We shall see in section
\ref{cosmo} that, if $C$, and hence $B$, are at least approximately
the same for all particle species, then cosmological bounds on
chameleon theories will require that $\left\vert\beta B'(\beta
\phi / M_{pl}) \phi / M_{pl}\right \vert < 0.1$ everywhere since
the epoch of nucleosynthesis. We preempt this requirement and use
it to justify the linearisation of $B(\beta \phi/M_{pl})$: \be
B\left(\frac{\beta \phi}{M_{pl}}\right) \approx B(0) + \frac{\beta
B'(0) \phi}{M_{pl}}. \ee For this to be a valid truncation we
require $(B''(0)/B'(0)) \beta \phi /M_{pl} \ll 1$. So long as
$\vert B''(0)\vert  < 10 \vert B'(0) \vert$, the cosmological
bounds on $\phi$ will then ensure that the above truncation of the
expansion of $B$ is a valid one.  The only forms of $B$ that
are excluded from this analysis are the ones where $\vert
B''(0)\vert \gtrsim 10 \vert B'(0)\vert$; we generally expect $B^{\prime \prime}(0) \sim {\mathcal{O}}(B^{\prime}(0))$.

Provided $B'(0) \neq 0$,
we can use the freedom in the definition of $\beta$ to set
$B'(0)=1$. When this is done, $\beta$ quantifies the strength of the chameleon-to-matter coupling.

For example: a particular choice for $C$ that has had some favour
in the literature, \cite{cham1,chameleoncosmology}, is $C=e^{k
\phi /M_{pl}}$ for some $k$. It follows that $B = k \phi / M_{pl}$, and so we choose $\beta
= k$ which ensures $B'(0)=1$, $B''(0)=0$.

We wish to construct our chameleon theories to be compatible with
Einstein's conception of gravity. By this we mean that we wish
them to display diffeomorphism and Poincar\'{e} invariance at the
level of the Lagrangian.  A natural consequence of Poincar\'{e}
invariance is that the chameleon couples to matter in a Lorentz
invariant fashion.   For a perfect fluid this implies that $\phi$
will generally couple to some linear combination of $\rho$ and the
fluid pressure ($P$) i.e $\rho + \omega P$.  The simplest way for
the chameleon to couple to matter in a relativistically invariant
fashion is for it to couple to the trace of the energy momentum
tensor; in this case $\omega = -3$. This said, apart from in the
early universe and in very high density objects such as neutron
stars, $P/\rho \ll 1$, and so the precise value of $\omega$ is not
of great importance. Apart from where such an assumption would be
invalid, we will take $P/\rho \ll 1$ and set $P=0$.

\subsection{A Lagrangian for Chameleon Theories}
It is possible to couple the chameleon to matter in a number of
different ways, and as such it is possible to construct many different actions for chameleon theories.  A reasonably general example of how the chameleon
can couple to trace of the energy momentum tensor is given by
following Lagrangian density \be \mathcal{L}= \sqrt{-g}
\left[-\frac{M_{pl}^2}{16 \pi} R(g) -
\frac{1}{2}\partial_{\mu}\phi\partial^{\mu}\phi + V(\phi)\right] +
\mathcal{L}_{m}(\psi^{(i)},g_{\mu \nu}^{(i)}), \ee where ${\mathcal{L}}_{m}$ is the Lagrangian density for normal matter. This Lagrangian
was first proposed in ref. \cite{chama}. The index $i$ labels the
different matter fields, $\psi^{(i)}$, and their chameleon
coupling.  The metrics $g_{\mu \nu}^{(i)}$ are conformally related
to the Einstein frame metric $g_{\mu \nu}$ by $g_{\mu \nu}^{(i)} =
\Omega_{(i)}^2 g_{\mu \nu}$ where $\Omega_{(i)} =
C_{(i)}(\beta^{(i)} \phi /M_{pl})$. The $C_{(i)}(\cdot)$ are model
dependent functions of $\beta^{(i)}\phi/M_{pl}$.  The
$\beta^{(i)}$ are chosen so that $B_{(i)}^{\prime}(0) := (\ln
C_{(i)})^{\prime}(0) = 1$.  $R(g)$ is the Ricci-scalar associated
with the Einstein frame metric.  For simplicity we will restrict
ourselves to a universal matter coupling i.e $g^{(i)}_{\mu \nu} =
\tilde{g}_{\mu \nu}$, $C_{(i)} = C$ and $\beta^{(i)} = \beta$. We
define $T^{\mu \nu} = (2/\sqrt{-\tilde{g}}) \delta
\mathcal{L}_{m}/\delta \tilde{g}_{\mu \nu}$. It follows that $T =
T^{\mu \nu} \tilde{g}_{\mu \nu} = \rho-3P$, where $\rho$ is the
physical energy density and $P$ is the sum of the principal
pressures. In general $\rho$ and $P$ are $\phi$-dependent.  With
respect to this action, the chameleon field equation is \be
-\square \phi = V_{,\phi}(\phi) + \frac{\beta
B^{\prime}(\beta\phi/M_{pl})(\rho-3P)}{M_{pl}}. \ee As mentioned
above, $\vert\beta \phi / M_{pl}\vert < 0.1$ is required for the theory to
be viable and so it is acceptable to approximate
$B^{\prime}(\beta\phi/M_{pl})$ by $B^{\prime}(0)$. We then scale
$\beta$ so that $B^{\prime}(0)=1$. The requirement, $\vert\beta \phi /
M_{pl}\vert < 0.1$, also ensures that $\rho$ and $P$ are independent of
$\phi$ to leading-order. The field
equation for $\phi$ is therefore \be -\square \phi = V_{,\phi}(\phi) + \frac{\beta
B^{\prime}(\beta\phi/M_{pl})(\rho-3P)}{M_{pl}}. \ee The above
Lagrangian should \emph{not} be viewed as specifying the
\emph{only} way in which $\phi$ can couple to matter. When one
considers varying constant theories, the matter coupling often
results from quantum loop effects \cite{shaw:2004}.  However,
despite the fact that many different Lagrangians are possible, it
is almost always the case that the field equation for $\phi$ takes
a form very similar to the one given above.

\subsection{Intrinsic Chameleon Mass Scale}
$\beta$ quantifies the
strength of the chameleon coupling. $M_{\phi}:=M_{pl}/\beta$ can
then be viewed as the intrinsic mass scale of the chameleon.
Although precise calculations of scattering amplitudes fall outside
the scope of this work, we expect that chameleon particles would be
produced in large numbers in particle colliders that operate at
energies of the order of $M_{\phi}$ or greater.

It is generally seen as `natural', from the point of view of
string theory, to have $M_{\phi} \approx M_{pl}$. When this
happens the chameleon has the same energy-scale as gravity. It has
also been suggested that the chameleon field arises from the
compactification of extra dimensions, \cite{morecham3}, if this is
the case then there is no particular reason why the true Planck
scale (i.e. of the whole of space time including the
extra-dimensions) should be the same as the effective
4-dimensional Planck scale defined by $M_{pl}$.  Indeed having the
true Planck scale being much lower than $M_{pl}$ has been put
forward as a means by which to solve the Hierarchy problem (e.g. the ADD scenario
\cite{HP,HP1,HP2}). In string-theory too, there is no particularly reason
why the string-scale should be the same as the effective
four-dimensional Planck scale.  It is also possible that the
chameleon might arise as a result of new physics with an
associated energy scale greater than the electroweak scale but
much less than $M_{pl}$. In light of these considerations it would
be pleasant if $M_{\phi}= M_{pl}/\beta \ll M_{pl}$, say of the GUT
scale, or, if we hoped to find traces of it at the LHC, maybe even
the TeV-scale.

A positive detection of a chameleon field with such a
sub-Planckian energy scale could provide us with the first
evidence for new physics beyond the standard model, but below the
Planck scale.

As pleasant as it might be to have $M_{\phi} \ll M_{pl}$, it is
generally agreed that the current experimental bounds on the existence
of light scalar fields rule out this possibility \cite{uzan,
bounds,damour1,damour2}. Indeed, in the absence of a chameleon mechanism similar to
that proposed in \cite{chameleoncosmology,chama}, bounds on the
violation of the weak equivalence principle (WEP) coming from
Lunar Laser Ranging (LLR), \cite{LLR,LLR1}, limit $\vert \beta \vert
\leq 10^{-5}$ for a light scalar field.  This implies $M_{\phi}
\gg M_{pl}$.  If the Planck scale is supposed to be associated with some fundamental maximum energy, such a large value of $M_{\phi}$ seems highly unlikely. Even if a (non-chameleon) scalar field has a mass of
the order of $1 {\mathrm{mm}}^{-1}$, then we must have $\beta <
10^{-1}$, \cite{EotWash}.

One of the major successes of the proposal of chameleon field by
Khoury and Weltman, \cite{cham1,chama}, was that chameleon fields
can, by attaining a large mass in high density environments such
as the Earth, Sun and Moon, evade the experimental limits coming
from LLR and other laboratory tests of gravity.  In this way, it
has been shown the scalar fields in theories that possess a
chameleon mechanism can couple to matter with the strength of
gravity, $\beta \sim \mathcal{O}(1)$ and still coexist with the
best experimental data currently available. Even though $\beta
\sim \mathcal{O}(1)$ has been shown to be possible, $\beta \gg 1$
is still generally assumed to be ruled out.

In this paper,
however, we challenge this assumption and show that it is indeed
feasible for $\beta$ to be very large. Moreover $M_{\phi} \approx
M_{GUT} \sim 10^{15}\,{\mathrm{GeV}}$ or $M_{\phi}\sim 1\,{\mathrm{TeV}}$
are allowed.   Tantalisingly the experimental precision required
to detect such a sub-Planckian chameleon theory is already within
reach.  Large matter couplings are allowed in normal scalar field
theories but only if the scalar field has a mass greater than
$(0.1\,{\mathrm{mm}})^{-1}$. This is \emph{not} the case
for chameleon theories.  We shall show that the mass of the
chameleon in the cosmos, or the solar system, can be, and generally
is, much less than $1\,{\mathrm{m}}^{-1}$.

\subsection{Initial Conditions}
Even though the term \emph{chameleon field} sounds rather exotic,
in a general scalar field theory with a matter coupling and
arbitrary self-interaction potential, there will generically be some values of $\phi$ about which the field theory
exhibits a chameleon mechanism. Whether or not $\phi$ ends up in such a
region will depend on its
cosmological evolution and one's choice of initial conditions. The
importance of initial conditions was discussed in
\cite{chameleoncosmology}. In that paper the potential was chosen
to be of runaway form $V \propto \phi^{-n}$, $n>0$.  We will
review what is required of the initial conditions for such
potentials in section \ref{cosmo} below. We shall see that the
larger $\beta$ is, the less important the initial conditions
become.  We will also see that the \emph{stronger} the coupling, the
\emph{stronger} the chameleon mechanism and so the more likely it becomes
that a given scalar field theory will display chameleon-like
behaviour.  This is one of the reasons for wanting to have a large value of $\beta$.

\subsection{The Importance of Non-Linearities}
Chameleon field theories necessarily involve highly non-linear
self-interaction potentials for the chameleon.  These
non-linearities make analytical solution of the field equations
much more difficult, particularly when the background matter
density is highly inhomogeneous.  Most commentators therefore
linearise the equations of chameleon theories when studying their
behaviour in inhomogeneous backgrounds
\cite{morecham1,morecham2,morecham3,morecham4}.  Such an
approximation may mislead theoretical investigations and result in
erroneous conclusions about  experiments which probe fifth force
effects.  In this paper we shall show, in detail,
that this linearisation procedure is indeed very often invalid.
When the non-linearities are properly accounted for, we will see
that the chameleon mechanism becomes much stronger.  It is this
strengthening of the chameleon mechanism that opens up the
possibility of the existence of light cosmological scalars that couple
to matter much \emph{more} strongly than gravity ($\beta \gg 1$).

\subsection{The Chameleon Potential}
The key ingredient of a chameleon field
theory, in addition to the chameleon-to-matter coupling, is a
non-linear and non-quadratic self-interaction potential $V(\phi)$.
It has been noted previously that $V(\phi)$ could play the role of
an effective cosmological constant \cite{chameleoncosmology}.
There are obviously many choices one could make for $V(\phi)$, and
whilst we wish to remain suitably general in our study, we must go
some way to specifying $V(\phi)$ if we are to make progress.  One
quite general form that has been widely used in the literature is
the Ratra-Peebles potential, $V(\phi)= M^4(M/\phi)^{n}$
\cite{ratra}, where $M$ is some mass scale and $n > 0$; chameleon
fields have also been studied in the context of $V(\phi)= k\phi^4
/ 4!$ \cite{gubser}.  In this paper we will consider both
of these types and generalise a little further. We take: \be
V(\phi) = \lambda M^4 (M/\phi)^{n}, \ee where $n$ can be positive
or negative and $\lambda > 0$.  If $n \neq -4$ then we can scale
$M$ so that without loss of generality $\lambda = 1$. When $n=-4$,
$M$ drops out and we have a $\phi^4$ theory. When $n>0$ this is
just the Ratra-Pebbles potential.

\subsection{Chameleon Field Equation}
With these assumptions and requirements, the chameleon field, $\phi$,
obeys the following conservation equation:
\begin{eqnarray}
- \square \phi = -n \lambda M^3 \left(\frac{M}{\phi}\right)^{n+1}
  +\frac{\beta (\rho+\omega P)}{M_{pl}}.
\label{micro}
\end{eqnarray}
For this to be a chameleon field we need the potential gradient
term, $V_{,\phi} = -n\lambda M^3 (M/\phi)^{n+1}$, and the matter
coupling term, $\beta (\rho + \omega P) / M_{pl}$, to be of
opposite signs. It is usually the case that $\beta > 0$ and
$P/\rho \ll 1$. If $n > 0$ we must therefore have $\phi > 0$.  In
theories with $n < 0$ we must have $\phi < 0$ and $n=-2p$ where
$p$ is a positive integer. We must also require that the effective
mass-squared of the chameleon field, $ m_{c}^2 = V_{,\phi \phi}$, be
positive, non-zero and depend on $\phi$. These conditions mean
that we must exclude the region $-2 \geq n \geq 0$.  If $n=-2$,
$n=-1$ or $n=0$ then the field equations for $\phi$ would be
linear.

\subsection{Natural Values of $M$ and $\lambda$}
When $n\neq-4$, one might imagine that our choice of potential has
arisen out of an expansion, for small $(\hat{M}/\phi)^n$, of
another potential $W(\phi) = \hat{M}^4 f((\hat{M}/\phi)^n)$ where
$f$ is some function. We could then write: \be W \sim \hat{M}^4
f(0) + \hat{M}^4 f'(0) \left(\frac{\hat{M}}{\phi}\right)^{n}, \ee
where $\hat{M}$ is some mass-scale. We define $M$ so that the
second term on the right hand side of the above expression reads
$M^4 (M/\phi)^n$. The first term on the right hand then plays the
r\^{o}le of a cosmological constant $\hat{M}^4f(0) \approx
\rho_{\Lambda}$. Assuming that both $f(0)$ and $f'(0)$ are
$\mathcal{O}(1)$, we would then have $M \approx \hat{M} \approx
\left(\rho_{\Lambda}\right)^{1/4} \approx (0.1\, {\mathrm{mm}})^{-1}$.
It is for this reason that one will often find $(0.1\,
{\mathrm{mm}})^{-1}$ referred to as a `natural' value for $M$,
\cite{chameleoncosmology,chama}.  When $n=-4$ we naturally expect
$\lambda \approx 1/4!$, \cite{phi4}.

\section{One body problem \label{sing}}
In this section, we consider the perturbation to the chameleon
field generated by a single body embedded in background of uniform
density $\rho_b$. For simplicity we shall model the body to be
both spherical and of uniform density $\rho_c$.  This analysis
will prove vital when we come to calculate the force between two
bodies that is induced by the chameleon field. We assume space-time to be Minkowski (at least to leading order) and we also assume that everything is static.  Under these assumptions $\square \rightarrow -\nabla^2$ where $\nabla^2 \phi = r^{-2} (r^2 \phi^{\prime})^{\prime}$; $\phi^{\prime} = {\mathrm{d}}\phi / {\mathrm{d}}r$.
% $\rho_c$.
%We label the radius of the
%body by $R$.
Whilst this problem has been considered elsewhere in the
literature \cite{chama,cham1,chameleoncosmology}, most commentators
have chosen to linearise the chameleon field equation, eq.
(\ref{micro}), before solving it. This linearisation is, however,
often invalid. In this section, we begin by briefly reviewing what
occurs when it is appropriate to linearise eq. (\ref{micro}), and,
in doing so, note where the linear approximation breaks down.  In
some cases, even though it is not possible to construct a
linearised theory that is valid everywhere, we shall demonstrate,
using the method of matched asymptotic expansions, how to
construct multiple linearisations of the field equation, each
valid in a different region, and then match them together to find
an asymptotic approximation to $\phi$ that is valid everywhere.
When this is possible, the chameleon field, $\phi$, will behave as
if these were the solution to a consistent, everywhere valid,
linearisation of the field equations; for this reason we deem this
method of finding solutions to be the \emph{pseudo-linear
approximation}. If a body is large enough, however, both the
linear and pseudo-linear approximations will fail. We shall see
that, when this happens, $\phi$ behaves in a truly non-linear
fashion near the surface of the body.  The onset of non-linear
behaviour is related to the emergence of a \emph{thin-shell} in
the body. The linear approximation is discussed in section
\ref{singlin} whilst the pseudo-linear approximation is considered
in section \ref{singpseu}.  We discuss the non-linear regime in
section \ref{singnon}.

We take the body that we are considering to be spherical with radius $R$
and uniform density $\rho_{c}$.  Assuming spherical symmetry, inside the body ($r < R$), $\phi$
obeys:
\begin{eqnarray}
\frac{d^2 \phi}{dr^2} + \frac{2}{r}\frac{d \phi}{dr} = -n\lambda M^3
\left(\frac{M}{\phi}\right)^{n+1} +\frac{\beta \rho_c}{M_{pl}},
\label{inbody}
\end{eqnarray}
and outside the body, $(r>R)$, we have:
\begin{eqnarray}
\frac{d^2 \phi}{dr^2} + \frac{2}{r}\frac{d \phi}{dr} = -n\lambda
M^3 \left(\frac{M}{\phi}\right)^{n+1} + \frac{\beta
\rho_b}{M_{pl}}. \label{outbody}
\end{eqnarray}
The right hand side of eq. (\ref{inbody}) vanishes when $\phi =
\phi_c$ where $$ \phi_{c} = M\left(\frac{\beta \rho_c}{n\lambda
M_{pl}M^3}\right)^{-\frac{1}{n+1}}. $$ This value of $\phi$
corresponds to the minimum, of the effective potential of the
chameleon field, inside the body. Similarly, the right hand side of
eq. (\ref{outbody}) vanishes when $\phi=\phi_b$ where $$ \phi_{b}
= M\left(\frac{\beta \rho_b}{n\lambda
M_{pl}M^3}\right)^{-\frac{1}{n+1}}. $$ This value of $\phi$
corresponds to the minimum, of the effective potential of the
chameleon field, outside the body. For large $r$ we must have $\phi
\approx \phi_b$.
 Associated with
every value of $\phi$ is an effective chameleon mass,
$m_{\phi}(\phi)$, which is the mass of small perturbations about that
value of $\phi$. This effective mass is given by:
\begin{eqnarray}
m^2_{\phi}(\phi) = V^{eff}_{,\phi\phi}(\phi) = n(n+1)\lambda M^2
\left(\frac{M}{\phi}\right)^{n+2}. \label{effmass}
\end{eqnarray}
We define $m_c = m_{\phi}(\phi_c)$ and $m_b = m_{\phi}(\phi_b)$. We shall see below that the larger
the quantity $m_c R$, the more
likely it is that a body will have a thin-shell.  In this section we
shall see both why this is so, and precisely how large $m_c R$ has to
be for a thin-shell to appear. Throughout this section we will require, as boundary conditions, that
\begin{eqnarray*}
\left.\frac{d\phi}{d r}\right\vert_{r=0} = 0\quad {\mathrm{and}} \quad \left.\frac{d\phi}{d r}\right\vert_{r=\infty} = 0.
\end{eqnarray*}
\subsection{Linear Regime \label{singlin}}
We assume that it is a valid approximation to linearise the equations
of motion for $\phi$ about the value of $\phi$ in the far background,
$\phi_{b}$. For this to be possible we must require that certain conditions,
which we state below, hold.  Writing $\phi = \phi_b + \phi_1$, the
linearised field equations are:
\begin{equation}
\frac{d^2 \phi_1}{d r^2} + \frac{2}{r}\frac{d \phi_1}{d r} = -n M^3
\left(\frac{M}{\phi_b}\right)^{n+1} + m^2_{b}\phi_1 + \frac{\beta
(\rho_{c}-\rho_{b})}{M_{pl}}H(R-r) + \frac{\beta \rho_b}{M_{pl}},
\end{equation}
where $H(R-r)$ is the Heaviside function: $H(x)=1$, $x\geq 0$, and
$H(x)=0$, $x<0$. For this
linearisation of the potential to be valid we need: $$
\frac{V_{,\phi \phi}(\phi_b)\phi_1}{V_{,\phi}(\phi_b)} < 1.$$ This translates to $\vert \phi_1/\phi_b \vert
< \vert n+1 \vert^{-1}$. Also,
for this linearisation to remain valid as $r \rightarrow \infty$,
we need $\phi_1 \rightarrow 0$, which implies that: $$ n M^3
\left(\frac{M}{\phi_b}\right)^{n+1} = \frac{\beta \rho_b}{M_{pl}}.
$$ Defining $\Delta \rho_c = \rho_c-\rho_b$, and solving the field
equations, we find that outside the body ($r>R$) we have: $$
\phi_1 = \frac{\beta \Delta \rho_c}{M_{pl}m_b^2}\frac{e^{m_b(R-
r)}}{m_{b}r}\left(\frac{\tanh(m_b R)-m_bR}{1+\tanh(m_bR)}\right).
$$ Inside the body ($r<R$), $\phi_1$ is given by $$ \phi_{1} =
-\frac{\beta \Delta \rho_{c}}{m_{b}^2 M_{pl}} + \frac{\beta \Delta
\rho_c}{M_{pl}m_b^2}\frac{\left(1+m_b R\right)e^{-m_bR}\sinh(m_b
r)}{m_b r}. $$ The largest value of $\vert\phi_1/\phi_b\vert$
occurs at $r=0$ and so, for this linear approximation to be valid,
we need: $\vert \phi_{1}(r=0)/\phi_{b} \vert < \vert n+1
\vert^{-1}$. This requirement is
equivalent to the statement that $$ \left\vert(1+m_b R)e^{-m_b
R}-1\right\vert \frac{\Delta \rho_{c}}{\rho_b} \sim
\frac{1}{2}m_b^2R^2 \frac{\Delta \rho_{c}}{\rho_b} = \frac{\Delta
\rho_c}{\rho_c}\left(\frac{\rho_b}{\rho_c}\right)^{\frac{1}{n+1}}
\frac{(m_cR)^2}{2}<1. $$ where
`$\sim$' means ``asymptotically in the limit $m_b R \rightarrow
0$". It is often the case that the background is much less dense
than the body i.e. $\rho_b \ll \rho_c$.  If this is the case then
it is clear, from the above expression, that there will be a
distinct difference between theories with $n>0$ and those with $n
\leq -4$.  In theories with $n > 0$, the lower the density of the
background, the better the linear theory approximation will hold,
whereas when $n \leq -4$ the opposite is true.  This can be
understood by considering the relation $\phi_b \propto
\rho_b^{-1/(n+1)}$.  If $n > 0$, the smaller $\rho_b$ becomes, the
larger $\phi_{b}$ will be. It is therefore possible for larger
perturbations in $\phi$ to be treated consistently in terms of the
linearised theory.  If, however, we have that $n\leq-4$ then
$\phi_b \rightarrow 0$ as $\rho_b \rightarrow 0$ and the opposite
is true.

We can, however, use the method of matched asymptotic expansions
to show that the region where behaviour, similar to that which
would be predicted by linearised theory, occur is significantly larger
than one would have guessed simply by requiring that the linear
approximation hold.

The results of this section, as well as those of sections
\ref{singpseu} and \ref{singnon}, are summarised in section
\ref{singsum} below.

\subsection{Pseudo-Linear Regime \label{singpseu}}
The defining approximation of the pseudo-linear regime (for both
positive and negative $n$) is that inside the body:
$$ \left(\frac{\phi_c}{\phi(r)}\right)^{n+1} \ll 1
$$ This is equivalent to $\beta \rho_c / M_{pl} \gg
nM^3(M/\phi(r))^{n+1}$.  When this holds we find $\phi \sim
\bar{\phi}(r)$ inside the body, where this defines $\bar{\phi}(r)$
and:
$$ \frac{1}{r}\frac{d^2(r\bar{\phi})}{dr^2} = \frac{\beta
\rho_c}{M_{pl}}.
$$ It follows that
$$ \phi \sim \bar{\phi} = \phi_0 + \frac{\beta \rho_c r^2}{6M_{pl}}.
$$ In this case `$\sim$' means ``asymptotically as
$(\phi_c/\phi(r))^{n+1} \rightarrow 0$". Outside the body, we can find
a similar asymptotic approximation:
$$ \phi \sim \bar{\phi}=\phi_0 + \frac{\beta \rho_c
R^2}{2M_{pl}}-\frac{\beta \rho_c R^3}{3M_{pl}r}.
$$

For this to be valid we must ensure that the neglected terms, in
the above approximation to $\phi$, are small compared to the
included ones; this requires that: $$ \frac{R^3}{3}\gg
\int^{r}_{0}d r' \int^{r'}_{0} dr^{\prime \prime} r^{\prime
\prime} \left(\frac{\phi_c}{\bar{\phi}(r^{\prime
\prime})}\right)^{n+1}. $$ For large $r$ we expect, as we did in
the previous section, that $\phi \rightarrow \phi_b$, and so: $$
\phi \sim \phi^{\ast}=\phi_b - \frac{Ae^{-m_b r}}{r}, $$ which
will remain valid whenever $Ae^{-m_br}{\phi_b r} \ll \vert
1/(n+1)$.  We shall refer to
$\bar{\phi}$ as the \emph{inner approximation} to $\phi$.
Similarly, $\phi^{\ast}$ is the \emph{outer approximation}.  So far
both $A$, and the value of $\phi_0$, remain unknown constants of
integration. In general, when $\phi \sim \phi^{\ast}$ we will
\emph{not} also have $\phi \sim \bar{\phi}$ (and vice versa). If,
however, there is some \emph{intermediate region} where both the
inner and outer approximations are simultaneously valid, then we
can match both expressions in that intermediate region and
determine both $\phi_0$ and $A$, \cite{hinch, hinch1}. A detailed
explanation of the use of matched asymptotic expansions with
respect to cosmological scalar fields is given in
\cite{shawbarrow1}.

For the moment we shall assume that such an intermediate region
\emph{does} exist. We check what is required for this assumption to
hold in appendix \ref{singpseuApp} and present the results of that
analysis below.  Given an intermediate region, we find:
\begin{eqnarray*}
A &=& \frac{\beta \rho_c R^3}{3 M_{pl}}, \\ \phi_0 &=& \phi_b -
\frac{\beta \rho_c R^2}{2M_{pl}} +\frac{\beta \rho_c m_b R^3}{3
M_{pl}}.
\end{eqnarray*}
The external field produced by a single body in the
pseudo-linear approximation is:
\begin{eqnarray}
\phi \sim \phi_b - \frac{\beta \rho_c R^3 e^{-m_br}}{3 M_{pl}r}
\Leftrightarrow \frac{\phi}{\phi_c} \sim \frac{\phi_b}{\phi_c} -
\frac{(m_cR)^2 e^{-m_br}}{3(n+1)(r/R)}, \label{phipseudo}
\end{eqnarray}
and the field inside the body is given by:
\begin{eqnarray}
\phi \approx \phi_b - \frac{\beta \rho_c R^2}{2M_{pl}} +\frac{\beta
\rho_c m_b R^3}{3 M_{pl}} + \frac{\beta \rho_c r^2}{6M_{pl}}.
\end{eqnarray}
In appendix \ref{singpseuApp}, we find that for the pseudo-linear
approximation to hold, we must require
\begin{subequations}
\label{pseudcon}
\begin{eqnarray}
m_c R
&\ll&\min\left((18)^{1/6}\left(\tfrac{m_c}{m_b}\right)^{\frac{n+4}{3(n+2)}},
\sqrt{2\vert n+1\vert} \left\vert
\left(\tfrac{m_{b}}{m_c}\right)^{\frac{2}{\vert n + 2
\vert}}-1\right\vert \right), \quad n < -4,\\ m_c R
&\ll&\min\left(\sqrt{3}\left(\tfrac{m_c}{m_b}\right)^{\frac{n+4}{3(n+2)}},
\sqrt{2\vert n+1\vert} \left\vert
\left(\tfrac{m_{c}}{m_b}\right)^{\frac{2}{\vert n + 2
\vert}}-1\right\vert \right), \quad n > 0, \\ m_c R &\ll& 1, \quad n =
-4.
\end{eqnarray}
\end{subequations}
When $n=-4$, we actually find a slightly different asymptotic behaviour
of $\phi$ outside the body, precisely:
\begin{eqnarray}
\phi \sim \phi_b - \sqrt{\frac{1}{2\lambda(y_0
+\ln(\min(r/R,1/m_b)))}} \frac{e^{-m_br}}{2r}, \label{phipseudo4}
\end{eqnarray}
where $y_0$ is an integration constant and: $$ \frac{(m_cR)^3}{9}
= \sqrt{\frac{3}{2y_0}} +\frac{1}{3}\left(\frac{3}{2
y_0}\right)^{3/2}. $$ The conditions given by eqs.
(\ref{pseudcon}a-c) ensure that the pseudo-linear approximation is
everywhere valid.  When these conditions fail, non-linear effects
begin to become important near the centre of the body.  As $m_c R$
is increased further the region where non-linear effects play a
r\^{o}le moves out from the centre of the body. Eventually, for
large enough $m_c R$, the non-linear nature of chameleon
potential, $V(\phi)$, is only important in a thin region near the
surface of the body; this is the thin-shell.  Since the emergence
of such a thin-shell is linked to non-linear effects becoming
important near the surface of the body, it must be the case that
the assumption that $\phi$ is given by equation (\ref{phipseudo})
(or by eq. (\ref{phipseudo4}) in $n=-4$ theories) breaks down for
some $r
> R$.  By this logic, we find, in appendix \ref{singpseuApp}, that a
thin-shell occurs when:
\begin{subequations}
 \label{thincond}
\begin{eqnarray}
m_c R
&\gtrsim&\min\left((18)^{1/6}\left(\tfrac{m_c}{m_b}\right)^{\frac{n+4}{3(n+2)}},
\sqrt{3\vert n+1\vert} \left\vert
\left(\tfrac{m_{b}}{m_c}\right)^{\frac{2}{\vert n + 2
\vert}}-1\right\vert \right), \quad n < -4, \\ m_c R
&\gtrsim&\min\left(\sqrt{3}\left(\tfrac{m_c}{m_b}\right)^{\frac{n+4}{3(n+2)}},
\sqrt{3\vert n+1\vert} \left\vert
\left(\tfrac{m_{b}}{m_c}\right)^{\frac{2}{\vert n + 2
\vert}}-1\right\vert \right), \quad n > 0, \\ m_c R &\gtrsim& 4, \quad
n = -4.
\end{eqnarray}
\end{subequations}
In both eqs. (\ref{pseudcon}a-b) and (\ref{thincond}a-b) the
second term in the $\min(\,\cdot\, , \,\cdot\,)$ is almost always
smaller than the first when $\rho_b / \rho_c \ll 1 \Leftrightarrow
m_b / m_c \ll 1$. The behaviour of $\phi$ both near to, and far
away from, a body with thin-shell is discussed below in section
 \ref{singnon}. The results of this section are summarised in section \ref{singsum}. Note that the thin-shell conditions , eqs. (\ref{thincond}a-c), necessarily imply that $m_c R \gg 1$.

\subsection{Non-linear Regime \label{singnon}}
We have just seen, in eqs. (\ref{thincond}a-c), that for
non-linear effects to be important, and the pseudo-linear
approximation to fail, we must have $m_c R \gg 1$.  In this regime
the body is, necessarily, very large compared to the length scale
$1/m_c$.  We expect that all perturbations in $\phi$ will die off
exponentially quickly over a distance of about $1/m_c$ and, as
such, $\phi\approx \phi_c$ will be almost constant inside the
body. Any variation in the chameleon field, that does take place,
will occur in a `thin-shell' of thickness $\Delta R \approx 1/m_c$
near the surface of the body. It is clear that $m_c R \gg 1$
implies $\Delta R / R \ll 1$.  In this section we will consider
both the behaviour of the field close to the surface of the body,
and far from the body.

\subsubsection{Close to the body \label{nonclose}}
We noted above that $m_c R\gg 1$ implies $\Delta R / R\ll 1$, we shall demonstrate this is a rigorous fashion below.  Given $\Delta R / R \ll 1$, when we consider the evolution of $\phi$ in the thin-shell region, we
can ignore the curvature of the surface of the body, to a good
approximation.

We therefore treat the surface of the body as being flat, with
outward normal in the direction of the positive $x$-axis. The
surface of the body defined to be at $x=0$ (i.e. $x=r-R$). Since
the shell is thin compared to the scale of the body, we are
interested in physics that occurs over length-scales that are very
small compared to the size of the body.  We therefore make the
approximation that the body extends to infinity along the $y$ and
$z$ axes and also along the negative $x$ axis. Given these
assumptions, we have that $\phi$ evolves according to

$$ \frac{d^2 \phi}{d x^2} = -n\lambda M^3 (M/\phi)^{n+1} + \beta
\rho_c / M_{pl}$$

As a boundary conditions (BCs) we have $\phi \rightarrow \phi_c$ and
$d \phi / d x \rightarrow 0$ as $x \rightarrow -\infty$.  With these
BCs, the first integral of the above equation is:
\begin{equation}
\frac{1}{2}\left(\frac{d \phi}{dx}\right)^2 \approx \lambda M^4
\left[(M/\phi)^n - (M/\phi_c)^n\right] + \frac{\beta \rho_c}{M_{pl}}
(\phi - \phi_c).
\label{phiin}
\end{equation}

Outside of the body, we assume that $\phi \rightarrow \phi_b$ as $x
\rightarrow \infty$, and that the background has density $\rho_{b}\ll
\rho_{c}$. Assuming that:
$$ \left\vert \frac{d^2\phi}{d x^2} \right\vert \gg
\left\vert\frac{2}{r}\frac{d \phi}{dx}\right\vert
$$ then we can ignore the curvature of the surface of the body and, in
$x>0$, we have:
\begin{eqnarray}
\frac{1}{2}\left(\frac{d \phi}{dx}\right)^2 = \lambda M^4(M/\phi)^{n}
- \lambda M^4(M/\phi_{b})^{n} + \frac{\beta \rho_b}{M_{pl}}(\phi -
\phi_{b}),
\label{phiout}
\end{eqnarray}
where $\phi_b$ is as we have defined above. Our assumption that
$\vert d^2 \phi / d x^2 \vert \gg (2/r)d \phi / d x$ then requires
that: $$
\frac{2\sqrt{2}}{m_{\phi}(\phi)r}\sqrt{\frac{n+1}{n}}\left(1-(n+1)\left(\frac{\phi}{\phi_b}\right)^{n}
+ n\left(\frac{\phi}{\phi_{b}}\right)^{n+1}\right)^{1/2} \ll 1. $$
Provided the pseudo-linear approximation breaks down, and that the
body has a thin-shell,  we expect that, near the surface of the
body, $\phi \sim \mathcal{O}(\phi_c)$. It follows that, whenever
$\rho_c \gg \rho_b$, $(\phi/\phi_b)^{n} \ll 1$ and
$(\phi/\phi_b)^{n+1} \ll 1$.  The above condition will therefore
be satisfied provided that $m_cR \gg
\sqrt{8(n+1)/n}$; this is generally a weaker condition than the
requirement that the body satisfy the thin-shell conditions, eqs
(\ref{thincond}a-c). On the surface at $x=0$, both $\phi$ and
${\mathrm{d}}\phi/{\mathrm{d}}x$ must be continuous. By comparing the
expressions for ${\mathrm{d}}\phi /{\mathrm{d}}x$ inside and outside the
body we have: $$ \frac{\phi(0)-\phi_c}{\phi_c} = \frac{1}{n}. $$

We now check that we do indeed have a thin shell i.e. $\Delta R
\ll R$. We expect that, near the surface of the body, almost all
variation in $\phi$ will concentrated into a shell of thickness
$\Delta R$. We define $m_{surf}$ by: $$ \int^{0}_{-\infty} dx
\frac{d\phi}{dx} = \phi(x=0)-\phi_c = \phi_c/n = m_{surf}
\frac{d\phi}{d x}(x=0). $$ $m_{surf}^{-1}$ is then, approximately,
the length scale over which any variation in $\phi$ dies off. As
it happens, $m_{surf}$ is also the mass of the chameleon field at
$x=0$. It follows that $\Delta R \approx m_{surf}^{-1}$. For this
shell to be thin, and for us to be justified in ignoring the
curvature of the surface of the body, we need $\Delta R / R \ll 1$
or equivalently $m_{surf}R \gg 1$. We find (assuming $\rho_b \ll
\rho_c$) that: $$ m_{surf}R \approx
\left(\frac{n}{n+1}\right)^{n/2+1}m_c R \sim \mathcal{O}(m_c R), $$ and so $m_{surf}R \gg
1$ follows from $m_{c}R \gg 1$, and $\Delta R \sim
\mathcal{O}(m_c^{-1})$. $m_{surf}R \gg 1$ will be automatically
satisfied whenever the thin-shell conditions eqs.
(\ref{thincond}a-c) hold.

Whenever $\rho_{c} \gg \rho_{b}$, eq. \ref{phiout} will, near
$r=R$, be well-approximated by: $$ \frac{1}{2}\left(\frac{d
\phi}{d x}\right)^2 \approx \lambda M^4 (M/\phi)^n. $$ Solving
this under the boundary conditions $\phi(x=0) = (1+1/n)\phi_c$ and
$\phi/\phi_c \rightarrow \phi_b/\phi_c \approx 0$ as $x
\rightarrow \infty$ we find
\begin{eqnarray}
 \frac{1}{m_{\phi}(\phi)} \sim \frac{\vert n+2 \vert (r-R)}{\sqrt{2n(n+1)}}  +\left(\frac{n+1}{n}\right)^{n/2+1}\frac{1}{m_{c}} \label{phiclosef}
\end{eqnarray}
This approximation will therefore break-down when $m_{\phi}(\phi)r\sim
\mathcal{O}(1)$,  which occurs when $r-R \sim \mathcal{O}(R)$. We
can see that, if $r-R \gg \sqrt{2n(n+1)}/(\vert n+2 \vert
m_{surf})$ then $m_{\phi}$, and hence also $\phi$, will be
independent of $m_{c}$ and hence also of $\phi_c$ and $\beta$ at
leading order.  Since $m_{surf} R \gg 1$, there will be some
region where eq. (\ref{phiclosef}) is both valid and, to leading
order, independent of $\beta$.

Although, in this approximation, we cannot talk about what occurs
for $(r-R) \gtrsim R$, it seems likely, in light of the behaviour
seen when $(r-R) \ll R$, that, whenever $r \gg 1/m_{surf} \approx
1/m_{c}$, the perturbation in $\phi$, induced by an isolated body
with thin-shell, will be independent of the matter coupling
$\beta$. We confirm this expectation in section \ref{farphithin}
below.

\subsubsection{Far field of body with thin-shell \label{farphithin}}
We found above that the emergence of a thin-shell was related to
non-linear effects being non-negligible near the surface of the
body. We noted that a thin-shell will exist whenever conditions
(\ref{thincond}a-c) hold. However, even when these conditions
hold, we do not expect non-linear effects to be important far from
the surface of the body.  Indeed, for large $r$ we should
expect that $\phi$ takes a functional form similar to that found
in the pseudo-linear approximation i.e. as given by eq.
(\ref{phipseudo}) (or eq. (\ref{phipseudo4}) if $n=-4$).   Although the functional form should be similar, in order to find the correct behaviour, one must replace $(m_c R)$ in equations (\ref{phipseudo}) and (\ref{phipseudo4}) by some other quantity $C$, say.   We show that, to leading order as $r \rightarrow \infty$, $C$ is \emph{independent} of the matter coupling $\beta$ and the density of the central body $\rho$. This confirms the expectation of section \ref{nonclose}
above.
%NEW UPDATE BEGINS HERE
The analysis for $n=-4$ is slightly more involved than it is for
other values of $n$.  We therefore consider the $n=-4$ case
separately below and in appendix \ref{singnon4App}. The analysis for
theories with runaway potentials that become singular at $\phi = 0$
(i.e $n >0$ theories) is much simpler than it is for theories where
the potential has a minimum at $\phi =0$ and which are non-singular
for all finite $\phi$ i.e. ($n < -4$ theories): we therefore
consider the $n < 0$ and $n > 0$ cases separately.
\\ \\
\\ \\
\noindent\textit{Runaway Potentials ($n >0$)}\\
\indent Away from the surface of the body we expect that non-linear effects will be negligible and as $r \rightarrow \infty$ we will have:
$$
\phi \sim \phi^{(0)} = \phi_b -  \frac{De^{-m_b r}}{r},
$$
for some $D$ where $\phi_b$ and $m_b$ are the values of the
chameleon and its mass in background.   It is clear from the field
equations however that $\nabla^2 \phi < \nabla^2 \phi^{(0)}$ and so,
given the boundary conditions at infinity, $\phi < \phi^{(0)}$
outside the body.  In $n>0$ theories there is a singularity of the
potential, and hence also of the field equations, at $\phi = 0$. It
is clear that this singularity cannot be reached in any physically
acceptable evolution and so we must always have $\phi > 0$, which in
turn implies $\phi^{(0)} > 0$ outside the body.  The minimum value
of $\phi^{(0)}$ outside the body occurs at $r=R$ and so we must
have:
$$
D < \phi_{b}e^{m_b R}R.
$$
In most cases of interest $m_b R \ll 1$ and so we have:
$$
D < \phi_b R.
$$
This upper bound on $D$ defines a critical form for the field outside the body:
$$
\phi_{crit} = \phi_b\left(1 - \frac{e^{m_b (R-r)}R}{r}\right).
$$
No matter what occurs inside the body ($r < R$) we must have $\phi > \phi_{crit}$ outside the body as $r \rightarrow \infty$.  This implies that:
$$
\left\vert \frac{d \phi}{d r} \right\vert < \frac{(1+m_b r)\phi_b e^{m_b (R-r)R}}{r^2} < \frac{\phi_b R}{r^2},
$$
as $r \rightarrow \infty$.  Ignoring non-linear effects, $\phi > \phi_{crit}$ is satisfied by all bodies that satisfy the conditions for the pseudo-linear approximation (eqs. (\ref{pseudcon}b)) but would be violated, in the absence of non-linear effects, by those that satisfy the thin-shell conditions (eqs. (\ref{thincond})).  We must therefore conclude that non-linear effects near the surface of body with thin-shells ensure $\phi > \phi_{crit}$ is always satisfied as $r \rightarrow \infty$. Furthermore, if $\phi \gg \phi_{crit}$ then
$$
\left\vert \frac{d \phi}{d r} \right\vert \ll \frac{\phi_b R}{r^2},
$$
and it follows from section \ref{singpseu} that the pseudo-linear approximation is valid for all $r$, which further implies that the body \emph{cannot} have a thin-shell.  Thus thin-shelled bodies must actually have $\phi$ being only greater less than $\phi_{crit}$ as $r \rightarrow \infty$. We are therefore justified in using $\phi_{crit}$ to approximate the far field of a body with a thin-shell.   In summary: in $n > 0$ theories, the far field of a body with a thin-shell has the following form:
$$
\phi \sim \phi_b - \phi_b e^{m_b (R-r)R}{r}.
$$
We note that this form, and the arguments with which we have derived it, do not depend, in any way, on the physics inside $r < R$.  The critical form of $\phi$ is determined entirely by the form of the potential and the background value of $\phi$.
\\ \\
\noindent\textit{Potentials with minimum ($n <0$)} \\
\indent For $n > 0$ theories the singularity of the potential at $\phi=0$
allowed us to determine asymptotic form of $\phi$ outside a body
with a thin-shell.  In $n < 0$ theories, however, the potential is
well-defined for all finite $\phi$ and so we cannot play the same
trick as we did above.  The $n=-4$ case is special and treated in
great detail in appendix \ref{singnon4App}. When $n=-4$, we find
that the far field of a body with thin-shell is given by: $$ \phi
\approx \phi_b -
\frac{e^{-m_br}}{r\sqrt{2\lambda(1+4\ln(\min(r/R,1/m_bR)))}} \sim
\phi_b - \frac{e^{-m_b r}}{2r\sqrt{2\lambda\ln(\min(r/R,1/m_bR))}}.
$$ A thin-shell is certainly present whenever $m_c R \gtrsim 4$. For
other negative values of $n$ we use a semi-analytical method.  We
saw when deriving the thin-shell conditions for $n < 0$ theories
that the background value of $\phi$ plays only a negligible r\^{o}le
since $\phi/\phi_b \gg 1$ near the body and, in most cases, $m_b R
\ll 1$. Assuming $m_b R \ll 1$, we simplify our analysis by setting
$\phi_b = 0$. Far from the body non-linear effects are sub-leading
order and we expect:
$$
\phi \sim \phi^{(0)} = -\frac{D}{r} + o(1/r).
$$
We now define a new coordinate $s = \sqrt{\vert n \vert A^{-(n+2)}}
Mr$ and $u = -\phi/AM$ for some constant $A$. With these definitions
the full field equation for $\phi$ outside the body (with $\phi_b =
0$) becomes:
$$
\frac{1}{s^2}\frac{d}{ds}\left(s^2\frac{d u}{ds}\right) = u^{-n-1},
$$
and as $s \rightarrow \infty$:
$$
u \sim \frac{D A^{-\frac{n+4}{2}}}{\sqrt{\vert n \vert}s} + o(1/s).
$$
We set $A^{n+4} = D^2/\vert n \vert$ so that $u \sim 1/s$ and define $t = 1/s$ so that the field equations become:
\begin{equation}
\frac{d^2u}{dt^2} = \frac{u^{-n-1}}{t^4}. \label{uteqn}
\end{equation}
The asymptotic form of $u$ as $r \rightarrow \infty$, $t \rightarrow
0$, requires that $u(t=0) = 0$ and $du/dt (t=0) = 1$ \emph{exactly}.
With these boundary conditions we numerically evolve eqn.
(\ref{uteqn}) towards larger $t$ (smaller $r$).  As one might expect
from such an elliptic equation, with these boundary conditions, a
singularity occurs at some finite $t$ which we label $t_{max}$. We
use our numerical evolutions to determine $t_{max}$ for each $n$.
For the evolution of $\phi$ to remain non-singular up to the surface
of the body, we need $r=R$ to correspond to a value of $t <
t_{max}$.  The limiting case is given by $t(R) = t_{max}$. This
limiting case determines a critical form for the $\phi$ field which
occurs when $A=A_{min}$ where:
$$
A_{min} = \vert n t_{max}^2 \vert^{\frac{1}{n+2}} (MR)^{\frac{2}{n+2}}.
$$
This corresponds to the following critical asymptotic form for $\phi$:
$$
\phi_{crit} \sim \phi_b - \left(\frac{t_{max}^{\vert n + 4\vert}}{\vert n \vert}\right)^{\frac{1}{\vert n +2 \vert}} (MR)^{\frac{n+4}{n+2}}\frac{e^{-m_b r}}{r},
$$
where we have reinserted the (almost always negligible) $\phi_b$ and
$m_b$ dependence.  We use numerical integration to calculate the
value of $\gamma(n) := t_{max}^{\vert n + 4 \vert}$ for different
values of $n$.  Our results are displayed in table \ref{TABgamma}.
\begin{table}[tbp]
\begin{center}
\caption{Values of $\gamma(n) = t_{max}^{\vert n + 4\vert}$ \label{TABgamma}}
\begin{tabular}{|c|c|}
\hline n & $\gamma(n)$ \\
\cline{1-2} -12 & 14.687 \\
\cline{1-2} -10 & 10.726 \\
\cline{1-2} -8 & 6.803  \\
\cline{1-2} -6 & 3.000  \\ \hline
\end{tabular}
\end{center}
\end{table}
Physically acceptable non-singular evolution implies that
asymptotically $\phi/\phi_{crit} < 1$. If $\phi/\phi_{crit} \ll 1$
then the conditions of the pseudo-linear approximation are satisfied
and so the body \emph{cannot} have a thin-shell.  Thin-shelled
bodies must therefore almost saturate this bound on $\phi$ and so
$\phi_{crit}$ provides a good approximation to the asymptotic
behaviour $\phi$ outside thin-shelled bodies.   We note that, as in
the $n >0$ case, the existence of a critical form for $\phi$ depends
in no way on the what occurs inside the body and, as such, is
independent of both $\beta$ and the density of the body.
%NEW UPDATE ENDS HERE
\\ \\
\noindent\textit{Critical Behaviour} \\
\indent The existent of a critical form for $\phi$ when $r \gg R$ implies
that, no matter how massive our central body, and no matter how
strongly it couples to the chameleon, the perturbation it produces
in $\phi$ for $r \gg R$ takes a universal value whenever the
thin-shell conditions, eqns (\ref{thincond}a-c), hold.

When $n \neq -4$, the critical form of the far field, depends only on $M$,
$n$, $R$ and on the chameleon mass in the background, $m_b$. When
$n = - 4$ the critical form for the far field depends only on
$\lambda$, $R$ and $m_b$.   For all $n$, the far field is,
crucially, found to be  \emph{independent} of the coupling,
$\beta$, of the chameleon to the isolated body.   This is one of
the main reasons why $\beta \gg 1$ is not ruled out by current
experiments. The larger $\beta$ becomes, the stronger the
chameleon mechanism and so the easier it is for a given body to
have a thin-shell. However, the far field of a body with a
thin-shell is independent of $\beta$, and so, in stark contrast to
what occurs for linear theories, larger values of $\beta$ do
\emph{not} result in larger forces between distant bodies.
%NEW UPDATE BEGINS HERE:  Just the change to betacrit for n <0
Defining the mass of our central body to be ${\mathcal{M}}=4\pi \rho_c R^3/3$ we
can express this critical behaviour of the far field in terms of an
effective coupling, $\beta_{eff}$, defined by:
$$ \phi \sim \phi_b - \frac{\beta_{eff} {\mathcal{M}}e^{-m_b r}}{4\pi
M_{pl}r},
$$ when $r \gg R$.  Assuming $\rho_{b}/\rho_c \ll 1$ we find that:
\begin{subequations}
\label{critbeta}
\begin{eqnarray}
\beta_{eff} &=& \frac{4\pi M_{pl}}{{\mathcal{M}}}
\left(\frac{\gamma(n)}{\vert n \vert}\right)^{\frac{1}{\vert n +
2\vert}}\left(MR\right)^{\frac{n+4}{n+2}}, \qquad n < -4, \\
\beta_{eff} &=& \frac{4\pi
M_{pl}}{{\mathcal{M}}}MR\left(\frac{n(n+1)M^2}{m_b^2}\right)^{\frac{1}{n+2}},
\qquad n > 0 \\ \beta_{eff}(r) &=& \frac{2\pi
M_{pl}}{\mathcal{M}}\left(2\lambda\ln(\min(r/R,1/m_bR))\right)^{-1/2},
\qquad n = -4.
\end{eqnarray}
\end{subequations}
The $\beta$ independence of $\beta_{eff}$ was first noted, in the
context of $\phi^4$ theory, in \cite{nelson}. However, the
authors were mostly concerned with region of parameter space
$\beta < 1$, $\lambda \ll 1$; in our analysis we go further:
considering a wider range of theories and also the possibility
that $\beta \gg 1$. %
% UPDATE STARTS HERE
$\beta$-independence was also present in the original work of Khoury
and Weltman \cite{chama,cham1} for $n>0$ theories. However, in those
works, the $\beta$ independence together with its important
implications for experiments, was not commented on. Especially
those that search for WEP
violations were not considered. As we shall see in section \ref{exper}
below, this $\beta$ independence means that if one uses test-bodies
with the same mass and outer dimensions then in chameleon theories, no
matter how much the weak equivalence principle is violated at a
particle level, there will be \emph{no} violations of WEP far from the
body. Simply because the far field is totally independence of both the
body's chameleon coupling and its density.

In this work, we have also shown that this $\beta$ independence is a
generic feature of all $V \propto \phi^{-n}$ chameleon theories and it
is not simply as artifact of the runaway ($n>0$) potentials considered
in \cite{chama,cham1}.  %
% UPDATE ENDS HERE
Indeed there are good reasons to believe that
similar behaviour will be seen in chameleon theories with other
potentials.  As we mentioned in the introduction, the field
equations for chameleon theories are necessarily non-linear.  It
is well-known that, that in non-linear theories, the evolution of
arbitrary initial conditions will generically be singular. If one
wishes to avoid singularities then tight constraints on the
initial conditions must be satisfied.  When considering the field
outside an isolated body, these conditions will, generally,
require that $\vert{\mathrm{d}} \phi / {\mathrm{d}}r \vert$ is smaller
than some critical, $r$-dependent, value. As a result, there will a
critical, or maximal, form that the field produced by a body can
take. This precisely what we have found for $\phi^{-n}$ theories.
The form of this critical far field will depend on the nature of
the non-linear potential, and possibly the coupling of $\phi$ to
any background matter, but, since we are outside the body, it \emph{cannot} depend on the coupling of the chameleon to the body itself.
Again, this is precisely what we have seen for $\phi^{-n}$
chameleon theories.

We can understand the $\beta$-independence, in a slightly
different way, as follows: just outside a thin-shelled body, the
potential term in eq. (\ref{micro}) is large and negative ($\sim
\mathcal{O}(-\beta \rho / M_{pl})$), and it causes $\phi$ to decay
very quickly. At some point, $\phi$ will reach a critical value,
$\phi_{crit}$, that is small enough so that non-linearities are no
longer important.  Since this all occurs outside the body,
$\phi_{crit}$ can only depend on the size of the body, the choice
of potential $(M, \lambda, n)$ and the mass of $\phi$ in the
background, $m_{b}$. This is precisely what we have found above.

We have seen above that the far field of a body with thin-shell is
independent of the microscopic chameleon-to-matter coupling,
$\beta$. This is one of the vital features that allows theories
with $\beta \gg 1$ to coexist with the current experimental
bounds.  It is also of great importance when testing for WEP
violations, since any microscopic composition dependence in $\beta$
will be invisible in the far field of the body.  We discuss these
issues further in section \ref{exper}, where we consider the
experimental constraints on $\beta$, $M$ and $\lambda$ in more
detail.

The results of this section are summarised in section
\ref{singsum} below.

%UPDATE ENDS HERE

\subsection{Summary \label{singsum}}
We have seen in this section that there are three important classes
of behaviour for $\phi$ outside an isolated body: the linear,
pseudo-linear and non-linear regimes.  In fact, although the
mathematical analysis differs, $\phi$ behaves in same way in both
the pseudo-linear and linear regimes. We have shown that linear, or
pseudo-linear, behaviour will occur whenever conditions
(\ref{pseudcon}a-c) on $m_c R$ hold.  As $m_c R$ is increased,
conditions (\ref{pseudcon}a-c) will eventually fail.  As $m_c R$
is increased still more, a thin-shell forms and we move into the
non-linear regime. A thin-shell will exist whenever the thin-shell
conditions, eqs. (\ref{thincond}a-c), hold; these are equivalent
to $m_c R > (m_c R)_{eff}$.  We have seen that, in the non-linear
regime, the far field is independent of the coupling of the
chameleon to the isolated body. The main results of this section
are summarised below.

We have been concerned with a spherical body of uniform
density $\rho_c$ and radius $R$.  The background has density
$\rho_b \ll \rho_c$. The chameleon in background ($r \gg R$) takes
the value $\phi_b$ and its mass there is $m_b=m_{\phi}(\phi_b)$.
We also have $$ \phi_{c} = M\left(\frac{\beta \rho_c}{n\lambda
M_{pl}M^3}\right)^{-\frac{1}{n+1}}, \qquad m_c = m_{\phi}(\phi_c).
$$
\subsubsection*{Linear and Pseudo-Linear Behaviour}
Non-linear effects are negligible when:
\begin{eqnarray*}
m_c R
&\ll&\min\left((18)^{1/6}\left(\tfrac{m_c}{m_b}\right)^{\frac{n+4}{3(n+2)}},
\sqrt{2\vert n+1\vert} \left\vert
\left(\tfrac{m_{b}}{m_c}\right)^{\frac{2}{\vert n + 2
\vert}}-1\right\vert \right), \quad n < -4,\\ m_c R
&\ll&\min\left(\sqrt{3}\left(\tfrac{m_c}{m_b}\right)^{\frac{n+4}{3(n+2)}},
\sqrt{2\vert n+1\vert} \left\vert
\left(\tfrac{m_{c}}{m_b}\right)^{\frac{2}{\vert n + 2
\vert}}-1\right\vert \right), \quad n > 0, \\ m_c R &\ll& 1, \quad n =
-4,
\end{eqnarray*}
and outside the body, $r> R$, $\phi$ behaves like:
\begin{eqnarray*}
\phi &\approx& \phi_b - \frac{(m_c R)^2 \phi_c Re^{-m_b r}}{3(n+1)r}, \qquad n \neq -4, \\
\phi &\approx& \phi_b - \sqrt{\frac{1}{2(y_0 +\ln(\min(r/R,1/m_b)))}}
\frac{e^{-m_br}}{2r}, \qquad n = -4,
\end{eqnarray*}
where $y_0$ is given by: $$ \frac{(m_cR)^3}{9} =
\sqrt{\frac{3}{2y_0}} +\frac{1}{3}\left(\frac{3}{2
y_0}\right)^{3/2}. $$ When non-linear effects are negligible a
body will certainly \emph{not} have a thin-shell.
\subsubsection*{Bodies with thin-shells}
A body with have a thin-shell when:
\begin{eqnarray*}
m_c R
&\gtrsim&\min\left((18)^{1/6}\left(\tfrac{m_c}{m_b}\right)^{\frac{n+4}{3(n+2)}},
\sqrt{3\vert n+1\vert} \left\vert
\left(\tfrac{m_{b}}{m_c}\right)^{\frac{2}{\vert n + 2
\vert}}-1\right\vert \right), \quad n < -4, \\ m_c R
&\gtrsim&\min\left(\sqrt{3}\left(\tfrac{m_c}{m_b}\right)^{\frac{n+4}{3(n+2)}},
\sqrt{3\vert n+1\vert} \left\vert
\left(\tfrac{m_{b}}{m_c}\right)^{\frac{2}{\vert n + 2
\vert}}-1\right\vert \right), \quad n > 0, \\ m_c R &\gtrsim& 4, \quad
n = -4.
\end{eqnarray*}
Outside the body there are two regimes of behaviour.  If $(r-R)/R
\ll 1$  and $m_{\phi}(\phi)/m_b \gg 1$ then $\phi$ is given by
\begin{eqnarray*}
 \frac{1}{m_{\phi}(\phi)} \sim \frac{\vert n+2 \vert (r-R)}{\sqrt{2n(n+1)}}  +\left(\frac{n+1}{n}\right)^{n/2+1}\frac{1}{m_{c}},
\end{eqnarray*}
and:
$$
\phi =  {\mathrm{sgn}}(n)M\left(\frac{n(n+1)\lambda M^2}{m_{\phi}^2(\phi)}\right)^{\frac{1}{n+2}}.
$$
If, however, $(r-R)/R \gtrsim 1$ then
$$ \phi \sim \phi_b - \frac{\beta_{eff} {\mathcal{M}}e^{-m_b r}}{4\pi
M_{pl}r},
$$
where $\mathcal{M}$ is the mass of the body, and $\beta_{eff}$ is the effective coupling
\begin{eqnarray*}
\beta_{eff} &=& \frac{4\pi M_{pl}}{{\mathcal{M}}}
\left(\frac{\gamma(n)}{\vert n \vert}\right)^{\frac{1}{\vert n +
2\vert}}\left(MR\right)^{\frac{n+4}{n+2}}, \qquad n < -4, \\
\beta_{eff} &=& \frac{4\pi
M_{pl}}{{\mathcal{M}}}MR\left(\frac{n(n+1)M^2}{m_b^2}\right)^{\frac{1}{n+2}},
\qquad n > 0 \\ \beta_{eff}(r) &\approx& \frac{2\pi
M_{pl}}{\mathcal{M}}\left(2\lambda\ln(\min(r/R,1/m_bR))\right)^{-1/2},
\qquad n = -4.
\end{eqnarray*}
We note that $\beta_{eff}$ is independent of coupling of the
chameleon to the body, and that $\phi$ is independent of the
body's mass, $\mathcal{M}$.  When $(r-R)/R \gtrsim 1$, $\phi$ only
depends on $r$, $R$, $M$, $\lambda$, $n$ and $m_b$.

%%%%%%%%%%%%%%%%%%%%%%%%%%%%%%%%%%%%%%%%%%%%%%%%%%%%%%%%%%%%%%%%

\section{Effective Macroscopic Theory \label{effmacr}}
Eq. (\ref{micro}) defines the microscopic, or particle-level,
field theory for $\phi$, whereas in most cases, which we wish to
study, we are interested in the large scale or coarse grained
behaviour of $\phi$. In macroscopic bodies the density is actually
strongly peaked near the nuclei of the individual atoms from which
it is formed and these atoms are separated from each other by
distances much greater than their radii. Rather than explicitly
considering the microscopic structure of a body, it is standard
practice to define an `averaged' field theory that is valid over
scales comparable to the body's size. If our field theory were
linear, then the averaged equations would be the same as the
microscopic ones e.g. as in Newtonian gravity. It is important to
note, though, that this is very much a property of \emph{linear
theories} and is not in general true of non-linear ones.
\begin{figure}[tbh]
\begin{center}
\includegraphics[width=7.3cm]{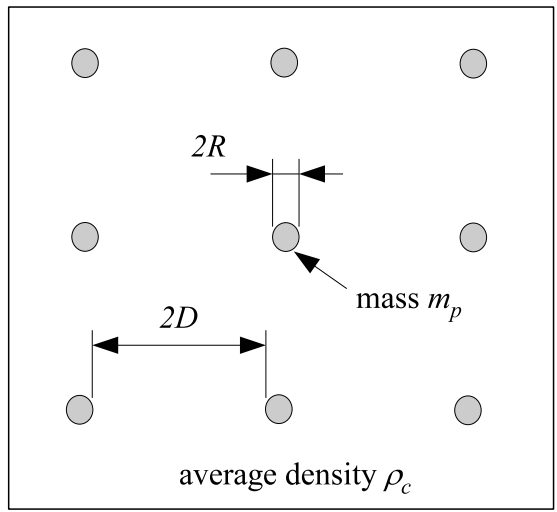}
\end{center}
\caption[Illustration of the model used for calculating the
macroscopic field theory.]{Illustration of the model for the
microscopic structure of the body considered in this section.  The
constituent particles are assumed to be spherical of radius $R$,
mass $m_{p}$ and of uniform density.  They are separated by an
average distance $2D$. The average density of the body is defined
to be $\rho_{c}$.} \label{FIGavgillu}
\end{figure}

Non-linear effects must therefore be taken into account.  Using
similar methods to those that were used in section \ref{sing} above, we derive an
effective theory that describes the behaviour of the course-grained or
macroscopic value of $\phi$ in a body with thin-shell.  We will
identify the conditions that are required for linear theory averaging
to give accurate results and consider what happens when non-linear
effects are non-negligible.

In this section, we derive an effective macroscopic theory
appropriate for use within bodies that possess a thin-shell and which
are made up of small particles, radius $R$ and mass $m_{p}$. These particles are
separated by an average distance $2D \gg R$.  The average density of the
body is $\rho_c$. We illustrate this set-up in FIG. \ref{FIGavgillu}.
A thin-shell, in this sense, means that the average value of
$\phi$ inside a sphere of radius $\gtrsim D$, will be
approximately constant, $\phi \approx \phi_c$, everywhere inside
the body apart from in a thin-shell close to the surface of the
body. Generally the emergence of a thin-shell is related to a
breakdown of linear theory on some level. The conditions for a
body to have a thin-shell are given by eqs. (\ref{thincond}a-c).
The outcome of this section will be to slightly modify these
conditions. Precisely, we will find that there is maximal, or
critical, value for the average chameleon mass, $m_c$.  Oddly, this critical, \emph{macroscopic} chameleon mass depends only on the \emph{microscopic} properties of the body.

\subsection{Averaging in Linear Theories}

We are concerned with finding an effective theory that will give
correct value of $\phi_c$. We have defined $m_{c} =
m_{\phi}(\phi_c)$. The microscopic field equations for $\phi$, as
given by eq. (\ref{micro}), are: $$ -\square \phi = -n\lambda M^3
\left(\frac{M}{\phi}\right)^{n+1} + \frac{\beta
\rho(\vec{x})}{M_{pl}}, $$ where the microscopic
matter density, $\rho(\vec{x})$, is strongly peaked
about the constituent particles of the macroscopic body but
negligible in the large spaces between them.  Before considering
what occurs in a non-linear theory, such as the chameleon theories
being studied here, we will review what would occur in the linear
case. For the field equation to be linear, the potential must be,
at worst, a quadratic in the scalar field $\phi$.  With the
potentials considered in this paper, a linear theory
emerges if $n=0, -1$ or $-2$. To examine why averaging, or coarse
graining, is not an issue if field equations are linear, and make
reference to what actually occurs in chameleon theories,  we shall
linearise eq. (\ref{micro}) about $\phi=\phi_0$ for some $\phi_0$.
It is important to note that we are performing this linearisation
only for the purpose of showing what occurs in linear theories; we
are \emph{not} claiming that a linearisation, such as this, is
actually valid. Defining $\phi = \phi_0 + \phi_1$, and
neglecting non-linear terms, we obtain:
\begin{eqnarray}
-\square \phi_1 = -n\lambda M^3 \left(\frac{M}{\phi_0}\right)^{n+1} +
 n(n+1)\lambda M^2 \left(\frac{M}{\phi_0}\right)^{n+2} \phi_1 +
 \frac{\beta \rho(\vec{x})}{M_{pl}}. \label{linavgeqn}
\end{eqnarray}
We will write the averaged, or coarse-grained, value of a quantity
$Q(\vec{x})$ as $< Q >(\vec{x})$ and define it by:
\begin{eqnarray}
< Q >(\vec{x}) = \frac{\int d^3 y \, Q(\vec{y})
\Theta(\vec{x}-\vec{y})}{\int d^3 y \Theta(\vec{x} - \vec{y})},
\label{coarseg}
\end{eqnarray}
where the function $\Theta(\vec{x}-\vec{y})$ defines the coarse
graining and the integral is over all space.  Different choices of
$\Theta$ will result in different coarse-grainings.  If we are
interested in averaging over a radius of about $D$ around the
point $\vec{x}$, then a sensible choice for $\Theta$ would
something like: $$ \Theta_{1}(\vec{x} - \vec{y}) = e^{-\frac{\vert
\vec{x}-\vec{y}\vert^2}{D^2}}, $$ or $$ \Theta_{2}(\vec{x} -
\vec{y}) = H\left(D-\vert\vec{x}-\vec{y}\vert\right), $$ where
$H(x)$ is the Heaviside function. For the coarse graining process
to be well-defined we must require that, whatever choice one makes
for $\Theta$, it vanishes sufficiently quickly as $\vert\vec{x} -
\vec{y}\vert \rightarrow \infty$ that the integrals of eq.
(\ref{coarseg}) converge. This will usually require $\Theta \sim
\mathcal{O}(\vert\vec{x}-\vec{y}\vert^{-3})$ as $\vert\vec{x} -
\vec{y}\vert \rightarrow \infty$.

Consider the application of the averaging procedure to the linear
field equation given by eq. (\ref{linavgeqn}).  It follows from
the assumed properties of $\Theta$ that $< \square \phi_1 > =
\square <\phi_1 >$ and so $$ \square <\phi_1 > = -n\lambda M^3
\left(\frac{M}{\phi_0}\right)^{n+1} + n(n+1)\lambda M^2
\left(\frac{M}{\phi_0}\right)^{n+2}< \phi_1 > + \frac{\beta < \rho
>(\vec{x})}{M_{pl}}. $$ This is the averaged field equation for
$\phi_1$.  It is clear from the above expression, that, although
the precise definition of the averaging operator depends on a
choice of the function $\Theta$, the averaged field equations are
\emph{independent} of this choice.  This independence is a
property of linear theories but it is not, in general, seen in
non-linear ones.  The averaged field equations for a non-linear
theory \emph{will}, generally, depend on ones choice of averaging.
In this section, we take our averaging function to be $\Theta_{2}$
as defined above; this is equivalent to averaging by volume in a
spherical region of radius $D$.  It is also clear that, for a
linear theory, the averaged field equation for $\phi_1$ is
functionally the same as the microscopic equation.  This, again,
would not be true if non-linear terms where present in the
equations; in general $$ < \phi^{n} > \neq < \phi >^{n} $$ unless
$n = 0$ or $1$ or $\phi$ is a constant.

\subsection{Averaging in Chameleon Theories}

Our aim, in this section, is to calculate the correct value of
$<\phi>$ and $<m_{\phi}(\phi)>$ inside a body with a thin-shell.
We have defined $\phi_{c} = <\phi>$ and $\rho_c = <\rho>$.
Although these calculations will implicitly depend on our choice
of averaging function, our results should also be approximately
equal, at least to an order of magnitude, to those that would be
found using any other sensible choice of coarse-graining
defined over length scales of about $D$ or greater.

If our chameleon theories were linear, we have seen that we would
expect $\phi = \phi_c^{(lin)}$ where $$ - \square \phi_c^{(lin)} =
-n \lambda M^3 \left(\frac{M}{\phi_c^{(lin)}}\right)^{n+1}
+\frac{\beta\rho_c}{M_{pl}}, $$ and so, for $\phi_c^{(lin)}
\approx const$, we have: $$ \phi_c^{(lin)} = M\left(\frac{\beta
\rho_c}{n \lambda M_{pl}M^3}\right)^{-1/(n+1)}. $$ In appendix
\ref{avgApp}, we show that, for some values of $R$, $D$ and
$m_{p}$, linearised theory will give the correct value of $\phi_c$
to a high accuracy i.e. $\phi_c \approx \phi_{c}^{(lin)}$.  This
happens when there either exists a consistent, everywhere valid,
linearisation of the field equations or we can construct a
pseudo-linear approximation along the same lines as was done in
section \ref{singpseu} . However, for some values of $R$, $D$ and
$m_{p}$, we find that non-linear effects are unavoidable.  When
$R$, $D$ and $m_{p}$ take such values, we will say that we are in
the \emph{non-linear regime}.  We find that, just as it did in
section \ref{singnon} above, the non-linear regime features
$\beta$-independent critical behaviour. The details of these
calculations can be found in appendix \ref{avgApp}.

We define:
\begin{eqnarray}
D_{c} &=& \left(n(n+1)\right)^{\frac{n+1}{n+4}}M^{-1}\left(\frac{3\beta
m_{p}}{4\pi M_{pl} \vert n \vert}\right)^{\frac{n+2}{n+4}},
\label{Dceqn} \\ D_{\ast} &=&
\left(\frac{n(n+1)}{MR}\right)^{\frac{n+1}{3}} M^{-1}\left(\frac{3\beta
m_{p}}{4\pi M_{pl} \vert n\vert}\right)^{\frac{n+2}{3}}, \notag
\end{eqnarray}
and note that $D_{\ast}/D_{c} = (D_{c}/R)^{(n+1)/3}$.  For the
linear approximation to be valid we need both $m_{c}D \ll 1$ and
$m_c^2 D^3/2R \ll 1$. This is equivalent to:
\begin{eqnarray*}
D_{c} &\ll& D \ll D_{\ast}, \qquad n <- 4 \\ \max(D_{c},D^{\ast})
&\ll& D, \qquad n > 0.
\end{eqnarray*}
When $n=-4$ we require $D \ll D_{\ast}$ and: $$
(12)^{3/2}\lambda^{1/2}\left(\frac{3\beta m_{p}}{4\pi n
  M_{pl}}\right) \ll 1.
$$ We can see that, for given $m_{p}$ and $R$, it is always
possible to find a $D$ such that the linear approximation is valid
when $n > 0$. However, when $n \leq -4$, it is possible that there
will exist \emph{no} value of $D$ for which the above conditions
hold. Whenever the linear approximation holds we have: $$ m_{c}
\approx m_{\phi}\left(\phi_{c}^{(lin)}\right). $$

We can construct a pseudo-linear approximation whenever:
\begin{subequations}
 \label{avgpseud}
\begin{eqnarray}
&\left(\frac{D_c}{R}\right)^{\frac{n+4}{n+1}} < 2\vert n + 1
\vert\left(1- \left(\frac{R}{D}\right)^{\frac{3}{\vert n + 1
\vert}}\right), \qquad n < -4& \\ &\frac{D}{D_c} > 1\quad \mathrm{and} \quad \frac{D_{\ast}}{D} <
\left[2(n+1)\left(1-\left(\frac{R}{D}\right)^{\frac{3}{n+1}}\right)\right]^{\frac{n+1}{3}},
\qquad n > 0 &\\ &m_{\phi}\left(\phi_{c}^{(lin)}\right)D \ll \left(\frac{243}{2\ln(D/R)}\right)^{1/6}, \qquad
n=-4.&
\end{eqnarray}
\end{subequations}
When the pseudo-linear approximation holds we again find:
$$ m_{c} \approx m_{\phi}\left(\phi_{c}^{(lin)}\right).
$$

As the inter-particle separation, $D$, is decreased we will
eventually reach a point where eqs. (\ref{avgpseud}a-c) fail to hold.
When this occurs it is because non-linear effects have become
important inside the individual particles that make up the body.  As
$D$ decreases still further these particles will eventually develop thin-shells of their own.  Non-linear effects become important when:
\begin{subequations}
\label{avgthin}
\begin{eqnarray}
\left(\frac{D_c}{R}\right)^{\frac{n+4}{n+1}} &>& 3\vert n + 1
\vert\left(1- \left(\frac{R}{D}\right)^{\frac{3}{\vert n + 1
\vert}}\right), \qquad n < -4 \\ \frac{D_{\ast}}{D} &>&
\left[3(n+1)\left(1-\left(\frac{R}{D}\right)^{\frac{3}{n+1}}\right)\right]^{\frac{n+1}{3}},
\qquad n > 0 \\ m_{\phi}\left(\phi_{c}^{(lin)}\right)D &\gtrsim& \left(\frac{243}{2\ln(D/R)}\right)^{1/6},
\qquad n=-4.
\end{eqnarray}
\end{subequations}
These conditions define the non-linear regime.  Between the
pseudo-linear, and fully non-linear regimes, there is, of course,
some intermediate region, however this has proven too difficult to
analyse analytically. We therefore leave the detailed analysis
of this intermediate behaviour to a later work.  This intermediate
region is, however, in some sense small and so we do not
believe it to have any great importance with respect to
experimental tests of chameleon theories.

\begin{figure}[tbh]
\begin{center}
\includegraphics[width=7.4cm]{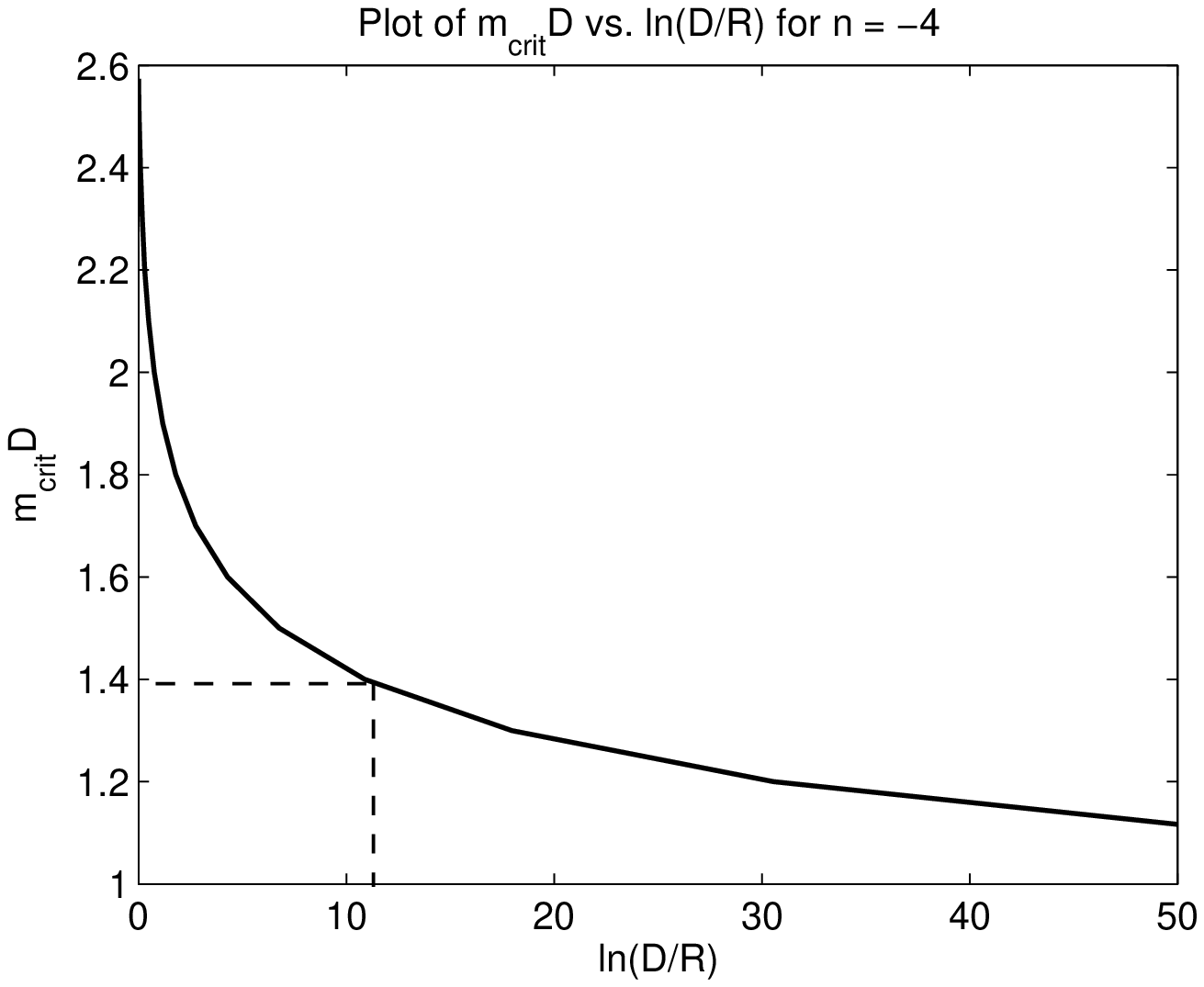}
\includegraphics[width=7.4cm]{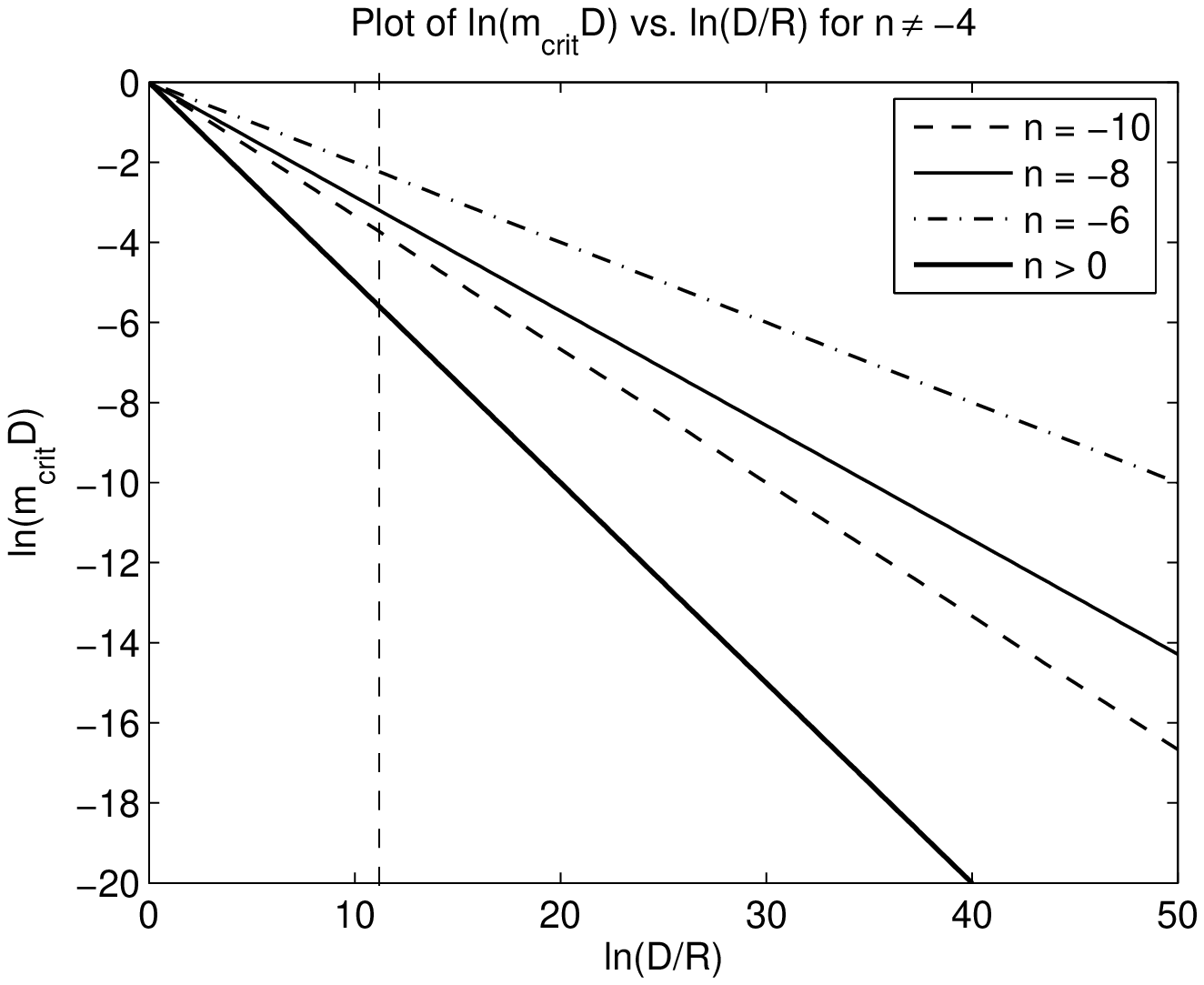}
\end{center}
\caption[Dependence of the critical chameleon mass on $D/R$.]{Dependence of the critical chameleon mass on $D/R$.  The
above plots show how $m_{crit} D$ depends on $\ln(D/R)$ for different
values of $n$.  The cases $n=-4$ and $n \neq -4$ are qualitatively
different and are therefore shown on separate plots.  $m_{crit}$ is the
maximal mass the chameleon can take inside a thin-shelled body. $2D$
is the average separation of the particles that comprise that body and
$R$ is the average radius of the constituent particles.  Typically we
find that $\ln(D/R) \approx 11$ for bodies with density $\rho \sim
1-10\,{\mathrm{g\,cm}}^{-3}$; $\ln(D/R) = 11$ is indicated on the plots.  Note
that when $n > 0$, $m_{crit}D$ is independent of $n$. Also note that
in $\phi^{4}$ theory $m_{crit}D \sim \mathcal{O}(1)$ whereas for other value of
$n$ it is generically much smaller.}
\label{FIGmaD}
\end{figure}

When the individual particles develop thin-shells, the $\phi$-field
external to the particles will be, by the results of section
\ref{sing}, independent of $\beta$. This ensures that the
chameleon mass far from the particles is also independent of
$\beta$. Therefore, whenever a body falls into the non-linear
regime, the average chameleon mass will take a critical value,
$m_{c} = m_{c}^{crit}$. This is defined in a similar way to which
$(m_{c}R)_{crit}$ was in section \ref{singnon}, i.e.
$m_{c}^{crit}$ is the maximal mass that the chameleon may have
when $r\sim {\mathcal{O}}(D)$ such that, when the microscopic
field equations are integrated, $(\phi_{c} / \phi)^{n}$ is finite
for all $r
> R$.  This definition implies a relationship between $m_{c}^{crit}$,
$R$ and $D$, however it does not depend on either $M$ or
$\lambda$. $m_{c}^{crit}$ is also found to depend on $n$; this is
because $n$ defines precisely how quickly $(\phi_{c}/\phi)^n$
blows up.  We derive expressions $m_{c}^{crit}$ in appendix
\ref{avgApp} finding:
\begin{eqnarray}
m_{c}^{crit} &\approx& \frac{\sqrt{3\vert n+1
\vert}}{D}\left(\frac{R}{D}\right)^{q(n)/2}S(n), \qquad n \neq -4
\label{mccrit} \\ m_{c}^{crit} &=& X/D, \qquad n=-4 \notag,
\end{eqnarray}
where $q(n) = \min((n+4)/(n+1),1)$, $S(n) = 1$ if $n > 0$ and $S(n) = (\gamma(n)/3)^{1/2\vert n+1\vert}$ if $n < 0$. $X$ is given by: $$
\frac{3\sqrt{3}}{\sqrt{2\ln(D/R)}} \approx X \cosh X - \sinh X. $$
We plot $m_{c}^{crit}D$ vs. $\ln(D/R)$ in figure \ref{FIGmaD}. For
an everyday body with density similar to water, we approximate $R$
and $m_{p}$ respectively by the radius and mass of carbon nucleus
(say) and find $\ln(D/R) \approx 11$, and so $m_{c}^{crit} \approx
1.4/D$ when $n=-4$. When $n \neq -4$, $m_c^{crit} D \ll 1$ follows
from $R/D \ll 1$.

It is interesting to note that, even though $m_{c}^{crit}$ is a
macroscopic quantity, it depends entirely on the details of the
microscopic structure of the body i.e. $R$ and $D$.  By combining
the results of this section, we find that the average mass of the
chameleon inside a body with thin-shell that is itself made out of
particles is given by: $$ m_{c} = \min \left( M\left(\frac{\vert
n\vert\lambda M_{pl}M^3}{\beta \rho_{c}}\right)^{\frac{n+2}{2(n+1)}},
m_{c}^{crit}(n,R,D)\right). $$ When evaluating the thin-shell
conditions, eqs. (\ref{thincond}a-c), it is therefore this value
of $m_c$ that should be used.

\section{Force between two bodies \label{forcetwo}}

In the previous two sections, we have considered how the chameleon
field, $\phi$, behaves both inside and outside an isolated body.
In this section we study the form that the chameleon field takes
when two bodies are present, and use our results to calculate the
resultant $\phi$-mediated force between those bodies. The results
of this sections will prove to be especially useful when we come
to consider the constraints on chameleon field theories coming
from experimental tests of gravity in section \ref{exper} below.

Chameleon field theories, by their very nature, have highly
non-linear field equations. This non-linear nature is especially
important when bodies develop thin-shells.  As a result of their
non-linear structure, one cannot solve the two (or many) body
problem by simply super-imposing the fields generated by two (or
many) isolated bodies, as one would do for a linear theory.

When the two bodies in question have thin-shells, we shall see that
the formula for the $\phi$-force is highly dependent on the
magnitude of their separation relative to their respective sizes.
We shall firstly consider the case where the separation between
the two bodies is small compared the radius of curvature of their
surfaces, and secondly look at the force between two distant
bodies. Finally we will consider the force between a very small
body and a very large body. We will also look at what occurs when
one or both of the bodies does \emph{not} have a thin-shell.

\subsection{Force between two nearby bodies \label{forclose}}
We consider the force between two bodies (hereafter body one and
body two) whose surfaces are separated by a distance $d$.  Both
bodies are assumed to satisfy the thin-shell conditions. The two
bodies are taken to be nearby in the sense that: $ d \ll R_{1},
R_{2}$ where $R_{1}$ and $R_{2}$ are respectively the radii of
curvature of the surface of body one and body two.  Since $d \ll
R_{1}, R_{2}$ we can ignore the curvature of the surfaces of bodies
to a first approximation.   With this simplification we treat the
bodies as being infinite, flat slabs and take body one to occupy the
region $x< 0$, and body two the region $x>d$. We use a subscript $1$
to refer to quantities that are defined for body one: e.g. the
density of body one is $\rho_{1}$ and the chameleon mass deep
inside body one is $m_1$, and a subscript $2$ for quantities
relating to body two.  Additionally a subscript or superscript $s$
is uses to refer to quantities that are defined on the surfaces of
the two bodies e.g. $m_{1}^{s}$ is the chameleon mass of the surface
of body one.   Subscript $0$ is used two label quantities defined at
that point between body one and body two where $d \phi / dx = 0$. We
assume also that the background chameleon mass, $m_{b}$, obeys $m_b
d \ll 1$, we discuss later what occurs if this is not the case.

We now consider the $\phi$-mediated force on body one due to body
two. With the above definitions, and $\phi$ obeys:
$$
\frac{d^2\phi}{dx^2} = V_{,\phi}(\phi)
$$
in $0 < x < d$ and
$$
\frac{d^2\phi}{dx^2} = V_{,\phi}(\phi) + \frac{\beta \rho_{1}}{M_{pl}}
$$
in $x < 0$.  Integrating these equations we find:
\begin{equation}
\left(\frac{d \phi}{d x}\right)^2 = 2(V(\phi)-V_0),
\label{phiinte}
\end{equation}
in $0 < x < d$, and in $x < 0$ we have:
$$
\frac{1}{2}\left(\frac{d \phi}{dx}\right)^2 = V(\phi)-V_1 + \frac{\beta \rho_{1} (\phi - \phi_1)}{M_{pl}}.
$$
Matching these expressions at $x = 0$ we have:
\begin{equation}
\phi_{1}^{s} - \phi_1 =  \frac{M_{pl}(V_{1} - V_{0})}{\beta \rho_{1}},
\label{phisurface}
\end{equation}
If the second body where \emph{not} present then $V_{0} = 0$ and $\phi_{1}^{s} = \bar{\phi}_{1}^{s}$ where:
$$
\bar{\phi}_{1}^{s} - \phi_1 =  \frac{M_{pl}V_{1}}{\beta \rho_{1}}.
$$
The attractive force per unit area of body one due to body two is therefore:
$$
\frac{F_{\phi}}{A} = \frac{\beta \rho_1}{M_{pl}}\left\vert\bar{\phi}_{1}^{s}-\phi_{1}^{s}\right\vert = V_{0}.
$$
This holds for all $V(\phi)$ not just the $\phi^{-n}$ potentials considered in this work.  To find $V_{0}$ we integrate eqn. (\ref{phiinte}) in the region $0 < x < d$ and find:
\begin{equation}
\sqrt{2}d = \frac{\sqrt{V_0}}{\vert V_{0,\phi} \vert}\left(\int^{y_1}_{1} \frac{dx}{W(x)\sqrt{x-1}} + \int^{y_2}_{1} \frac{dx}{W(x)\sqrt{x-1}}\right),
\end{equation}
where $y_1 = V_{1}^{s} / V_0$ and $y_2 = V_{2}^{s}/V_0$ and $W( x = V/V_0) = V_{\phi}/V_{0,\phi}$. We evaluate the integrals in the above expression in two important limits.
\subsubsection{Limit 1: $y_{1} = V_{1}^{s}/V_0 = 1+\delta$}
In this limit we assume $V_{1} \approx V_{1}^{s} \approx V_{0}$,  this would occur if $V_1 < V_2$ and $d$ is suitably small. In this limit:
$$
\int^{y_1}_{1} \frac{dx}{W(x)\sqrt{x-1}} = 2\int^{\delta^{1/2}}_{0} \frac{d z}{W(1+z^2)} \approx 2\delta^{1/2} + \mathcal{O}(\delta).
$$
We also define $\phi_{1}^{s}  - \phi_{1} = \phi_{1}^{s} \epsilon > 0$.  With this definition equation (\ref{phisurface}) becomes:
$$
\phi_{1}^{s} \epsilon = \frac{M_{pl}\left(V_{1} - V_{1}^{s}\right)}{\beta \rho_{1}} + \frac{M_{pl} V_{0} \delta}{M_{pl}} \approx \phi_{1}^{s} \epsilon + \frac{M_{pl}}{\beta \rho}\left(-\frac{1}{2}\phi_{1}^{s\,2} m_1^2 \epsilon^2 + V_{0}\delta + \mathcal{O}(\epsilon^3)\right),
$$
thus
$$
\phi_{1}^{s}\epsilon \approx \sqrt{\frac{2V_{0}\delta}{m_1^2}},
$$
and
$$
V_0 = V_1 - \frac{\beta \rho}{M_{pl}} \sqrt{\frac{2V_{1}\delta}{m_1^2}} + \mathcal{O(\delta)}.
$$

\subsubsection{Limit 2: $y_{1} = V_{1}^{s}/V_0 \gg 1$}
This limit occurs when either $d$ is suitably large or if $V_1 \gg V_2$. We take $1/y_1 = \delta \ll 1$.  We consider $V \propto \phi^{-n}$ potentials and so $W = x^{1 +1/n}$. In this limit we have:
$$
\int^{y_1}_{1} \frac{dx}{W(x)\sqrt{x-1}} = \int^{1}_{\delta} d z\,z^{\frac{1}{n} - \frac{1}{2}}(1-z)^{-\frac{1}{2}} \approx B\left(\frac{1}{n} + \frac{1}{2},\frac{1}{2}\right) - \left(\frac{2n}{n+2}\right) \delta^{\frac{1}{n} + \frac{1}{2}} + ...
$$
where $B(\,\cdot\,,\,\cdot\,)$ is the Beta function. To leading order in $\delta$ we have:
$$
\phi_{1}^{s} = \phi_{1} + \frac{M_{pl} V_{1}}{\beta \rho_{1}}.
$$
We are now in a position to evaluate $V_{0}(d)$ and hence $F_{\phi}/A$.  Without loss of generality we take $V_{1}\leq V_{2}$ and consider three limits:
\subsubsection{Large separations}
If
$$
m_{1}d, \, m_{2}d \gg \frac{2\sqrt{2}n}{n+2}\left(\frac{n}{n+1}\right)^{-\frac{n+1}{2}}B\left(\frac{1}{n} + \frac{1}{2},\frac{1}{2}\right).
$$
then we have $V_{2} \geq V_{1}\gg V_{0}$ and so:
$$
\sqrt{2\lambda}\vert n \vert M \left\vert \frac{M}{\phi_0} \right \vert^{\frac{n}{2}+1} d  = 2B\left(\frac{1}{n} + \frac{1}{2},\frac{1}{2}\right) + \mathcal{O}\left((V_0/V_1)^{\frac{1}{n}+\frac{1}{2}}, (V_0/V_2)^{\frac{1}{n}+\frac{1}{2}}\right).
$$
In this limit:
$$
\frac{F_{\phi}}{A} = V_{0} = \lambda^{\frac{2}{n+2}} M^4 \left(\frac{\sqrt{2}B\left(\frac{1}{n}+\frac{1}{2},\frac{1}{2}\right)}{\vert n \vert M d}\right)^{\frac{2n}{n+2}} = \lambda^{\frac{2}{n+2}}  K_{n} M^4 (Md)^{\frac{2n}{n+2}},
$$
where we have defined $$K_{n} = \left(\frac{\sqrt{2}B\left(\frac{1}{n}+\frac{1}{2},\frac{1}{2}\right)}{\vert n \vert}\right)^{\frac{2n}{n+2}}.$$

\subsubsection{Small separations: $m_{2} \approx m_{1}$}
If
$$
m_1 d > \left(\frac{n+1}{n}\right) \left(\left(\frac{m_2}{m_1}\right)^{n/(n+2)} - \left(\frac{m_1}{m_2}\right)^{n/(n+2)}\right), \quad m_2 d \ll 1
$$
then $V_{2} \approx V_{1} \approx V_{0}$ and:
$$
V_0 \approx V_{1}^{1/2}V_{2}^{1/2}\left[1-\left(\frac{n}{n+1}\right) \frac{V_{0}^{1/2}m_{0}d}{V_{1}^{1/2} + V_{2}^{1/2}}\right] \approx V_{1}^{1/2}V_{2}^{1/2}\left[1-\left(\frac{n}{n+1}\right) \frac{V_{1}^{1/2}m_{1}d}{V_{1}^{1/2} + V_{2}^{1/2}}\right].
$$
The force per area is therefore:
$$
\frac{F_{\phi}}{A} = V_{1}^{1/2}V_{2}^{1/2}\left[1-\left(\frac{n}{n+1}\right) \frac{V_{0}^{1/2}m_{0}d}{V_{1}^{1/2} + V_{2}^{1/2}}\right] \approx V_{1}^{1/2}V_{2}^{1/2}\left[1-\left(\frac{n}{n+1}\right) \frac{V_{1}^{1/2}m_{1}d}{V_{1}^{1/2} + V_{2}^{1/2}}\right].
$$
\subsubsection{Small separations: $m_{2} \gg m_{1}$}
When $m_{2} \gg m_{1}$ it is generally the case that $V_{2} > V_{1}^{s} \gg V_{0} > V_{1}$.  In this case we have:
$$
\frac{F_{\phi}}{A} = V_{0} = (n+1)V_{1}\left(1 - \left(\frac{m_1}{m_{2}} +  \left(\frac{n}{n+1}\right)^{\frac{n+2}{2}}\frac{(n+2)m_1d}{\sqrt{2n(n+1)}}\right)^{\frac{2}{n+2}}\right).
$$
This limit is valid provided that $V_{0} > 0$ and $V_{1}^{s} \gg V_{0}$, which requires:
$$
\frac{m_1}{m_{2}} +  \left(\frac{n}{n+1}\right)^{\frac{n+2}{2}}\frac{(n+2)m_1d}{\sqrt{2n(n+1)}} \ll 1.
$$

\noindent If $m_{b}d$ is \emph{not} $\ll 1$ then the $F_{\phi}/A$ given above are further suppressed by a factor of $\exp(-m_{b}d)$.
\subsection{Force between two distant bodies \label{fordist}}

We shall now consider the force between two bodies, with thin-shells, that are
separated by a `great distance', $d$.  By `great distance' we mean
$d \gg R_1, \, R_2$, where $R_1$ and $R_2$ are respectively the
length scales of body one and body two.  Given that $d \gg R_{1},
\, R_{2}$, then, to a good approximation, we can consider just the
monopole moment of the field emanating from the two bodies, and
model each body as a sphere with respective radii $R_1$ and $R_2$.

We expect that, outside some thin region close to the surface of
either body, the pseudo-linear approximation (with the field
taking its critical value i.e. $(m_c R) \rightarrow (m_c R)_{crit}
\approx (m_c R)_{eff}$) is appropriate to describe the field of
either body. In the region where pseudo-linear behaviour is seen,
we can safely super-impose the two one body solutions to find the
full two body solution.

Close to the surface of body one the mass of the chameleon induced by body one
will act to attenuate the perturbation to the chameleon field
created by body two.  This effect can be quite difficult to model
correctly.  We can predict the magnitude of the field, however, by
noting that the perturbation to $\phi$, induced by body two, near
to body one will be $$ \delta \phi_{2} = -\frac{\beta_{eff}^{(2)}
{\mathcal{M}}_2 e^{-m_b d}}{4 \pi M_{pl}d}. $$ From the results of
section \ref{singnon}, we know that
$\beta_{eff}^{(2)}{\mathcal{M}}_{2}/M_{pl}$ only depends on the
radius of body two and the theory dependent parameters, $M$,
$\lambda$ and $n$; $\beta_{eff}$ is as given in eqs.
 (\ref{critbeta}a-c); $m_b$ is the mass of the
chameleon field in the background i.e. far away from either of the
two bodies. ${\mathcal{M}}_{2}$ is the mass of body two.

We define the perturbation to $\phi$ induced by body one near to
body one in a similar manner $$ \delta \phi_{1} =
-\frac{\beta_{eff}^{(1)} {\mathcal{M}}_1 e^{-m_b d}}{4 \pi
M_{pl}d}. $$ From eqs. (\ref{critbeta}a-c) we know that $\delta
\phi_{1}$ is independent of $\beta$ and the mass of body one,
${\mathcal{M}}_{1}$. The force on body one due to body two will be
proportional to $\nabla \delta \phi_{2}$, however, since this must
also be the force on body two due to body one, it must also be
proportional to $\nabla \delta \phi_{1}$ evaluated near body two.
>From this we can see that the force on one body due to the other
must, up to a possible ${\mathcal{O}}(1)$ factor, be given by $$
F_{\phi} =
\frac{\beta_{eff}^{(1)}{\mathcal{M}}_{1}\beta_{eff}^{(2)}{\mathcal{M}}_{2}
(1+m_b d)e^{-m_b d}}{(4\pi M_{pl})^2 d^2}. $$ The functional
dependence of this force on $M$, $n$, $R_1$, $R_2$ and $\lambda$
depends on whether $n < -4$, $n=-4$ or $n >0$.  We consider these
three cases separately below.  In all cases the force is found to
be independent of $\beta$, ${\mathcal{M}}_1$ and
${\mathcal{M}}_2$.
\subsubsection{Case $n<-4$}
When $n<-4$, eq. (\ref{critbeta}a) gives $$
\frac{\beta_{eff}^{(1)}{\mathcal{M}}_{1}}{4\pi M_{pl}} =
\left(\frac{\gamma(n)}{\vert n\vert}\right)^{\frac{1}{\vert n +2\vert}}
\left(MR_{1}\right)^{\frac{n+4}{n+2}}. $$ The expression for
$\beta_{eff}^{(2)}$ is similar but with $1 \rightarrow 2$. The
force between two spherical bodies, with respective radii $R_1$
and $R_2$, separated by a distance $d \gg R_{1}, R_{2}$, is
therefore given by
\begin{eqnarray}
F_{\phi} =\left(\frac{\gamma(n)}{\vert n\vert}\right)^{\frac{2}{\vert n
+2\vert}}\frac{\left(MR_{1}\right)^{\frac{n+4}{n+2}}\left(MR_{2}\right)^{\frac{n+4}{n+2}}(1+m_bd)e^{-m_bd}}{d^2}. \label{FarForce1}
\end{eqnarray}
\subsubsection{Case $n>0$}
If $n > 0$ then $\beta_{eff}^{(1)}$  is given by eq.
(\ref{critbeta}b) to be $$
\frac{\beta_{eff}^{(1)}{\mathcal{M}}_{1}}{4\pi M_{pl}} =
MR_1\left(\frac{n(n+1)M^2}{m_b^2}\right)^{\frac{1}{n+2}}. $$ Again
the expression for $\beta_{eff}^{(2)}$ is similar. The force
between two distant bodies is therefore found to be
\begin{eqnarray}
F_{\phi} =
\left(\frac{n(n+1)M^2}{m_b^2}\right)^{\frac{2}{n+2}}\frac{M^2R_{1}R_{2}(1+m_b
d)e^{-m_b d}}{d^2}. \label{FarForce2}
\end{eqnarray}
\subsubsection{Case $n=-4$}
When $n=-4$, $\beta_{eff}^{(1,2)}$ are given by eq.
(\ref{critbeta}c) and they are actually weakly dependent on $r$.
Using eq. (\ref{critbeta}c) we find that
\begin{eqnarray}
 F_{\phi} = \frac{(1+m_bd)e^{-m_b d}}{8\lambda
\sqrt{\ln(d/R_1)\ln(d/R_2)}d^2}. \label{FarForce3}
\end{eqnarray}

\subsection{Force between a large body and a small body}

One subcase that is not included in the above results is the force
between a very large body with radius of curvature $R_{1}$, and a
very small body with radius of curvature $R_{2}$, that are
separated by an \emph{intermediate} distance $d$  i.e. $R_{2} \gg
d \gg R_{1}$. We assume that both bodies have thin-shells.  In
this case we find a behaviour that is half-way between the two
cases described above in sections \ref{forclose} and
\ref{fordist}. The magnitude of field produced by the large body
will be much greater than that of the small body.

If we ignore the small body and assume that the average mass of
the chameleon in the background obeys $m_{b} d \ll 1$, then the field produced by body one is
given by eq. (\ref{phiclosef}).  Using this equation, we find that
$$ \frac{d \phi}{dx} (x=d) \approx
\lambda^{\frac{1}{n+2}}\left(\frac{2}{(n+2)^2}\right)^{\frac{n}{2(n+2)}}
\left(Md\right)^{\frac{n+4}{n+2}}d^{-2}. $$ The effective coupling
of body two to this $\phi$-gradient will be $\beta_{eff}$ as it is
given by eqs. (\ref{critbeta}a-c). If $m_{b}d \gtrsim 1$ then this
gradient will be, up to an order $\mathcal{O}(1)$ coefficient,
attenuated by a factor of $(1+m_{b}d)\exp(-m_b d)$. The force
between the two bodies is therefore given by
\begin{subequations}
\begin{eqnarray}
F_{\phi} &=& \left(\frac{\gamma(n)}{\vert n \vert}\right)^{\frac{1}{\vert
n+2\vert}
}\left(MR_{2}\right)^{\frac{n+4}{n+2}}\left(\frac{2}{(n+2)^2}\right)^{\frac{n}{2(n+2)}}
\notag\\&&\left(Md\right)^{\frac{n+4}{n+2}}\frac{(1+m_bd)e^{-m_bd}}{d^2},
\qquad n<-4 \\ F_{\phi} &=&
MR_2\left(\frac{n(n+1)M^2}{m_b^2}\right)^{\frac{1}{n+2}}\left(\frac{2}{(n+2)^2}\right)^{\frac{n}{2(n+2)}}
\notag\\&&\left(Md\right)^{\frac{n+4}{n+2}}\frac{(1+m_bd)e^{-m_bd}}{d^2},
\qquad n>0 \\ F_{\phi} &=& \frac{1}{2\lambda
\sqrt{2\ln(d/R_2)}}\frac{(1+m_bd)e^{-m_bd}}{d^2}, \qquad n=-4.
\end{eqnarray}
\label{forceBigSmall}
\end{subequations}
As before, $d$ is the distance of separation. These formulae will
prove useful when we consider the $\phi$-force between the Earth
and a test-mass in laboratory tests for WEP violation in section
\ref{WEPsec}.

\subsection{Force between bodies without thin-shells}

If neither of the two bodies have thin-shells, $\phi$ behaves just
like a standard, linear, scalar field with mass $m_b$. The force
between the two bodies, with masses ${\mathcal{M}}_1$ and
${\mathcal{M}}_2$, is given by: $$ F_{\phi} = \frac{\beta^2
{\mathcal{M}}_1 {\mathcal{M}}_2 (1+m_b d)e^{-m_bd}}{4\pi M_{pl}^2
d^2}. $$ As above, $m_b$ is the mass of the chameleon in the
background. If one of the bodies has a thin-shell, body one say,
but the other body does not, then the force is given by $$ F_{\phi}
= \frac{\beta_{eff}^{(1)}\beta {\mathcal{M}}_1
{\mathcal{M}}_2(1+m_bd)e^{-m_bd}}{4\pi d^2}, $$ whenever $d \gg
R_1$ where $R_1$ is the radius of curvature of body one and
$\beta_{eff}^{(1)}$ is as given by eqs. (\ref{critbeta}). If $d
\ll R_{1}$ then $$ F_{\phi} =
\lambda^{\frac{1}{n+2}}\left(\frac{2}{(n+2)^2}\right)^{\frac{n}{2(n+2)}}\left(Md\right)^{\frac{n+4}{n+2}}
\frac{\beta {\mathcal{M}}_2 (1+m_b d)e^{-m_bd}}{M_{pl} d^2}. $$ As
above, $d$ is the distance of separation.

\subsection{Summary}
In this section we have considered the force that the chameleon
field, $\phi$, induces between two bodies, with masses
${\mathcal{M}}_1$ and ${\mathcal{M}}_2$ and radii $R_1$ and $R_2$,
separated by a distance $d$.  The chameleon mass in the far
background is taken to be $m_{b}$.  When both bodies have
thin-shells, we found that, to leading order, the $\phi$-force
between them is independent of the matter coupling, $\beta$,
provided that $m_1 d, m_2 d \gg 1$; $m_1$ is the mass that the
chameleon has inside body one, and $m_2$ is similarly defined with
respect to body two.  The force between two such bodies is also
independent ${\mathcal{M}}_1$ and ${\mathcal{M}}_2$ but does in
general depend on $R_1$ and $R_2$, as well as on $M$, $n$ and $\lambda$.
The main results of this section are summarised below.
\subsubsection*{Neither body has a thin-shell}
$$ F_{\phi} = \frac{\beta^2 {\mathcal{M}}_1 {\mathcal{M}}_2 (1+m_b
d)e^{-m_bd}}{4\pi M_{pl}^2 d^2}. $$
\subsubsection*{Body one has a thin-shell, body two does not}
If the two bodies are close together ($m_1^{-1} \ll d \ll R_{1}$)
then $$F_{\phi} =
\lambda^{\frac{1}{n+2}}\left(\frac{2}{(n+2)^2}\right)^{\frac{n}{2(n+2)}}\left(Md\right)^{\frac{n+4}{n+2}}
\frac{\beta {\mathcal{M}}_2 (1+m_b d)e^{-m_bd}}{M_{pl} d^2}.$$ If
the bodies are far apart $d \gg R_{1}$ then
\begin{eqnarray*}
F_{\phi} &=& \left(\frac{\gamma(n)}{\vert n \vert}\right)^{\frac{1}{\vert
n+2\vert}}\left(MR_1\right)^{\frac{n+4}{n+2}}\frac{\beta
{\mathcal{M}}_2(1+m_bd)e^{-m_bd}}{M_{pl}d^2}, \qquad n < -4, \\
F_{\phi} &=&
MR_1\left(\frac{n(n+1)M^2}{m_b^2}\right)^{\frac{1}{n+2}}\frac{\beta
{\mathcal{M}}_2(1+m_bd)e^{-m_bd}}{M_{pl}d^2}, \qquad n > 0, \\
F_{\phi} &=& (2\lambda\min(d/R,1/m_b R))^{-1/2}\frac{\beta
{\mathcal{M}}_2(1+m_bd)e^{-m_bd}}{2M_{pl}d^2}, \qquad n = -4.
\end{eqnarray*}
This force is independent of the mass of body one, and the
chameleon's coupling to it.
\subsubsection*{Both bodies have thin-shells}
If the two bodies of close by and $m_1^{-1}, m_2^{-1} \ll d \ll R_1,
R_2$ then the force per unit area is
\begin{eqnarray*}
\frac{F_{\phi}}{A} \approx \lambda^{\frac{2}{2+n}} M^4
\left[\frac{\sqrt{2}B\left(\frac{1}{2},\frac{1}{2}+\frac{1}{n}\right)}{\vert
n \vert Md}\right]^{\frac{2n}{n+2}}.
\end{eqnarray*}
Different formulae for $F_{\phi}/A$ apply if either $m_{1}d < 1$ or $m_2 d < 1$, these are given at the end of section \ref{forclose} above.
\begin{eqnarray*}
F_{\phi} &=&\left(\frac{\gamma(n)}{\vert n\vert}\right)^{\frac{2}{\vert n
+2\vert}}\frac{\left(MR_{1}\right)^{\frac{n+4}{n+2}}\left(MR_{2}\right)^{\frac{n+4}{n+2}}(1+m_bd)e^{-m_bd}}{d^2},
\quad n < -4 \\ F_{\phi} &=&
\left(\frac{n(n+1)M^2}{m_b^2}\right)^{\frac{2}{n+2}}\frac{M^2R_{1}R_{2}(1+m_b
d)e^{-m_b d}}{d^2}, \quad n >0 \\
 F_{\phi} &=& \frac{(1+m_bd)e^{-m_b d}}{8\lambda
\sqrt{\ln(d/R_1)\ln(d/R_2)}d^2}, \quad n = -4,
\end{eqnarray*}
where $\gamma(n)$ is given in table \ref{TABgamma}. If body one is much larger than body two and they are at an
intermediate separation: $R_2, m_1^{-1} \gg d \gg R_1$ then
\begin{eqnarray*}
F_{\phi} &=& \left(\frac{\gamma(n)}{\vert n \vert}\right)^{\frac{1}{\vert
n+2\vert}
}\left(MR_{2}\right)^{\frac{n+4}{n+2}}\left(\frac{2}{(n+2)^2}\right)^{\frac{n}{2(n+2)}}
\notag\\&&\left(Md\right)^{\frac{n+4}{n+2}}\frac{(1+m_bd)e^{-m_bd}}{d^2},
\qquad n<-4 \\ F_{\phi} &=&
MR_2\left(\frac{n(n+1)M^2}{m_b^2}\right)^{\frac{1}{n+2}}\left(\frac{2}{(n+2)^2}\right)^{\frac{n}{2(n+2)}}
\notag\\&&\left(Md\right)^{\frac{n+4}{n+2}}\frac{(1+m_bd)e^{-m_bd}}{d^2},
\qquad n>0 \\ F_{\phi} &=& \frac{1}{2\lambda
\sqrt{2\ln(d/R_2)}}\frac{(1+m_bd)e^{-m_bd}}{d^2}, \qquad n=-4.
\end{eqnarray*}

\section{Laboratory Constraints\label{exper}}

The best bounds on corrections to General Relativity  come from laboratory experiments such as
the E\"{o}t-Wash experiment, \cite{EotWash} and Lunar Laser
Ranging tests for WEP violations, \cite{LLR,LLR1}. At very small distances $d \lesssim 10\mu m$, the best bounds on the strength of any fifth force come from measurements of the Casimir force.

In this section we will
consider, to what extent, the results of these tests constrain the class of
chameleon field theories considered here. We will find the rather startling result that $\beta \gg 1$ is \emph{not} ruled out for chameleon theories.

One of original reasons for studying chameleon theories with $n>0$
potentials, \cite{chama}, was that the condition for an object to
have a thin-shell, eq. (\ref{thincond}b), was found to depend on
the background density of matter.  It is clear from eq.
(\ref{thincond}b), that the smaller $\rho_{b}$ is, the larger $m_c
R$ must be for a body to have a thin-shell.  This property lead
the authors of ref. \cite{chama} to conclude that the thin-shell
suppression of the fifth-force associated with $\phi$ would be
weaker for tests performed in the low density vacuum of space
$\rho_b \sim 10^{-25}\,{\mathrm{g\,cm}}^{-3}$, than it is in the
relatively high density laboratory vacuum $\rho_b \sim
10^{-17}\,{\mathrm{g\,cm}}^{-3}$.  As a result, it is possible that,
if the same experimental searches for WEP violation, which were
performed in the laboratory in \cite{WEP,WEP1,WEP2,willbook,willbook1}, were to be repeated in
space, they would find equivalence principle violation at a level
\emph{greater} than that already ruled out by the laboratory-based tests.
It is important to note that this is very much a property of $n>0$
theories. It is clear from eqs. (\ref{thincond}a) and
(\ref{thincond}c) that, when $n \leq -4$, the thin-shell condition,
for a body of density $\rho_c$, is only very weakly dependent on
the background density of matter when $\rho_b \ll \rho_c$.  As a
result, space-based searches for WEP violation will \emph{not}
detect any violation at a level that is already ruled by lab-based
tests for $n \leq -4$ theories.  Planned space-based tests such as
STEP \cite{STEP}, SEE \cite{SEE}, GG \cite{GG} and MICROSCOPE
\cite{MICRO} promise a much greater precisions than their lab-based
counterparts. MICROSCOPE is  due to be launched in 2007.  This improved
precision will, in all cases, provide us with better bounds on
chameleon theories.

\subsection{E\"{o}t-Wash experiment}
The University of Washington's E\"{o}t-Wash experiment,
\cite{EotWash,Eotnew}, is designed to search for deviations from the
$1/r^2$ drop-off gravity predicted by General Relativity. The
experiment uses a rotating torsion balance to measure the torque
on a pendulum.   The torque on the pendulum is induced by an
attractor which rotates with a frequency $\omega$.  The attractor
has 42 equally spaced holes, or `missing masses', bored into it.
As a result, any torque on the pendulum, which is produced by the
attractor, will have a characteristic frequency which is some integer multiple of $21 \omega$. This characteristic frequency allows any torque due to background
forces to be identified in a straightforward manner. The torsion
balance is configured so as to factor out any background forces.
The attractor is manufactured so that, if gravity drops off as
$1/r^2$, the torque on the pendulum vanishes.

The experiment has been run with different separations between the
pendulum and attractor. The Eot-Wash group recently announced some
new results which go a long way towards better constraining the
parameter space of chameleon theory, \cite{Eotnew}.  The experiment
has been run for separations, $55 \mu m \leq d \leq 9.53 mm$.   Both
the attractor and the pendulum are made out of molybdenum with a
density of about $\rho_{Mb} \sim 10\,{\mathrm{g\,cm}}^{-3}$
and are $0.997\,{\mathrm{mm}}$ thick. Electrostatic forces are
shielded by placing a $10\, \mu{\mathrm{m}}$ thick, uniform BeCu
sheet between the attractor and pendulum.  The density of this sheet
is $\rho_{BeCu} \sim 8.4\,{\mathrm{g\,cm}}^{-3}$. The
r\^{o}le played by this sheet is crucial when testing for chameleon
fields.  If $\beta$ is large enough, the sheet will itself develop a
thin-shell.  When this occurs the effect of the sheet is not only to
shield electrostatic forces, but also to block any chameleon force
originating from the attractor.

The force per unit area between the attractor and pendulum plates due to a scalar field with matter coupling $\beta$ and constant mass $m$, where $1/m \ll 0.997mm$ is:
\begin{eqnarray}
\vert \frac{F}{A} \vert = \frac{\alpha e^{-md} 2\pi G
\rho_{Mb}^2}{m^2} \nonumber \end{eqnarray} where $\alpha =
\beta^2/4\pi$ and $d$ is the separation of the two plates.  The
strongest bound on $\alpha$ coming from the Eot-Wash experiment is
$\alpha < 2.5\times 10^{-3}$ for $1/m = 0.4\,--\,0.8 mm$.

Depending on the values of $\beta$, $M$ and $\lambda$ there are three
possible situations:
\begin{itemize}
\item The pendulum and the attractor have thin-shells, but the BeCu sheet does not
\item The pendulum, the attractor \emph{and} the BeCu sheet all have thin-shells.
\item Neither the test masses nor the BeCu sheet have thin-shells.
\end{itemize}

In the first case the $\phi$-mediated force per unit area in a
perfect vacuum is given by one of the equations derived in section
\ref{forclose} (depending on the separation $d$).  In reality the
vacuum used in these experiments is not perfect actually has a
pressure of $10^{-6}\,\mathrm{Torr}$ which means that the chameleon
mass in the background, $m_{b}$, is non-zero and so $F_{\phi}/A$ is
suppressed by a factor of $\exp(-m_{b} d)$. Fortunately however
$m_{b} d \ll 1$ for all but the largest $\beta$. A further, and far
more important, suppression occurs when the BeCu sheet has a
thin-shell. If $m_{BeCu}$ is the chameleon mass inside the BeCu
sheet and $d_{BeCu}$ the sheet's thickness, the existence of a
thin-shell in the electromagnetic shield causes the chameleon
mediated force between the pendulum and attractor to be suppressed
by a further factor of $\exp(-m_{BeCu}d_{BeCu})$.  The thin-shell
condition for the BeCu sheet implies that $m_{BeCu}d_{BeCu} \gg 1$,
and so this suppression all but removes any detectable chameleon
induced torque on the pendulum due to the attractor. The BeCu sheet
will itself produce a force on pendulum but, since the sheet is
uniform, this force will result in no detectable torque.  If $M$ and
$\lambda$ take natural values the electromagnetic sheet develops a
thin-shell for $\beta \gtrsim 10^{4}$: as a result of this the
Eot-Wash experiment can only very weakly constrain large $\beta$
theories.

If neither the pendulum or the attractor have thin-shells then we
must have $m_{b}d \ll 1$ and the chameleon force is just
$\beta^2/4\pi$ times the gravitational one.  Since this force drops
of as $1/r^2$ it will be undetectable from the point of view this
experiment. In this case, however, $\beta$ is constrained by other
experiments such as those that search for  WEP violation
\cite{LLR,LLR1,WEP,WEP1,WEP2} or those that look for Yukawa forces
with larger ranges \cite{Irvine}.

\begin{figure}[tbh]
\begin{center}
\includegraphics[width=7.4cm]{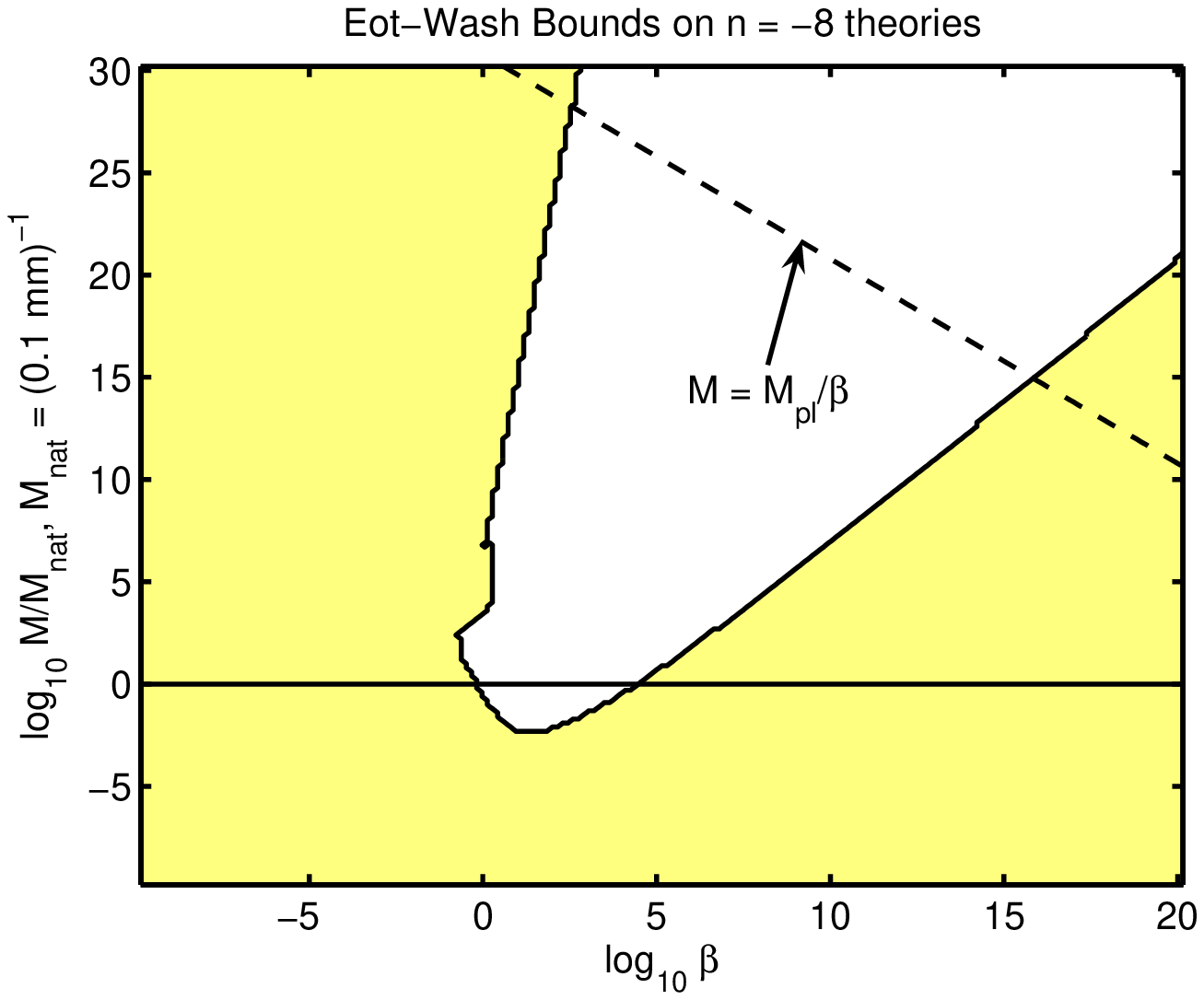}
\includegraphics[width=7.4cm]{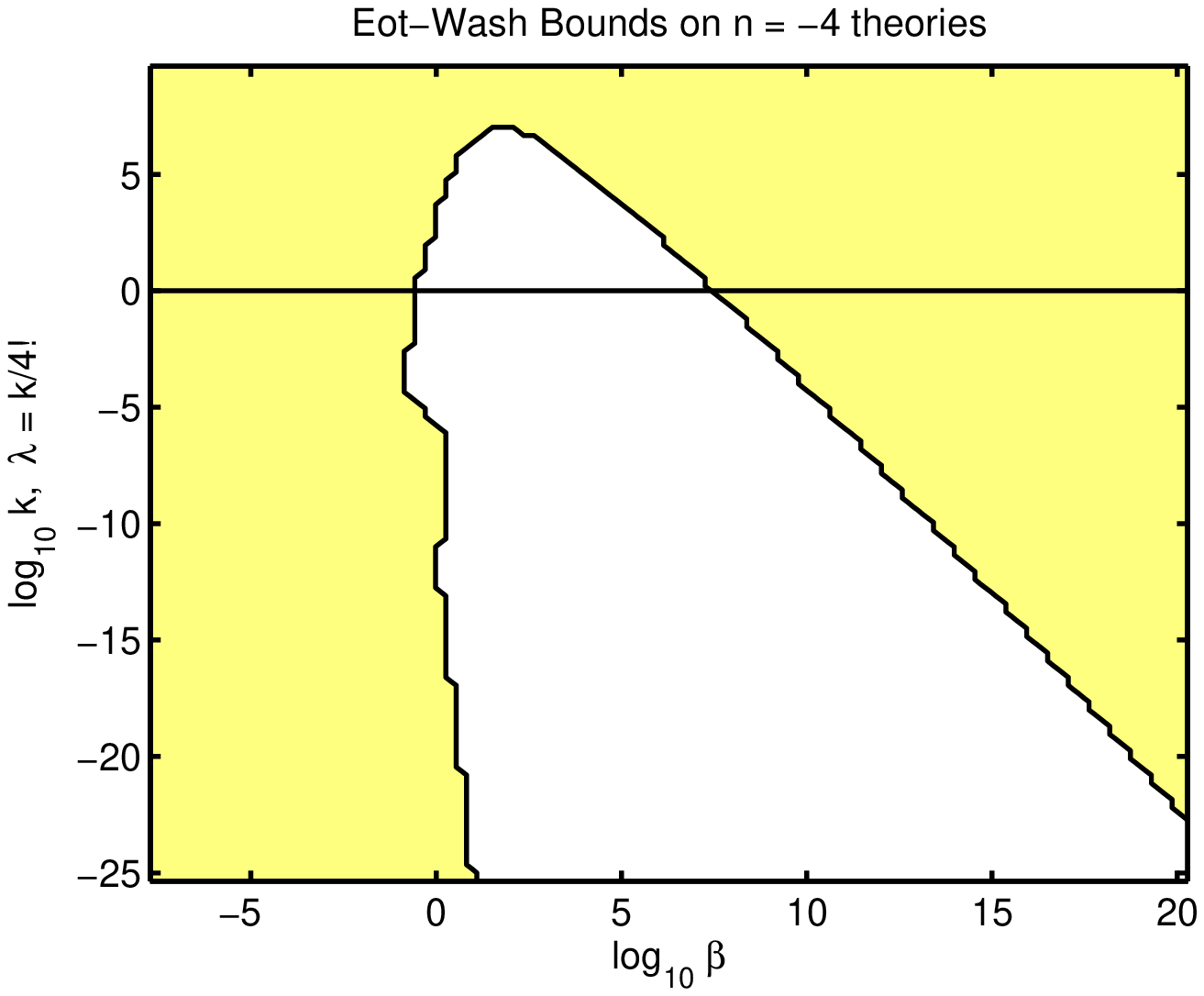}
\includegraphics[width=7.4cm]{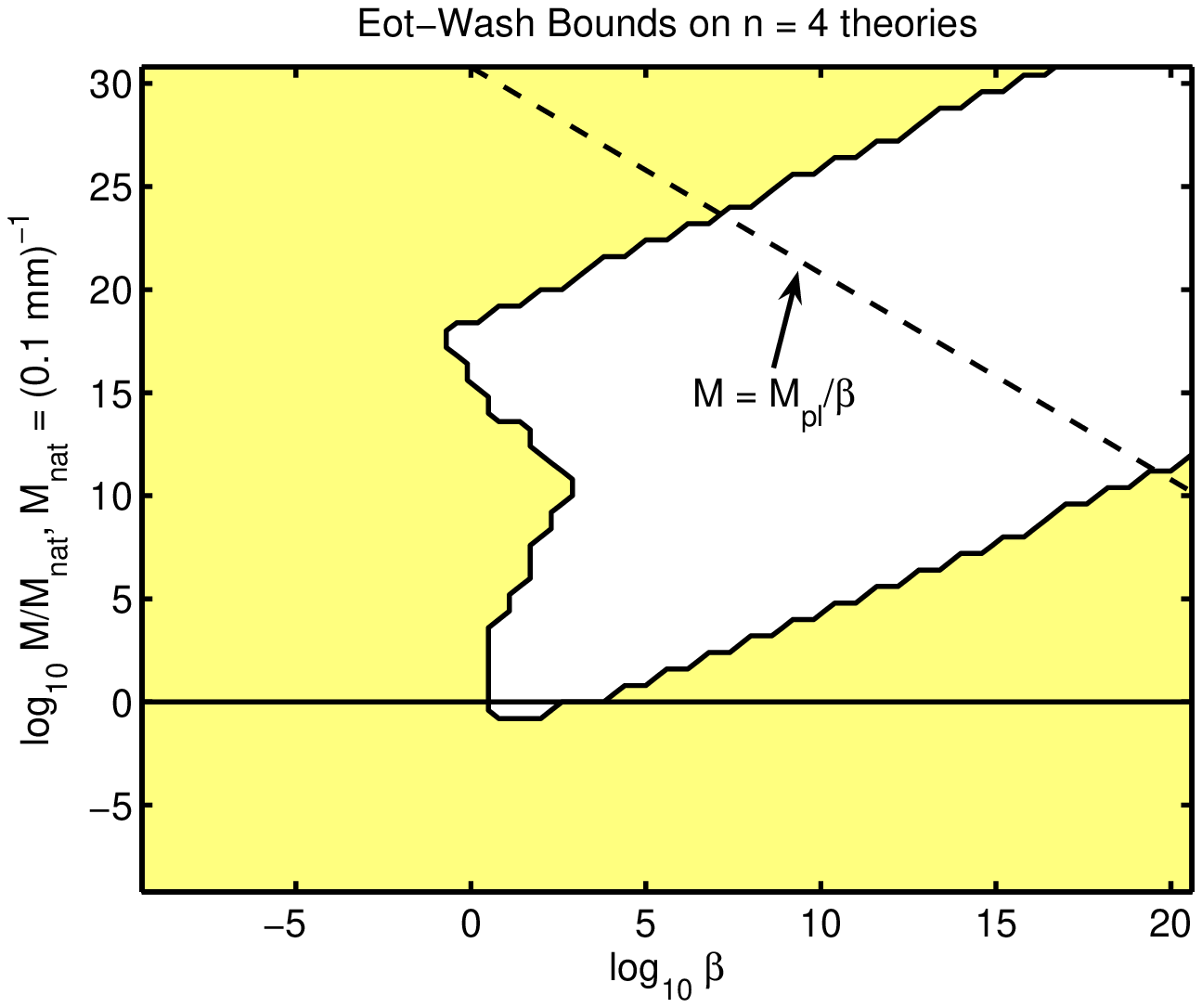}
\end{center}
\caption[E\"{o}t-Wash constraints on chameleon theories.]{[Colour Online] Constraints
on chameleon theories coming from the E\"{o}t-Wash bounds on
deviations from Newton's law. The shaded area shows the
regions of parameter space that are allowed by the current data.  The solid black
lines indicate the cases where $M$ and $\lambda$ take `natural
values'.  The dotted-black line indicates when $M = M_{\phi} :=
M_{pl}/\beta$ i.e. when the mass scale of the potential is the
same as that of the matter coupling.  Other $n < -4$ theories are
similar to the $n = - 8$ case, whilst the $n=4$ plot is typical of
what is allowed for $n>0$ theories.  The amount of \emph{allowed}
parameter space increases with $\vert n \vert$.} \label{eotbounds}
\end{figure}

Given all of the considerations mentioned above, we used the formulae given in section \ref{forclose} to evaluate the latest E\"{o}t-Wash constraints on the parameter space of chameleon theories. Our results are shown in  FIG \ref{eotbounds}.
In these plots the shaded region is \emph{allowed} by the
current bounds.

When $\beta$ is small, the chameleon mechanism present in
these theories becomes very weak, and from the point of view of
the E\"{o}t-Wash experiment, $\phi$ behaves like a normal
(non-chameleon) scalar field.  When $\beta \gg 1$, the $\phi$-force is independent of the coupling of the chameleon to attractor or the pendulum, but
does depend on the mass of the chameleon in the BeCu sheet,
$m_{BeCu}$.  The larger $m_{BeCu}$ is, the weaker the E\"{o}t-Wash
constraint becomes. Larger $\beta$ implies a larger $m_{BeCu}$,
and this is why the allowed region of parameter space increases as
$\beta$ grows to be very large.

When $n=-4$, we can see that a natural value of $\lambda$ is ruled
out for $10^{-1} \lesssim \beta \lesssim 10^{4}$, but is
permissible for $\beta \gtrsim 10^{4}$.  This is entirely due to
the that BeCu sheet has a thin-shell, in $n=-4$ theories with
$\lambda = 1/4!$, whenever $\beta \gtrsim 10^{4}$.

It is important to stress that, despite the fact that the
E\"{o}t-Wash experiment is currently unable to detect $\beta \gg 1$,
this is not due to a lack of precision. One pleasant feature, of the
$\beta$-independence of the $\phi$-force, is that if you can detect, or rule out, such a force for  one value of
$\beta \gg 1$, then you will be able to detect it, or rule out, \emph{all} such $\beta \gg 1$ theories.    If design of the experiment can be altered so that electrostatic forces are compensated without using a thin-sheet then the experimental precision \emph{already} exists to  detect, or rule out, almost all $\beta \gg 1$, $\phi^{4}$ theories with
$\lambda \approx 1/4!$.

In conclusion, an experiment, along the same lines of the E\"{o}t-Wash test, could detect, or rule out, the existence of sub-Planckian, chameleon fields with natural values of $M$ and $\lambda$ in the near future, provided it is designed to do so.

\subsection{Casimir force experiments}
Short distance tests of gravity fail to constrain strongly coupled
chameleon theories as a result of their use of a thin metallic sheet
to shield electrostatic forces. However, experiments designed to
detect the Casimir force between two objects, control electrostatic
effects by inducing an electrostatic potential difference between
the two test bodies.  By varying this potential difference and
measuring the force between the test masses, it is possible to
factor out electrostatic effects.  As a result, Casimir force
experiments provide an excellent way in which to bound chameleon
fields where the scalar field is strongly coupled to matter.

Casimir force experiments measure the force per unit area between
two test masses separated by a distance $d$. It is generally the
case that $d$ is small compared to the curvature of the surface of
the two bodies and so the test masses can be modeled, to a good
approximation, as flat plates.

In section \ref{forclose} we evaluated the force per unit area between two flat, thin-shelled slabs with densities $\rho_1$ and $\rho_{2}$.  The Casimir force is between two such plates is:
$$
\frac{F_{Cas}}{A} = \frac{\pi^2}{240 d^4}.
$$
Whilst a number of experimental measurement of the Casimir force
between two plates have been made, the most accurate measurements of
the Casimir force have been made using one sphere and one slab as
the test bodies. The sphere is manufactured so that its radius of
curvature, $R$, is much larger than the minimal distance of
separation $d$.  In this case the total Casimir force between the
test masses is:
$$
F_{Cas} = 2\pi R \left(\frac{1}{3}\frac{\pi^2}{240}\frac{1}{d^3}\right) = 3.35\left(\frac{R}{d^3} \frac{(\mu m)^3}{cm} \right) \mu \mathrm{dyn}.
$$
In all cases, apart from when $n=-4$ and $m_1 d, m_2 d \gg 1$, the
chameleon force per area grows more slowly than $d^{-4}$ as $d
\rightarrow 0$.  When $n=-4$ and $m_1 d, m_2 d \gg 1$ we have
$F_{\phi}/A \propto d^{-4}$.  It follows that the \emph{larger} the
separation, $d$, used, the better Casimir force searches constrain
chameleon theories.  Additionally, these tests provide the best
bounds when the test masses \emph{do} have thin-shells as this
results in a strongly $d$ dependent chameleon force. Large test
masses are therefore preferable to small ones.

Note that if the background chameleon mass is large enough that
$m_{b}d \gtrsim 1$ then $F_{\phi}$ is suppressed by a factor of
$\exp(-m_b d)$.  The smaller the background density,
$\rho_{b}$, is, the smaller $m_{b}$ become.  Since small
$m_{b}$ is clearly preferably, the best bounds come from experiments
that use the lowest pressure laboratory vacuum.
\begin{figure}[tbh]
\begin{center}
\includegraphics[width=7.4cm]{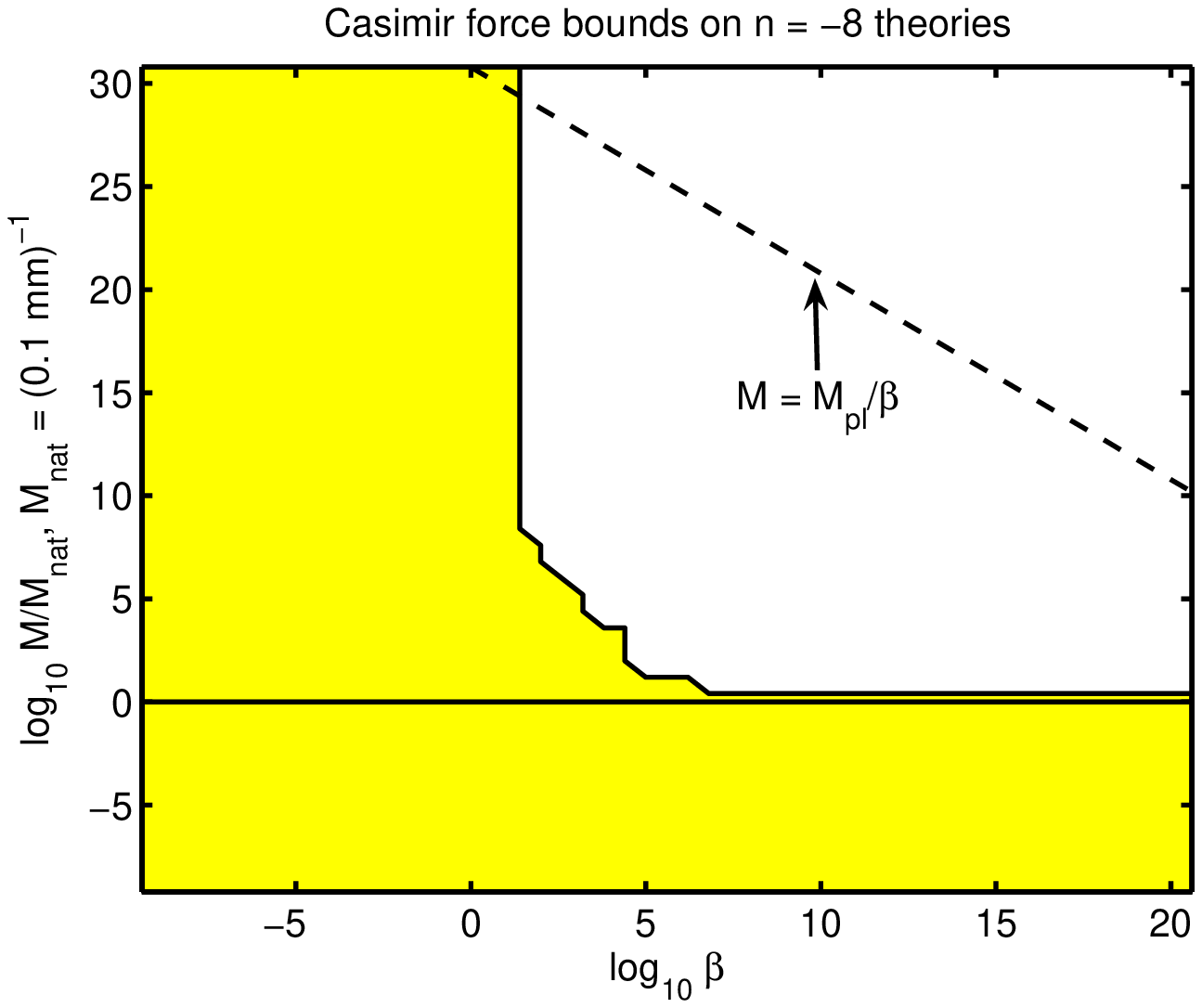}
\includegraphics[width=7.4cm]{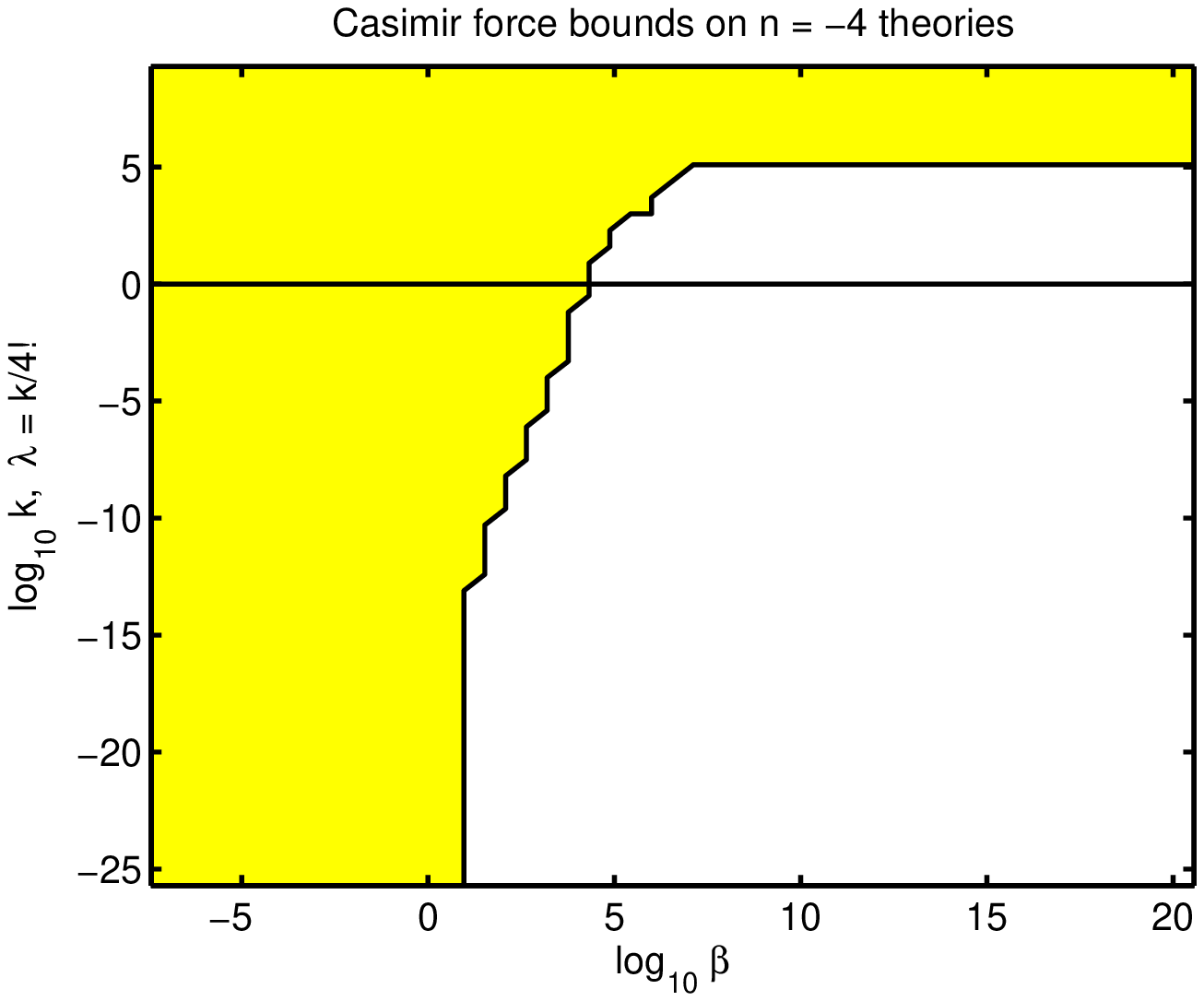}
\includegraphics[width=7.4cm]{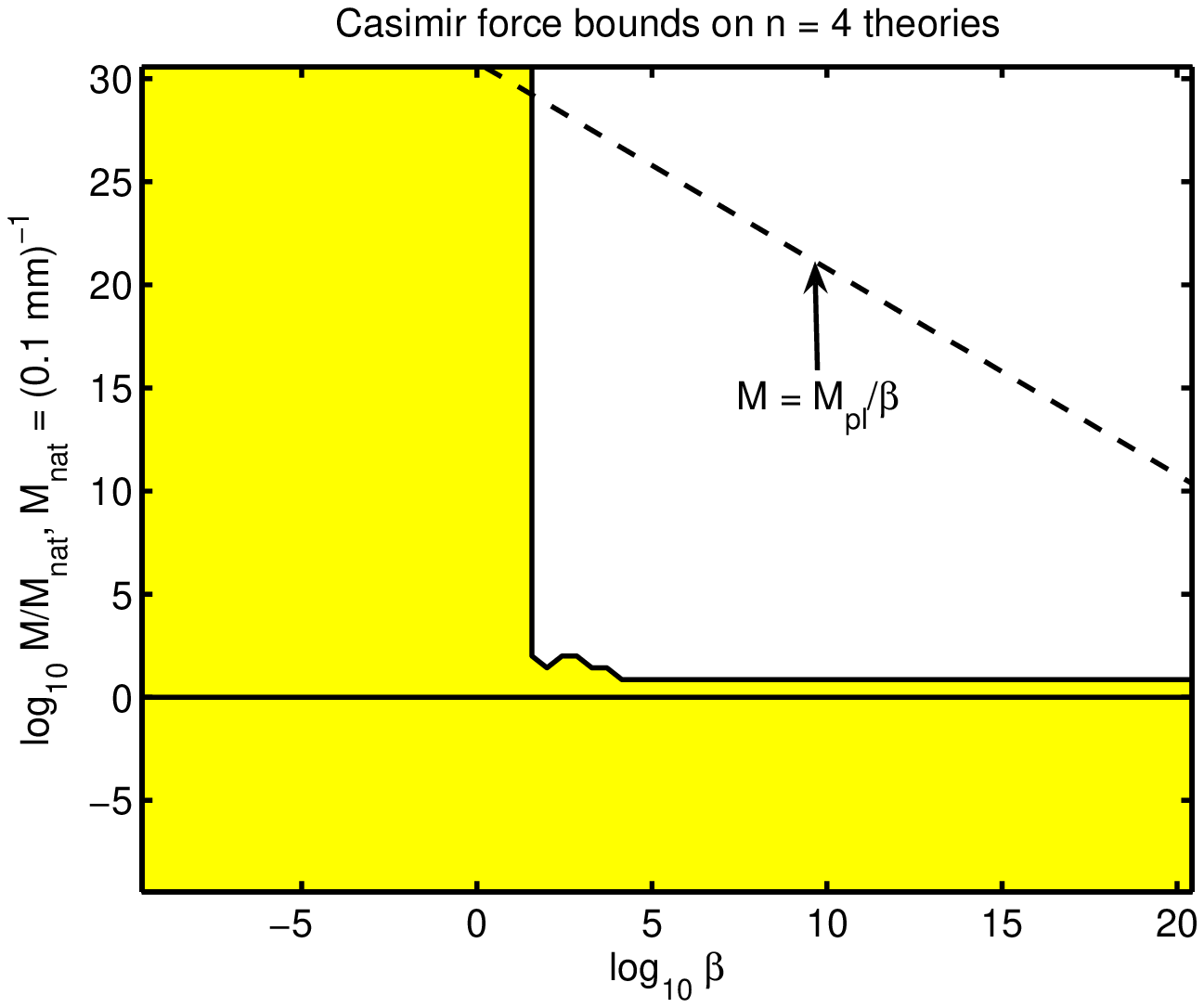}
\end{center}
\caption[Casimir force experiment constraints on chameleon theories.]{[Colour Online] Constraints
on chameleon theories coming from experimental searches for the Casimir force. The shaded area shows the
regions of parameter space that are allowed by the current data.  The solid black
lines indicate the cases where $M$ and $\lambda$ take `natural
values'.  The dotted-black line indicates when $M = M_{\phi} :=
M_{pl}/\beta$ i.e. when the mass scale of the potential is the
same as that of the matter coupling. Other $n < -4$ theories are
similar to the $n = - 8$ case, whilst the $n=4$ plot is typical of
what is allowed for $n>0$ theories.  The amount of \emph{allowed}
parameter space increases with $\vert n \vert$ i.e. as the potential becomes steeper.} \label{Casfigure}
\end{figure}

In \cite{Lamcas}, Lamoreaux reported the measurement of the Casimir
force using a torsion balance between a sphere, with radius of
curvature $12.5\mathrm{cm} \pm 0.3\mathrm{cm}$ and diameter of
$4\mathrm{cm}$, and a flat plate. The plate was $0.5\mathrm{cm}$
thick and $2.54\mathrm{cm}$ in diameter. The apparatus of placed in
a vacuum with a pressure of $10^{-4}\mathrm{mbar}$.  Distances of
separation of $6 - 60 \mu\mathrm{m}$ we used and in the region $d
\approx 7  - 10 \mu\mathrm{m}$ it was found that:
$$
\vert F^{measured}-F_{Cas}^{theory}\vert \lesssim 1 \mu\,\mathrm{dyn}.
$$
Another measurement, this time using a microelectromechanical
torsional oscillator, was performed by Decca \emph{et al.} and
reported in \cite{Decca}.  In this experiment, the sphere was much
smaller than that used by Lamoreaux, being on $296 \pm 2 \mu
\mathrm{m}$ in radius;  the plate was made of $3.5 \mu \mathrm{m}$
thick, $500 \times 500 \mu \mathrm{m}^2$ polysilicon.   The
smallness of these test masses means that they will only have
thin-shells when $\beta$ is very large.   In the region $d \approx
400 n\mathrm{m} - 1200 n\mathrm{m}$, $\vert
F^{measured}-F_{Cas}^{theory} \vert \lesssim 7.5 \times 10^{-2} \mu
\mathrm{dyn}$.  We show how these experiments constrain the
parameter space of $\phi^{-n}$ chameleon theories in FIG
{\ref{Casfigure}}. Other Casimir force tests (e.g. \cite{othercas} )
are less suited to constraining chameleon theories such as those
considered here.  As in the previous plots, the shaded area is
allowed, the solid black line is $M \sim
(\rho_{\Lambda})^{1/4}$ or $\lambda = 1/4!$, and the dotted
black line is $M = M_{pl}/\beta$.   We note that Casimir force
experiments provide very tight bounds on $\lambda$ and $M$ when
$\beta \gg 1$.  A natural value of $\lambda$  when $n=-4$ is ruled
out for all $\beta \geq 10^{4}$.  When this is combined with the
latest E\"{o}t-Wash data we can rule out $\lambda = 1/4!$ for all
$\beta > 1$.  For other $n$, we see that we cannot have $M$ much
larger than its `natural' value $(\rho_{\Lambda})^{1/4}$ for
large $\beta$.  If the bounds on extra forces at $d \sim 1- 10\mu m$
can be tighten by roughly an order of magnitude then a natural value
of $M$ can be ruled out for all large $\beta$. Casimir force tests
provided by far the best bounds on $M$ and $\lambda$ for large
$\beta$. We note that $n = -4$ theories are the most tightly
constrained by Casimir force experiments, this is not surprising
since $F_{\phi}/F_{Cas} \sim const$ and is $> 1$ in the region when
$m_1d, m_2 d \gg 1$ in this model, whereas in all the other theories
$F_{\phi}/F_{Cas}$ decreases as $d$ is made smaller.

More generally, the steeper the potential in a given theory is, the more slowly $F_{\phi}/A$ increases as $d \rightarrow 0$ and, as a result, the weaker Casimir force bounds on the theories parameter space are.

\subsection{WEP violation experiments \label{WEPsec}}
The weak equivalence principle (WEP) is the statement that the
(effective) gravitational and inertial mass of a body are equal. If it is
violated then the either the strength of gravitational force on a
body depends on its composition, or there is a composition
dependent `fifth-force'.  Since we believe that gravity is
geometric in nature, most commentators, ourselves included, would
tend to interpret any detection of a violation of WEP in terms of
the latter option. The existence of light scalar fields that
couple to matter usually results in WEP violations.   As we mentioned in the introduction to this thesis, the
experimental bounds on WEP violation are exceeding strong,
\cite{LLR,LLR1,willbook,willbook1,fischbach,WEP,WEP1,WEP2}, and, at present, they represent the strongest
bounds on the parameters of, non-chameleon, scalar-tensor theories
such as Brans-Dicke theory, \cite{uzan, BD}.   A number of planned
satellite missions promise to increase the precision to which we
can detect violations of WEP by between 2 to 5 orders of magnitude
\cite{MICRO,SEE,STEP,GG}.   The precision that is achievable in
laboratory based tests also continues to increase at a steady
rate.

Experiments that search for violations of the weak equivalence
principle generally fall into two categories: laboratory based experiments, which often employ a modified torsion balance, \cite{WEP,WEP1,WEP2}, and solar system tests such as lunar laser
ranging (LLR) \cite{LLR,LLR1}.

The laboratory based searches use a modified
version of the E\"{o}t-Wash experiment mentioned above.  In these experiments the test masses are manufactured to have different compositions.  The aim is then to detect, and measure, any difference in the acceleration of test-masses
towards an attractor, which is usually the Earth, the Sun or the Moon.  In some versions of the
experiment a laboratory body is used as the attractor.

If the test masses have thin-shells then the $\phi$-force pulling
them towards the Earth is given by equations
(\ref{forceBigSmall}a-c).  The chameleon force towards the Moon or
Sun is given by eqs. (\ref{FarForce1})-(\ref{FarForce3}). If the
attractor is a laboratory body then, depending on the separations
used, the force is given by either eqs.
(\ref{FarForce1})-(\ref{FarForce3}) or by eq. (\ref{forclose}).

We label the
mass and radius of the attractor by ${\mathcal{M}}_{3}$ and $R_{3}$
respectively, and take the mass and radius (or size) of the two
test-masses to be given by $\{{\mathcal{M}}_{1},\,R_{1}\}$ and $\{{\mathcal{M}}_{2},\, R_{2}\}$.  We
define $\alpha_{13}$ to be the relative strength, compared to gravity, of $\phi$-force
between the attractor (body three) and the first of the test masses
(body one).  $\alpha_{23}$ is defined similarly as a measure of the $\phi$-force between body two and body three.  The difference between
the acceleration of the two test masses towards the attractor is quantified by the E\"{o}tvos
parameter, $\eta$, where
$$
\eta = \frac{2\vert \alpha_{13} - \alpha_{23} \vert}{\vert 2 + \alpha_{13}+\alpha_{23}\vert} \approx \vert\alpha_{13}-\alpha_{23}\vert.
$$
When the test-bodies have thin-shells, we found, in section \ref{forcetwo}, that the $\phi$-force is independent of the masses of the
test-bodies, the mass of the attractor and the coupling of the test-masses and attractor to
the chameleon. The only property of the attractor and test bodies, which the $\phi$-force does depend on, is their respective radii.  Since the gravitational force between the test-masses and the attractor \emph{does} depend on the masses of the bodies, it follows that $\alpha_{13}$ only depends on ${\mathcal{M}}_{1}$, ${\mathcal{M}}_{3}$, $R_1$, $R_3$, $M$ (or $\lambda$), $m_b$ and $n$, where $m_b$ is the chameleon mass in the background.  It does \emph{not} depend on the chameleon's coupling to the test-mass, $\beta$.   The situation with $\alpha_{23}$ is very similar.

Since the $\phi$-force is independent of the coupling, $\beta$, any microscopic composition dependence in $\beta$ will be
hidden on macroscopic length scales.  The only `composition'
dependence in $\alpha_{13}$ is through the masses of the bodies and
their dimensions ($R_1$ and $R_3$).

Taking the third body to be the Earth, the Sun or the Moon,
experimental searches for WEP violations have, to date, found that
$\eta \lesssim 10^{-13}$ \cite{WEP,WEP1,WEP2}.  Future satellite tests
promise to be able to detect violations of WEP at between the
$10^{-15}$, \cite{MICRO}, and the $10^{-18}$ level, \cite{STEP}.
It also is claimed that future laboratory tests will be able to see
$\eta \sim 10^{-15}$, however, whilst the precision to detect at
such a level is achievable, there are a number of systematic effects
that need to be compensated for, before an accurate measurement
can be made.

In most of these
searches, although the composition of the test-masses is different,
they are manufactured to have the \emph{same} mass (${\mathcal{M}}_1={\mathcal{M}}_2$) and the
\emph{same} size ($R_1 = R_2$).  Therefore, if the test-masses have
thin-shells, we will have $\alpha_{13}=\alpha_{23}$ and so $\eta =
0$ identically.  As a result, a chameleon field will produce \emph{no} detectable WEP violation in these experiments. The only implicit dependence of this
result on $\beta$ is that, the \emph{larger} the coupling is, the more
likely it is that the test-masses will satisfy the thin-shells
conditions.

If one wishes to detect the chameleon using WEP violations searches, then one must either ensure that test-masses do not satisfy the
thin-shell conditions, or that they have different masses and/or
dimensions.

We have just argued that \emph{all} of the chameleon theories considered here
will automatically satisfy all laboratory bounds on WEP violation,
provided the test-masses have thin-shells.  This occurs entirely as a result of the design of those experiments.

Let us consider then a putative experiment, which
could be conducted, that would, in principle, be able to detect the
chameleon through a violation of WEP. In this experiment the test
masses are of different densities ($\rho_{1}$ and $\rho_{2}$) but of
the same mass, ${\mathcal{M}}_{test}$. Crucially the radii (size) of the
two bodies are taken to be different: $R_{1}$ and $R_{2}$. We now
calculate the E\"{o}tvos parameter, $\eta$, taking the attractor to be either  the Earth, the Sun or the Moon.

\subsubsection{Attractor is the Earth}
If the attractor is the Earth then we obtain
\begin{eqnarray*}
\eta &=& \left(\frac{M_{pl}^2(1+m_b d)e^{-m_b
d}}{{\mathcal{M}}_{E}{\mathcal{M}}_{test}}\right)\left(\frac{3}{\vert n
\vert}\right)^{\frac{1}{\vert
n+2\vert}}\left\vert\left(MR_1\right)^{\frac{n+4}{n+2}}-\left(MR_2\right)^{\frac{n+4}{n+2}}\right\vert\\
&&\left(\frac{2}{(n+2)^2}\right)^{\frac{n}{2(n+2)}}\left(Md\right)^{\frac{n+4}{n+2}},
\qquad n < -4, \\ \eta &=& \left(\frac{M_{pl}^2(1+m_b d)e^{-m_b
d}}{{\mathcal{M}}_{E}{\mathcal{M}}_{test}}\right)\left(\frac{n(n+1)M^2}{m_{b}^2}\right)^{\frac{1}{n+2}}\left\vert
M(R_1-R_2)\right\vert\\&&
\left(\frac{2}{(n+2)^2}\right)^{\frac{n}{2(n+2)}}\left(Md\right)^{\frac{n+4}{n+2}},
\qquad n > 0, \\ \eta &=& \left(\frac{M_{pl}^2(1+m_b d)e^{-m_b
d}}{4\sqrt{2}\lambda
{\mathcal{M}}_{E}{\mathcal{M}}_{test}}\right)\left\vert\frac{1}{\sqrt{\ln(d/R_{1})}}-\frac{1}{\ln(d/R_{2})}\right\vert,
\qquad n = -4,
\end{eqnarray*}
where $m_{b}$ is the mass of the chameleon in the background region between the
test masses and the surface of the Earth; $d$ is the distance between the test masses and the surface of the Earth. ${\mathcal{M}}_{Earth}$ is the mass of the Earth, and $R_{Earth}$ its radius.

Current experimental precision bounds $\eta \lesssim 10^{-13}$. We shall assume that our putative experiment, if conducted, would find
$\eta \lesssim 10^{-13}$.  However, even if this is the case, we are still only able to recover very weak bounds on $\{\beta, M, \lambda\}$.  The bounds on $\beta$ are especially weak due to the
$\beta$-independence of the $\phi$-force whenever the test-bodies have
thin-shells. The only real bound on $\beta$ comes out of requirement that it be large enough for the test-masses to have thin-shells.

\begin{figure}[tbh]
\begin{center}
\includegraphics[width=7.4cm]{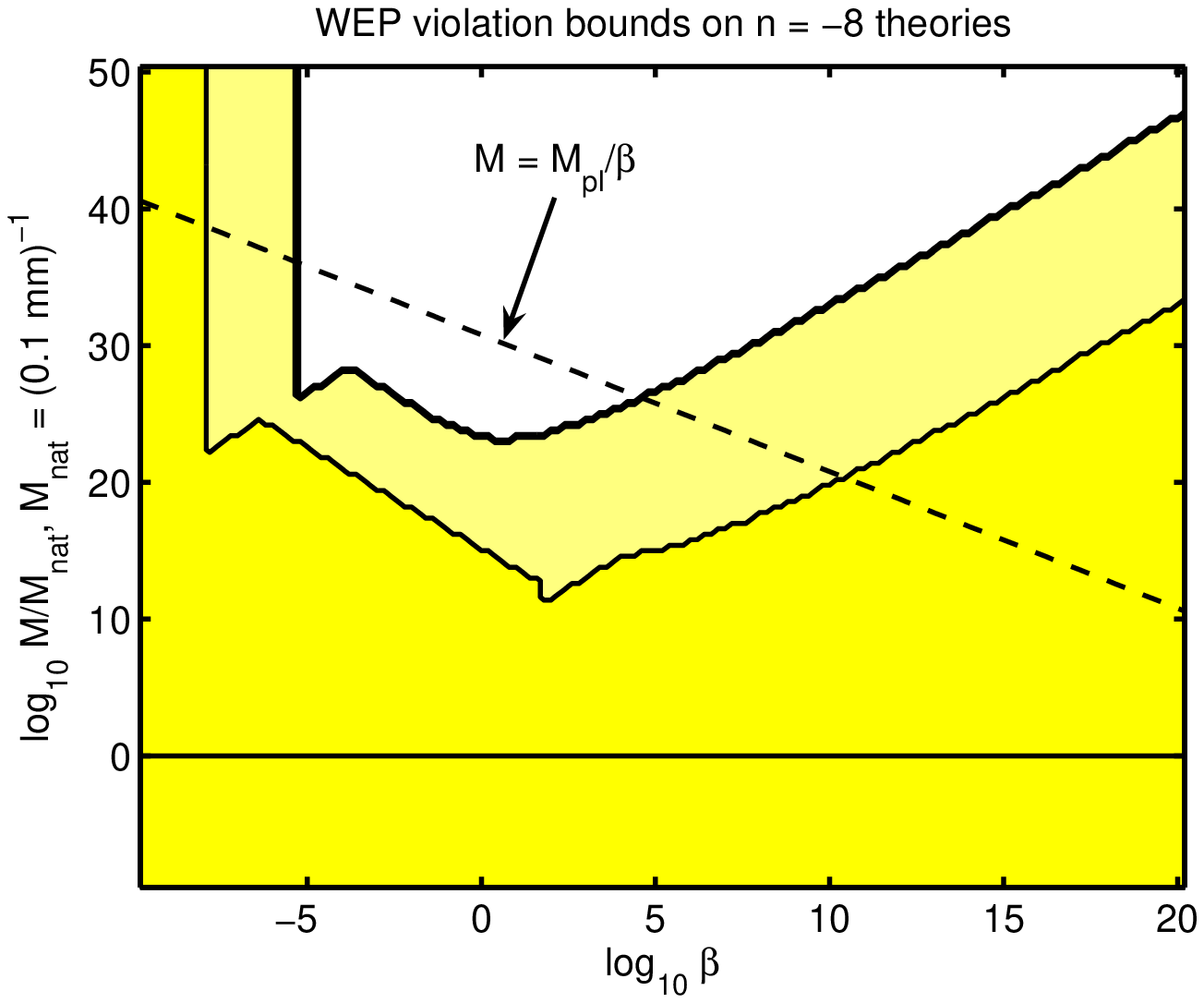}
\includegraphics[width=7.4cm]{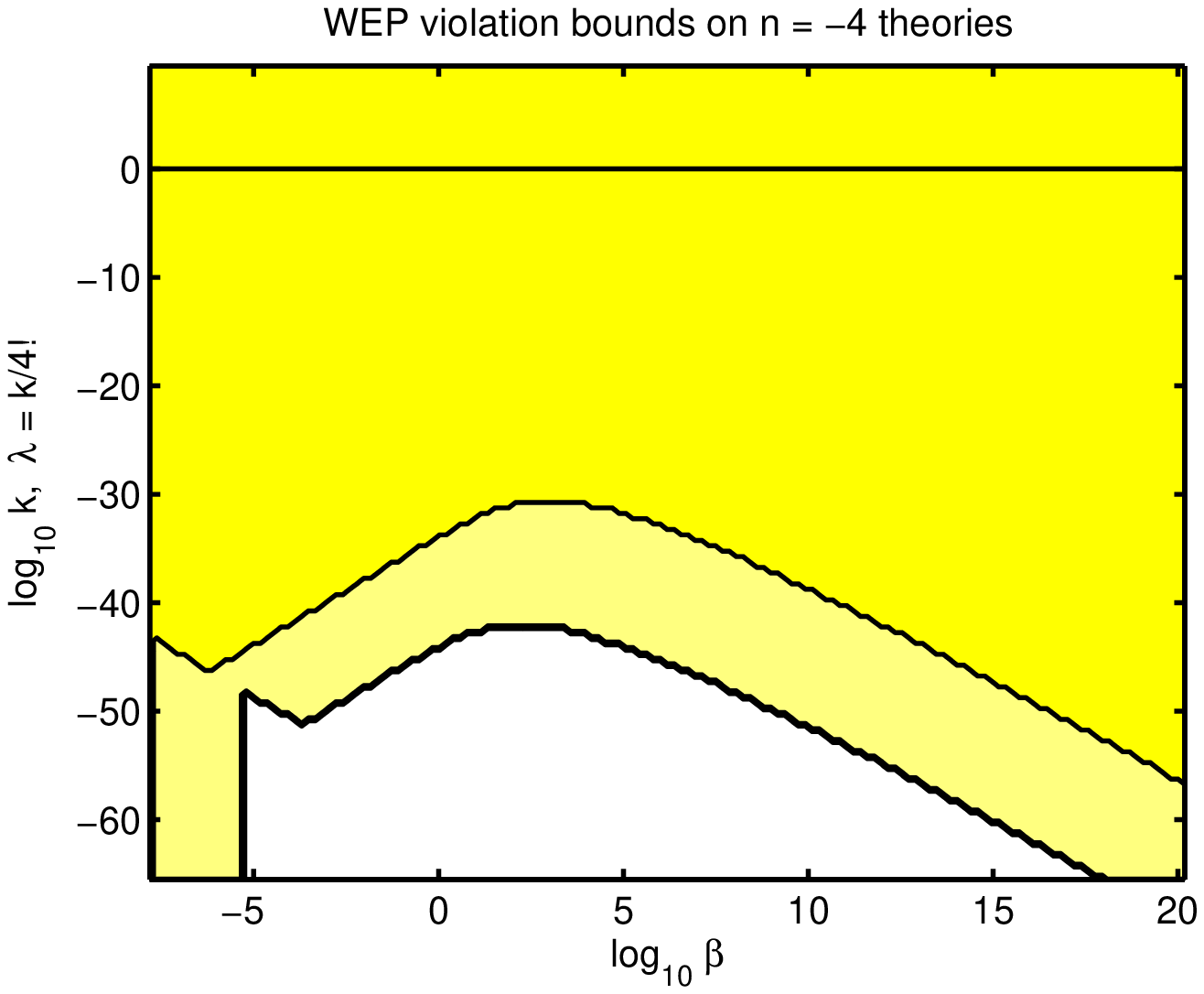}
\includegraphics[width=7.4cm]{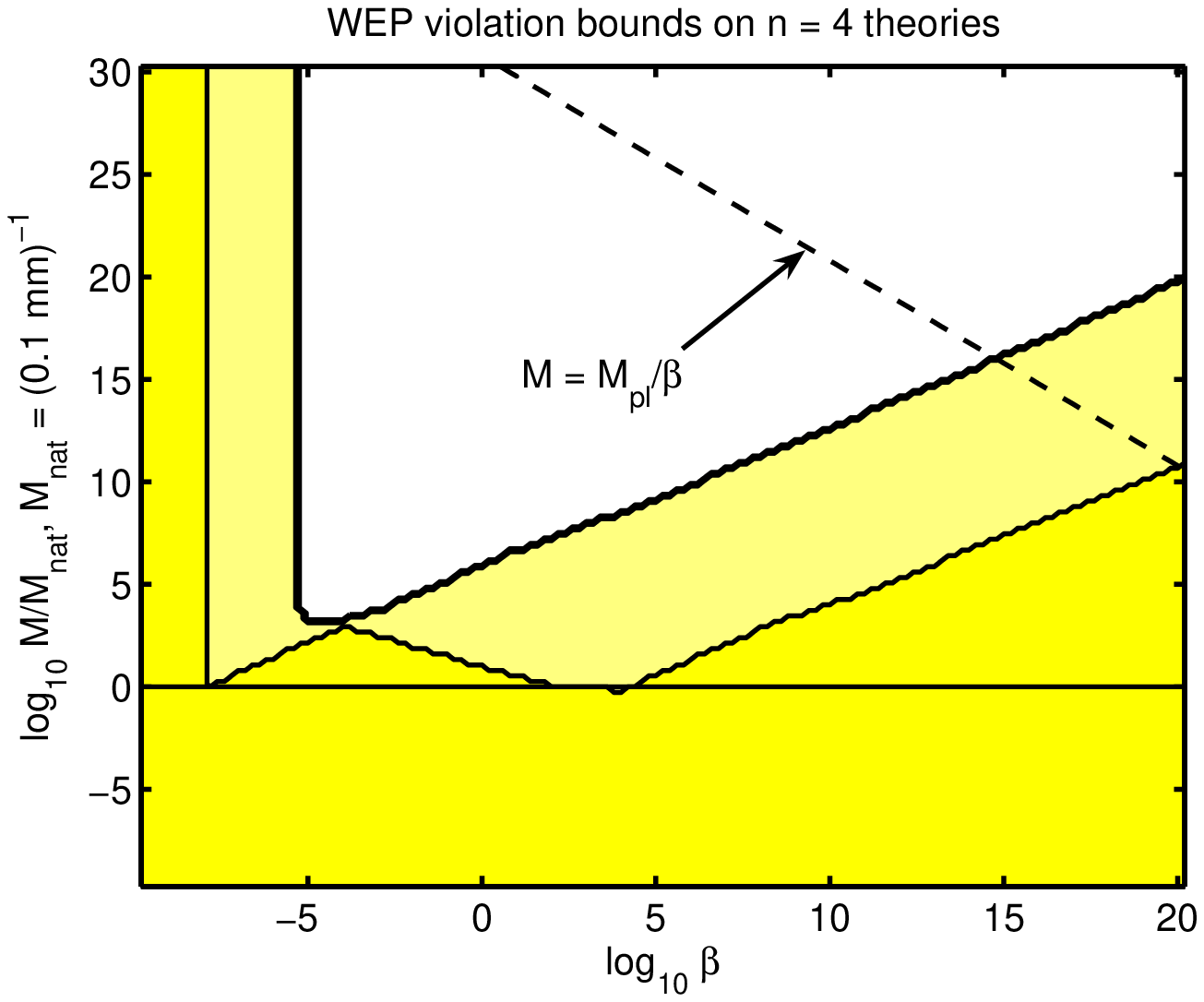}
\end{center}
\caption[WEP violation constraints on chameleon
theories]{[Colour Online] Constraints on chameleon theories coming from WEP
violation searches. The whole of shaded area shows the regions of
parameter space that are allowed by the current data.  Future
space-based tests could detect the more lightly shaded region.
The solid black lines indicate the cases where $M$ and
$\lambda$ take `natural values'.  The dotted-black line indicates
when $M = M_{\phi} := M_{pl}/\beta$ i.e. when the mass scale of
the potential is the same as that of the matter coupling.  Other
$n < -4$ theories are similar to the $n = - 8$ case, whilst the
$n=4$ plot is typical of what is allowed for $n>0$ theories.  The
amount of \emph{allowed} parameter space increases with $\vert n
\vert$.} \label{WEPbounds}
\end{figure}

For definiteness, we take the test-masses to be spherical,
with a mass of ${\mathcal{M}}_{test}=10\,g$. We assume that one of them is made entirely of copper
and the other from aluminum.  If the test-masses have thin-shells then,
even if $m_b d \ll 1$, finding $\eta <10^{-13}$ would,
when $n=-4$, only limit
$$ \lambda \gtrsim 10^{-30}.
$$
For theories with $n<-4$, $\eta < 10^{-13}$ is easily satisfied
provided that: $M < 10^{10}\,{\mathrm{mm}}^{-1}$.  When $n > 0$,
the resultant bound is on a combination of $M$ and $m_{b}^2$.  The WEP bounds on the parameter space of $n > 0$ theories are generally stronger than those for other $n$.  We
plot the effect of these bounds on the parameter space of our
chameleon theories in FIG \ref{WEPbounds}.

\subsubsection{Attractor is the Sun or the Moon}
Constraints on chameleon theories can also be found by considering the differential
acceleration of the test masses towards the Moon or the Sun, rather than towards the Earth.  The
analysis for both of these scenarios proceeds along the much same lines.
Since, for the reasons we explain below, the lunar bound will be by
far the stronger, we will only explicitly consider the case where the
third body is the Moon. In this case, the force between the test mass
and the Moon is given by equations (\ref{FarForce1}),
(\ref{FarForce2}) and (\ref{FarForce3}) for $n< -4$, $n>0$ and $n=-4$ respectively.

We define $m_{b}$ to be the average chameleon mass in the region
between the Earth and the Moon, and $m_{atm}$ to be the mass of
the chameleon in the Earth's atmosphere; $m_{lab}$ is the
background mass of the chameleon in the laboratory. $R_{a}$ is the
thickness of the Earth's atmosphere.   $d$ is the distance of
separation between the laboratory apparatus and the Moon.
Evaluating $\eta$ we find:
\begin{eqnarray}
\eta&=&\left(\frac{M_{pl}^2(1+m_b d)e^{-m_b
d-m_aR_a}}{{\mathcal{M}}_{Moon}{\mathcal{M}}_{test}}\right)\left(\frac{3}{\vert
n\vert}\right)^{\frac{2}{\vert n
+2\vert}}\\ &&\left(MR_{Moon}\right)^{\frac{n+4}{n+2}}\left\vert\left(MR_{1}\right)^{\frac{n+4}{n+2}}-\left(MR_{2}\right)^{\frac{n+4}{n+2}}\right
\vert, \qquad n < -4\notag \\ \eta&=&\left(\frac{M_{pl}^2(1+m_b d)e^{-m_b
d-m_aR_a}}{{\mathcal{M}}_{Moon}{\mathcal{M}}_{test}}\right)\\&&\left(\frac{n(n+1)M^2}{m_b
m_{lab}}\right)^{\frac{2}{n+2}}MR_{Moon}\vert M(R_1-R_2)\vert, \qquad
n > 0,\notag
\end{eqnarray}
\begin{eqnarray}
\eta &=& \left(\frac{M_{pl}^2(1+m_b d)e^{-m_b
d-m_aR_a}}{{8\lambda\mathcal{M}}_{Moon}{\mathcal{M}}_{test}}\right)\\ &&\left\vert\frac{1}{\sqrt{\ln(d/R_1)\ln(d/R_{Moon})}}-\frac{1}{\sqrt{\ln(d/R_2)\ln(d/R_{Moon})}}\right\vert,
\end{eqnarray}
%\begin{eqnarray}
% \eta &=& \left(\frac{M_{pl}^2(1+m_b d)e^{-m_b
%d-m_aR_a}}{{\mathcal{M}}_{Moon}{\mathcal{M}}_{test}}\right)\frac{1}{8\lambda
%\sqrt{\ln(d/R_1)\ln(d/R_2)}}, \qquad n = -4.
%\end{eqnarray}
where ${\mathcal{M}}_{Moon}$ is the mass of the Moon and $R_{Moon}$ its radius. If the Sun is used as the attractor then $\eta$ is given by a similar set of equations to those shown above only with ${\mathcal{M}}_{Moon} \rightarrow {\mathcal{M}}_{Sun}$ and $R_{Moon} \rightarrow R_{Sun}$.  $M_{Sun}$ and $R_{Sun}$ are respectively the mass and the radius of the Sun. If the attractor is taken to have fixed density, then we can see that $\eta$ dies off at least as quickly as $1/R^2$, where $R$ is the radius of the attractor. It follows that the predicted value $\eta$, induced by chameleon, is always smaller when the Sun is used as the attractor than when the Moon is. This is because $R_{Sun} \gg R_{Moon}$.
The predicted values of $\eta$ have a similar dependence on the radii of the test-masses i.e. dying off at least as quickly as $1/R^2$.  The corollary of this result is that if we are unable to detect $\phi$ in lab-based, micro-gravity experiments where the radii of the test-masses and the attractor are both of the order of $10\,{\mathrm{cm}}$, or smaller, then the $\phi$-force between larger (say
human-sized) objects would also be undetectably small.  For this
reason, in the context of chameleon theories, measurements of the
differential acceleration of the Earth and Moon towards the Sun,
e.g. lunar laser ranging \cite{LLR,LLR1}, are not competitive with the
bounds on WEP violation found in laboratory-based searches.

\subsubsection{Summary}
We show how bounding the E\"{o}tvos parameter by $\eta <
10^{-13}$, with the attracting body being either the Moon or the
Earth, constrains chameleon theories in FIG. \ref{WEPbounds}. As
with the E\"{o}t-Wash plots: the whole of the shaded area is
currently allowed, whilst the more lightly shaded area is that
which could be detected by proposed space-based tests of gravity
such as SEE, STEP, GG and MICROSCOPE.  It is claimed,
\cite{STEP,SEE,MICRO,GG}, that these experiments will be able to
detect $\eta$ down to $10^{-18}$.  This improved precision is
responsible for most of the increased ability of spaced based
tests to detect a chameleon field.  When $n > 0$, the thin-sell
condition is stronger in the low density background of space than
it is in the relatively higher density background of the
laboratory; this effect accounts for some of extra ability
that future space-based experiments have to detect the chameleon.
We note that WEP violation
searches only have any real hope of detecting the chameleon field, if
$M$ take a `natural' value i.e. $M \sim (0.1{\mathrm{mm}})^{-1}$, in $n > 0$ theories.

\subsection{Discussion}
In FIG. \ref{WEPbounds}, we plot how all of the bounds on WEP violation
mentioned in refs. \cite{WEP,WEP1,WEP2,LLR,LLR1}, as well as the putative bound
resulting from the modified WEP violation test we considered above,
constrain the parameter space of our chameleon theories.  The E\"{o}t-Wash bounds are shown in FIG. \ref{eotbounds}, and the Casimir bounds in FIG. \ref{Casfigure} These E\"{o}t-Wash, Casimir and WEP violation bounds are also included in the plots of section \ref{allbounds}.

It is important to note that, the larger $\beta$ is, the stronger the
chameleon mechanism becomes.  A strong chameleon mechanism results in larger chameleon masses, and larger chameleon masses in turn result in weaker chameleon-mediated forces.  A stronger chameleon mechanism also increases the likelihood of the test masses, used in these experiments, having thin-shells.   Large values of $\beta$ cannot therefore be detected at present by the E\"{o}t-Wash and WEP tests, if $\lambda$ and $M$ take natural values. If the matter coupling is very small, $\beta \ll 1$, then the chameleon mechanism is very weak and $\phi$ behaves as a normal (non-chameleon) scalar field.  The current precision of laboratory tests of gravity prevent them from seeing even non-chameleon theories with $\beta < 10^{-5}$.  Experiments that search for the Casimir force are better able to detect large $\beta$ theories primarily due to the way in which they cancel electrostatic forces. The $\beta \sim O(1)$ region is, however, currently inaccessible to Casimir force experiments.

Casimir force experiments provide upper bounds on $1/\lambda$ and
$M$ but do not directly rule out large $\beta$. Upper bounds on
$\beta$  do arise, however, from astrophysical considerations.  It
so happens that for $\beta \lesssim 10^{20}$, these astrophysical
constraints are weaker than those coming from Casimir force
searches, however they are important since they constrain how the
chameleon field can behave in much lower (and higher) density
backgrounds than those that are easily accessible in the laboratory,
and thus effectively probe a different region of the chameleon
potential. These bounds are discussed in sections \ref{compact} and
\ref{cosmo} below.

In conclusion:  contrary to most expectations laboratory tests of
gravity do \emph{not} rule out scalar field theories with a large
matter couplings, $\beta \gg 1$, provided that they have a strong
enough chameleon mechanism. When $n=-4$, a natural value of $\lambda$ is ruled out for all $\alpha = \beta^2/4\pi < 10^{-2}$, and even $\lambda \sim O(1)$ requires $\beta \sim 10^{7}$, i.e $\alpha \sim 10^{13}$.  When $n \neq -4$, large $\beta$ theories with natural values of $M$ are allowed for all $n \neq -6$.  Chameleon theories, with natural values of $M$, could well be detected or ruled out by a number of future experiments provided they are properly designed to do so.

\section{Implications for Compact Bodies \label{compact}}

In this section we will consider the effect that the fifth force
associated with the chameleon field has upon the physics of
compact bodies such as white dwarfs and neutron stars.  In the
preceding analysis, we have shown that the $\phi$-force is only
comparable in strength to gravity over very small scales.   In
neutron stars and white dwarfs, however, the average
inter-particle separations are very small, about
$10^{-13}{\mathrm{cm}}$ and $10^{-10}{\mathrm{cm}}$ respectively, and the
physics of such compact objects can therefore be very sensitive to
additional forces that only become important on small scales. The
stability of white dwarf and neutron stars involves a delicate
balancing act between the degeneracy of, respectively, electrons
or neutrons  and the effect of gravity.   If the presence of the
chameleon field were to alter this balance significantly, one
might find oneself predicting that such compact objects are always
unstable.  If chameleon theories were to make such a prediction,
for some values of the parameters $\{\beta, M, \lambda\}$, then we
could, obviously, rule out those parameter choices.   As well as
the issue of stability, we must also consider potential
astrophysical observables such as the Chandrasekar mass limit, and
the mass-radius relationship.

In 1930, Chandrasekar made the important discovery that white
dwarfs had a \emph{maximum} mass $\sim 1.4 M_{\odot}$,
\cite{chandrasekar,chandrasekar1}.  The precise value depends on the composition
of the star.   A similar maximum mass was found for neutron stars
by Landau \cite{landau}.  It was additionally noted that the mass,
$M_{star}$, of a white dwarf or neutron star would depend on its
radius, $R$, in a very special manner.  This is the mass-radius
relationship.  It is possible to extrapolate both $M_{star}$ and
$R$ from astronomical data, for example see \cite{Provencal,
Shipman, Shipman2}. In all of those works, and others like them,
the mass-radius relationship, as predicted by General Relativity ,
has been found to be in good agreement with the data.  It is,
therefore, important that the addition of a chameleon field should
not greatly alter this relationship.   The quoted $1\sigma$ error
bars on most of the determinations of $M_{star}$ and $R$ are
between about 3 and 10\% of the central value. It would certainly be fair to say then that any new theory,
which predicts deviations in the value of $M_{star}(R)$ from the
GR value of less than about 10\%, is consistent with all current
data.  Much greater deviations, are however ruled out.  We shall
use this criterion to bound the parameters of our chameleon
theories.

We firstly consider how the presence of a chameleon alters the
Chandrasekar mass limit and the mass-radius relation for both
white dwarfs and neutron stars. Our analysis proceeds along the
same lines as that presented in \cite{bholes}.  We will start by
considering a white dwarf and then note how our results carry over
to neutron stars. We suppose that the white dwarf contains $N$
electrons.  Charge neutrality then implies that there are $N$
protons.  There will also be neutrons present (approximately $N$
of them) but for this calculation we will merely group the protons
and neutrons together into $N$ nucleons where each nucleon
contains one proton. We denote the mass of a nucleon by $m_{u}$
and take it to be the atomic mass unit, $m_{u}=1.661 \times
10^{-24}g$.  White dwarfs are kept from collapsing by the pressure
of degenerate electrons, whilst their gravitational potential comes
almost entirely from the nucleons (as $m_{u} \gg m_{e}$).  We
model the white dwarf as being at zero temperature.  If should be
noted that we are not interested, so much, in accurately
determining the mass-radius relationship or Chandrasekar mass
limit, as we are in seeing to what extent they deviate from the
general relativistic form.

In the limiting cases of non-relativistic, $\Gamma = 5/3$, and
relativistic, $\Gamma = 4/3$, behaviour, the equation of state for
the electrons can be written in polytropic form: $$ P = K
\rho^{\Gamma} $$ with $K$ a constant.  For relativistic electrons:
$$ K = \frac{3^{1/3}
\pi^{2/3}}{4}\frac{1}{m_{u}^{4/3}\mu_{e}^{5/3}}, $$ where $\mu_{e}
\approx 2$ is the chemical potential for the electrons in the
white-dwarf, \cite{bholes}.  We require that the white dwarf be in
hydrostatic equilibrium. Ignoring general relativistic
corrections, this implies $$ \vec{\nabla}P = -\rho \vec{\nabla}
\phi_{N} - \frac{\beta \rho \vec{\nabla} \phi}{M_{pl}}, $$ where
the last term is the additional element that comes from the
chameleon field. $\phi_{N}$ is the Newtonian gravitational
potential: $\nabla^2 \phi_{N} = 4\pi \rho / M_{pl}^2$.  In most
realistic scenarios the density inside the white dwarf will only
change very slowly over length scales comparable to the inverse
chameleon mass. We can therefore take $$ \phi(x) \approx
M\left(\frac{M_{pl}M^3 n \lambda}{\beta
\rho(\vec{x})}\right)^{\frac{1}{n+1}}, $$ to hold inside the
white dwarf.  It is standard practice, \cite{bholes, brane}, to
solve for hydrostatic equilibrium by minimizing an appropriately
chosen energy functional.  In the absence of a chameleon field
this is: $$ \bar{E} = U+W, $$ where $W$ is the gravitational
potential energy: $$ W=-\int d^{3}x \frac{1}{2} \phi_{N}\rho, $$
and $U$ is the internal energy of the white dwarf.  It is shown in
\cite{bholes} that when one perturbs the density in such a way
that the total mass is conserved ($\delta\rho = - \vec{\nabla}
\cdot (\rho \vec{\xi})$ for some vector field $\vec{\xi}$) we
have: $$ \delta \bar{E}= \delta U + \delta W = \int
\left(\vec{\nabla}P + \rho \vec{\nabla}\phi_{N}\right)\cdot
\vec{\xi} d^3 x. $$ It follows that $\delta \bar{E}$ vanishes for
a star in hydrostatic equilibrium.  To solve for hydrostatic
equilibrium in the presence of a chameleon field, we need to minimise
the following energy functional: $$ E = \bar{E} +W_{\phi} =U+W
+W_{\phi}, $$ where: $$ W_{\phi} = \frac{n+1}{n} \int d^3 x
\frac{\beta\phi}{M_{pl}}\rho. $$ To see that this is the correct
expression we consider $\delta W_{\phi}$:
\begin{eqnarray*}
\delta W_{\phi} &=& \frac{n+1}{n}\int d^3x \frac{\beta \delta(\phi
\rho)}{M_{pl}} = \int d^3x \frac{\beta \phi}{M_{pl}}\delta \rho \\ &=&
-\int d^3 x \frac{\beta \phi}{M_{pl}}\vec{\nabla}\cdot(\rho \vec{\xi})
= \int d^3 x \left(\frac{\beta \rho \vec{\nabla}\phi}{M_{pl}}\right)
\cdot \xi.
\end{eqnarray*}
With this definition $\delta E=0$ is seen to be equivalent to
requiring hydrostatic equilibrium:
$$ \delta E = \int \left(\vec{\nabla}P + \rho \vec{\nabla}\phi_{N} +
\frac{\beta \rho \vec{\nabla}\phi}{M_{pl}}\right)\cdot \vec{\xi} \,
d^3 x = 0.
$$
\subsection{The Mass-Radius Relation}
Schematically we have $W_{\phi} \propto \frac{n+1}{n}
(\beta{<\phi>}/{M_{pl}})M_{star}$ where $<..>$ indicates an average
and $M_{star}=m_u N$ is the mass of the star. We note that $W_{\phi}
\sim \rho^{-1/(n+1)}$, and that it is negative for $n \leq -4$ and positive for
$n > 0$. To study the effect of the chameleon upon the Chandrasekar
mass limit, and the mass-radius relationship, we assume that the density of
the white dwarf is uniform. Whilst this is not at all accurate, it
is sufficient to see when the chameleon does, or does not, have a
significant effect. Also, whilst not being accurate, this
approximation still gives the mass-limit and mass-radius relationship
up to an $\mathcal{O}(1)$ factor. This said, we shall consider a more accurate
model later when we look at general relativistic corrections. The total internal energy of the white dwarf is given by:
$$ U = N\left(\left(p_{F}^2 + m_e^2\right)^{1/2}-Nm_{e}\right) > 0,
$$ where $p_{F}=N^{1/3}/R$ is the Fermi momentum of degenerate
electrons, \cite{bholes,brane}.  For $W$ we find:
$$ W = -\frac{3m_{u}^2 N^2}{5M_{pl}^2 R}.
$$ Evaluating $W_{\phi}$ in this approximation yields:
$$ W_{\phi} = \alpha_{\phi} \frac{n+1}{n} \frac{\beta
\phi(\rho)}{M_{pl}}m_{u}N.
$$
We have included a numerical factor $\alpha_{\phi}$ in the definition of $W_{\phi}$ given above. Although in the uniform density approximation we have
$\alpha_{\phi} = 1$, we chosen to leave $\alpha_{\phi}$ in the above
equation so that our results can be more easily reassessed in light of the more accurate evaluation of $W_{\phi}$ performed in
appendix \ref{alphaphiApp}.  The total energy is then:
$$ E(R) = N\left(\left(N^{2/3}/R^{2} +
m_e^2\right)^{1/2}-Nm_{e}\right) -\frac{3m_{u}^2 N^2}{5M_{pl}^2 R} +
\alpha_{\phi} \frac{n+1}{n} \frac{\beta \phi(\rho) m_{u}N}{M_{pl}},
$$ and $\rho \propto 1/R^{3}$.  We will have equilibrium when $d
E(R)/dR = 0$ which implies:
$$ \frac{N^{1/3}}{\left(N^{2/3} + m_e^2R^2\right)^{1/2}} =
\frac{3m_{u}^2N^{2/3}}{5M_{pl}^2} + \frac{3\alpha_{\phi}}{\vert n
\vert}\left\vert \frac{\beta \phi(\rho)}{M_{pl}}\right\vert
m_{u}RN^{-1/3}.
$$ We note that the term on the left-hand side of the above expression
is always less than $1$, and that both terms on the right hand side
are positive definite.  This implies that there is a maximal value of
$N$. This maximal value is found by setting the left hand side to $1$
and solving for $N$.  Because both terms on the right hand side are
positive definite, the maximum value of $N$, with a chameleon field
present, will be \emph{less than or equal to} the value it takes in pure General Relativity . We can see that, both with an without a chameleon, we have:
$$ N< N_{max} = \left(\frac{5M_{pl}^2}{3m_u^2}\right)^{3/2}.
$$
Following what was done for the braneworld corrections to gravity
in \cite{brane} we define:
$$ R_{0} = N^{1/3}/m_{e},
$$ and $x = R/R_{0}$.  We also define $\rho_0 = 3m_{u}N/4\pi R_0^3 =
3m_{u}m_{e}^3/4\pi$, and $Y = (N/N_{max})^{2/3}$.  Hydrostatic
equilibrium, $dE/dR = 0$, is therefore equivalent to:
$$ \frac{1}{\sqrt{1+x^2}} = Y + \frac{3\alpha_{\phi}}{\vert n
\vert}\left\vert \frac{\beta \phi(\rho_0)}{M_{pl}}\right\vert
\frac{m_{u}}{m_{e}} x^{\frac{n+4}{n+1}}.
$$ This is the mass-radius relationship for a white-dwarf star. The
Chandrasekar mass-limit follows from setting $x=0$.  We can see that it is unchanged by the presence of a chameleon field.  The chameleon field will,
however, alter the mass-radius relationship. We shall assume, and later require, that the second term on the right hand side, i.e. the
chameleon correction, is small compared to the first. Solving perturbatively under
this assumption we find:
$$ x \approx \sqrt{\frac{1}{Y^2}-1} - \frac{3 \alpha_{\phi}}{\vert n
\vert} \left\vert \frac{\beta \phi(\rho_0)}{M_{pl}}\right\vert
\frac{m_{u}}{m_{e}}\frac{\left(\frac{1}{Y^2}-1\right)^{\frac{n+4}{2(n+1)}}}{Y^2\sqrt{1-Y^2}}.
$$
The maximum value of $x$ for relativistic white-dwarfs ($p_{F} \geq m_{e}$) is $x=1$.  In order for our assumption that the effect of the
chameleon field could be treated perturbatively to be valid, we need:
$$ \frac{3\alpha_{\phi}}{\vert n \vert}\left\vert \frac{\beta
\phi(\rho_0)}{M_{pl}}\right\vert \frac{m_{u}}{m_{e}}
\sqrt{1+x^2}x^{\frac{n+4}{2(n+1)}} \ll 1.
$$
The right hand side is clearly increasing with $x$, and so we evaluate it at $x=1$. For corrections to the mass-radius relationship to be smaller than 10\% we must therefore require:
$$ \frac{3\sqrt{2} \alpha_{\phi}}{\vert n \vert} \left\vert
\frac{\beta \phi(\rho_0)}{M_{pl}}\right\vert \frac{m_{u}}{m_{e}} < 0.1.
$$
Evaluating this expression, for a white dwarf, we find
$$ \left(5.065 \times 10^{-41}\right) \alpha_{\phi}
\left(\frac{3.19\beta \times 10^{8}}{\vert n
\vert}\right)^{\frac{n}{n+1}} \left(\frac{M}{1
\,{\mathrm{mm}}^{-1}}\right)^{\frac{n+4}{n+1}}\lambda^{\frac{1}{n+1}} < 0.1,
$$ The above expression provides us with an upper-bound on $\beta$.
For natural values of $M$ and $\lambda$, this upper-bound is strongest for $n=-4$ theories.

For the corrections to the
mass-radius relationship to be smaller than 10\% when $n=-4$ we need
$$ \beta < 6.70\lambda^{-1/4} \times 10^{18}.
$$ Alternatively, for the corrections to be smaller than 1\% we require $\beta <
1.19\lambda^{-1/4} \times 10^{18}$.  As the data improves it might, in future, to be able to limit any such corrections to being smaller than 0.1\%. In this case we would need $\beta < 2.18\lambda^{-1/4} \times 10^{17}$. In evaluating these limits, we have
used the accurate value for $\alpha_{\phi}(n=-4)$ found in appendix
\ref{alphaphiApp}: $\alpha_{\phi}(n=-4)=0.58$.  Despite the fact that these represent some of the best upper bounds on $\beta$ for chameleon fields, it is clear that $\beta \gg 1$ is still allowed. Even if the astronomical data improves to the point where we can rule out corrections at the 0.1\% level, we would still be unable to rule out $M_{pl}/\beta \gtrsim 1\,{\mathrm{TeV}}$.

The calculation for a neutron star proceeds along similar lines.
In a neutron star, the neutrons provide both the degeneracy
pressure and the gravitational potential. We must therefore
replace both $m_u$ and $m_e$ by $m_n$.  For the correction for the
mass-radius relationship to be less than 10\% we must require: $$
\left(8.15 \times 10^{-51}\right) \alpha_{\phi}
\left(\frac{1.98\beta \times 10^{18}}{\vert n
\vert}\right)^{\frac{n}{n+1}} \left(\frac{M}{1
\,{\mathrm{mm}}^{-1}}\right)^{\frac{n+4}{n+1}}\lambda^{\frac{1}{n+1}}
< 0.1. $$ The left hand side of the above expression is a factor
of $(m_e/m_n)^{\frac{n+4}{n+1}}$ smaller than the equivalent
expression for a white dwarf. It follows that, for all $n=-4$, a
weaker bound on the parameters results.  When $n=-4$ we find the
same bound.  In FIG.
\ref{whplots} we have plotted the white-dwarf bounds on $\beta$,
$M$ and $\lambda$. We have, conservatively, assumed that
corrections to the mass radius relationship are smaller than 5\%.
The plots are similar to those done for the E\"{o}t-Wash and WEP
bounds.  The whole of the shaded region is allowed and the plots
for other theories with $n < -4$ are similar to those with $n=-8$.
Similarly, the plots for other $n>0$ theories looks much the same
as the $n=4$ plot does.

These white dwarf bounds are included in the plots of section \ref{allbounds}, where all the bounds on these chameleon theories are collated.

\subsection{General Relativistic Stability}
We have already derived conditions for the effect of the chameleon to
be small compared to that of the Newtonian potential and thus produce only a
negligible change to the mass-radius relation. It also is necessary to
consider how the inclusion of a chameleon affects the stability of
white dwarfs and neutron stars. This requires the inclusion of general relativistic effects.

A compact body in hydrostatic
equilibrium (${\mathrm{d}}E/{\mathrm{d}}R=0$) will be stable against small perturbations
whenever ${\mathrm{d}}^2E/{\mathrm{d}}R^2 > 0$, i.e. we are at a minima of the energy. For a
proof of this result see \cite{bholes,brane}. The onset of instability
occurs when ${\mathrm{d}}^2 E / {\mathrm{d}}R^2 = 0$. For Newtonian gravity this occurs when
the star becomes relativistic i.e. $\Gamma = 4/3$.  Ignoring general
relativistic effects but including chameleon effects we find that ${\mathrm{d}}^2
E / {\mathrm{d}}R^2$ is given by:
$$ \frac{{\mathrm{d}}^2 E}{{\mathrm{d}}R^2} = -\frac{2}{R}\frac{{\mathrm{d}}E}{{\mathrm{d}}R} +
\frac{N^{4/3}}{R^2} \left(\frac{N^{1/3}m_{e}^2 R^2}{(N^{2/3} +
m_{e}^2R^2)^{3/2}} +
\left(\frac{n+4}{n+1}\right)\frac{3\alpha_{\phi}}{\vert n \vert} \left
\vert \frac{\beta \phi(\rho)}{M_{pl}}\right \vert m_{u}N^{-1/3}
R\right).
$$ When ${\mathrm{d}}E /{\mathrm{d}}R=0$, the contribution from the chameleon field to the
right hand side of this equation is positive. It follows that the effect of the
chameleon field is to increase the stability of white dwarfs and
neutron stars i.e. it makes ${\mathrm{d}}^2 E/{\mathrm{d}}R^2$ more positive.

It is well known, \cite{bholes}, that General Relativity  alters
the stability of white dwarfs and neutrons stars. When GR effects
are included gravity is generally stronger. As a result, it tends
to have a destabilising effect on configurations that are stable
when studied in Newtonian physics.  In the absence of chameleon
corrections, but including GR effects (assuming $GM_{star}/R
\ll 1$) the criterion for stability is roughly: $$ \Gamma -4/3 >
\kappa \frac{M_{star}}{R M_{pl}^2}, $$ where $\kappa \sim
{\mathcal{O}}(1)$, \cite{bholes}.

Whilst GR destabilises white-dwarf stars, we have just noted that the chameleon force acts to stabilise them. In this section we will see how the leading
order general relativistic effects balance out
against the chameleon force, and study their cumulative effect on the
stability of compact objects.  The assumption that
general relativistic effects are small means that these results will
be more accurate for white dwarfs than they will for neutron stars.
A full derivation of the potential energies associated with the
leading order general relativistic effects can be found in
\cite{bholes}. For the sake of brevity we shall not repeat that
analysis here but merely quote the results.

We shall assume that the electrons in the white dwarfs are
approximately relativistic and so satisfy $P=K\rho^{4/3}$. We shall
also assume spherical symmetry. Just as we do in appendix \ref{alphaphiApp},
we evaluate the different contributions to the energy of the
white-dwarf under the assumption that the fluid satisfies the
Newtonian equation of hydrostatic equilibrium at leading order. We then solve the resulting Lane-Emden equation for $P$ numerically.  This
procedure is valid if one only wishes, as we do, to calculate the
GR and chameleon field corrections to leading order.

The internal energy of a $n=3$ polytrope is:
$$ U = k_{1}K \rho_{c}^{1/3}m_u N,
$$ and the gravitational potential of the star is given by:
$$ W = -k_{2}\rho_{c}^{1/3} M_{pl}^{-2} (m_u N)^{5/3},
$$ where $k_{1}$ and $k_{2}$ are found in ref. \cite{bholes} to be (for
$\Gamma=4/3$):
$$ k_{1} = 1.75579, \qquad k_{2} = 0.639001.
$$ In addition to the correction coming from General Relativity  (which
we will consider shortly) we also need to account for the fact that
the electrons are not entirely relativistic.  At leading order, this gives the following correction to the internal energy:
$$ \Delta U = k_{3}m_{e}^2/(\mu_{e}m_{u})^{2/3} m_{u}N\rho_{c}^{-1/3},
$$ where $k_{3} = 0.519723$, \cite{bholes}. Finally, the leading order general
relativistic contribution to the energy is found to be of the form:
$$ \Delta W_{GR} = -k_{4} M_{pl}^{-4} (m_u N)^{7/3} \rho_{c}^{2/3}
$$ where $k_4 =0.918294$, \cite{bholes}.  Including the effect of the chameleon, the
energy of the white dwarf is given by:
$$ E = U + W+\Delta U+\Delta W_{GR} +W_{\phi}.
$$
\begin{figure}[tbh]
\begin{center}
\includegraphics[width=7.4cm]{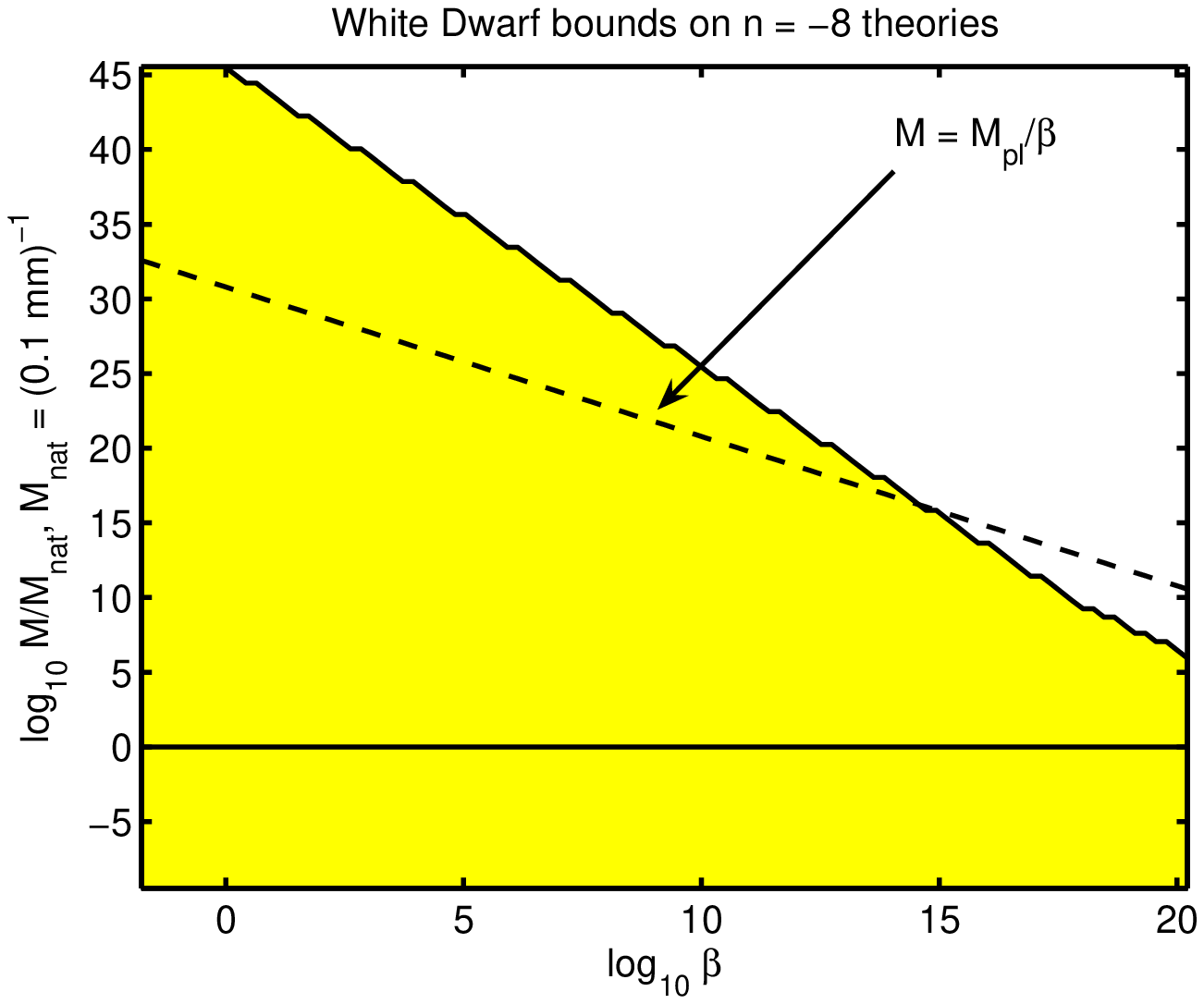}
\includegraphics[width=7.4cm]{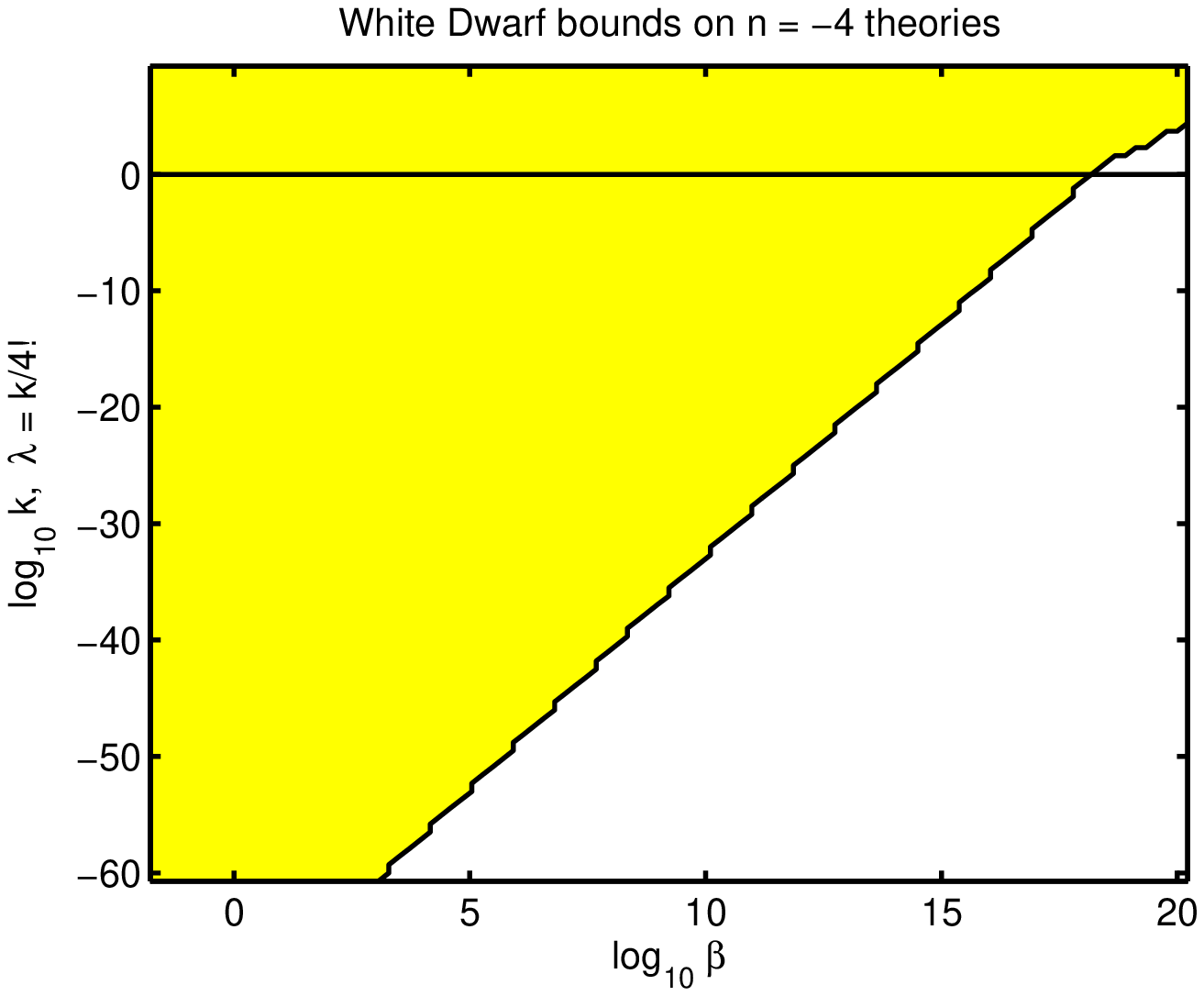}
\includegraphics[width=7.4cm]{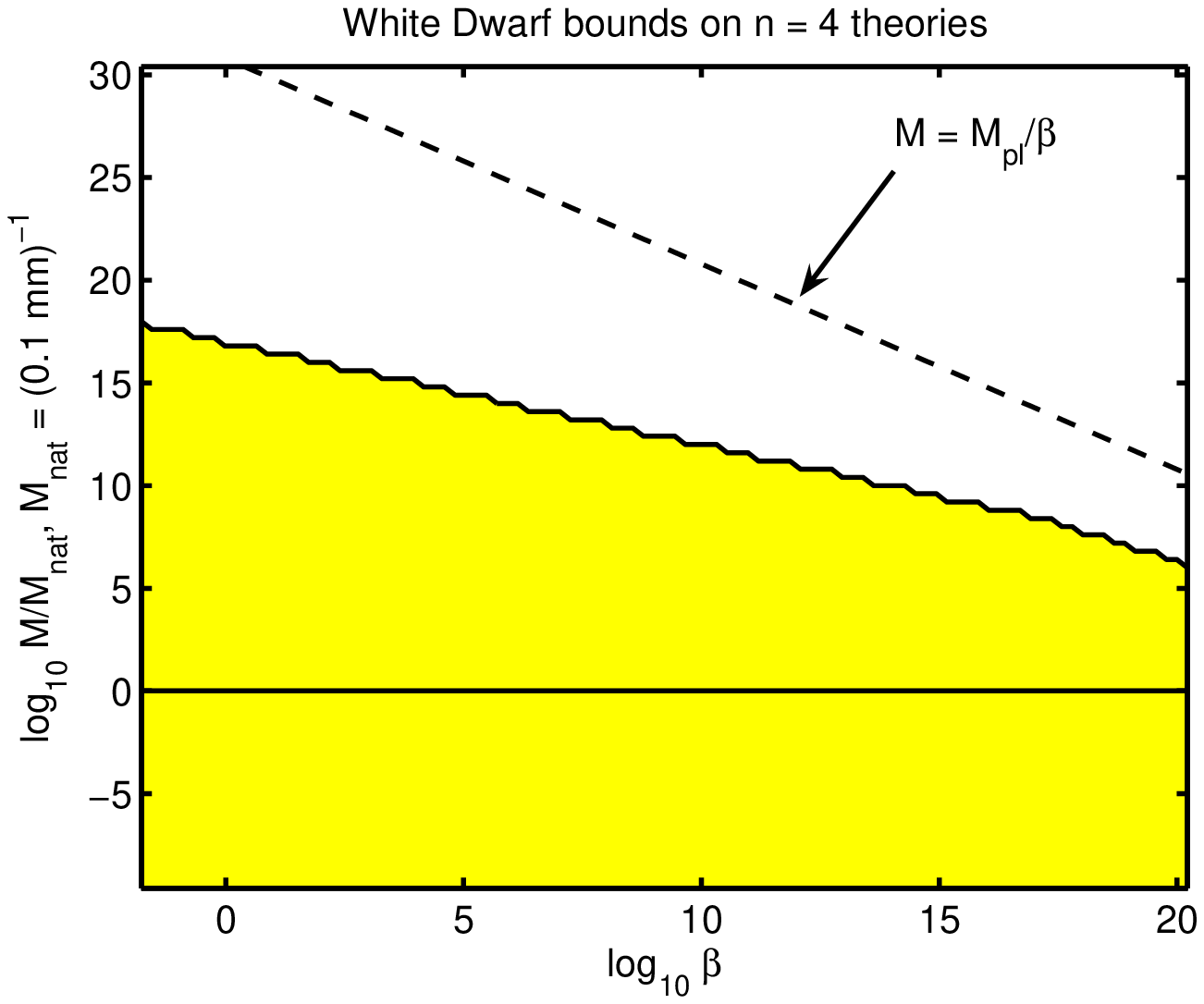}
\end{center}
\caption[Compact object constraints on chameleon
theories]{[Colour Online] Constraints on chameleon theories coming from white
dwarfs and neutron stars.  The shaded area shows the regions of
parameter space that are allowed assuming that any alterations to
the white-dwarf mass-radius relationship are at the $5\%$ level or smaller and
that the chameleon only induces changes that are smaller than
$10\%$ in the maximum white-dwarf density, $\rho_{crit}$.  Neutron
star bounds are \emph{not} competitive with the white-dwarf
constraints. The solid black lines indicate the cases where
$M$ and $\lambda$ take `natural values'.  The dotted-black line
indicates when $M = M_{\phi} := M_{pl}/\beta$ i.e. when the mass
scale of the potential is the same as that of the matter coupling.
Other $n < -4$ theories are similar to the $n = - 8$ case, whilst
the $n=4$ plot is typical of what is allowed for $n>0$ theories.
The amount of \emph{allowed} parameter space increases with $\vert
n \vert$. \label{whplots}}
\end{figure}
We define $B = k_{1}K$, $C=k_2 M_{pl}^{-2}$, $D=
k_{3}m_e^2/m_u^{2/3}$ and $F=k_4 M_{pl}^{-4}$. We also define $H =
\alpha_{\phi}/\vert{n}\vert \vert \beta \phi(\rho_c) M_{pl}^{-1}
\vert \rho_{c}^{1/(n+1)}$. As defined, $H$ is actually independent
of $\rho_c$.  With these definitions, the energy is extremised
when: $$ \frac{dE}{d \rho_c} = \tfrac{1}{3}(BM_{star} -
CM_{star}^{5/3})\rho_c^{-2/3} -
\tfrac{1}{3}DM_{star}\rho_{c}^{-4/3} -
\tfrac{2}{3}FM_{star}^{7/3}\rho_{c}^{-1/3} -
HM_{star}\rho_{c}^{-\frac{n+2}{n+1}} = 0. $$ At leading order, we
drop the terms proportional to $D$, $F$ and $H$ and recover the
standard Chandrasekar limit for the mass of a white dwarf: $$
M_{star}= \left(\frac{A}{B}\right)^{3/2} =
1.457\left(\frac{\mu_{e}}{2}\right)^{-2} M_{\odot}, $$ where
$M_{\odot}$ is the mass of the Sun.  Instability begins to occur
when ${\mathrm{d}}^2 E / {\mathrm{d}}^2 \rho_c = 0$ which, given that
${\mathrm{d}}E/{\mathrm{d}}\rho_c =0$, is equivalent to:
\begin{eqnarray} \frac{2}{3} D\rho_c^{-4/3} - \frac{2}{3}FM^{4/3}_{star}\rho_{c}^{-1/3}
+ \frac{n+4}{n+1}H\rho_c^{-\frac{n+2}{n+1}} = 0. \label{rhocritW}
\end{eqnarray} Solving this equation for $\rho_{c}$ gives a critical density,
$\rho_{crit}$, at which instability occurs. We find that, for all
$n \neq -4$, there \emph{will be} a chameleon induced correction
to $\rho_{crit}$.  When $n \neq -4$, we must either have $n\leq-6$
or $n > 0$ and so then $(n+2)/(n+1) > 4/5 > 1/3$. This observation
means that eq. (\ref{rhocritW}) still has solutions. The effect of
the chameleon is to raise the value of $\rho_{crit}$.  Even in
pure General Relativity  it turns out that this critical density is
so high that it will only be important for white dwarfs in which
the core is ${}^{4}{\mathrm{He}}$, \cite{bholes}. In all other cases,
except for that where the core is ${}^{12}{\mathrm{C}}$, $\rho_{crit}$
is greater than the neutronisation threshold, and so such
high-density white dwarfs will not occur.  In the case of carbon
white dwarfs, the central density is in fact limited by
pyconuclear reactions, \cite{bholes}.  Since the addition of a
chameleon field \emph{raises} $\rho_{crit}$, this
change can only be potentially important for Helium white dwarfs.

If the effect of the chameleon is small then, at leading order, $\rho_{crit} =
CB^2/DA^2=2.65 \times 10^{10} \,{\mathrm{g\,cm}}^{-3}$ for ${}^4{\mathrm{He}}$ stars.  The leading order chameleon correction to $\rho_{crit}$ is
\begin{eqnarray}
\frac{\delta \rho_{crit}}{\rho_{crit}} &=&
\frac{3(n+4)}{2(n+1)}\left(\frac{\alpha_{\phi}
(2m_{u})^{2/3}\rho_{crit}^{1/3}}{\vert n \vert m_{e}^2 k_{3}}\right)
\left\vert \frac{\beta \phi(\rho_{crit})}{M_{pl}} \right \vert \\ &=&
1.33 \times
10^{-44}\frac{(n+4)}{(n+2)}\alpha_{\phi}\left(\frac{1.22\beta \times
10^{12}}{\vert n
\vert}\right)^{\frac{n}{n+1}}\left(\frac{M}{1\,{\mathrm{mm}}^{-1}}\right)^{\frac{n+4}{n+1}}
\notag.
\end{eqnarray}
If we wish to require that the presence of a chameleon do little
to alter the stability properties of white dwarfs in general
relativity, we will need $\delta \rho_{crit}/\rho_{crit} \ll 1$.
This gives us another upper bound on $\beta$. In general, however,
it is not competitive with the white-dwarf mass-radius relation
bound on $\beta$.  The requirement $\delta \rho_{crit}/\rho_{crit}
< 0.1$ is included in the plots of FIG. \ref{whplots}.

\subsection{Discussion}
It should be noted that, in this section, the bounds that have been
derived on $\beta$ have been found under the assumption that the
chameleon field couples to relativistic matter in the same way as it
does to normal matter i.e. it just couples to the rest mass energy
density of matter.  As we noted in the introduction, however, it is usually the case that the chameleon in fact couples
to some linear combination of the energy density and the pressure
e.g. $\rho + \omega P$. In the simplest models $\omega=-3$.  The total energy density, $\rho_{tot}$, of the star with
equation of state $P=K\rho_{0}^{\Gamma}$ is given by:
$$ \rho_{tot} = \rho_{0} + \frac{P}{\Gamma -1} = \rho_{0} + p P
$$ where $\rho_{0}$ is the rest mass energy density and $\Gamma = 1+1/p$.  In calculations presented above, we have implicitly
assumed that $\rho_{tot} + \omega P = \rho_0 + (p+\omega) P \approx \rho_{0}$.
The bounds that we have derived come from the sector where the matter
in the star is relativistic i.e. $p=3$.  If the chameleon couples to matter through the trace of the energy momentum
tensor i.e. $\omega=-3$ then we do, in fact, have $\rho_{tot}=\rho_{0}$, just as we have assumed.

In the non-relativistic case, $P \ll \rho_{0}$, and so $\rho_{tot} + \omega P
\approx \rho_{0}$ is always true.  Even in the relativistic case, since
$P/\rho=K\rho_{0}^{1/3} \ll 1$ for white dwarfs, and $P/\rho=K\rho_{0}^{1/3} \sim
\mathcal{O}(1)$ for neutron stars, we always have $P\lesssim \mathcal{O}(\rho)$, and so different values of $\omega$ will only alter our bounds by at most an $\mathcal{O}(1)$ quantity. There is one caveat: if $\omega$ is chosen so that
$\rho_{tot} + \omega P$ can become negative, then the $n>0$ chameleon
field theories will cease to display chameleon behaviour.  This would immediately rule them out for all $\beta \gtrsim 1$.

\section{Cosmological and Other Astrophysical Bounds \label{cosmo}}

\subsection{Nucleosynthesis and the Cosmic Microwave Background}

%Radiation Era Constraints}
The  compact object bounds present above constrain a chameleon field behaves in very high density backgrounds whereas cosmological bounds on chameleon theories constrain how the behaviour of the chameleon field in low-density backgrounds. We have assumed that the chameleon
couples to the energy density and pressure of matter in the
combination: $$ \rho + \omega P. $$ In the radiation era $P\approx
3\rho$.  Provided then that $\rho(1 + \omega/3) > 0$,
i.e. $\omega > -3$, and $\beta$ is large enough, the chameleon
will simply stay at the minima of its effective potential, which
is itself slowly evolving over time.  For this to be the case it
is required that: $$ \left\vert \square \phi_{c} \right\vert \ll
\left\vert\frac{\beta\rho(1+\omega/3)}{M_{pl}}\right
\vert, $$ where $\phi_{c}$ is the value of $\phi$ at its effective
minima: $$\phi_{c} = M\left(\frac{M^3M_{pl}\lambda n}{\beta
\rho(1+\omega/3)}\right)^{\frac{1}{n+1}}, $$ and so:
\begin{eqnarray}
 \square \phi_c &=& -\ddot{\phi}_c - 3H \dot{\phi}_c =
 -\frac{4(n+5)}{(n+1)^2}H^2 \phi_{c} \\ &=&
 -\frac{4(n+5)}{(n+1)^2}\left(\frac{8\pi}{3\beta^2(1+\omega/3)}\right)\left(\frac{\beta
 \phi_{c}}{M_{pl}}\right) \left(\frac{\beta\rho(1+\omega/3)}{M_{pl}}\right),\nonumber
\label{boxphic}
\end{eqnarray}
 where we have used $H^2 = 8\pi \rho/3M_{pl}^2$ and $\dot{\rho} = -4H
 \rho$ as appropriate for the radiation era.

We shall show below that we must require that $\vert \beta \phi_c
/M_{pl} \vert < 0.1$ since the epoch of nucleosynthesis. When $\beta
\gtrsim 1$, it follows from eq. (\ref{boxphic}) that this
requirement is enough to ensure that $\vert \square \phi_c \vert
\ll \beta\rho(1+\omega/3)/M_{pl}$ provided that
$\omega > -3$.

The simplest, and perhaps the most natural way, for the chameleon to
interact with matter in a relativistically invariant fashion, however,
is for it to couple to the trace of the energy momentum tensor
i.e. $\omega = -3$. When $\omega = -3$ the above analysis does not
apply.  The strongest bounds on the parameters
of chameleon theories arise in $\omega = -3$ case.

When $\omega = -3$ we must evaluate $\rho - 3P$.  Although $\rho
\approx 3P$ in the radiation era, $\rho - 3P$ is not identically
zero. Following \cite{chameleoncosmology}, we find, for each particle species
$i$:
$$ \rho_{i} - 3P_{i} = \frac{45}{\pi^4} \frac{H^2 M_{pl}^2}{8\pi}
\frac{g_{i}}{g_{\ast}(T)} \tau(m_{i}/T),
$$ where $g_{\ast} = \sum_{bosons} g_{i}^{boson}(T_{i}/T)^4 +
(7/8)\sum_{fermions} g_{i}^{fermion} (T_{i}/T)^4$ is the standard
expression for the total number of relativistic degrees of freedom;
$g_{i}$ and $T_{i}$ are respectively the degrees of freedom and temperature of the
$i$th particle species. The function $\tau$ is defined by:
$$ \tau(x) = x^2 \int_{x}^{\infty} du \frac{\sqrt{u^2 - x^2}}{e^{u}
\pm 1}
$$ with the $+$ sign for fermions and the $-$ for bosons. This
function goes like $x^2$ when $x \ll 1$ and $e^{-x}$ when $x \gg 1$,
but it is $\mathcal{O}(1)$ when $x\sim \mathcal{O}(1)$.  The case
$\beta \sim \mathcal{O}(1)$ and $n>0$ was analysed very thoroughly
in \cite{chameleoncosmology}. We will therefore restrict ourselves
to looking at the $n \leq -4$ cases. We will also consider, for all
$n$, what new features emerge when we take $\beta \gg 1$.  In both
cases we will see that theories with $n \leq -4$ and/or $\beta \gg 1$ are much less susceptible to different initial
conditions than those with $n > 0$ and $\beta \sim \mathcal{O}(1)$.

We consider the cases $n \leq -4$ and $n>0$ separately below. Before we do so, we note some similarities between the two
cases. Whatever the sign of $n$, the effective potential will have a
minima at:
$$ \phi_{min}(T) = M\left(\frac{M_{pl}M^3 n \lambda}{\beta
\hat{\rho}(T)}\right)^{\frac{1}{n+1}},
$$ where we define:
$$ \hat{\rho}(T) = \sum_{i} \left(\rho_{i} - 3P_{i}\right).
$$ From the form of $\rho_{i} - 3P_{i}$, it is clear that $\hat{\rho}$
will be dominated by the heaviest particle species that satisfies
$m_{i} \lesssim T$. When $m_{i} \ll T$, $\tau \propto (m_{i}/T)^2$ and
so the largest value of $m_{i}$ dominates, whereas the contribution
from species with $m_{i} \gg T$ are exponentially suppressed.  The
function $\tau(m_{i}/T)$ is peaked near $m_{i}/T =1$.

If the chameleon
is not at $\phi_{min}$, and $(\phi_{min}/\phi)^{n+1} \ll 1$, then this
peak will result in the value of $\phi$ being `kicked' towards
$\phi_{min}(T)$.  We label the distance that $\phi$ moves due to this
`kick' by $(\Delta \phi)_{i}$ where
$$ (\Delta \phi)_{i} \approx - \frac{\beta g_{i} M_{pl}}{8\pi
g_{\ast}(m_{i})}\left(\begin{array}{l} \tfrac{7}{8}\\
1\end{array}\right).
$$ The $7/8$ is for fermions and the $1$ for bosons. This formula is valid so long as $\vert \phi_{min}(T)-\phi\vert > \vert \Delta\phi_{i} \vert$ i.e. so long as the kick is not large enough to move $\phi$ to its minimum.

The largest jump of this sort will occur for the smallest value of
$g_{i}/g_{\ast}$. It will therefore occur when electrons decouple from
equilibrium at $T \approx 0.5\,{\mathrm{MeV}}$. If, however, $\vert \phi -
\phi_{min}(T) \vert$ is smaller in magnitude than this above quantity,
then $(\Delta\phi)_{i} = \phi_{min}-\phi$.  Whether or not $\phi$ will
stay at $\phi_{min}(T)$, as it evolves with time, will depend on the mass
of the chameleon at $\phi_{min}(T)$.  If $m_{\phi}^2 \gg H^2$ then it
will stay at the minimum. Otherwise it will tend to slowly evolve
towards values of $\phi$ for which $(\phi_{min}/\phi)^{n+1} < 1$.

If
$(\phi_{min}/\phi)^{n+1} > 1$ then $\phi$ will very quickly (in under
one Hubble time) roll down the potential. It will either overshoot
$\phi_{min}$, or if the mass of the chameleon at $\phi_{min}$ is large
enough, stick at $\phi_{min}$.  We can therefore assume that our
initial conditions are such that, in the far past, $\phi$ is either at
$\phi_{min}$ or $(\phi_{min}/\phi )^{n+1} < 1$.

\subsubsection{Case $n \leq -4$}
We note that when $n \leq -4$, before any jump, we have $\phi_{min} /
\phi > 1$ and so $\vert \phi_{min} - \phi \vert \leq \vert \phi_{min}
\vert$. It follows that:
$$ \frac{\phi_{min}(T \approx m_{i})}{(\Delta \phi)_{i}} \approx
\frac{45(n+1)}{\pi^4} \frac{H^2(T \approx
m_{i})}{m_{\phi}^2(\phi_{min})}.
$$ This seems to suggest that, if $(\Delta \phi)_{i}$ is large enough to move
$\phi$ to $\phi_{min}(T)$, then we will necessarily have
$m_{\phi}^2(\phi_{min}) > H^2$, and so $\phi$ will stay at
$\phi_{min}(T)$ in the subsequent evolution.  However, this is not
quite the case. As $T$ drops below $m_{i}$, the $i$th species
decouples and its energy density decreases exponentially.  This causes
$\hat{\rho}$, and consequently $m_{\phi}(\phi_{min}(T))$, to decrease quickly.
Roughly, $\hat{\rho}$ shortly after decoupling will be a factor of
$(m_{j}/m_{i})^2$ smaller than it was before, where the $j$th species
is the most massive species of particle obeying $m_{j} < m_{i}$.

If $\phi$ reaches $\phi_{min}(T\approx m_{i})$ with the $i$th kick,
then as $T$ decreases the chameleon will roll quickly down to the
potential towards $\phi_{min}(T < m_{i})$; $\vert \phi_{min}(T < m_{i}) / \phi_{min}(T \approx
m_{i}) \vert < 1$.  For $\phi$ to stick at $\phi_{min}$, for $m_{j} < T <
m_{i}$, and not overshoot it, we need:
$$ \frac{H^2(T)}{m_{\phi}^2(\phi_{min}(T))} \approx
\left(\frac{T}{m_{i}}\right)^{\frac{2n}{n+1}}
\left(\frac{m_{i}}{m_{j}}\right)^{\frac{2(n+2)}{n+1}}\frac{H^2(m_i)}{m_{\phi}^2(\phi_{min}(m_{i}))}
\ll 1.
$$ Since $T < m_{i}$ in this region, and $n/(n+1) > 0$, it is
sufficient to require:
$$
\left(\frac{m_{i}}{m_{j}}\right)^{\frac{2(n+2)}{n+1}}\frac{H^2(m_i)}{m_{\phi}^2(\phi_{min}(m_{i}))}
\ll 1.
$$
We know that $m_{\phi} \propto \beta^{(n+2)/(n+1)}$ and so the above condition is
more likely to be satisfied for larger values of $\beta$ than for
smaller ones.

We note that, even when $\phi \neq \phi_{min}$, we cannot
have $\phi / \phi_{min} \gg 1$. If this were the case initially, when
$\phi = \phi_{i}$ say, then the gradient in the $\phi$ potential would
be very steep and in one Hubble time $\phi$ would move a distance
$\Delta \phi$ where:
$$ \frac{\Delta \phi}{\phi_{i}} \sim -\beta
\left(\frac{m_{i}}{T}\right)^2
\left(\frac{M_{pl}}{\phi_{min}}\right)\left(\frac{\phi_{min}}{\phi_{i}}\right)^{(n+2)}.
$$ It is clear from this expression that for $\beta \gtrsim \mathcal{O}(1)$,
$\phi /\phi_{min}$ will decrease very quickly and pretty soon
$\phi/\phi_{min} \lesssim 1$.

When $\beta \gtrsim \mathcal{O}(1)$ it is therefore valid to assume that,
for almost of all of the radiation era evolution, $\phi(T)/\phi_{min}(T)\lesssim 1$.  The larger $\beta$ becomes, the
greater the extent to which this assumption holds true.

The main purpose of the above discussion is to illustrate that, for
theories with $n \leq -4$, the late time behaviour of $\phi$ will be
virtually, independent of one's choice of initial
conditions. The same is not true, or at least not true to the same
extent, of theories in which $n > 0$.  The corollary of this result is
that $\vert\beta \phi / M_{pl}\vert$ could be very large at the beginning of the
radiation epoch and still be less than 0.1 from the epoch of nucleosynthesis onwards.  This would correspond to the masses of particles during
the early radiation era being very different from the values they have
taken since the epoch of nucleosynthesis until the present day.  The larger
$\beta$ is, the easier it is to support large changes in the particle
masses.  This is one reason why $\beta \gg 1$ is theoretically very interesting.
In chameleon theories with $\beta \gg 1$, the constants
of nature that describe the physics of the very early universe
(i.e. pre-nucleosynthesis) could take very different values from the
ones that they do today. This could have some interesting implications for baryogenesis at the electroweak scale.

The best radiation-era bounds on $\beta, M$ and $\lambda$ come from
big-bang nucleosynthesis (BBN) and the isotropy of the cosmic
microwave background (CMB). As noted in \cite{chameleoncosmology}, a
variation in the value of $\phi$ between now and the epochs of BBN and
recombination would result in a variation of the particle
masses (relative to the Planck mass) of about:
$$ \left\vert\frac{\Delta m}{m}\right\vert \sim \frac{\beta\vert
\Delta \phi \vert}{M_{pl}}.
$$ BBN constrains the particle masses at nucleosynthesis to be within
about 10\% of their current values, \cite{chameleoncosmology}.  This
requires:
$$ \vert \phi_{BBN} \vert \lesssim 0.1\beta^{-1} M_{pl}.
$$

We have argued above that $\phi_{BBN}/\phi_{min}(T_{BBN}) \lesssim 1$
and so the above condition will be satisfied whenever:
$$ \vert \phi_{min}(T_{BBN}) \vert \lesssim 0.1\beta^{-1} M_{pl}.
$$

Nucleosynthesis occurs at temperatures between $0.1\,{\mathrm{MeV}}$ and $1.3\,{\mathrm{MeV}}$.  Since $\phi_{min}(T)$ decreases with temperature, we
conservatively evaluate the above condition with $T_{BBN}=2\,{\mathrm{MeV}}$. At
such temperatures the largest contribution to $\hat{\rho}$ will come
from the electrons (with $m_{e}(today) = 0.511\,{\mathrm{MeV}}$) and:
$$ \hat{\rho} \approx \frac{g_{e}T^2m_{e}^2}{24} \approx
\frac{g_{e}({\mathrm{MeV}})^4}{24},
$$ where $g_{e} = 4$ ($2$ from the electrons and $2$ from the
positrons). The BBN constraint on $\beta$, $\lambda$ and $M$ for $n
\leq -4$ is therefore:
$$\lambda \vert n \vert\left(\frac{1.8\beta \times 10^6}{\lambda \vert
n \vert}\right)^{\frac{n}{n+1}}
\left(\frac{M}{1\,{\mathrm{mm}}^{-1}}\right)^{\frac{n+4}{n+2}} \lesssim 1.1
\times 10^{37}.
$$ This is, however, a weaker bound on $\{\beta, M, \lambda\}$ than the white dwarf mass-radius relation constraint discussed in section \ref{compact} above.

Another important restriction on these chameleon theories comes out of considering the
isotropy of the CMB \cite{wmap}.  As is mentioned in
\cite{chameleoncosmology} a difference between the value of $\phi$ today and the value it had during the epoch of recombination would mean that the electron mass
at that epoch differed from its present value $\Delta m_{e} / m_{e}
\approx \beta \Delta \phi / M_{pl}$.  Such a change in $m_{e}$ would,
in turn, alter the redshift at which recombination occurred, $z_{rec}$:
$$ \left \vert \frac{\Delta z_{rec}}{z_{rec}} \right\vert \approx
\left \vert \frac{\beta\Delta \phi}{M_{pl}}\right \vert.
$$ WMAP bounds $z_{rec}$ to be within 10\% of the value that has been
calculated using the present day value of $m_{e}$,
\cite{chameleoncosmology}. We define $\hat{\rho}_{rec}$ and $\hat{\rho}_{BBN}$ to be, respectively, the value that $\hat{\rho}$ takes at the recombination and BBN epochs. Now $\hat{\rho}_{rec} \gg
\rho_{today}$, where $\rho_{today}$ is the cosmological energy density,
and $\phi \propto \rho^{-1/(n+1)}$, it follows that $\vert \Delta \phi \vert
\approx \vert \phi_{rec} \vert$ for $n \leq -4$.  $\phi_{rec}$ is the value of $\phi$ had during the epoch of recombination.  However, since
$\hat{\rho}_{rec} < \hat{\rho}_{BBN}$, we also have $\phi_{rec}/\phi_{BBN} < 1$, and so this CMB bound is always
weaker than the one coming from BBN. For this reason we do not evaluate the CMB constraint here.

\subsubsection{Case $n>0$}

When we considered theories with $n \leq -4$, the analysis was made
easier by the fact that $V(\phi)$ had a minimum.  This allowed us to
bound the magnitude of $\phi$ to be approximately less than that of
$\phi_{min}(T)$.  However, when $n > 0$, the potential is of runaway
form and we cannot bound the magnitude of $\phi$ in such a way.  Many
of the issues involved with $n>0$ potentials were discussed very
thoroughly in ref. \cite{chameleoncosmology}. In that work it was
assumed that $\beta \sim 1$ and so, generically $m_{\phi}(\phi_{min})
\ll H$ in the radiation era, due the fact that $\hat{\rho} \ll \rho$.
However, if $\beta \ll 1$, it is possible for $m_{\phi}$ to be large
compared to $H$.

As with the $n\leq -4$ case, it is not necessary to require that $\phi$
is at its effective minima during the whole of the radiation
era. All that is really needed is for $\phi$ be sufficiently close to
the minima at the epochs of BBN and recombination that we are able to
satisfy their constraints.  It was shown in \cite{chameleoncosmology}
that the total sum of all of the kicks that occur before BBN will move
the chameleon a distance of approximately: $\beta M_{pl}$.  BBN requires that:
$$ \vert \Delta \phi_{BBN} \vert = \phi_{today}-\phi_{BBN} \lesssim 0.1\beta^{-1} M_{pl}.
$$ Provided that, at the beginning of the radiation era, $\phi \ll
\beta M_{pl} / 8\pi$, then $\phi$ at BBN will easily satisfy the above
bound provided that $\beta \phi_{today}/M_{pl} < 0.1$.

The first of these requirements is a statement about initial
conditions. It is clear that the larger $\beta$ becomes, the less
restrictive this condition is.  Indeed for $\beta \gg 1$, it is
quite possible to have had $\beta \phi / M_{pl} \sim 0.1
\mathcal{O}(\beta^2) \gg 1$ at the beginning of the radiation era
and still satisfy this bound.  The larger the matter coupling is,
the less important the initial conditions become, and the greater
the scope for large changes in the values of the particle masses
and other constants to have occurred between the pre-BBN universe
and today.

Recombination enforces a similar bound to BBN:
$$ \frac{\Delta z_{rec}}{z_{rec}} \approx \frac{\beta (\phi_{today} -
\phi_{rec})}{M_{pl}} \lesssim 0.1.
$$ If the requirements on the initial conditions are satisfied, we will
have $\phi_{today} \gg \phi_{rec},\,\phi_{BBN}$ and so, in both cases,
$\Delta \phi \approx \phi_{today}$.  We must therefore require that
$$\beta \phi_{today} / M_{pl} \lesssim 0.1.$$

$\phi_{today} \propto \rho_{c}^{-1/(n+1)}$ where
$\rho_{c}$ is that part of the cosmological energy density of
matter that the chameleon couples to.  For the most part, we have,
in this paper, assumed that the chameleon couples to all forms of
matter with the same strength.  However, up to now, we have only
been concerned with baryonic matter.  Even if the chameleon to
baryon coupling is virtually universal, there is not necessarily any
reason to expect the chameleon to couple to dark matter with the
same strength. It is possible that the chameleon does not interact
with dark matter at all.  Since it clear that $\phi_{today}$ is a
very important quantity when it comes to bounding $n>0$ chameleon
theories, it is crucial to know what fraction of the cosmological
energy density the chameleon \emph{actually} couples to. The smaller
$\rho_{c}$ is, the larger $\phi_{today}$ will be. The larger
the value of $\phi_{today}$, the more restrictive the resultant
bound on $\{\beta, M, \lambda\}$. This implies that cosmological
bounds on a chameleon theory that couples only to baryonic matter
will be \emph{stronger} than those on a chameleon that also couples
to dark matter. Not knowing how the chameleon couples to dark
matter, we chose to be cautious, and err on the side of specifying a
bound that is perhaps slightly too tight, rather than too loose.  We
therefore work on the assumption that the chameleon only couples to
baryonic matter, and so $\rho_c \approx 0.42 \times
10^{-30}\,{\mathrm{g\,cm}}^{-3}$, \cite{wmap}. Under this
assumption, we find: $$ 8.34\times 10^{-3} \vert n \vert
\left(\frac{1.93\times 10^{-29}
\beta}{n}\right)^{\frac{n}{n+1}}\left(\frac{M}{1\,\mathrm{mm}^{-1}}\right)^{\frac{n+4}{n+1}}
< 1.$$

If the chameleon to dark matter coupling is similar in magnitude to the baryon coupling then $\rho_{c}
\approx 2.54 \times 10^{-30}\,{\mathrm{g\,cm}}^{-3}$ and we have the less restrictive bound:
$$ 1.39\times 10^{-3} \vert n \vert \left(\frac{1.17\times 10^{-28}
\beta}{n}\right)^{\frac{n}{n+1}}\left(\frac{M}{1\,\mathrm{mm}^{-1}}\right)^{\frac{n+4}{n+1}}
< 1.$$
The bound that we have just found has come about from the requirement that the particle masses at BBN and recombination are within 10\% of the values they take in regions of cosmological density i.e. $\rho_{today} \sim 10^{-30} \,{\mathrm{g\,cm}}^{-3}$. However, all cosmological
determinations of the particle masses, and indeed of the other constants of
nature, have come from analysing measurements made in regions with densities much \emph{greater} than the cosmological one.  For instance, recent cosmological determinations of $m_{p}/m_{e}$ have employed the absorption and emission spectra of
dust clouds around QSOs \cite{reinhold,webbmore}. These dust clouds
have typical densities of the order of $\rho \sim 10^{-25} -
10^{-24}\,{\mathrm{g\,cm}}^{-3}$.  If we take $\rho_{c}$ to be
$10^{-25}\,{\mathrm{g\,cm}}^{-3}$, then the BBN and CMB bounds would only require:
$$ 3.52\times 10^{-8} \vert n \vert \left(\frac{4.59\times 10^{-24}
\beta}{n}\right)^{\frac{n}{n+1}}\left(\frac{M}{1\,\mathrm{mm}^{-1}}\right)^{\frac{n+4}{n+1}}
< 1.$$

\subsubsection{Summary}
\begin{figure}[tbh]
\begin{center}
\includegraphics[width=7.4cm]{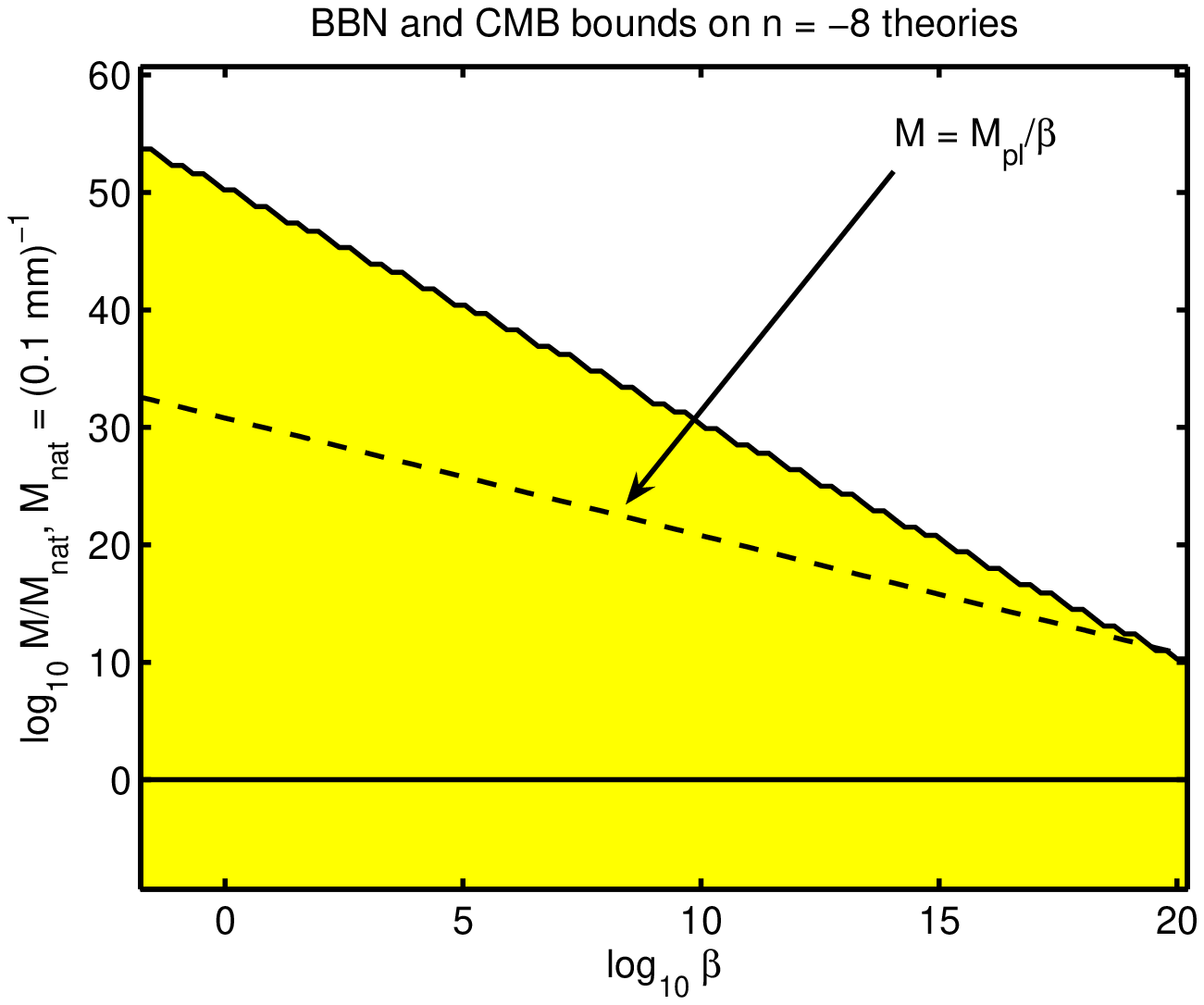}
\includegraphics[width=7.4cm]{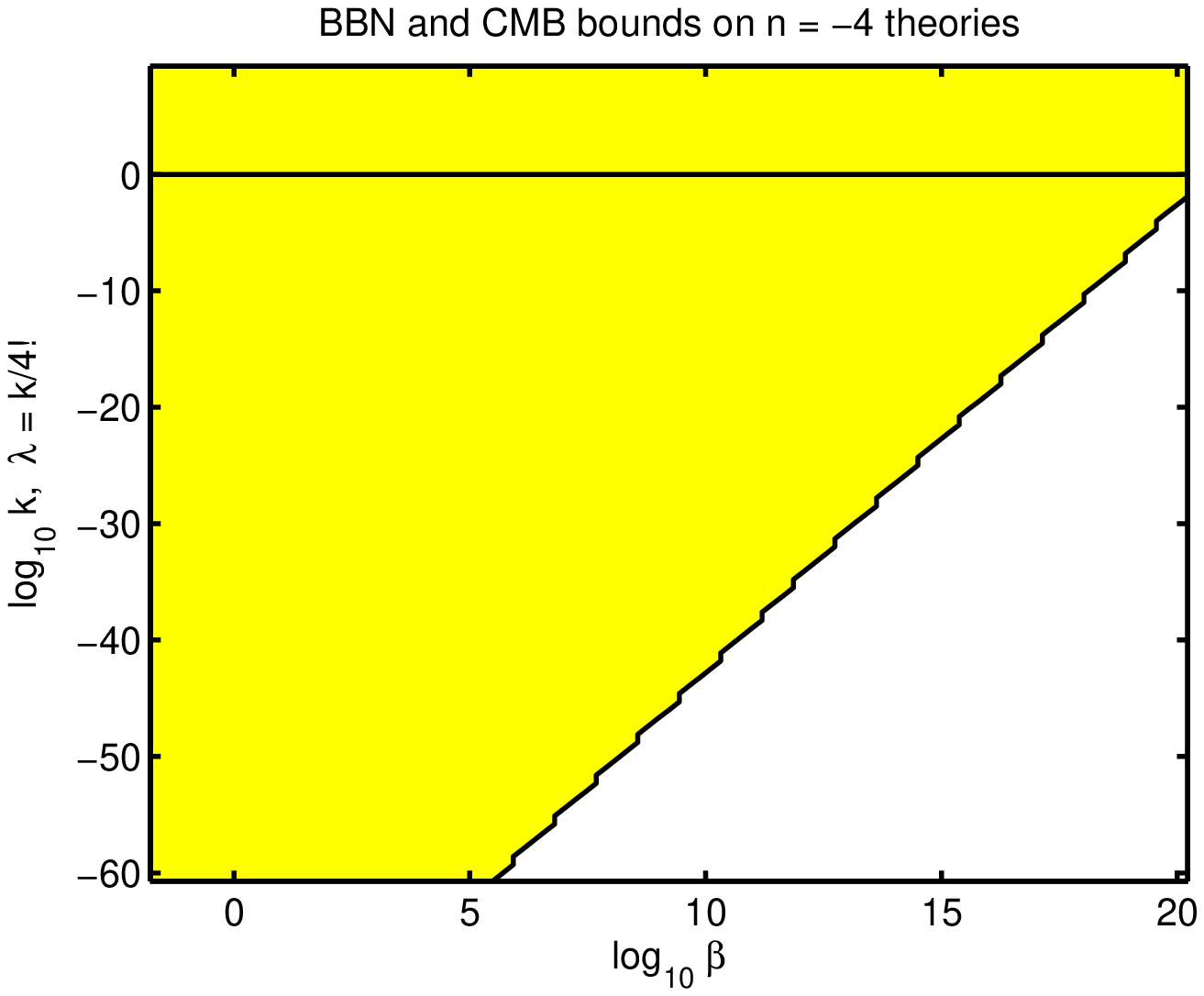}
\includegraphics[width=7.4cm]{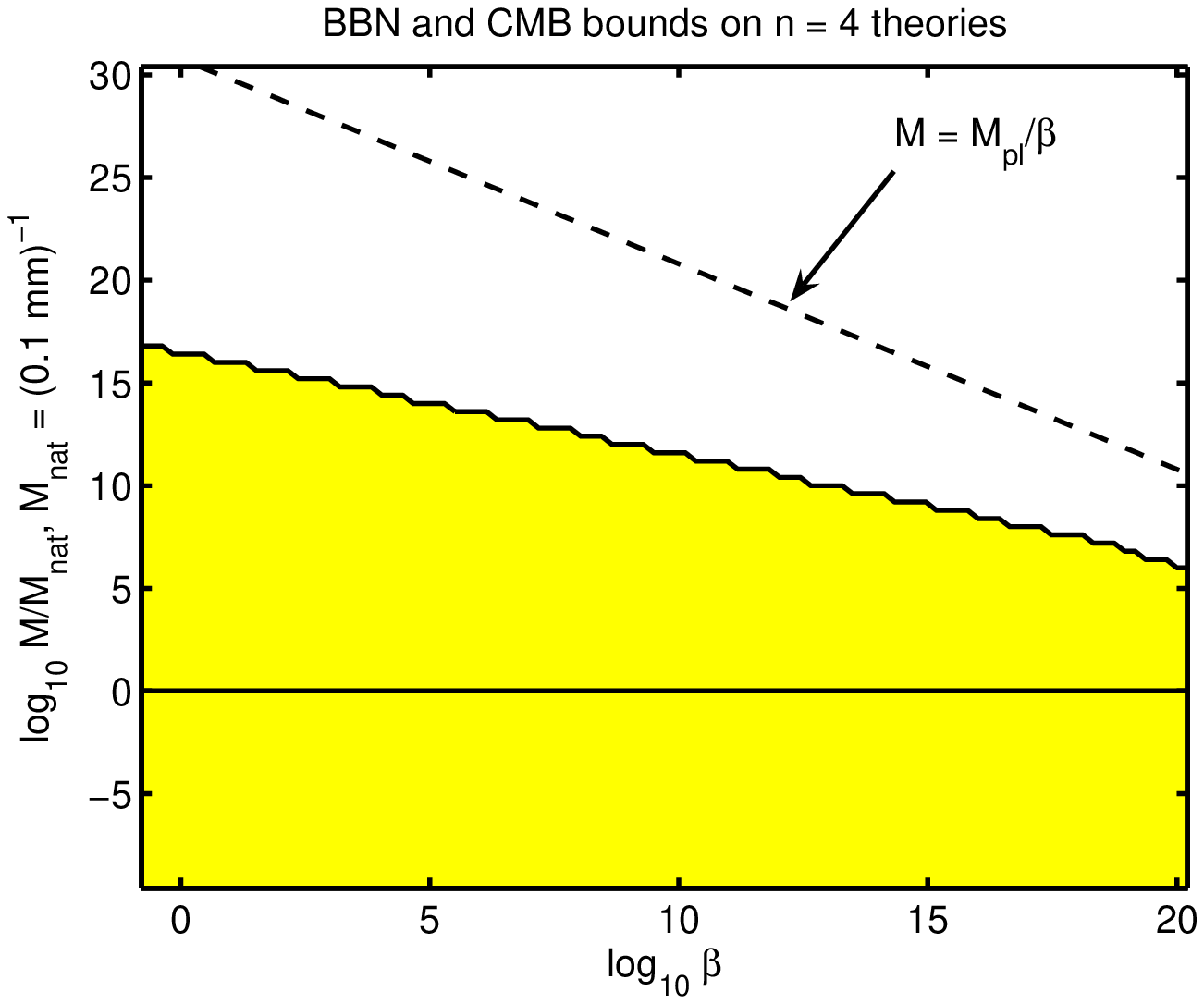}
\end{center}
\caption[Cosmological (BBN and CMB) constraints on chameleon
theories]{[Colour Online] Constraints on chameleon theories coming from the
particle masses at big-bang nucleosynthesis and the constraints
the redshift of recombination.   The shaded area shows the regions
of parameter space that are allowed by the current data.  The
solid black lines indicate the cases where $M$ and $\lambda$
take `natural values'.  The dotted-black line indicates when $M =
M_{\phi} := M_{pl}/\beta$ i.e. when the mass scale of the
potential is the same as that of the matter coupling.  Other $n <
-4$ theories are similar to the $n = - 8$ case, whilst the $n=4$
plot is typical of what is allowed for $n>0$ theories.  The amount
of \emph{allowed} parameter space increases with $\vert n \vert$.
In these plots we have assumed that the chameleon couples only to
baryons.  Slightly \emph{weaker} constraints result if the
chameleon additionally couples to dark matter.} \label{bbnplots}
\end{figure}

We have plotted the BBN and CMB constraints on $\{\beta, M \lambda\}$ in FIG. \ref{bbnplots}. We have plotted what occurs in the most restrictive case i.e. when the chameleon couples only to baryons.  The whole of the shaded region is currently allowed.

Another class of potentially important cosmological bounds can be
derived by employing astronomical bounds on the allowed variation of
the fundamental constants of nature during the matter era.  We discuss
this further below.

\subsection{Variation of fundamental constants}

Any variation in a chameleon field will lead to a variation in the
masses of the particle species to which the chameleon couples. This
variation is relative to the fixed Planck mass, $M_{pl}=G^{-1/2}$.
If the chameleon couples to all matter particles in the same way,
then all the fundamental particle masses will vary in the same
fashion and so their ratios will remain constant.  It is also
feasible to construct theories where a variation in the chameleon
leads to a variation in some other fundamental `constants' of
nature. For instance one might propose a theory where the
fine-structure constant, $\alpha_{em}$, is given by $\alpha_{em} =
f(\beta \phi / M_{pl})$ for some function $f$.  If this is the case then, in
addition to bounds on any allowed variation in the particle masses,
we would also have to apply the whole range of very stringent bounds
on the possible variation of $\alpha_{em}$ mentioned above.

For the purposes of this section, however, we assume that $\alpha_{em}$
is a true constant.  It should be noted that even if $\alpha_{em}$
does vary with $\phi$ at the same level as the particle masses do,
the resulting bounds on the parameters of the theory are only
slightly more stringent than those already found.

The best matter era bounds on the variation of the particle masses
come from measurements of the ratio $\mu = m_{p}/m_{e}$,
\cite{reinhold, webbmore, uzan}. We shall assume that the chameleon
couples to protons with strength $\beta_{p}$ and to electrons with
strength $\beta_{e}$. The relative change in the proton and electron
masses is then given by:
\begin{eqnarray*}
\frac{\Delta m_{p}}{m_{p}} &\approx& \frac{\beta_{p} \Delta
\phi}{M_{pl}}, \\ \frac{\Delta m_{e}}{m_{e}} &\approx& \frac{\beta_{e}
\Delta \phi}{M_{pl}}.
\end{eqnarray*}
Without any a priori knowledge about the magnitude, or sign, of $\beta_{e} -
\beta_{p}$ it is difficult to derive any bounds on chameleon
theories simply by considering $\Delta \mu / \mu$, where $\mu = m_{p}/m_{e}$.
The simplest assumption one could make about the matter coupling, $\beta$, is that it is universal i.e. $\beta_{p}=\beta_{e}=\beta$. If this is the case then  $\Delta \mu =0$ identically. An alternative, but still very reasonable, assumption
about the chameleon coupling, which would
produce $\Delta \mu \neq 0$, is that the
chameleon couples to all \emph{fundamental} particles with the same universal
coupling, $\beta_{U}$, but that the QCD scale, $\Lambda_{QCD}$, is
independent of $\phi$.  When the quark masses vanish the proton mass
is proportional to $\Lambda_{QCD}$.  The masses of the three lightest
quarks, $m_{u}$, $m_{d}$ and $m_{s}$, are considerably smaller than
$\Lambda_{QCD}$ and as a result contribute only a small correction to the
proton mass (at approximately the 10\% level). If $\Lambda_{QCD}$ is $\phi$-independent, we expect the proton mass to depend only weakly on $\phi$,
through the quark masses, and so $\beta_{p} \sim \mathcal{O}(\beta_{U}/10)$
whereas $\beta_{e} = \beta_{U}$. Since the mass of the electron is so
much smaller than the proton mass, the coupling of the chameleon to
baryons is approximately given by $\beta_{p}$, and so it is
$\beta_{p}$ that is constrained by the experiments mentioned in sections
\ref{exper} and \ref{compact}. However, the BBN and CMB requirements constrain $\beta_{e}=\beta_{U}$. A change in $\phi$ of $\Delta \phi$ would therefore induce a change in $\mu$ of:
\begin{eqnarray}
\frac{\Delta \mu}{\mu} \approx -9 \frac{\beta \Delta
\phi}{M_{pl}}. \label{deltamu}
\end{eqnarray}
As we reported above, the Reinhold \emph{et al.}, \cite{reinhold}, result is
consistent with a difference between the laboratory value of $\mu$ and
the one measured in such a dust cloud at the $3.5\sigma$ level: $ \Delta \mu/\mu = 2.0\pm0.6\times 10^{-5}$  where $\Delta \mu = \mu_{dust}-\mu_{lab}$.  It should be noted that
in the context of the chameleon theories considered here $\Delta \phi
= \phi_{dust}-\phi_{lab}$ is \emph{always positive} and so $\Delta \mu
/ \mu < 0$ if $\beta_{p} =\beta_{e}/10$ as we have assumed.   Under this assumption, it is not possible to reproduce the result of Reinhold \emph{et al.}. If we had alternatively assumed that the fundamental particle masses are true constants but that $\Lambda_{QCD}$ is $\phi$-dependent, then we would be able to have $\Delta \mu/\mu > 0$.

We interpret the Reinhold
result conservatively i.e. as limiting any variation in $\mu$ to be at
or below the $2 \times 10^{-5}$ level.  We shall also assume that the
laboratory experiments that measure $\mu$ are performed close enough
to the Earth's surface that the background value of $\phi$ for these
experiments is approximately the value the chameleon takes inside the
Earth.  This is last assumption is also a conservative one i.e. it
will result in a tighter bound on the chameleon theory parameters.
Taking the density of Earth to be $\rho_{Earth} \approx
5.5\,{\mathrm{g\,cm}}^{-3}$ and the density of the dust clouds from which
the absorption spectra come to be $\rho_{dust} \sim
10^{-25}\,{\mathrm{g\,cm}}^{-3}$ we find that we must require:
\begin{eqnarray*}
2.88&\times& 10^{-29}\vert n \vert\left(\frac{253\beta}{\vert n
  \vert}\right)^{\frac{n}{n+1}}\left(\frac{M}{1\,\mathrm{mm}^{-1}}\right)^{\frac{n+4}{n+1}}\lambda^{\frac{1}{n+1}}
  < 1, \qquad n \leq -4 \\ \beta &<& 1.42 \lambda^{-1/4} \times
  10^{19}, \qquad n=-4, \\ 1.58&\times& 10^{-3} \vert n \vert
  \left(\frac{4.59\times 10^{-24}
  \beta}{n}\right)^{\frac{n}{n+1}}\left(\frac{M}{1\,\mathrm{mm}^{-1}}\right)^{\frac{n+4}{n+1}}
  < 1, \qquad n > 0.
\end{eqnarray*}
For fixed $M$ and $\lambda$ this places an upper-bound on $\beta$. When $n>0$, the bounds
coming from varying-$\mu$ are competitive with the other cosmological
constraints and they provide a weak upper bound on $\beta$.  When $n \leq -4$, however, the white dwarf bounds of
section \ref{compact} still provide the best upper bound on $\beta$.  It should be noted that the bounds on $\{\beta, M, \lambda\}$ deduced from measurements  of $\Delta \mu /\mu$ are highly model-dependent.  For this reason we do not plot them here.

\section{Combined bounds on chameleon theories \label{allbounds}}
\begin{figure}[tbh]
\begin{center}
\includegraphics[width=7.4cm]{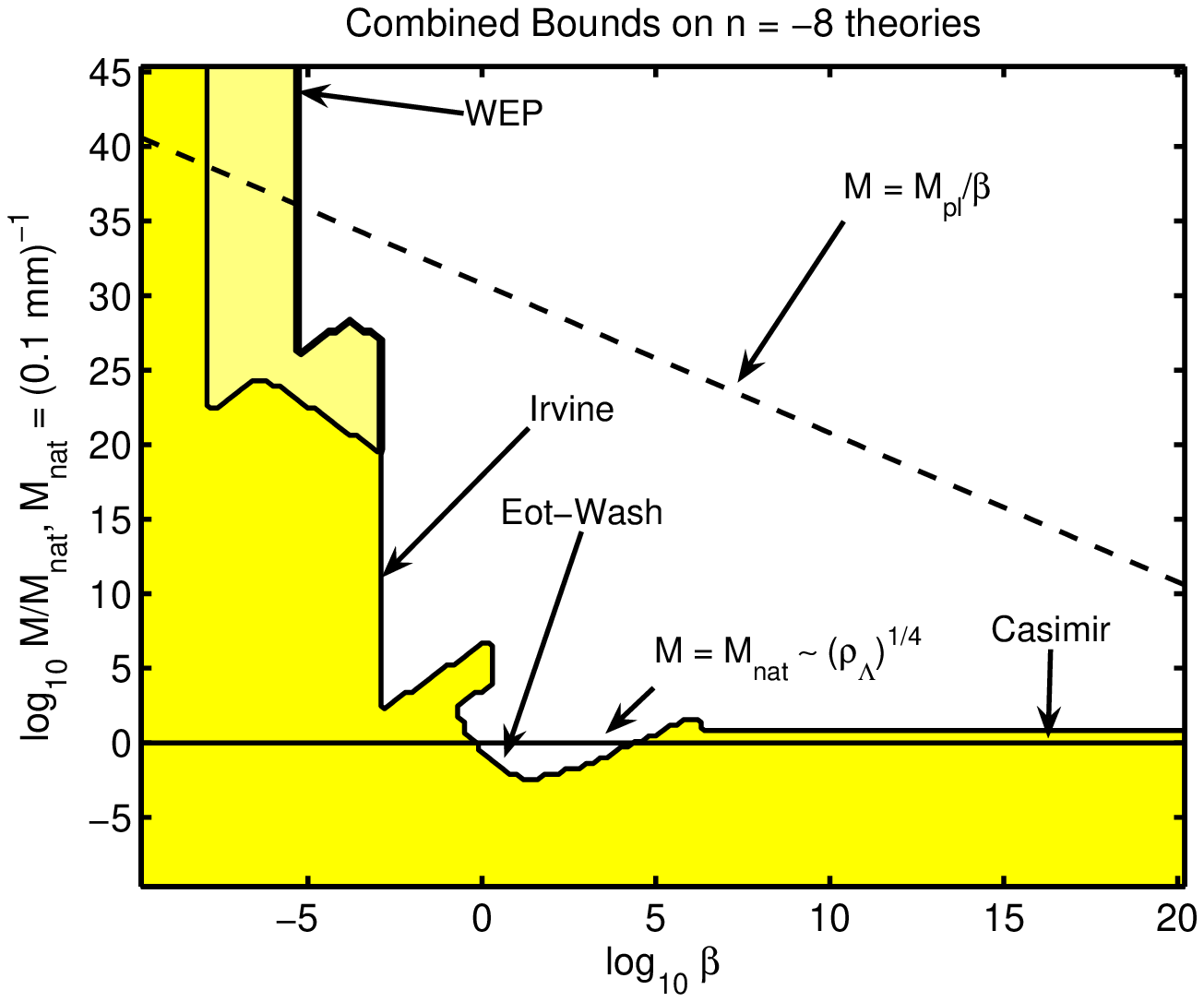}
\includegraphics[width=7.4cm]{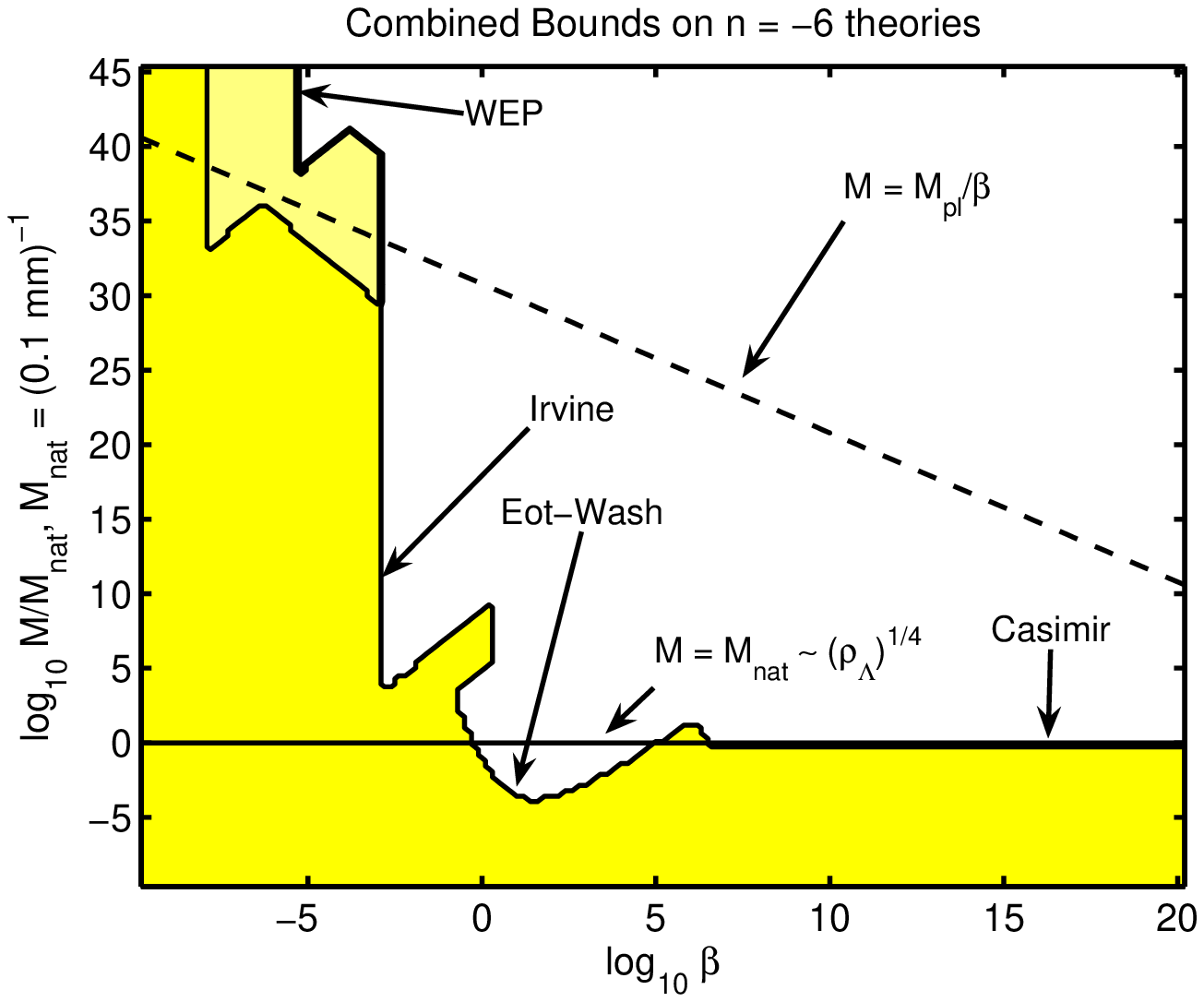}
\includegraphics[width=7.4cm]{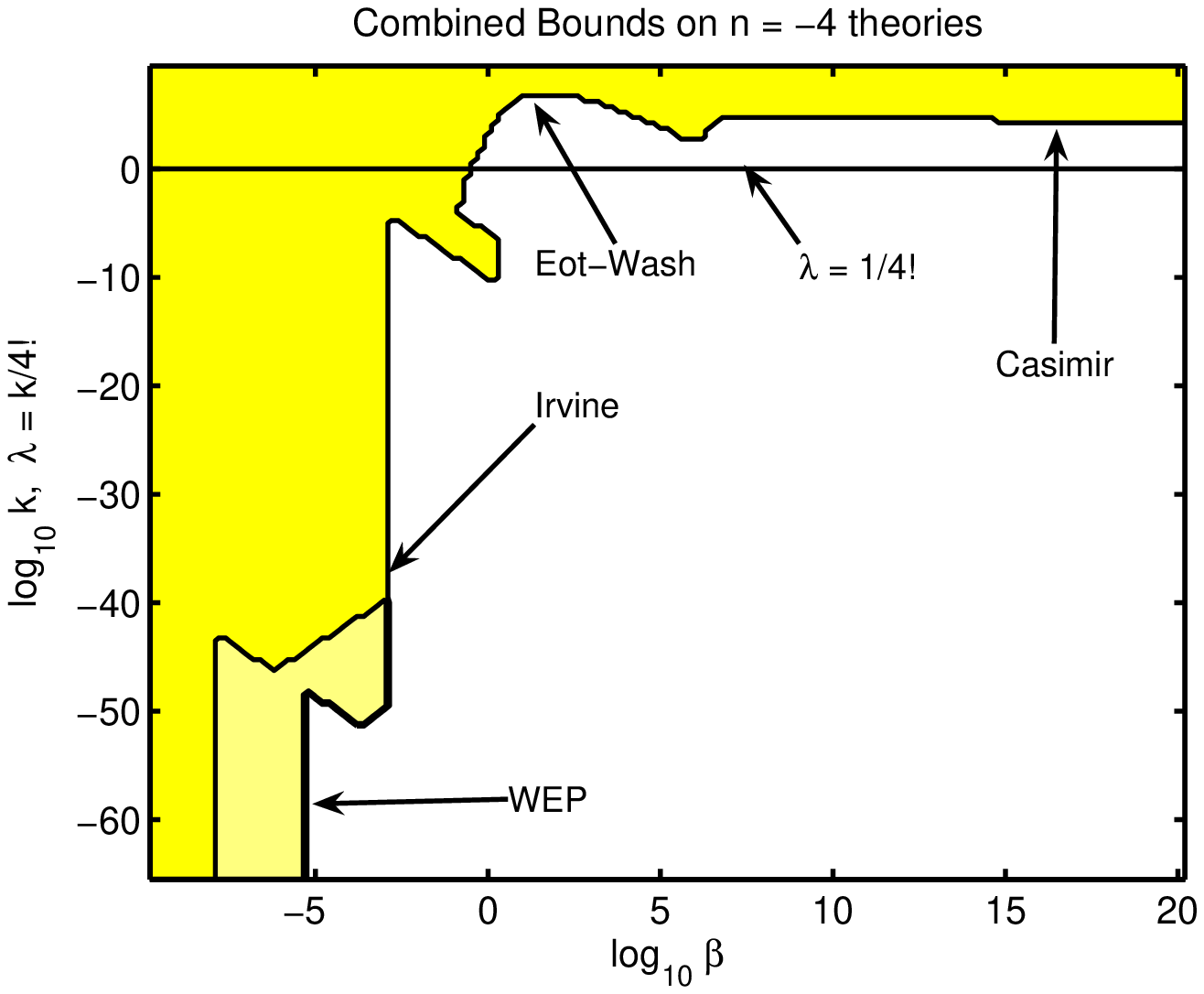}
\includegraphics[width=7.4cm]{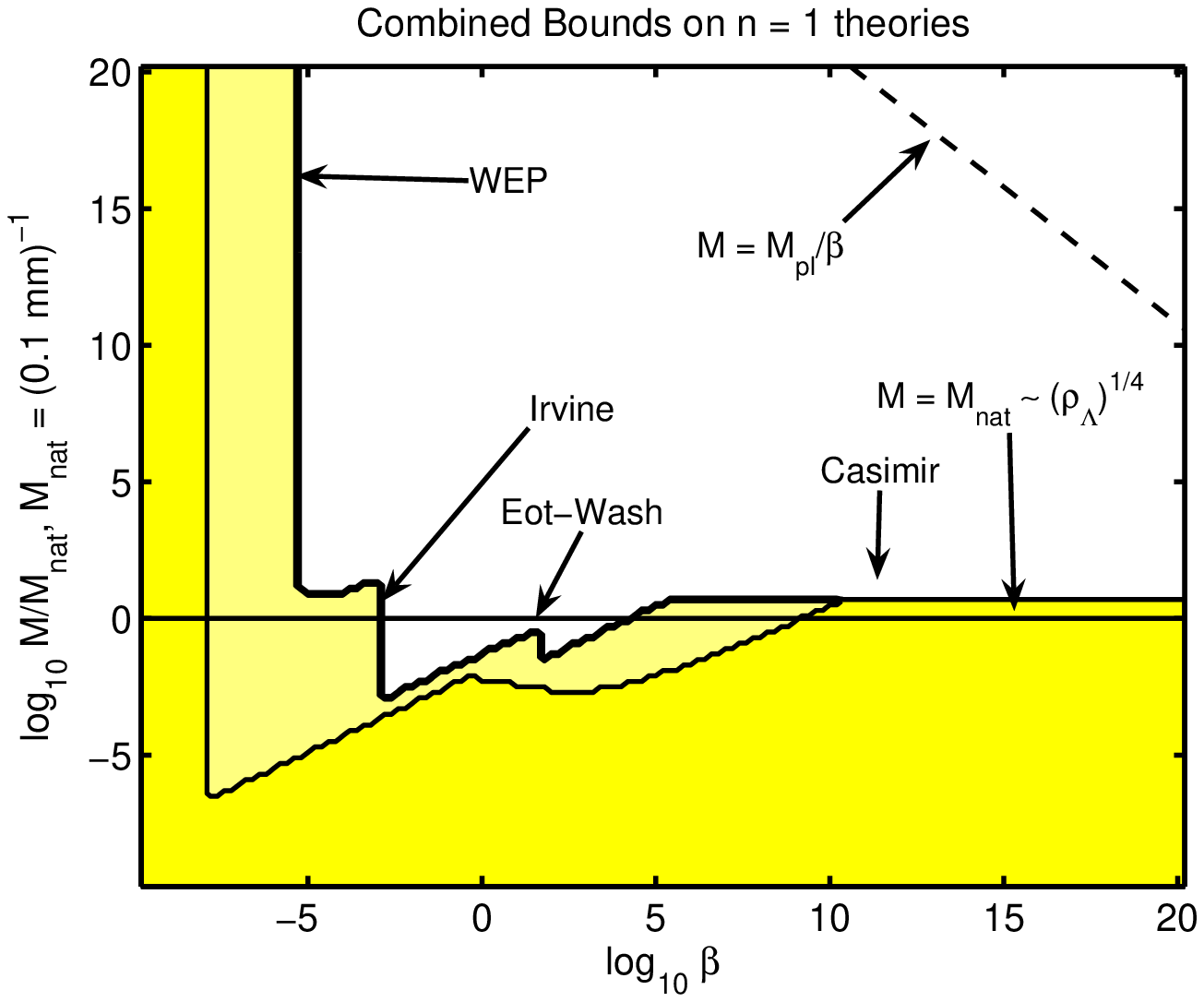}
\includegraphics[width=7.4cm]{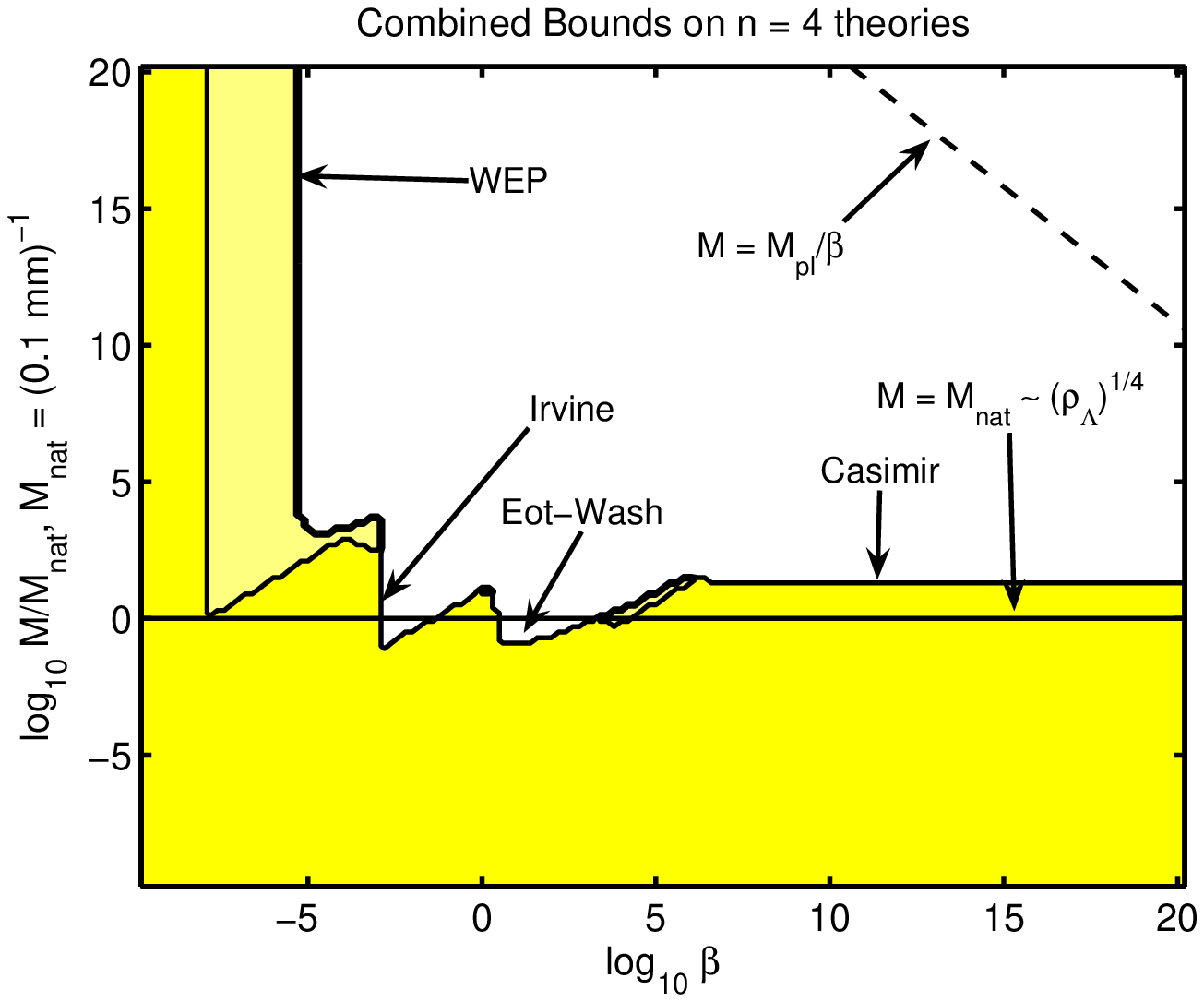}
\includegraphics[width=7.4cm]{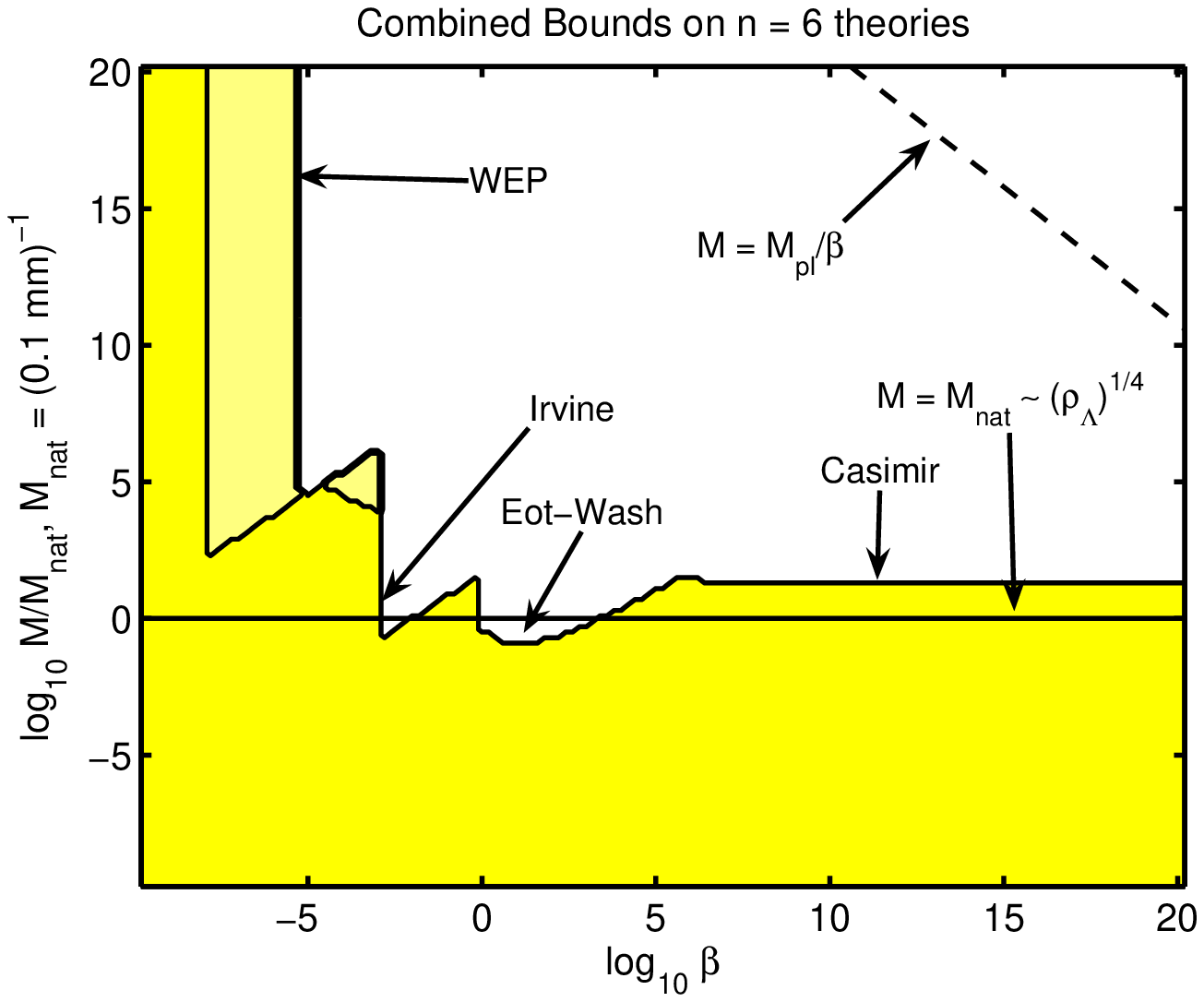}
\end{center}
\caption[Combined constraints on chameleon theories]{[Colour Online] Combined
constraints on chameleon theories. The whole of shaded area shows
the regions of parameter space that are allowed by the current
data.  Future space-based tests could detect the more lightly
shaded region.  The solid black lines indicate the cases
where $M$ and $\lambda$ take `natural values'.  For $n \neq -4$, a
natural value for $M$ is required if the chameleon is to be dark
energy. The dotted-black line indicates when $M = M_{\phi} :=
M_{pl}/\beta$ i.e. when the mass scale of the potential is the
same as that of the matter coupling.  Other $n < -4$ theories are
similar to the $n = - 6$ and $n = -8$ cases, whilst the $n=4$ and
$n=6$ plots are typical of what is allowed for $n>0$ theories.
The amount of \emph{allowed} parameter space increases with $\vert
n \vert$.  \label{allboundsFIG}}
\end{figure}

All of the chameleon theories considered in this work have a
two-dimensional parameter space, spanned either by $M$ and $\beta$
($n\neq -4$), or by $\lambda$ and $\beta$ ($n=-4$).  We combine the
constraints found in sections \ref{exper}-\ref{cosmo} to bound the
values of $\beta$ and $M$ (or $\lambda$) for different $n$. We
plotted the constraints for $n=-8,\,-6,\,-4,\,1\,4$ and $n=6$ in
FIG. \ref{allboundsFIG}. In these figures we have included all the
bounds coming from the E\"{o}t-Wash experiment, \cite{EotWash,
Eotnew}, as well as those coming from Casimir force searches,
\cite{Lamcas, Decca}. We also include the bounds (labeled Irvine)
coming from another search for Yukawa forces, \cite{Irvine}.   We
also show how \emph{current} WEP violation experiments constrain
these theories  - i.e. experiments that have actually been done as
opposed to the putative WEP violation experiment described in
section \ref{WEPsec}.  The white-dwarf and BBN constraints are also
included in the plots, however, for these theories, they are always
weaker than those laboratory tests (when $\beta \lesssim 10^{20}$).
The plots for other $n < -4$ theories are very similar to the
$n=-8$ and $n=-6$ plots, whilst the $n=4$ and $n=6$ cases are
representative of $n>0$ theories.  In general, the larger $\vert n
\vert$ is, the larger the region of allowed parameter space.  This
is because, in a fixed density background, the chameleon mass,
$m_c$, scales as $\vert n+ 1\vert^{1/2}/\vert n \vert^{1/2(n+1)}$,
and so $m_c$ increases with $\vert n\vert$.  The larger $m_c$ is, in
a given background, the stronger the chameleon mechanism, and a
stronger chameleon mechanism tends to lead to looser constraints.

We have indicated on each plot how the variety of different
bounds, considered above, combine to constrain the theory. In each
plot, the whole of the shaded area indicates the allowed values of
$M$ and $\beta$ (or $\lambda$ and $\beta$).
Three satellite experiments (SEE~\cite{SEE}, STEP~\cite{STEP} and
GG~\cite{GG}) are currently in the proposal stage, whilst a fourth
one (MICROSCOPE~\cite{MICRO}) will be launched in 2007.  These
experiments will be able to detect WEP violations down to $\eta =
10^{-18}$. The more lightly shaded region on the plots indicates
those regions of parameter space that could be detected by these
experiments - we have assumed that their WEP violation experiments to be along the lines of the putative experiment described in section \ref{WEPsec}.

In $n>0$ theories, the increased precision, promised by these
satellite tests, is not the only advantage that they offer over
their lab-based counterparts.  In space, the background density is
much lower than in the laboratory.  As a result, the thin-shell
condition, eq,. (\ref{thincond}b), is generally more restrictive
for bodies in the vacuum of space than it is for the same bodies
here on Earth.  It is therefore possible for test-bodies, that
had thin-shells in the laboratory, to lose them when they are
taken into space \cite{chama,cham1,chameleoncosmology}.   If such a
thing occurs for the test-masses used in the aforementioned
satellite experiments,  then it is possible that they might see
WEP violations in space at a level that had previously been
thought ruled out by Earth-based tests i.e. $\eta > 10^{-13}$.
This interesting feature of $n>0$ chameleon theories was first
noted in \cite{chama}.  It is important to stress that this effect
will \emph{not} occur if $\beta$ is so large that the satellites
themselves develop thin-shells \cite{chama,cham1}.
In chameleon theories where the potential
has a minimum, i.e. $n \leq -4$, the thin-shell condition, eqs.
(\ref{thincond}a \& c), is only weakly dependent on the background
density of matter.  As a result, $n \leq -4$ theories will
generally be oblivious to the background in which the experimental
tests of it are conducted.  It is for this reason that future
space-based tests will be better able to constrain $n>0$ theories
than they will $n\leq-4$ ones.

In all of the plots, we show $\beta$ running from $10^{-10}$ to
$10^{20}$.  $\beta < 10^{-10}$ will remain invisible to even the
best of the currently proposed space-based tests, and $\beta
> 10^{20}$ corresponds to $M_{pl}/\beta < 500\,{\mathrm{MeV}}$. The region in which $M_{pl}/\beta \lesssim 200\,{\mathrm{GeV}}$
is, in fact, probably already \emph{ruled out}.  If $\beta$ were
so large that $M_{pl}/\beta < 200\,{\mathrm{GeV}}$, then we would
probably have already seen some trace of the chameleon in particle
colliders. This said, without a quantum theory for the
chameleon it is hard to say  how chameleon theories behave at high
energies.  A result a detailed calculation of the chameleon's
effect on scattering amplitudes is not possible at this stage. A
full quantum mechanical treatment of the chameleon is very much
beyond the scope of this work, but remains one possible area of
future study.

The chameleon mass ($m_c$), in a background of fixed density,
scales as $\lambda^{\frac{n}{n+1}}M^{-\frac{n+4}{2(n+1)}}$.  As we
mentioned above, the larger $m_c$, is the easier it is to satisfy
the thin-shell conditions, eqs. (\ref{thincond}a-c), and the
stronger the chameleon mechanism becomes.  Since
$(n+4)/(2(n+1))\geq 0$ and $n/(n+1)
> 0$, for all theories considered here, the chameleon mechanism becomes \emph{stronger} as $M \rightarrow 0$, or
$\lambda\rightarrow \infty$, and all of the constraints are more
easily satisfied in these limits. It is for this reason that we
truncate our plots for some small $M$ and, when $n=-4$, for a
large value of $\lambda$.  Values of $M$ that are smaller than those shown,  or values of $\lambda$ that are larger, are still allowed.

The upper limit on $M$ (and lower limit on $\lambda$) has been
chosen so as to show as much of the allowed parameter space as
possible.   When $\beta$ is very small, the chameleon mechanism is
so weak that, in all cases, $\phi$ behaves like a standard
(non-chameleon) scalar field.   When this happens, the values of
$M$ and $\lambda$ become unimportant, and the bounds one finds are
on $\beta$ alone.  This transition to non-chameleon behaviour
can be seen to occur towards the far left of each of the plots.

It is clear from FIG. \ref{allboundsFIG} that $\beta \gg 1$ is,
rather unexpectedly, very much allowed for a large class of
chameleon theories. We can also see that, rather disappointingly,
future space-based searches for WEP violation, or corrections to
$1/r^2$ behaviour of Newton's law, will only have a small effect
in limiting the magnitude of $\beta$. If $M_{pl}/\beta \sim
1\,{\mathrm{TeV}}$ is feasible, pending a detailed calculation of
chameleon scattering amplitudes, that chameleon particles might be
produced at the LHC.

The solid black line on each of the plots indicates the `natural' values of
$M$ and $\lambda$ i.e. $M \sim (0.1\,{\mathrm{mm}})^{-1}$ and $\lambda
= 1/4!$. The E\"{o}t-Wash experiment and measurements of the Casimir force rule out $\lambda = 1/4!$ in $\phi^{4}$, however (expect when $n=-6$) $M\sim (0.1 mm)^{-1}$ is allowed for all $10^{4} \lesssim \beta \lesssim 10^{18}$. In
particular, $M_{pl}/\beta \sim 10^{15}\,{\mathrm{GeV}} \approx
M_{GUT}$, i.e. the GUT scale, and $M_{pl}/\beta \sim
1\,{\mathrm{TeV}}$ are allowed.

The dotted black line indicates the cases where $M = M_{pl}/\beta$
i.e. when there is only one mass scale associated with the
chameleon theory.   It is clear, however, that no such theories
are allowed if $\beta \lesssim 10^{20}$. In all cases we must
require $M \ll M_{pl}/\beta$.   As noted in \cite{chama, cham1},
this requirement introduces a hierarchy problem in the chameleon
theory itself. This problem is not, however, present in $\phi^4$
($n=-4$) theory.

The larger $\beta$ is the stronger the chameleon mechanism
becomes.  A strong chameleon mechanism results in larger chameleon
masses, and larger chameleon masses in turn result in weaker
chameleon-mediated forces. A stronger chameleon mechanism also
increases the likelihood of the test masses, used in the
experiments consider in section \ref{exper}, having thin-shells.
Large values of $\beta$ cannot, therefore be detected, at present
by the E\"{o}t-Wash and WEP tests, if $\lambda$ and $M$ take
natural values.  It is for this reason that, as can be seen in
FIG. \ref{allboundsFIG}, the E\"{o}t-Wash experiment and the WEP
violation searches place the greatest constraints on the parameter
space in a region about $\beta \sim {\mathcal{O}}(1)$.  Casimir force tests are much better able to detect large $\beta$, but they can ultimately only place an upper-bound on $M$ or $1/\lambda$.

For large $\beta$, Casimir force test succeed where the E\"{o}t-Wash experiment fails because in the former does \emph{not} use a thin metal sheet to cancel electrostatic effects. If it were not for the presence of
the BeCu sheet, the E\"{o}t-Wash test would be able to detect, or
rule out, almost \emph{all} $\beta \gg 1$ theories with $M \sim (\rho_{\Lambda})^{1/4}$.
\begin{figure}[tbh]
\begin{center}
\includegraphics[width=10cm]{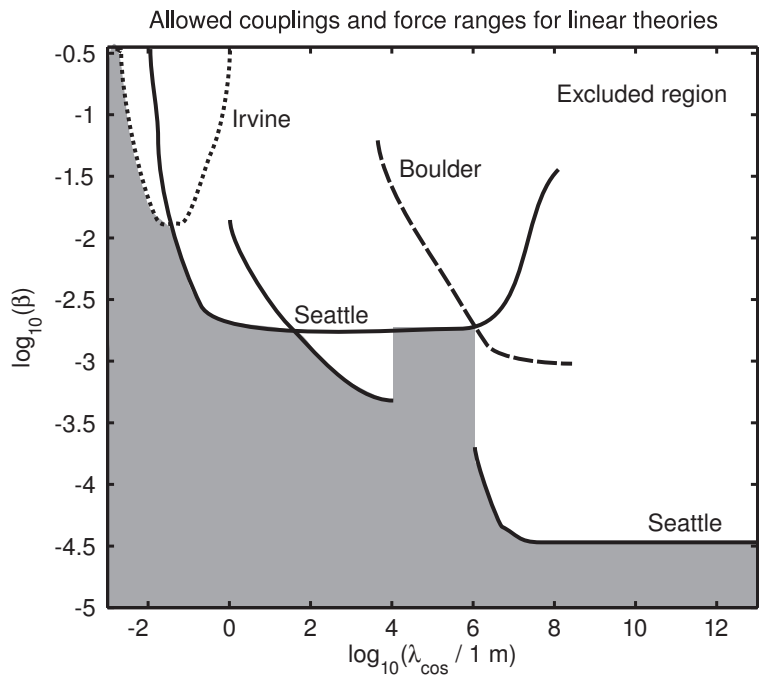}
\end{center}
\caption[Allowed couplings and force ranges for scalar fields for linear theories]{Allowed couplings and force ranges for scalar fields which couple to baryon number and have approximately linear field equations. The shaded area shows the parameter space that is allowed by the current bounds on scalar field theories with linear field
equations. $\lambda = 1/m$ is the range of the force, and $m$ is
the mass of the scalar field. $\beta$ is the matter coupling of the
scalar.  $\beta > 1$ is ruled out for all but the smallest ranges
(currently $\lambda < 10^{-4}{\mathrm{m}}$). We refer the reader to
\cite{EotWash} for a plot of the allowed parameter space for
$\lambda < 10^{-3}{\mathrm{m}}$. Irvine, Seattle and Boulder refers to
\cite{Irvine}, \cite{WEP,WEP1,WEP2} and \cite{Boulder} respectively. }
\label{FIGlin}
\end{figure}

\begin{figure}[tbh]
\begin{center}
%\FIGURE{ \epsfig{file=allranm8cos.eps,width=7.4cm}
\includegraphics[width=7.4cm]{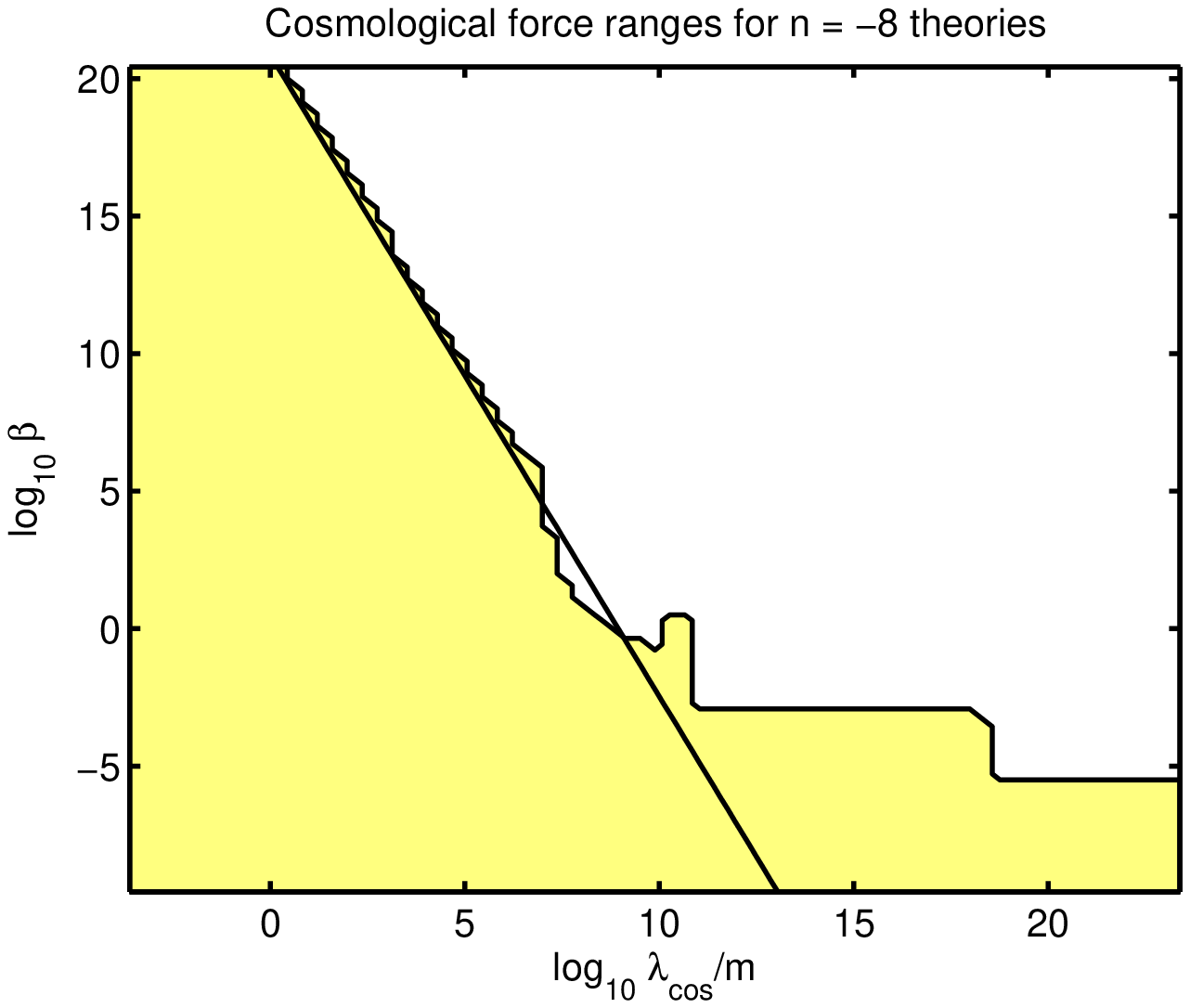}
\includegraphics[width=7.4cm]{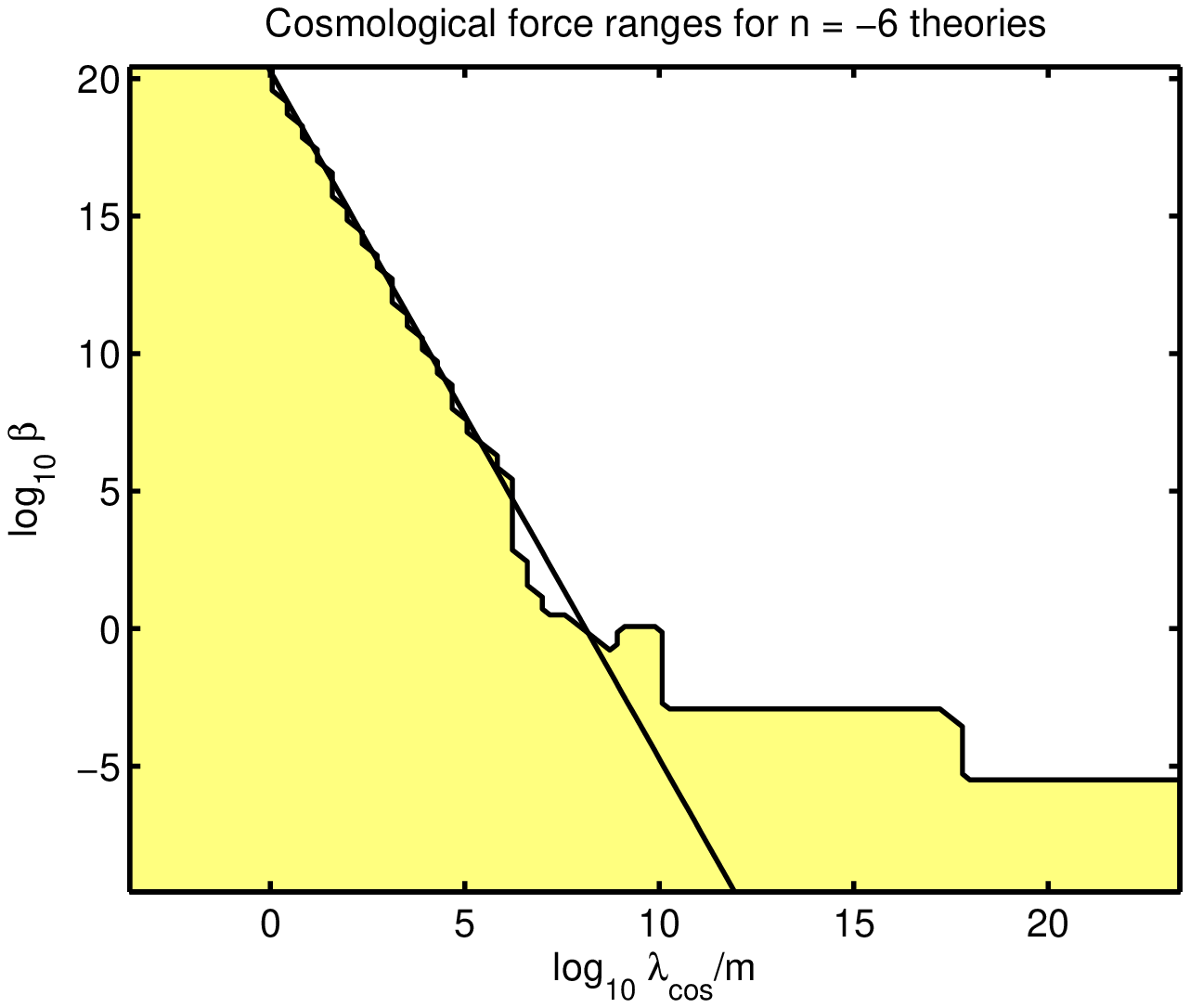}
\includegraphics[width=7.4cm]{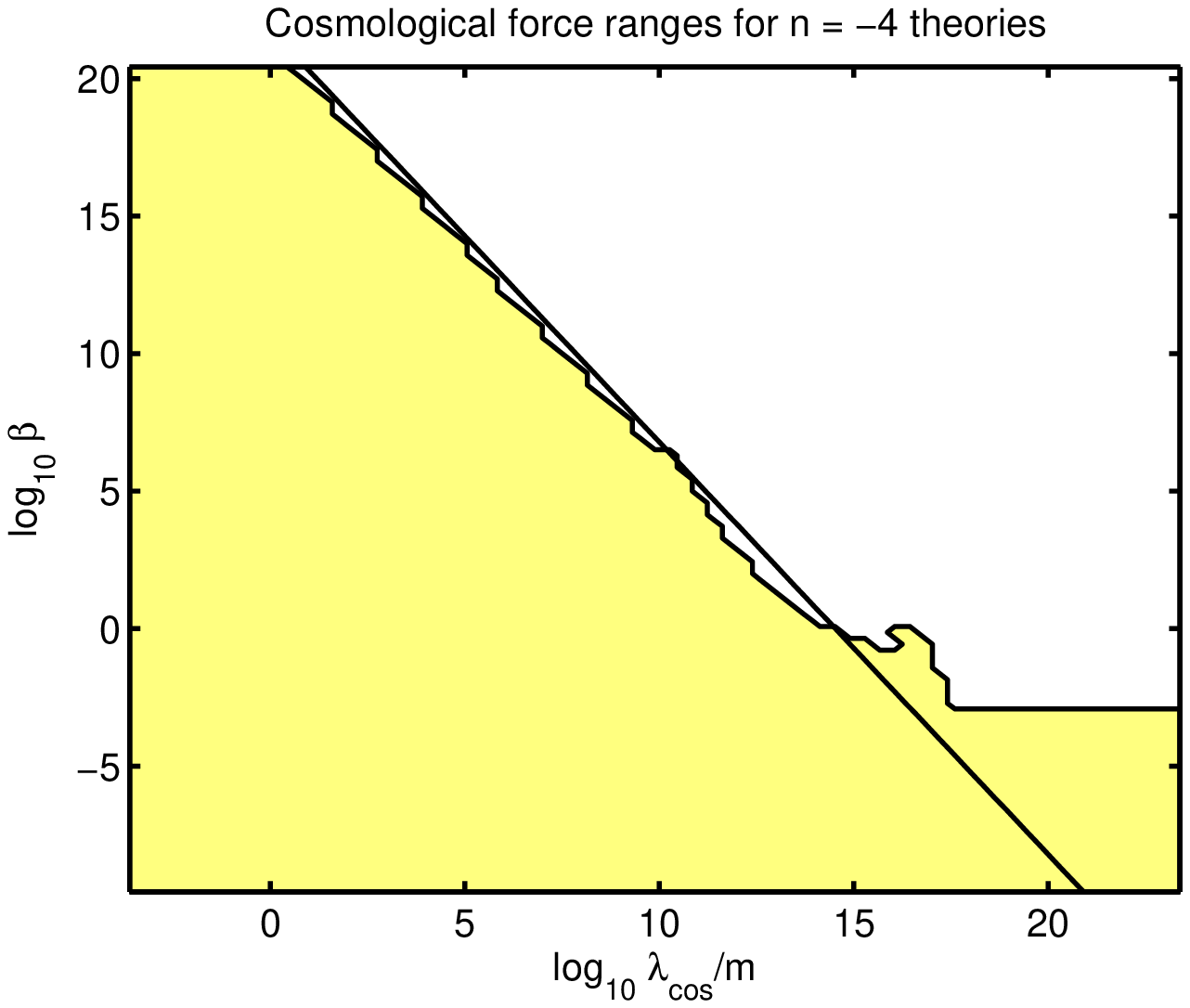}
\includegraphics[width=7.4cm]{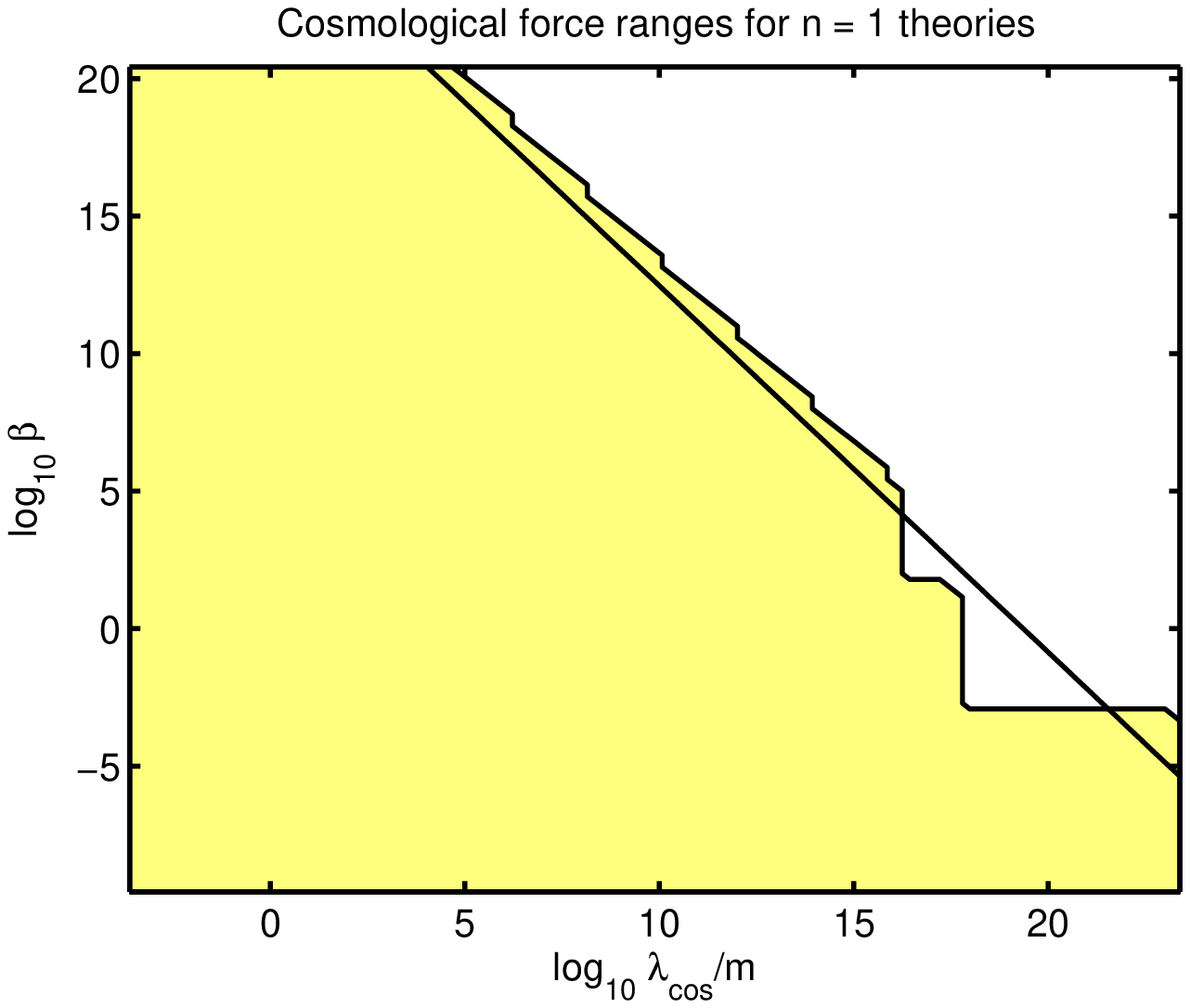}
\includegraphics[width=7.4cm]{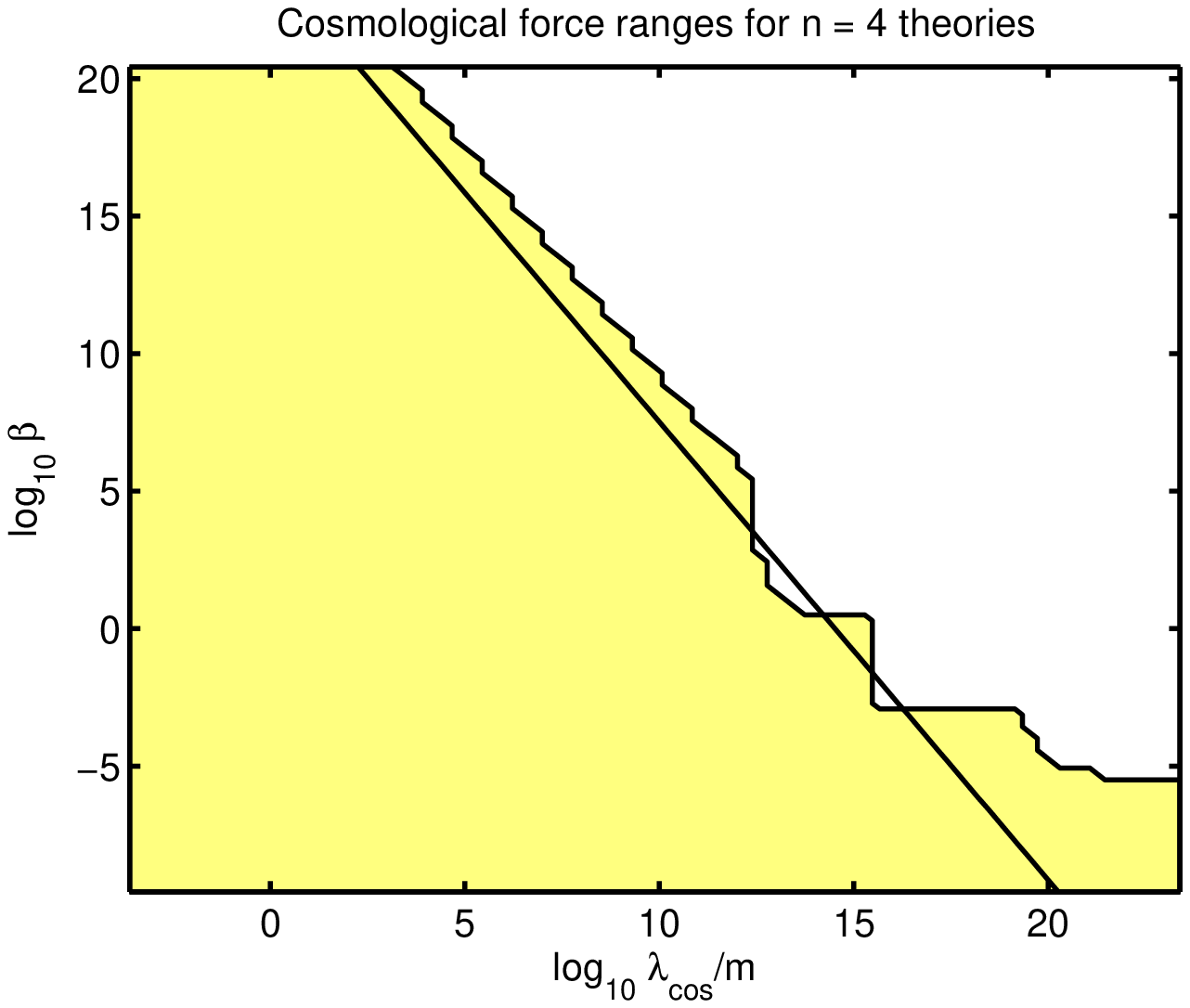}
\includegraphics[width=7.4cm]{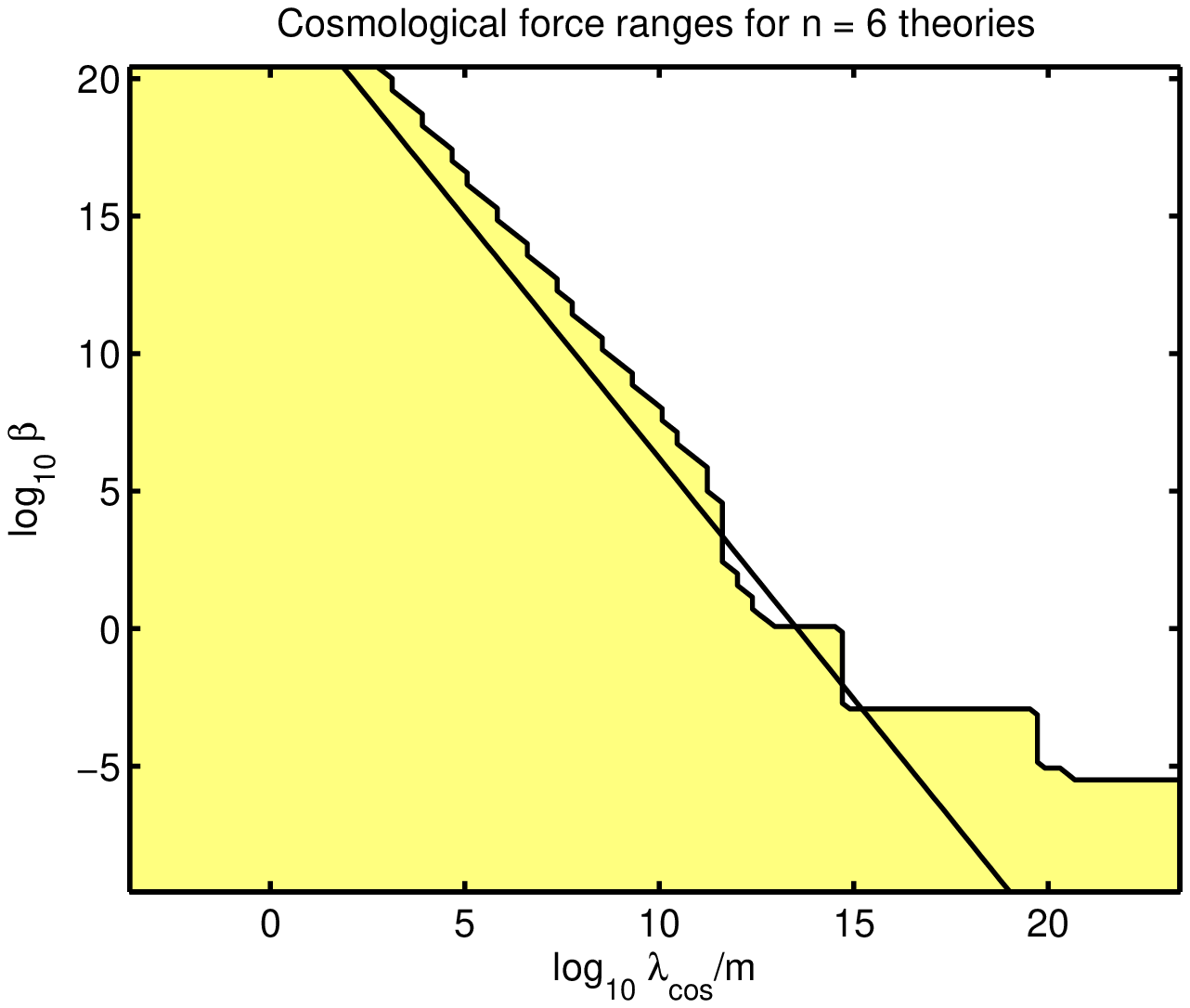}
\end{center}
\caption[Allowed couplings and cosmological force ranges for chameleon theories.]{[Colour Online] Allowed couplings and cosmological force ranges for chameleon theories.  The shaded area shows the allowed parameter space
with all current bounds.  $\lambda_{cos}$ is the range of the
chameleon force in the cosmological background with density $\rho \sim 10^{-29}\,{\mathrm{g\,cm}}^{-3}$. It is related to the
cosmological mass of the chameleon, $m_{c}^{cos}$, by $\lambda_{cos} =
1/m_{c}^{cos}$. The solid black lines indicate the cases where $M$ and $\lambda$ take `natural' values. Plots for theories with $n < -4$ or $n > 0$ are
similar to the cases $n=-8,\,-6$ and $n=4,\,6$ respectively.}
\label{FIGcompcos}
\end{figure}
\begin{figure}[tbh]
\begin{center}
\includegraphics[width=7.4cm]{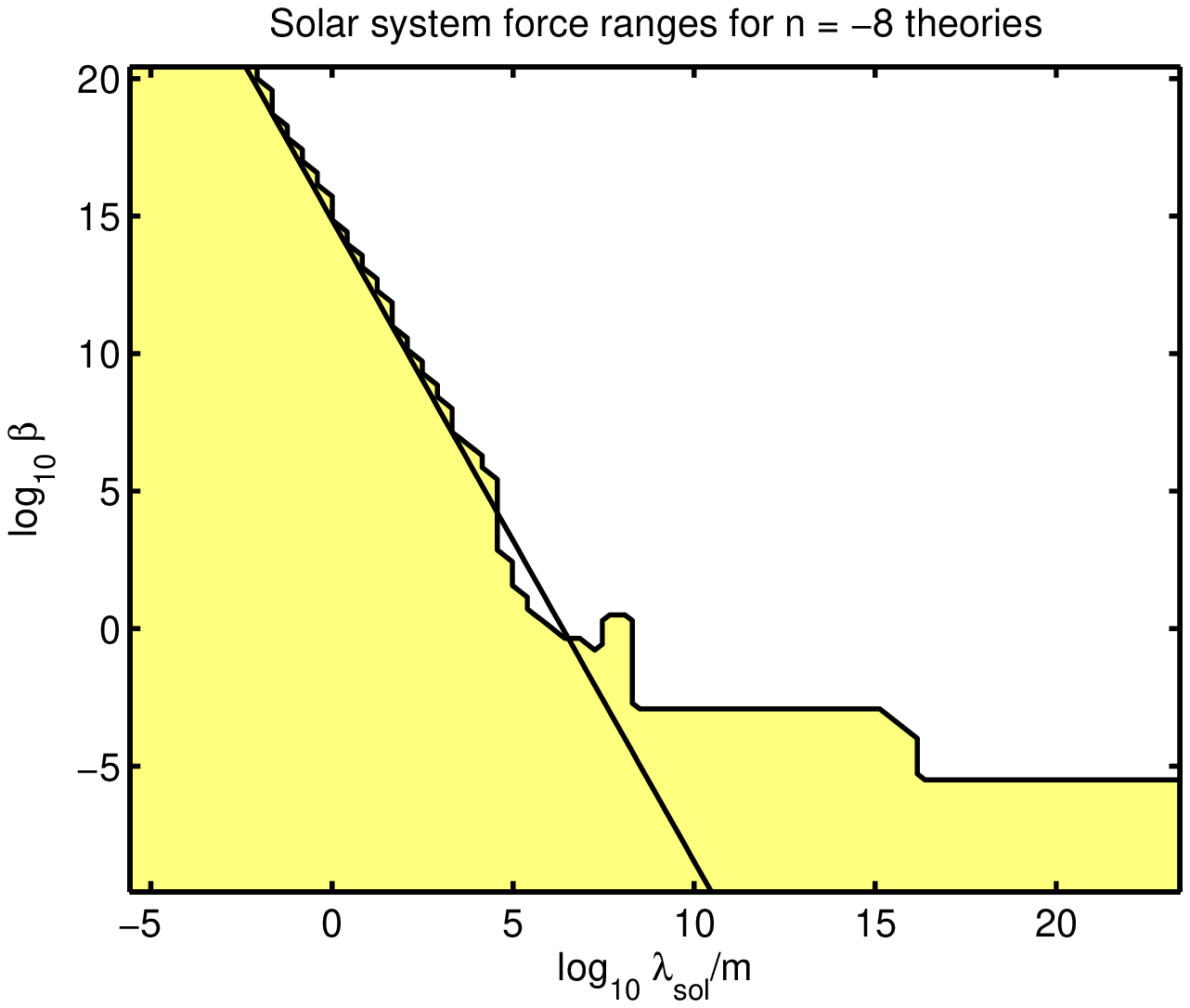}
\includegraphics[width=7.4cm]{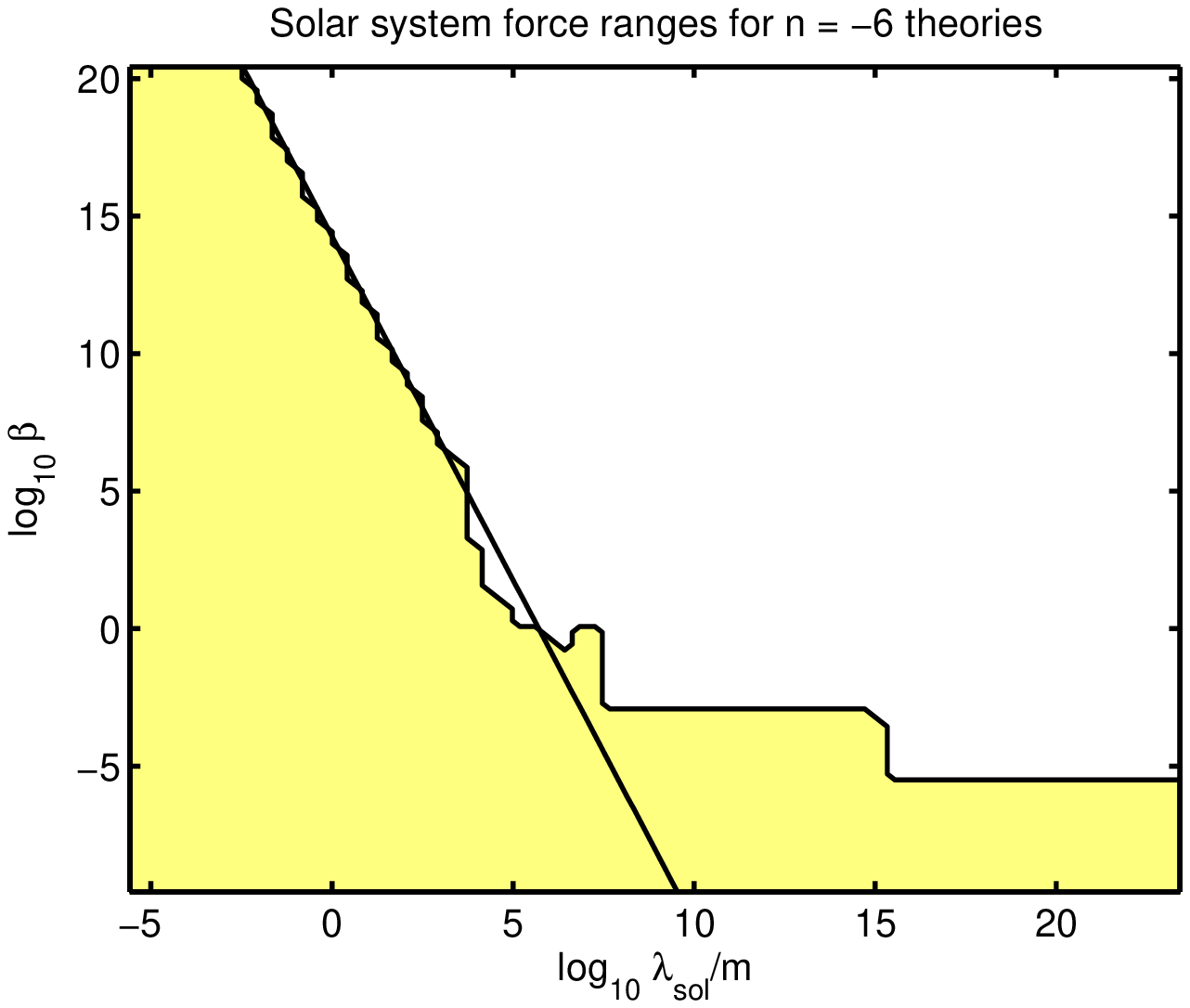}
\includegraphics[width=7.4cm]{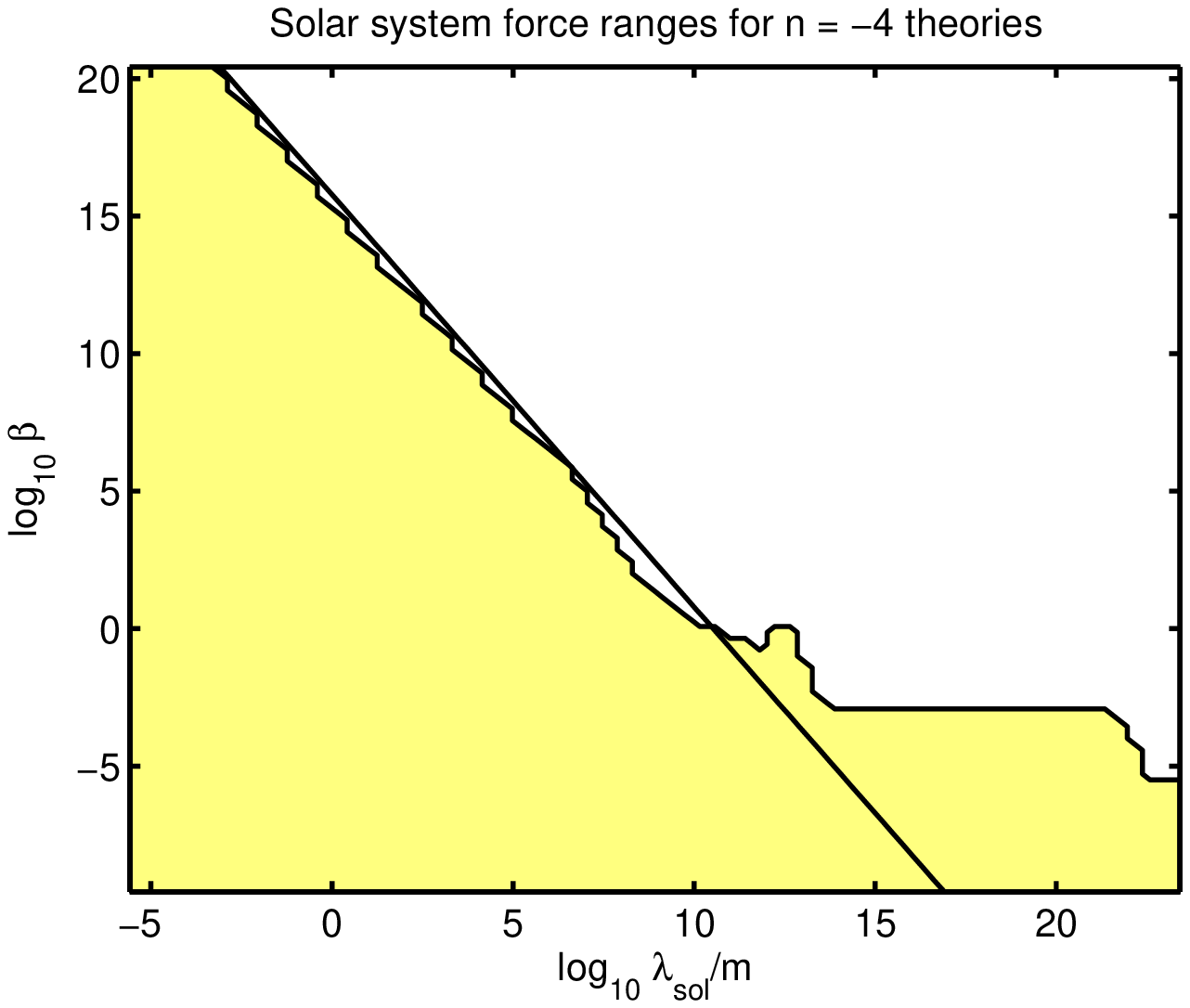}
\includegraphics[width=7.4cm]{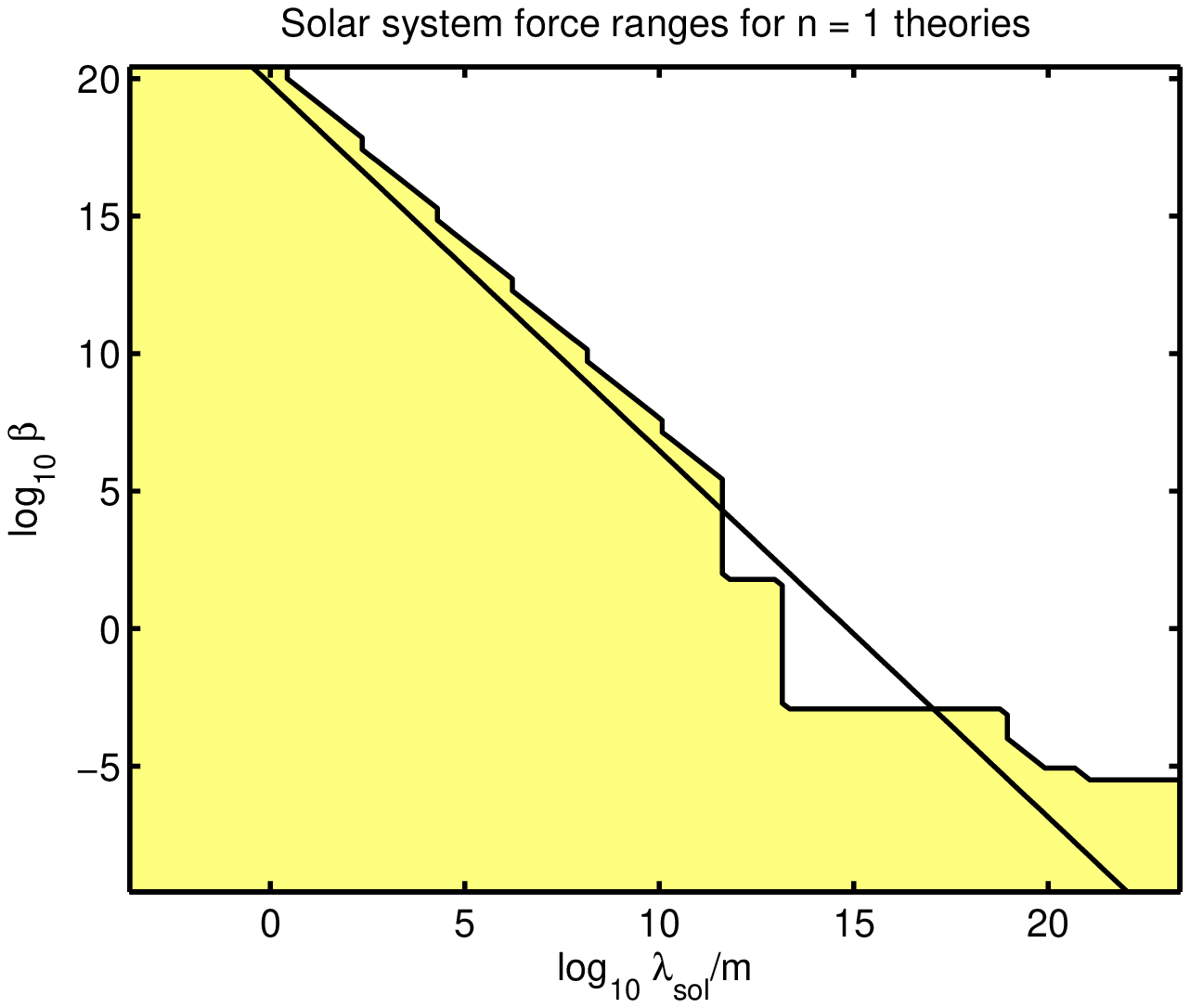}
\includegraphics[width=7.4cm]{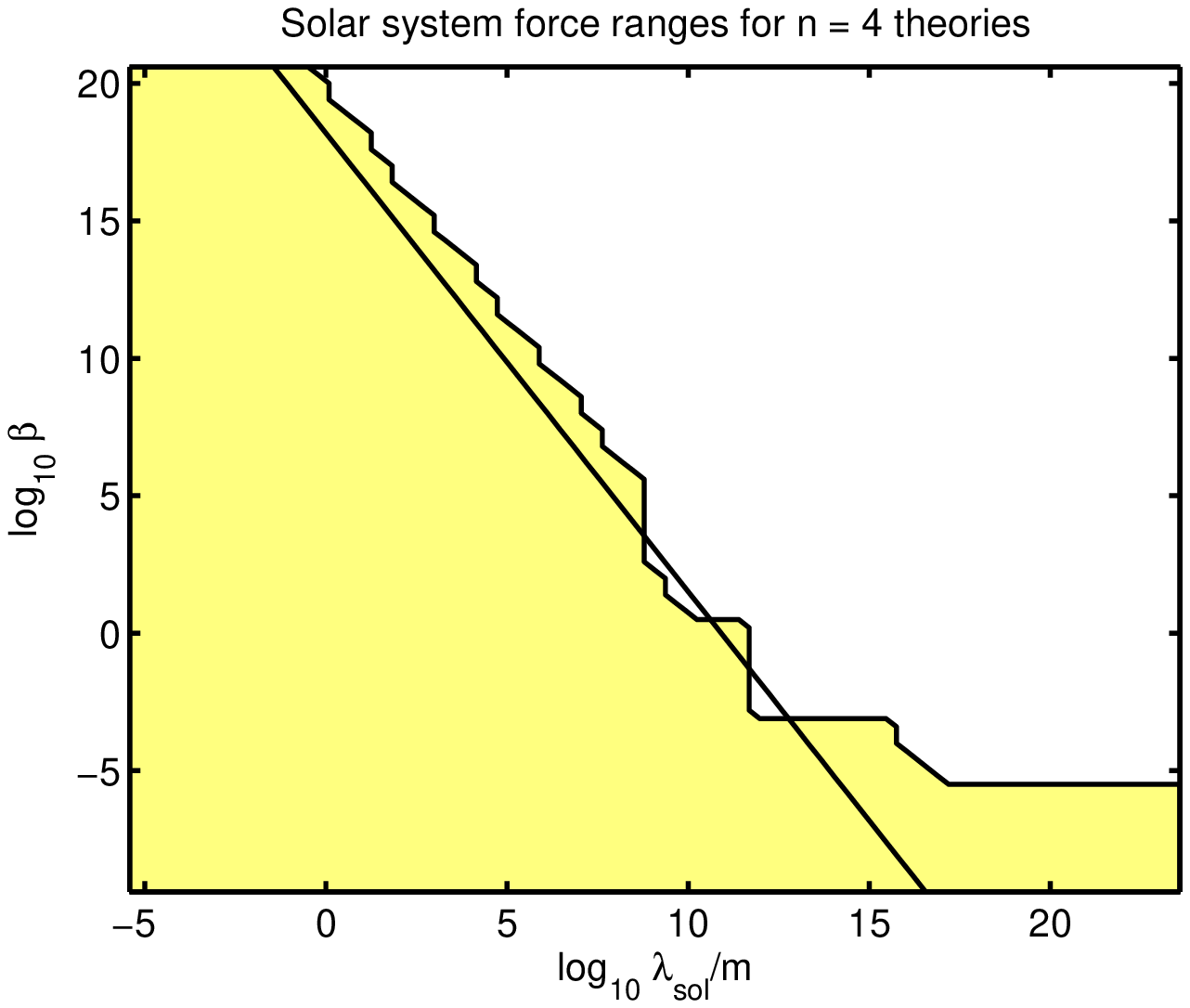}
\includegraphics[width=7.4cm]{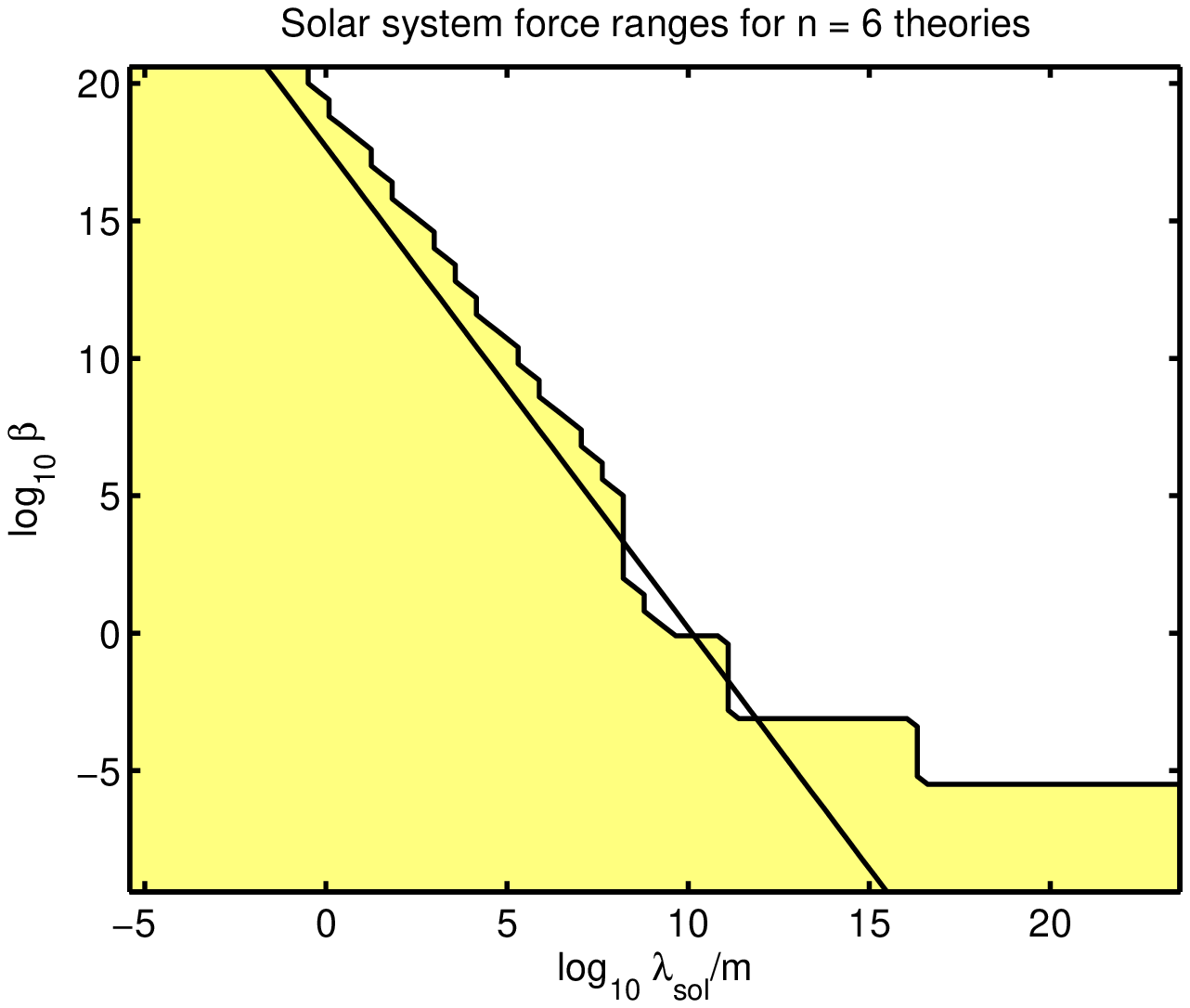}
\end{center}
\caption[Allowed couplings and solar system force ranges for chameleon theories.]{[Colour Online] Allowed couplings and cosmological force ranges for chameleon theories.  The shaded area shows the allowed parameter space
with all current bounds.  $\lambda_{sol}$ is the range of the
chameleon force in the solar system, where the average density of matter is $\rho \sim 10^{-24}\,{\mathrm{g\,cm}}^{-3}$. It is related to the
cosmological mass of the chameleon, $m_{c}^{sol}$, by $\lambda_{sol} =
1/m_{c}^{sol}$. The solid black lines indicate the cases where $M$ and $\lambda$ take `natural' values. Plots for theories with $n < -4$ or $n > 0$ are
similar to the cases $n=-8,\,-6$ and $n=4,\,6$ respectively.}
\label{FIGcompsol}
\end{figure}

The fact that we can have $\beta \gg 1$, in scalar theories with a
chameleon mechanism, is entirely due to the non-linear nature of
these theories.  Almost all the quoted bounds on the coupling to
matter are for scalar field theories with linear field equations.
In such theories $\phi$ evolves according to: $$ -\square \phi = m_c^2 \phi + \frac{\beta
\rho}{M_{pl}}, $$ where the field's mass, $m$, is
constant (i.e. not density dependent).  The $\phi$-force between
two bodies with masses ${\mathcal{M}}_1$ and ${\mathcal{M}}_2$, which are
separated by a distance $d$, takes the Yukawa form
$$ F_{12} = \frac{\beta^2 (1+d/\lambda)e^{-d/\lambda}{\mathcal{M}}_1
{\mathcal{M}}_2}{M_{pl}^2d^2}. $$ where $\lambda = 1/m_c$ is the
\emph{range} of the force.  The best limits of $\lambda$ and
$\beta$ come from WEP violation searches, \cite{WEP,WEP1,WEP2, LLR,LLR1}, and
searches for corrections to the $1/r^2$ behaviour of gravity,
\cite{EotWash}.  The 95\% confidence limits on $m$ and $\beta$ for
such a linear theory, where the field couples to baryon number, are plotted in FIG. \ref{FIGlin} with the allowed regions shaded and the excluded regions left white. It is clear that $\beta > 1$ is ruled out for all but the smallest
ranges (currently $\lambda \lesssim 10^{-4}\,{\mathrm{m}} =
0.1\,{\mathrm{mm}}$).

To make the comparison with the linear case more straightforward,
we have replotted the allowed parameter space for chameleon
theories with $n=-8,\,-6,\,-4,\,4$ and $n=6$ in terms of its
coupling to matter, $\beta$, and the range of the chameleon force
cosmologically, $\lambda_{cos}$, and in the solar system,
$\lambda_{sol}$, in FIGS. \ref{FIGcompcos} and \ref{FIGcompsol}.
FIG. \ref{FIGcompcos} shows the cosmological range, whilst
$\lambda_{sol}$ is shown in FIG. \ref{FIGcompsol}. The solid black
line, in each of these plots, indicates the case where $M$ and
$\lambda$ take their `natural' values.  We can clearly see that,
in stark contrast to the linear case, chameleon theories are
easily able to accommodate both $\beta \gg 1$ and $\lambda \gg
1\,{\mathrm{m}}$. This
underlines the extent to which non-linear scalar field theories
are different from linear ones, and the important r\^{o}le that is
played by the chameleon mechanism.

\section{ Conclusions and Discussion \label{conclude}}

In this article we have investigated scalar field theories in which the field is strongly coupled to matter. 
In particular, we have studied the so called chameleon scalar fields. A scenario presented by Khoury and Weltman
\cite{cham1} that employed self-interactions of the scalar-field to
avoid the most restrictive of the current bounds on such fields and its coupling.
In the models that they
proposed, these fields would couple to matter with gravitational
strength, in harmony with general expectations from string theory,
whilst, at the same time, remaining very light on cosmological
scales.   
In this work we went  much further and show, contrary to most expectations, that the scenario presented in \cite{cham1} allows scalar fields, which are very light on cosmological scales, to couple to matter much \emph{more} strongly than gravity does, and yet still satisfy \emph{all} of the current experimental and observational constraints.

Previous investigations on such scenarios, \cite{chama, chameleoncosmology,
nelson,phi4}, noted that an important feature of chameleon field
theories is that they make unambiguous and testable predictions
for near-future tests of gravity in space. This is timely as there
are currently four satellite experiments either in the proposal
stage or due to be launched shortly (SEE~\cite{SEE},
STEP~\cite{STEP}, GG~\cite{GG} and MICROSCOPE~\cite{MICRO}).  A
reasonably sized region of the parameter space of the chameleon
theories considered here will be visible to these missions.
Theories with very large couplings $\beta \gg 1$ will, however,
remain undetectable.   The ability of these planned missions to
detect large $\beta$ theories could, however, be exponentially
increased if the experiments they carry were to be redesigned slightly in
the light of our findings.

Previous studies claimed that typical test masses in the above
satellite experiments do not have a thin shell. Therefore, the
extra force is comparable to their gravitational interaction. The
chameleon model hence predicts that MICROSCOPE, STEP and GG could
measure violations of the weak equivalence principle that are
stronger than currently allowed by laboratory experiments.
Furthermore, the SEE project could measure an effective Newton's
constant different, by order of unity, from that measured on
Earth.  We have seen, in this paper, that both of these
features are very much properties of chameleon theories with
runaway ($n>0$) potentials. They will not, in general, occur if
the chameleon potential has a minimum (e.g. $n<-4$ theories).
These features are also very much associated with a gravitational
strength chameleon coupling i.e. $\beta \sim {\mathcal{O}}(1)$,
and will not occur if $\beta \gg 1$, or $\beta \ll 1$.

The major result, presented in this work, is that current
experiments do \emph{not} limit the coupling of the chameleon to
matter, $\beta$, to be order ${\mathcal{O}}(1)$ or smaller.
Indeed, if we wish to have a `natural' value of $M$  in a
$V\propto \phi^{-1}$ theory then we must require $\beta \gtrsim 10^{4}$ (or $\beta \lesssim 10^{-3}$.  If
$\beta \gg 1$, the test-masses in the planned satellite
experiments will still have thin-shells. As such SEE, STEP, GG and
MICROSCOPE, as they are currently proposed,  will be \emph{unable}
to detect the chameleon and place an upper-bound on $\beta$.

We have shown that upper-bounds on the matter coupling, $\beta$, can be derived from astrophysical and cosmological considerations. Also, if $\beta$ is very large, of the order of $10^{17}$ or greater, then it might even be possible to detect the effect on the chameleon on scattering amplitudes in particle colliders.  This possibility is one avenue that requires further in depth study.

We noted in the introduction that $\beta \gg 1$ could be seen as
being pleasant in light of the hierarchy problem \cite{HP,HP1,HP2}.
If a chameleon with a large $\beta$ where detected, it would imply
that new physics emerges at a sub-Planckian energy-scale:
$M_{\phi} = M_{pl}/\beta$.  

A large value of the matter coupling is also preferable to an
order unity value in that it leads to the late time behaviour of
the chameleon being much more \emph{weakly} dependent on the
initial conditions, than it would be if $\beta \lesssim
{\mathcal{O}}(1)$.

The magnitude of $\beta (\Delta\phi = \phi_1 - \phi_2)/M_{pl}$
quantifies the relative amount by the particle masses differ between a
region where $\phi = \phi_1$ and one where $\phi=\phi_2$. The
larger the coupling is then, the easier it is for there to have
been a very large difference between the current values of
particle masses, and the values that they had in the very early
universe (i.e. pre-BBN). If the particle masses were very
different from their present values at, say, the epoch of the
electroweak phase transition, then the predictions of electroweak
baryogenesis could be significantly altered.

In this paper have taken the chameleon potential to have a
power-law form i.e $V\propto \phi^{-n}$.   This is certainly
\emph{not} the only class of chameleon potentials that it is
possible to have. In general, any potential that satisfies $\beta
V_{,\phi} < 0$, $V_{,\phi\phi} > 0$ and
$V_{,\phi\phi\phi}/V_{,\phi} > 0$ in a region near some $\phi =
\phi_{0}$ will produce a chameleon theory.  In fact, a generic
potential might contain many different regions in which chameleon
behaviour is displayed. In some of these regions, the potential
may appear to have a runaway form, and so behave qualitatively as
an $n>0$ theory. In other regions, the potential might have a
local minimum, leading to $n \leq -4$ type behaviour.   The
existence of the matter coupling provides one with a mechanism,
along the lines of that considered by Damour and Polyakov
\cite{pol}, by which the scalar field $\phi$ can, during the
radiation era, be moved into a region where it behaves like a
chameleon field.   The larger $\beta$ is, the more effective this
mechanism becomes. Given this mechanism, one important avenue for
further study is to see precisely how general late-time chameleon
behaviour is of a generic scalar field theory with a strong
coupling to matter.

In this paper we have avoided the temptation to linearise
the chameleon field equation, eq. (\ref{micro}), when it is not
valid to do so.  We have, instead, combined matched asymptotic
expansions with approximate analytical, and exact numerical,
solution of the full non-linear equations to study the behaviour
of chameleon field theories in more detail.

The main results of this analysis were:
\begin{itemize}
\item We found the conditions under which a body would have a
thin-shell, and noted that the development of a thin-shell is related to
the onset of non-linear behaviour.
\item We have shown that the far field of a body with a thin-shell is
\emph{independent} of the coupling strength $\beta$: this is a generic property of all the chameleon field
theories with a power-law potential.  This $\beta$-independence
was seen, in section \ref{exper}, to have important consequences
for the design of experiments that search for WEP violations, and
was seen to be vital in allowing theories with $\beta \gg 1$ to be
compatible with the current experimental bounds. The $\beta$ independence of the far-field of thin-shelled bodies was seen to cause the $\phi$-force between two
such bodies to be $\beta$-independent also. 
\item For $\beta \sim O(1)$ the best bounds on $M$ and $\lambda$ currently come from the E\"{o}t-Wash experiment \cite{EotWash,Eotnew}  For $\beta \gg 1$, the best bounds on $M$ and $\lambda$ were found to come from measurements of the Casimir force.
\item Non-linear effects were shown to limit the magnitude of
the average chameleon mass in a thin-shelled body to be smaller
than some critical value, $m^{crit}_c$. Intriguingly, $m^{crit}_c$ is independent of $\beta$, $M$ and $\lambda$, and
depend only on $n$ and the \emph{microscopic} properties of the
thin-shelled body.
\item The experimental constraints on the coupling of chameleon fields to matter are much weaker than those on non-chameleon fields.  In fact when $n \neq -4$, the constraints on large $\beta$ theories are weaker than those in which the scalar field couples to matter with gravitational strength ($\beta \sim O(1)$). Almost paradoxically, strongly-coupled scalar fields are actually harder to detect than weakly coupled ones.
\item Perhaps the most important result, though, is that the ability of table-top gravity tests to see strongly
coupled, chameleon fields could be exponentially increased if
certain features of their design could be adjusted in the
appropriate manner.  The detection, or exclusion, of chameleon
fields with $\beta \gg 1$ represents a significant but
ultimately, we believe, achievable challenge to experimentalists. Searches for large $\beta$ chameleon fields represent one way in which table-top tests of gravity could be used to probe for new physics beyond the standard model. Whether, or not, chameleon fields actually exist, it is important to note those areas into which our current experiments cannot see and, if possible, design experiments to probe those areas. 
\end{itemize}

We have shown, in this paper, that scalar field theories
that couple to matter much more strongly than gravity are not only
viable but could well be detected by a number of future
experiments provided they are properly designed to do so.  This
result opens up an altogether new window which might lead to a
completely different view of the r\^{o}le played by light scalar
fields in particle physics and cosmology.

\begin{acknowledgments} We would like to thank J. Barrow, T. Clifton,
T. Dent, M. Doran, H. Gies, J. Khoury, N. Nunes, A. Upadhye and C. Wetterich and  for helpful discussions and
comments.  DFM acknowledges support from the Research Council of
Norway through project number 159637/V30, the Humboldt Foundation and
of the Perimeter Institute for Theoretical Physics.  DJS acknowledges PPARC.
\end{acknowledgments}
\appendix
% ********** Appendix 1 **********
\section{Pseudo-Linear regime for single-body problem \label{singpseuApp}}

In the pseudo-linear approximation, we assume that non-linear
effects are, locally, everywhere sub-leading order. The
cumulative, or integrated, effect of the non-linearities is not,
necessarily small. This means that, whilst we assume that there
always exists at least one self-consistent linearisation of the
field equations about every point, we do not require there to be a
linearisation that is \emph{everywhere} valid. Instead we aim to
contrast two linearisations of the field equations: the
\emph{inner} and the \emph{outer approximations} to $\phi$.

The \emph{inner approximation} is intended to be an asymptotic
approximation to the chameleon that is valid both inside an
isolated body, and close to the surface of that body.  We take the
isolated body to be spherically symmetric, with uniform density
$\rho_{c}$ and radius $R$.  Far from the body, $r \gg R$, the
inner approximation will, in general, break down.

The \emph{outer approximation} is an asymptotic approximation to
$\phi$ that is valid for large values of $r$. We require that it
remains valid as $r \rightarrow \infty$.  In general the outer
approximation will not be valid for $r \sim \mathcal{O}(R)$.

The boundary conditions on the evolution of $\phi$ are:
$$ \left.\frac{d \phi}{dr}\right\vert_{r=0} = 0, \qquad \left.\frac{d
\phi}{dr}\right\vert_{r=\infty} = 0.
$$ The first of these conditions is defined at $r=0$.  We will
generally find that \emph{only} the inner approximation is valid at $r=0$. As a result we
cannot apply the $r=0$ boundary condition to the outer-approximation.
Similarly the $r=\infty$ boundary condition will be applicable to the
outer-approximation but not to the inner one.

Since we cannot directly apply all the boundary conditions to both
approximations, there will generally be undefined constants of
integration in both the inner and outer expansions for $\phi$.
This ambiguity in both expansions can, however, be lifted if there
exists some \emph{intermediate} range of values of $r$ ($r_{out}<
r < r_{in}$ say) where both the inner \emph{and} outer
approximations are valid.

Asymptotic expansions are locally unique \cite{hinch, hinch1}. Thus, if both the outer and inner approximation are simultaneously valid in some intermediate region, then they must equal to each other in that region.  By appealing to this fact, we can match the inner and outer approximations in the intermediate region. In this way, we effectively apply \emph{all} of the boundary conditions to \emph{both} expansions. This \emph{method of
matched asymptotic expansions} is described in more detail in \cite{shawbarrow1}.

\subsection{Inner approximation}
Inside the body, $0\leq r \leq R$, the chameleon obeys:
\begin{eqnarray}
\frac{d^2 \phi}{dr^2} + \frac{2}{r}\frac{d \phi}{dr} = -n\lambda
M^3\left(\frac{M^3}{\phi}\right)^{n+1} + \frac{\beta
\rho_c}{M_{pl}}. \label{phiAppev}
\end{eqnarray}
The inner approximation is defined by the assumption $$ n\lambda
M^3\left(\frac{M^3}{\phi}\right)^{n+1} \ll \frac{\beta
\rho_c}{M_{pl}}. $$ Defining $$ \phi_c = M\left(\frac{\beta
\rho_c}{n \lambda M_{pl}M^3}\right)^{-\frac{1}{n+1}}, $$ \noindent
we see that the above assumption is equivalent to: $$ \delta(r) :=
\left(\frac{\phi_c}{\phi(r)}\right)^{n+1} \ll 1 $$ We define the
inner approximation by solving eq. (\ref{phiAppev})
 for $\phi$ as an asymptotic expansion in the small parameter
$\delta(0)$; we shall see below that $\delta(r)< \delta(0):=\delta$.

Whenever the inner approximation is valid we have:
$$ \phi \sim \phi_0 + \frac{\beta \rho_c r^2}{6 M_{pl}} + \mathcal{O}(\delta),
$$ where the order $\delta$ term is:
$$ \phi_{\delta}(r) = \frac{\delta\beta \rho_c}{r M_{pl}}\int_{0}^{r}
dr^{\prime} \int_{0}^{r^{\prime}}dr^{\prime \prime}\, r^{\prime
\prime} \left(\frac{\phi_0}{\bar{\phi}(r^{\prime
\prime})}\right)^{n+1}.
$$
\noindent We have defined $\bar{\phi}(r) := \phi_0 + \beta \rho_c r^2 / 6M_{pl}$
for $r<R$. $\phi_0$ is an undefined constant of integration.  It
will be found by the matching of the inner approximation to the outer
one. For the inner approximation to remain valid inside the body we
need:
$$ \frac{\phi_{\delta}(r)}{\bar{\phi}(r)} \ll 1.
$$
Outside the body, $r > R$, $\phi$ obeys:
$$ \frac{d^2 \phi}{dr^2} + \frac{2}{r} \frac{d \phi}{dr} = -n \lambda
M^3 \left(\frac{M^3}{\phi}\right)^{n+1}.
$$
Whenever $\delta(r) < 1$, we can solve the above equation in the
inner approximation finding:
$$ \phi \sim \bar{\phi}(r) + \phi_{\delta}(r).
$$
Outside the body, $r > R$, we define $\bar{\phi}(r)$ to be
\begin{eqnarray}
\bar{\phi}(r) = \phi_0 + \frac{\beta \rho_c r^2}{2 M_{pl}} -
\frac{\beta \rho_c R^3}{3 M_{pl} r}. \label{phibar}
\end{eqnarray}
The order $\delta$ term, $\phi_{\delta}(r)$, is given by the same expression as it was for $r < R$.

The inner approximation will therefore be valid, both inside and
outside the body, provided that: $$
\frac{\phi_{\delta}(r)}{\bar{\phi}(r)} \ll 1. $$ In general, this
requirement will only hold for $r$ less than some finite value of
$r$, $r=r_{in}$ say. To fix the value of $\phi_0$, and properly
evaluate the above condition, we must now consider the outer
approximation.

\subsection{Outer Approximation}
When $r$ is very large, we expect that the presence of the body
should only induce a small perturbation in the value of $\phi$.
Assuming that, as $r \rightarrow \infty$, $\phi \rightarrow
\phi_b$, where $\phi_b$ is the value of $\phi$ in the background,
then the outer approximation is defined by the assumption $\vert
(\phi - \phi_b) / \phi_b \vert < 1/\vert n+1 \vert.$ We may
therefore write: $$ -n \lambda
M^3\left(\frac{M^3}{\phi}\right)^{n+1} \sim -n \lambda
M^3\left(\frac{M^3}{\phi_b}\right)^{n+1} + m_b^2 (\phi-\phi_b) +
O\left((\phi/\phi_b -1)^2\right) $$ where $$ m_b^2 = \lambda
n(n+1) M^2 \left(\frac{M}{\phi}\right)^{n+2}, $$ is the mass of
the chameleon in the background.  The assumption that $\vert (\phi
- \phi_b) / \phi_b \vert < 1/\vert n+1 \vert$ is essentially the
same assumption as was made in the linear approximation in section
\ref{singlin}.  In the linear approximation, however, this
assumption was required to hold all the way up to $r=0$. All that
is required for the pseudo-linear approximation to work, is that
the outer approximation be valid for all $r > r_{out}$, where
$r_{out}$ is any value of $r$ less than $r_{in}$.  This is to say
that, we need there to be some intermediate region where both the
inner and outer approximations are simultaneously valid.

Outside of the body $\phi$ obeys: $$ \frac{d^2 \phi}{dr^2} +
\frac{2}{r} \frac{d \phi}{dr} = -n \lambda M^3
\left(\frac{M^3}{\phi}\right)^{n+1} + \frac{\beta\rho_b}{M_{pl}}.
$$ For the outer approximation to remain valid as $\{r \rightarrow
\infty,\, \phi \rightarrow \phi_b\}$  we need $$ n \lambda M^3
\left(\frac{M^3}{\phi_b}\right)^{n+1} =
\frac{\beta\rho_b}{M_{pl}}. $$ Solving for $\phi$ in the outer
approximation, we find $\phi \sim \phi^{\ast}$ where:
\begin{eqnarray}
\phi^{\ast} = \phi_b - \frac{A e^{-m_b r}}{r}. \label{phiast}
\end{eqnarray}
\noindent $A$ is an unknown constant of integration.  It will be
determined through the matching procedure.

\subsection{Matching Procedure}
We assume that there exists an intermediate region, $r_{out} < r <
r_{in}$, where both the inner and outer approximations are valid.
This region does not need to be very large. All that is truly
needed is for there to exist an open set, about some point $r=d$,
where both approximations are valid.  We shall consider what is
required for such an open set to exist in section \ref{pseudval}
below. For the moment we shall assume that it does exist, and
evaluate $\phi_0$ and $A$. In the intermediate region we must have
$$ \phi \sim \bar{\phi} \sim \phi^{\ast}, $$ by the uniqueness of
asymptotic expansions. The requirement that $\delta(r) \ll 1$
ensures that $m_b r \ll 1$ in any intermediate region.  Expanding
$\phi^{\ast}$ to leading order in $m_b r$ and equating it to
$\bar{\phi}$ we find that
\begin{eqnarray*}
\phi_0 &=& \phi_b - \frac{\beta \rho_c}{2M_{pl}}, \\ A &=& \frac{\beta
\rho_c}{3M_{pl}}.
\end{eqnarray*}
Now that the previously unknown constants of integration, $A$ and
$\phi_0$, have been found, we can evaluate the conditions under which an
intermediate region actually exists.

\subsection{Conditions for Matching \label{pseudval}}
For the inner approximation to be valid we must certainly require that
$(\bar{\phi}(r=0)/\phi_c)^{-(n+1)} \ll 1.$ This is equivalent to:
\begin{eqnarray}
(m_c R)^2 \ll 2\vert n+1 \vert\left((\rho_c/\rho_b)^{1/(n+1)}
-1\right). \label{inhold}
\end{eqnarray}
It is also interesting to note what is required for the pseudo-linear approximation to be valid outside the body i.e. without requiring it to be valid for $r<R$.  This gives the weaker condition:
$$ (m_c R)^2 \ll 3\vert n+1\vert\left\vert(\rho_c/\rho_b)^{1/(n+1)}
-1\right\vert.
$$
If this weaker
condition also fails then, irrespective of what occurs inside the body, we cannot even have a solution where $\phi \sim \bar{\phi}$ outside of the body.  As we show below, the condition that a body have a thin-shell is equivalent to the requirement that this weaker condition fail to hold.

For an intermediate matching region to exist, there must, at the
very least, exist some $d$ such that, in an open set about $r=d$,
both the inner and outer approximations are valid.  We must also
require that the inner approximation be valid for all $r < d$, and
that the outer approximation hold for all $r > d$.

For the outer approximation to hold for all $r>d$ we need:
$$ \left\vert\frac{(n+1)(\phi^{\ast} - \phi_b)}{\phi_b}\right \vert
\ll 1.
$$ Using $A= \beta \rho_c / 3 M_{pl}$ and eq. (\ref{phiast}) we can see
that this is equivalent to:
\begin{eqnarray}
(m_c R)^2 \frac{R}{3d} \ll
\left(\frac{\rho_c}{\rho_b}\right)^{1/(n+1)}. \label{mccond}
\end{eqnarray}
Using this condition, we define $d_{min}$ to be the value of $d$
for which the left hand side and right hand side of the above
expression are equal: $(m_cR)^2R/3d_{min} =
(\rho_b/\rho_c)^{-1/(n+1)}$. For the outer approximation to be
valid at $d$ we require $d > d_{min}$.

For the inner approximation, $\phi \sim \bar{\phi}$, to hold in
the intermediate region we require that, for all $R< r < d$: $$
\frac{R^3}{3}\gg \int^{r}_{0}d r' \int^{r'}_{0} dr^{\prime \prime}
r^{\prime \prime} \left(\frac{\phi_c}{\bar{\phi}(r^{\prime
\prime})}\right)^{n+1}. $$ We must also have that
$(\phi_c/\bar{\phi}(r=0))^{n+1}$ i.e. that condition
(\ref{inhold}) holds. Provided that this is the case then, for all
$r$ in $(R,d)$, we need
\begin{eqnarray}
\frac{R^3}{3}\gg \int^{d}_{R}d r' \int^{d}_{r'} dr^{\prime \prime}
r^{\prime \prime} \left(\frac{\phi_c}{\bar{\phi}(r^{\prime
\prime})}\right)^{n+1}. \label{intexr}
\end{eqnarray}
If both eqs. (\ref{inhold}) and (\ref{intexr}) hold then the inner
approximation will be valid for all $r < d$. \noindent We evaluate
eq. (\ref{intexr}) approximately as follows: For $n \leq -4$ we
define $r=d'$ by $\bar{\phi}-\phi_b = \phi_b$. For all $r$ in
$(d',d)$ we approximate $\bar{\phi}$ in the above integral by
$\phi_b$, and for all $r$ in $(R,d')$ we approximate $\bar{\phi}$
by $(\bar{\phi}-\phi_b)$.

When $n > 0$, eq. (\ref{inhold}) implies that $\phi_b
>(\bar{\phi}-\phi_b) \gg \phi_c$ and so such no $d'$ exists.   When $n>0$ and condition (\ref{inhold}) holds, we can therefore find a good estimate for the validity of the pseudo-linear regime by setting $\phi=\phi_b$ in the above integral.  We consider the cases $n \leq -4$, $n=-4$ and $n>0$ separately below.

\subsubsection{Case: $n < -4$}
We shall deal with the subcases $d' \leq R$ and $d' > R$ separately.
\\ \\
\noindent\textit{Subcase:} $d' \leq R$ \\
\indent The subcase where $d' < R$ includes those circumstances where the
linear regime is valid (see section \ref{singlin}).  Since we
already found, in section \ref{singlin}, how $\phi$ behaves when
linear approximation holds, we shall only consider what occurs
when the linear approximation fails. If the linear approximation
fails, then the outer approximation must break down outside the
body i.e. $d_{min}/R > 1$.

Evaluating eq. (\ref{intexr}), we find that we must, at the very
least, require that: $$ \left(\frac{d_{min}}{R}\right)^3
\frac{\rho_b}{\rho_c}< 1, $$ which, for all $n$, is equivalent to:
\begin{eqnarray*}
m_c R &<&
\sqrt{3}\left(\frac{\rho_c}{\rho_b}\right)^{\frac{n+4}{6(n+1)}},
\\ m_c R &<& \sqrt{3\vert n+1
\vert}\left(\frac{\rho_c}{\rho_b}\right)^{1/2\vert n+1\vert}.
\end{eqnarray*}
The second criteria is just the statement that $d' < R$. When $\rho_b \ll
\rho_c$, this latter condition is more restrictive than the former.
\\ \\
\noindent\textit{Subcase:} $d' > R$ \\
\indent From the definition of $d'$ we find $d'/R = (m_cR)^2/(3\vert
n+1\vert)\left(\rho_c/\rho_b\right)^{1/\vert n+1 \vert}$ and $d' =
d_{min}/\vert n+1 \vert$. It follows that $d \geq d_{min} > d'$.  When $d'>R$, an
intermediate region will exist so long as:
\begin{eqnarray*}
\frac{3d^3}{2R^3}\frac{\rho_b}{\rho_c} +
\frac{3}{(n+4)(n+3)}\left(\frac{(m_c R)^2}{3\vert
n+1\vert}\right)^{\vert n+1 \vert} \ll 1.
\end{eqnarray*}
The smaller $d$ is, the more likely it is that this condition will be satisfied; we therefore evaluate the condition at $d=d_{min}$. Both of the terms on the left hand side are positive, and so for an intermediate region to exist we must have:
\begin{eqnarray*}
m_cR &<& (18)^{1/6}\left(\frac{\rho_c}{\rho_b}\right)^{(n+4)/6(n+1)}
\approx 1.6\left(\frac{\rho_c}{\rho_b}\right)^{(n+4)/6(n+1)}, \\ m_cR
&<& \sqrt{3\vert n+1\vert}\left(\frac{(n+4)(n+3)}{3}\right)^{1/2\vert
n+1\vert}.
\end{eqnarray*}
For $n <-4$, the second of these conditions is usually the more
restrictive. However, since $n < -4$ implies $n \leq -6$, this second
condition is itself, in general, less restrictive than requiring either
$(\phi_c/\bar{\phi}(r=0))^{n+1} \ll 1$ or $(\phi_c/\bar{\phi}(r=R))^{n+1} \ll 1$.
\\ \\
\noindent\textit{Conditions}\\ 
\indent Putting together all of the conditions found above, we see that,
for the pseudo-linear approximation to be valid all the way from
$r = \infty$ to $r=0$ for $n <- 4$, we must, at the very least,
have:
%\begin{widetext}
$$ m_c R <
\min\left((18)^{1/6}\left(\frac{m_c}{m_b}\right)^{(n+4)/3(n+2)}
, \sqrt{2\vert n+1\vert}\left\vert(m_b/m_c)^{2/\vert n+2\vert}
-1\right\vert^{1/2} \right),
$$
%\end{widetext}
where we have used $\rho_c / \rho_b =  (\phi_b/\phi_c)^{n+1} = (m_c/m_b)^{2(n+1)/(n+2)}$.
%NEW UPDATE BEGINS HERE
\\ \\
\noindent\textit{Thin-Shell Condition} \\
\indent If the above condition fails to hold, then it is a sign that
non-linear effects have become important.   However, even when
non-linear effects are important, we only expect them to be so in
a region very close to the surface of the body itself.  In section \ref{singnon} we assumed that our body had a thin-shell and considered the behaviour of the field close to the surface of the body.  We found there to be pronounced non-linear behaviour near the surface of such bodies.  This implies that the pseudo-linear approximation must break down for  $r>R$ for bodies with thin-shells.   We further found that the assumption that the body had a thin-shell required:
$$
\left(\frac{n}{n+1}\right)^{n/2 + 1} m_c R \gg 1.
$$
For a body to have a thin-shell we must therefore require that this condition hold and that the pseudo-linear approximation would break down for $r > R$: these are the thin-shell conditions.  It is almost always the case that the latter of these two conditions is the more restrictive.  In section \ref{singpseu} we found this condition to be:
$$ (m_c
R) >
\min\left((18)^{1/6}\left(\frac{m_{c}}{m_b}\right)^{(n+4)/3(n+2)}
, \sqrt{3\vert
n+1\vert}\left\vert\left(\frac{m_b}{m_{c}}\right)^{1/\vert
n+2\vert} -1\right\vert^{1/2} \right). $$
Since this condition implies $m_c R \gg 1$, it is the \emph{thin-shell} condition for $n < 0$ theories.
%NEW UPDATE ENDS HERE

\subsubsection{Case $n > 0$}
The case of $n>0$ is actually slightly simpler than the $n \leq
-4$ one because we cannot have $d' > R$.  Other than that, the
analysis proceeds in much that same way as it does for the $n<-4$
case. For this reason, we will not repeat the details of the
calculations here.
\\ \\
\noindent\textit{Conditions} \\
\indent For the pseudo-linear
approximation to be valid all the way up to $r=0$ we must, at the very
least, have:
\begin{eqnarray*}
m_c R <
\min\left((18)^{1/6}\left(\frac{\rho_c}{\rho_b}\right)^{\frac{n+4}{6(n+1)}}
, \sqrt{2(n+1)}\left\vert(\rho_c/\rho_b)^{\frac{1}{n+1}}
-1\right\vert^{1/2} \right).
\end{eqnarray*}
When $\rho_b \ll \rho_c$, the most restrictive bound comes from
the second term on the right hand side.
%NEW UPDATE BEGINS HERE
\\ \\
\noindent\textit{Thin-Shell Condition} \\ 
\indent As in $n < 0$ theories, a body with have a thin-shell provided that $m_c R \gg 1$ and non-linear effects are important near the surface of the body which implies that the pseudo-linear approximation breaks down outside the body.  This latter condition in fact implies the former and is therefore the \emph{thin-shell} condition for $n > 0$ theories.  This condition reads:
\begin{eqnarray*}
m_c R >
\min\left(\sqrt{3}\left(\frac{m_c}{m_b}\right)^{(n+4)/6(n+2)} ,
\sqrt{3(n+1)}\left\vert(m_c/m_b)^{1/(n+2)} -1\right\vert^{1/2}
\right),
\end{eqnarray*}
where the second term on the right hand side is usually the more restrictive
when $\rho_{b} \ll \rho_c$.
%NEW UPDATE ENDS HERE

\subsubsection{Case $n=-4$}
The $n=-4$ case, i.e. $\phi^{4}$ theory, requires a more involved
analysis. The reason for this is that, unlike the $n < -4$
theories, there does \emph{not} exist a solution to this theory
where $\phi \sim -A/r$ as $r \rightarrow \infty$. If we propose
such a leading order behavior for the inner approximation, then it
can be easily checked that the next-to-leading order term dies off
as $\ln(r)/r$, i.e. more slowly than the leading order one. This
means that, for some finite $r$, the next-to-leading order term
will dominate over the leading order one. When this happens the
inner approximation will break down. It can also be checked that
higher order terms will always die off more slowly than the terms
of lower order.  This complication will only manifest itself,
however, when the conditions for the pseudo-linear approximation
fail, or almost fail, to hold i.e when $(m_c R)$ is large.
\\ \\
\noindent\textit{Inner approximation and matching} \\
\indent To avoid these difficulties, we shall use a different form for the
leading order behaviour for $\phi$ when $n=-4$.  We write: $$
\frac{\phi}{\phi_c} \sim \frac{\phi_b}{\phi_c} + \frac{\alpha(r)
 e^{-m_b r}}{m_c r}. \label{phi4leada}
$$ To leading order in the inner approximation we neglect terms of
$\mathcal{O}(m_b r)$ and smaller.  The field equation for $\phi$
is then found to be equivalent to: $$ \frac{d^2 \alpha}{d y^2} -
\frac{d \alpha}{d y} \sim \frac{\alpha^3}{3} $$ where $y = y_0
+\ln(r/R)$. The outer approximation is still given by $\phi \sim
\phi^{\ast}$.  To perform the matching we need to know the large
$r$ behaviour of the inner approximation, and the small $r$
behaviour of the outer one.  Solving for $\alpha$ we find that: $$
\alpha \sim \sqrt{\frac{3}{2 y}}. $$ The next-to-leading
correction to $\alpha$ is: $$
\frac{\ln(y)}{2}\left(\sqrt{\frac{3}{2 y}}\right)^3. $$ It is
clear that, as we would wish, the next-to-leading order term dies
off faster than the leading order one. We shall see below that,
for this approximation to be valid near the body, we need $y_0 \gg
0.634$. This approximation also breaks down when $m_b r \gg 1$.
Matching the inner and outer approximations in the same manner as
we did before, we find for $r > R$
\begin{eqnarray}
 \frac{\phi}{\phi_c} \approx \frac{\phi_b}{\phi_c} +
\frac{\alpha(\min(r,1/m_b)) e^{-m_b r}}{m_c r}. \label{phi4lead}
\end{eqnarray}
Inside the body, $r < R$, we assume that, at leading order,
$\phi/\phi_c$ behaves in the same way as if did for all the other
values of $n$: $$ \frac{\phi}{\phi_c} \sim \frac{\phi_0}{\phi_c} +
\frac{(m_cR)^2 r^2}{6(n+1)R^2}. $$ By requiring $\phi$ to be
$C_{1}$ continuous at $r=R$, we find $y_0$ to be
\begin{eqnarray}
\frac{(m_cR)^3}{9} = \sqrt{\frac{3}{2 y_0}} +
\frac{1}{3}\left(\sqrt{\frac{3}{2 y_0}}\right)^3 \label{y0eqn1}
\end{eqnarray} and so $$ \frac{\phi_0}{\phi_c} =
\frac{\phi_b}{\phi_c} + \frac{(m_cR)^2}{18} +
\frac{1}{m_cR}\sqrt{\frac{3}{2y_0}}. $$
\\ \\
\noindent\textit{Conditions} \\
\indent For this approximation to be a valid approximation, we must
firstly require that the next-to-leading order correction to
$\alpha$ is always small compared to the leading order one.  This
implies $$ y_0 \gg 0.508, \quad m_c R \ll 3.13. $$ We must also require
$\phi(r=0)=\phi_0 \ll \phi_c$ which gives (assuming $\rho_b /
\rho_c \ll 1$): $$ y_0 \gg 0.634, \quad m_c R \ll 2.915. $$ This
is the stronger of the two bounds.

When the field equations are solved numerically we find that the form for $\phi$ given above is an accurate approximation whenever $m_c R < 1$.
%NEW UPDATE BEGINS HERE
\\ \\
\noindent\textit{Critical Far Field} \\
\indent The form of the far field for bodies with thin-shells in $n=-4$ theories is examined
in appendix \ref{singnon4App} below.  However, it is interesting
to note that, even in the  pseudo-linear approximation for $n=-4$,
there is already the first hint of $\beta$-independent behaviour in the far field. When $m_b r \ll 1$, we found that
\begin{eqnarray*}
\phi \approx \phi_b -
\frac{(y_0 + \ln(r/R))^{-1/2}e^{-m_b r}}{2\sqrt{2\lambda}r}.
\end{eqnarray*}
Thus, when $\ln(r/R) \gg y_0$ (provided $m_{b} r \ll 1$), we have, to
leading order:
\begin{eqnarray*}
\phi \sim \phi_b -
\frac{e^{-m_b r}}{2\sqrt{2\lambda \ln(r/R)}r},
\end{eqnarray*}
which is manifestly independent of $\phi_c$ and hence also of
$\beta$. In $n \neq -4$ theories, $\beta$-independent critical
behaviour was reserved for bodies with thin-shells. When $n=-4$, we
can see that leading order, $\beta$-independent behaviour in the
far field ($r \gg R$) can occur for \emph{all} $y_0$ i.e.
\emph{all} values of $(m_c R)$. However if $(m_c R) \ll 1$, this
critical behaviour will only be seen for exponentially large
values of $r/R$.  If, however, $m_c R \gtrsim 2$ then $\phi$ will
be $\beta$-independent at leading order for all $r/R \gtrsim 10$.
%NEW UPDATE ENDS HERE

\section{Far field in $n=-4$ theory for a body with thin-shell \label{singnon4App}}
Using eq. (\ref{phiclosef}) and the other results of section
\ref{singnon}, we find that, if a body has a thin-shell, then for
$0 < (r-R)/R \ll 1$ we have $$ \phi \approx
\frac{1}{-\sqrt{2\lambda}(r-R) + \frac{4}{3\phi_c}}, $$ when
$n=-4$. In appendix \ref{singpseuApp}, we saw that, far from the
body and when $m_b r \ll 1$, we have:
\begin{eqnarray*}
\phi \approx \phi_b -
\frac{(y_0 + \ln(r/R))^{-1/2}e^{-m_b r}}{2\sqrt{2\lambda}r}.
\end{eqnarray*} We can find the value that $y_0$ takes for
a thin-shelled body, and hence determine the behaviour of the far
field, by matching the leading order large $r$ behaviour of the
first expression to the leading order behaviour of the second
expression as $r \rightarrow R$.  This gives: $$ 4y_0 = 1
\Rightarrow y_0 = 1/4.$$ Numerical simulations show that this
tends to be a slight over-estimate of the true far-field.  We
find therefore that far from a body with thin-shell, the
chameleon field is approximately given by:
\begin{eqnarray*}
\phi \approx \phi_b -
\frac{(1 + 4\ln(r/R))^{-1/2}e^{-m_b r}}{\sqrt{2\lambda}r}.
\end{eqnarray*}
Indeed, since we are far from the body, it is almost always that
case that $r/R \gg 1.3$ and so $4\ln(r/R) \gg 1$. As a result, it
is usually a very good approximation to take
\begin{eqnarray*}
\phi \sim \phi_b -
\frac{e^{-m_b r}}{2\sqrt{2\lambda \ln(r/R)}r}.
\end{eqnarray*}
We note that, whenever the pseudo-linear approximation holds, $y_0
= 1/4$ is equivalent, by eq. (\ref{y0eqn1}), to an effective value
of $m_c R$ of $(m_c R)_{eff} = 4.04$. Numerical simulations
confirm that there is pronounced, thin-shell behaviour whenever
$m_c R \gtrsim 4$.  When $n=-4$, the condition for a body to have
a thin-shell is therefore $m_c R \gtrsim 4$.  This condition is
shown to be a sufficient condition for thin-shell behaviour in
section \ref{singnon} above.

It is clear that, far from the body, $\phi$ is independent of both
$\beta$ and the mass of the body, $\mathcal{M}$.  There is a very
weak, log-type, dependence on the radius of the body, $R$.  The
strongest parameter dependence is on $\lambda$. We can use the
form of $\phi$ given above to define an effective coupling,
$\beta_{eff}$.  $\beta_{eff}$ is defined so that $$ \phi \approx
\phi_b - \frac{\beta_{eff}(r/R) \mathcal{M}e^{-m_b r}}{4\pi M_{pl}
r}. $$ It follows that, for $r \gtrsim 2R$: $$ \beta_{eff} \approx
\frac{2 \pi M_{pl}}{\mathcal{M} \sqrt{2\lambda
\ln(\min(r/R,1/m_bR))}}. $$

\section{Effective macroscopic theory \label{avgApp}}

The analysis of the effective field theory for a macroscopic body
proceeds along the same lines as the analysis that was performed
to find the field about an isolated body in section \ref{sing},
and appendices \ref{singpseuApp} and \ref{singnon4App}.  A
detailed discussion of precisely what is meant by `effective', or
`averaged', macroscopic theory is given section \ref{effmacr}
above.   For the purposes of this appendix, our aim is to find the
average value of the chameleon mass inside a body in a thin-shell.

We firstly consider what is required for linearisation of the
field equations to be a good approximation. We then introduce a
pseudo-linear approximation.  Finally, we consider what we expect
to see when non-linear effects are strong. We assume that the
macroscopic body has a thin-shell, and is composed of spherical
particles of mass $m_p$ and radius $R$.  The average
inter-particle separation is taken to be $2D$. The average density
of the macroscopic body is: $\rho_{c} = 3m_{p}/4\pi D^3$. The
density of the particles is $\rho_{p} = 3m_{p}/4\pi R^3$.  We
label the average (i.e. volume averaged over a scale $\gtrsim D$)
value of $\phi$ deep inside the body by $<\phi>$. We shall assume
that the effect of the other particles, on a particle at $r=0$, is
sub-leading when $r \ll D$.  In general, the surfaces on which $d
\phi / dr =0$ will not be spherical, but their shape will depend
on how the different particles are packed together.  It will,
however, make the calculation much simpler, and easier to follow,
if we assume that the field about every particle is approximately
spherically symmetric for $0 < r < D$, and that at $r=D$, $d\phi /
d r = 0$.  We take everything to be approximately symmetric
about $r=D$. We define $\phi(r=D) = \phi_c$; $m_c =
m_{\phi}(\phi_c)$. We argue below that $<\phi> \approx \phi_c$,
and that the average chameleon mass is approximately equal to
$m_c=m_{\phi}(\phi_c)$. Our aim therefore is to find $\phi_c$
and $m_c$.

We shall see below that it is possible, for all values of the
parameters $\{D, R\}$, to construct an outer approximation that is
valid near $r = D$.  The outer approximation will be valid so long
as $(\phi - \phi_c)/\phi_c \ll 1/\vert n+1 \vert$.  In the linear
regime, this outer approximation will be valid everywhere. In the
pseudo-linear and non-linear regimes, however, the outer
approximation will only be valid for $r > d_{min}$ where $$
d_{min} = \frac{m_c^2 D^3}{3}. $$

When $n \neq -4$, we shall find that $R \ll D$ implies that we always
have $m_c D \ll 1$. When $n = -4$, $R \ll D$ implies that $m_c R \lesssim \sqrt{3}$ is always true. It follows that $(d_{min}/D)^3 \ll 1$. The
volume in which the outer approximation holds is: $$
\mathcal{V}_{out} = 4\pi (D^3-d_{min}^3)/3. $$ Since
$(d_{min}/D)^3 \ll 1$, $\mathcal{V}_{out} \approx 4\pi D^3/3$ i.e.
the entire volume of the region $0 <r < D$. This means that the
volume averaged value of $\phi$, and $m_{\phi}(\phi)$, will be
dominated by the value $\phi$ takes in the outer expansion. Since
$\phi \approx \phi_c$ in the outer expansion, and $m_{\phi}
\approx m_c$, it follows that $<\phi> \approx \phi_c$ and
$<m_{\phi}> \approx m_c$.

Throughout this appendix, we shall therefore refer to $\phi_c$
as the average value of $\phi$, and $m_c$ as the average chameleon
mass. The averaged behaviour of $\phi$, in a body with a
thin-shell, is entirely determined by $\phi_c$ and $m_c$.  Our aim
of finding an effective macroscopic theory is therefore equivalent
to calculating $\phi_c$ and $m_c$.

If linear theory holds inside the body then $\phi_c = \phi_c^{(lin)}$ where:
$$ \phi_c^{(lin)} = M\left(\frac{\beta \rho_{c}}{n \lambda
M_{pl}M^3}\right)^{-1/(n+1)},
$$ We also define:
$$\phi_p = M\left(\frac{\beta \rho_{p}}{n \lambda
M_{pl}M^3}\right)^{-1/(n+1)}.$$

\subsection{Linear Regime}
We write $\phi = \phi_c + \phi_1$, where $\vert \phi_1/\phi_c
\vert \ll 1$, and $\phi(r=D)=\phi_c$. Linearising the equations
about $\phi_0$, one obtains:
\begin{eqnarray}
\frac{d^2 (\phi_1/\phi_c)}{d r^2} + \frac{2}{r}\frac{d
(\phi_1/\phi_c)}{dr} = -\frac{m_c^2}{n+1} + m_c^2(\phi_1/\phi_c) +
\frac{3\beta M_{b}}{4\pi R^3 \phi_c}H(R-r),
\end{eqnarray}
and $$ \frac{3\beta M_{b}}{4\pi R^3 \phi_c} = \frac{m_c^2
D^3}{(n+1)R^3}\left(\frac{\phi_c}{\phi_c^{(lin)}}\right)^{n+1}. $$
For this linearisation of the potential to be valid we need, just
as we did in section \ref{singlin}, $\vert \phi_1 / \phi_c \vert
\ll 1/\vert n+1 \vert$. Solving these equations is straightforward
and we find that for $r
> R$:
$$ (n+1)\frac{\phi_1}{\phi_c} = 1- \frac{\cosh(m_c(r-D))D}{r} -
\frac{\sinh(m_c(r-D))}{m_c r}.
$$ Inside the particle, $r < R$, we have:
\begin{eqnarray*}
(n+1)\frac{\phi_1}{\phi_c} &=& 1-
\frac{D^3}{R^3}\left(\frac{\phi_c}{\phi_c^{(lin)}}\right)^{n+1}\left(1-\frac{\sinh(m_cr)R}{\sinh(m_cR)r}\right)-\\
&-& \frac{(\cosh(m_c(R-D))m_c
D+\sinh(m_c(R-D)))\sinh(m_cr)}{\sinh(m_cR)m_cr}.\nonumber
\end{eqnarray*}
For $d \phi / dr$ to be continuous at $R=d$, we need:
\begin{eqnarray}
\left(\frac{\phi_c}{\phi_c^{(lin)}}\right)^{n+1} =
\frac{R^3}{D^3}\frac{m_cD\cosh(m_c D)-
\sinh(m_cD)}{m_cR\cosh(m_cR)- \sinh(m_cR)}. \label{C22}
\end{eqnarray}
 The largest value of
$\vert \phi_1 / \phi_c \vert$ occurs when $r=0$. For this
linearisation to be valid everywhere we must therefore require:
\begin{eqnarray}
1&-&\frac{m_c D\cosh(m_c D)- \sinh(m_cD)}{m_cR\cosh(m_cR)-
\sinh(m_cR)}\left(1-\frac{m_c R}{\sinh(m_c R)}\right) \nonumber\\ &-&
\frac{(\cosh(m_c(R-D))m_c D+\sinh(m_c(R-D)))}{\sinh(m_cR)} \ll 1.
\end{eqnarray}
We shall see below that $m_c R \ll 1$.  Given that $m_c R$ is small,
the left hand side of the above condition becomes:
$$ \frac{m_c^3 D^3}{2m_c
R}\left[\frac{3(\sinh(m_cD)-m_cD\cosh(m_cD))}{m_c^3 D^3}\right] + 1 +
\sinh(m_c D)m_c D- \cosh(m_c D) + \mathcal{O}(m_cR)
$$ For this quantity to be small compared with $1$, we need both $m_c D \ll 1$, which implies $m_c R \ll 1$,  and $m_c^2 D^3 /
2R \ll 1$.  For all $n \neq -4$ we define:
\begin{eqnarray}
D_{c} &=& \left(n(n+1)\right)^{\frac{n+1}{n+4}}\left(\frac{3\beta
m_{p}}{4\pi M_{pl} \vert n \vert}\right)^{\frac{n+2}{n+4}}, \\
D_{\ast} &=& \left(\frac{n(n+1)}{MR}\right)^{\frac{n+1}{3}}
\left(\frac{3\beta m_{p}}{4\pi M_{pl} \vert
n\vert}\right)^{\frac{n+2}{3}},
\end{eqnarray}
we note that $D_{\ast}/D_c = (D_c / R)^{(n+1)/3}$. With these
definitions the requirement that $(m_c D)^2 \ll 1$ is equivalent
to $D \gg D_c$.  When $n < -4$ we need $D \ll D_{\ast}$ for $m_c^2
D^3 / R \ll 1$, whereas when $n > 0$ we need $D \gg D_{\ast}$ for
the same condition to hold.  Therefore, when $n > 0$ we need $D \gg D_{c}, D^{\ast}$, whereas for $n < -4$ we require $D_{c} \ll D \ll D^{\ast}$.

When $n > 0$, no matter what value $R$ takes, there will always be
some range of $D$ for which the linear approximation holds.  If $n
< -4$, however, we must require that $D_{\ast} \gg D_{c}$ for
there to exist \emph{any} value of $D$ for which the linear
approximation is valid.  It follows that, for the linear
approximation to be valid for \emph{any} $D$ (when $n < -4$) we
need $R \gg D_{c}$, i.e. $m_\phi(\phi_p) R \ll 1$.

When $n=-4$ we need both $D \ll D_{\ast}$ and: $$
(12)^{3/2}\lambda^{1/2}\left(\frac{3\beta m_{p}}{4\pi
  nM_{pl}}\right) \ll 1.
$$
This second condition implies both that $m_{c} D \ll 1$ and $m_{\phi}(\phi_p)
R \ll 1$.

We conclude that, for large enough particle separations, it is
always possible to find some region where the linear approximation
holds when $n > 0$. However, when $n \leq -4$ we must also require
that $m_{\phi}(\phi_p) R \ll 1$ for there to be \emph{any} value
of $D$ for which the linear approximation is appropriate.

 Whenever
the linear approximation holds, it follows from $m_c D \ll 1$ and
eq. (\ref{C22}) that we must have: $$\phi_c \approx
\phi_c^{(lin)}\quad \Leftrightarrow \quad m_{c} \approx
m_{\phi}(\phi_c^{(lin)}). $$

\subsection{Pseudo-Linear Regime}
The pseudo-linear approximation proceeds in much the same way as
it did in section \ref{singpseu}, and appendix \ref{singpseuApp},
for an isolated body. Near each of the particles we can use the
same inner approximation as was used for a single body.  This is
because, near any one particle, the other particles are
sufficiently far away that their effect is very much sub-leading
order.
\\ \\
\noindent\textit{Inner and Outer Approximations} \\
\indent The inner approximation (for $n \neq -
4$) is therefore:
$$ \phi/\phi_c \sim \bar{\phi}/\phi_c = A -
\frac{m_c^2D^3}{3(n+1)r}\left(\frac{\phi_c}{\phi_c^{(lin)}}\right)^{n+1},
$$ where $A$ is to be determined by the matching procedure. We will
deal with the $n=-4$ case separately later.

Previously, the outer approximation was defined so that it
remained valid as $r \rightarrow \infty$.  In this model, however,
we are assuming that everything is symmetric about $r=D$, and so
we need only to require that the outer approximation remain valid up to
$r=D$. Near $r = D$, we assume that $\phi \approx \phi_c$ and
linearise about $\phi_c$. Requiring that $d\phi/dr = 0$ when
$r=D$, we find: $$ \phi/\phi_c \sim 1 + \frac{1}{n+1}\left(1-
\frac{\cosh(m_c(r-D))D}{r} - \frac{\sinh(m_c(r-D))}{m_c r}\right).
$$ The outer approximation, as defined above, is also good for $n
\neq -4$.
\\ \\
\noindent\textit{Matching} \\
\indent For the pseudo-linear approximation to work we need, just as we
did in appendix \ref{singpseuApp}, there to exist an intermediate
region where both the inner and outer approximations are
simultaneously valid.  We will discuss what conditions this
requirement imposes shortly, but, before we do, we shall assume
that such a region \emph{does} exist, and match the inner and
outer approximations.  Matching the inner and outer approximations
we find:
\begin{eqnarray}
A &=& 1 + \frac{1}{1+n}\left(1-\cosh(m_c D) + m_cD\sinh(m_cD)\right),
\\ \left(\frac{\phi_c}{\phi_c^{(lin)}}\right)^{n+1} &=&
\frac{3(\cosh(m_c D)m_c D - \sinh(m_c D))}{(m_cD)^3},
\label{Mavg1}
\end{eqnarray}
\\ \\
\noindent\textit{Conditions for matching} \\
\indent
For the outer approximation to remain valid in the intermediate
region, where $r \approx d$ say, we require that, for all
$r\in(d,D)$: $$ \left \vert 1-\cosh(m_c D) + m_c D \sinh(m_c D) -
\frac{m_c^2 D^3}{3 d}\frac{3(\cosh(m_c d)m_c D - \sinh(m_c
D))}{(m_cD)^3} \right \vert \ll 1. $$ This is equivalent to $m_c D
\ll 1 \Rightarrow \phi_{c} \approx \phi_{c}^{(lin)}$.  We must
also require $$ \frac{m_c^2 D^3}{3 d} \ll 1. $$ We define
$d_{min}$ to be the smallest value of $d$ for which the above
condition holds.

For the inner approximation, $\phi \sim \bar{\phi}$, to hold in
the intermediate region, we require that conditions, similar  to those
that were found in the isolated body case, hold (see section
\ref{singpseu} and appendix \ref{singpseuApp}).
Specifically, we require that for all $r$ in $(R,d)$
\begin{eqnarray}
\frac{R^3}{3}\gg \int^{d}_{R}d r' \int^{d}_{r'} dr^{\prime \prime}
r^{\prime \prime} \left(\frac{\phi_p}{\bar{\phi}(r^{\prime
\prime})}\right)^{n+1}. \label{intexr2}
\end{eqnarray}
We also require that $(\phi_c/\phi(r=0))^{n+1} \ll 1$.  We note that
$(\phi_{c}^{(lin)}/\phi_{p})^{n+1} = \rho_{p}/\rho_{c} =
D^3/R^3$. We consider the subcases $n < - 4$, $n > 0$ and $n=-4$
separately.

\subsubsection{Case $n<-4$}
The analysis proceeds in the same way as it did for an isolated body
in appendix \ref{singpseuApp}. We find that we must require
$$ m_c^2 R^2 = \left(\frac{D_c}{R}\right)^{\frac{n+4}{n+1}} < 2\vert
n+1 \vert\left(1- \left(\frac{R}{D}\right)^{3/\vert n+1 \vert}\right).
$$
If we only wish for the pseudo-linear approximation to remain valid up to $r=R$, then we must require the weaker condition
$\bar{\phi}(r=R)/\phi < 1$.  This is equivalent to
$$ m_c^2 R^2 = \left(\frac{D_c}{R}\right)^{\frac{n+4}{(n+1)}} < 3\vert
n+1 \vert\left(1- \left(\frac{R}{D}\right)^{3/\vert n+1 \vert}\right).
$$
Provided that either of these conditions hold, the conditions for the outer approximation to be valid in the intermediate region are automatically satisfied.

Whenever the first of the
above requirements holds, the pseudo-linear approximation will give
accurate results.

When the latter (and weaker) of the two conditions fails, we
expect pronounced non-linear behaviour near the particles. When
this happens the far field induced by
each particle becomes $\beta$-independent. We will discuss how
this affects the values of $\phi_{c}$ and $m_{c}$ in section
\ref{C33} below.

$m_c D \ll 1$ implies $\phi_c \approx \phi_{c}^{(lin)}$ via eq.
\ref{Mavg1}.  It follows that $m_c \approx
m_{\phi}(\phi^{(lin)})$. The resulting macroscopic theory,
therefore, looks precisely like it did in the linear regime.
\subsubsection{Case $n > 0$}
The analysis for the $n>0$ case proceeds in the much same way as
it did for a single body (see appendix \ref{singpseuApp}). For the
outer approximation to hold, we require that: $$ D/ D_c
> 1, $$ which implies $m_c D \ll 1$.  We also need $$
\frac{D_{\ast}}{D} <
\left[2(n+1)\left(1-\left(\frac{R}{D}\right)^{\frac{3}{n+1}}\right)\right]^{\frac{n+1}{3}}.
$$ If we only wish to require that the pseudo-linear approximation
hold up to $r=R$, then we can relax this second condition to: $$
\frac{D_{\ast}}{D} <
\left[3(n+1)\left(1-\left(\frac{R}{D}\right)^{\frac{3}{n+1}}\right)\right]^{\frac{n+1}{3}}.
$$ When the first of the two conditions holds, the pseudo-linear
approximation gives accurate results.  Whenever the second
condition fails, we expect pronounced non-linear behaviour near
the particles.  When this happens, we expect that the far field
(in $r \gg R$) will attain a critical form. We discuss the
consequences of this in section \ref{C33} below.

As in the $n < - 4$ case, when the pseudo-linear approximation holds we have $m_c D \ll 1$ and so $\phi_c \approx \phi_{c}^{(lin)}$. This implies $m_c \approx m_{\phi}(\phi^{(lin)})$.

\subsubsection{Case $n=-4$}
The case of $n=-4$ is, as it was in the one particle case, the
most complicated to study.  However, the analysis proceeds almost
entirely along the same lines as it did in the one particle case.
We find, using the results of appendix \ref{singpseuApp}, that the
field near the particles will behave like:
\begin{eqnarray}
\frac{\phi}{\phi_c} \sim A + \frac{\sqrt{\frac{3}{2y}}R}{m_c r},
\label{phi4int}
\end{eqnarray}
where $y = y_0 + \ln(r/R)$, and, provided non-linear effects are
small i.e. $y_0 \gtrsim 0.6$, we find $y_0$ to be given by: $$
\frac{(m_\phi(\phi_p)R)^3}{9} = \sqrt{\frac{3}{2 y_0}} +
\frac{1}{3}\left(\sqrt{\frac{3}{2 y_0}}\right)^3. $$ When $n=-4$,
we have that $m_{\phi}(\phi^{lin})D = m_{\phi}(\phi_p) R$.

Using this, and matching the inner approximation to the outer one,
we find:
\begin{eqnarray}
A &=& 1 + \frac{1}{1+n}\left(1-\cosh(m_c D) + m_cD\sinh(m_cD)\right),
\\ \sqrt{\frac{3}{2 (y_0 +\ln(D/R))}} &=& \left(\cosh(m_c D)m_c D -
\sinh(m_c D)\right)/3. \label{phi4match2}
\end{eqnarray}
where $y(D) = y_0 + \ln(D/R)$.

Whenever non-linear effects are small we have $y_0 \gtrsim 0.6$,
which implies $$\left(\sqrt{\frac{3}{2y_0}}\right)^3/3 \ll
\sqrt{\frac{3}{2y_0}}$$ and so $$ \sqrt{\frac{3}{2y_0}} \approx
\frac{(m_{\phi}(\phi_p)R)^2}{9} =
\frac{(m_{\phi}(\phi^{(lin)}D))^3}{9}.$$ Since $y_0 > 0.6$ implies
$\sqrt{\frac{3}{2y_0}} < 1.6$ and so it follows eq.
(\ref{phi4match2}) that: $$ (m_c D)^3 \approx
(m_{\phi}(\phi^{(lin)}D))^3\sqrt{\frac{1}{1+\ln(D/R)/y_0}}. $$
The above approximation is very accurate when it predicts $m_c D <
1$, and even when $m_c D \gtrsim 1$ it gives a good estimate for
$m_c D$.

For macroscopic, everyday, bodies with densities of the order of
$1\,-\,10\,{\mathrm{g\,cm}}^{-3}$, we tend to find $\ln(D/R) \approx 11$.
If linear theory is to give a good estimate of $m_c$, we need:
$$ \ln(D/R)/y_0 \ll 1 \Rightarrow y_0 \gg 11 \Rightarrow
m_c(\phi^{(lin)}) \ll 1.5.$$  More generally, linear theory gives
a good approximation to $m_c$ whenever: $$
2(m_{\phi}(\phi^{(lin)})D)^6 \ln(D/R) / 243 \ll 1. $$ Therefore, unless
$D/R$ is improbably large ($D \gtrsim R e^{243/2}$), we will have we
will have $m_c \approx m_c^{(lin)}$ whenever $m_c^{(lin)}D \lesssim 1$.
When $m_c(\phi^{(lin)}) \gtrsim 1.5$, we actually move into the a
regime of $\beta$-independent, critical behaviour, more about
which shall be said below.

\subsubsection{Summary}
When the pseudo-linear approximation holds (and
$m_{\phi}(\phi^{(lin)})D \lesssim 1$), we have found that $m_c
\approx m_{\phi}(\phi^{lin})$.  As a result, despite the fact that
there does \emph{not} exist an everywhere valid linearisation of
the field equations, linear theory actually gives the correct
value of $m_c$, at least to a good approximation.

\subsection{Non-linear Regime \label{C33}}
When the pseudo-linear approximation fails it is because
non-linear effects have become important near the surface of the
particles, and they have developed thin-shells of their own.  Far from
the particles we expect that the field will take its critical form
as given by eqs. (\ref{critbeta}a-c). We can find the value of
$\phi_{c}$, in this case, by matching our the outer approximation
for $\phi$ to the critical form of the far field around the
particles.

When $n < -4$ we will have critical behaviour whenever: $$
m_{\phi}^2(\phi_p) R^2 =
\left(\frac{D_c}{R}\right)^{\frac{n+4}{(n+1)}} > 3\vert n+1
\vert\left(1- \left(\frac{R}{D}\right)^{3/\vert n+1 \vert}\right).
$$ When this happens the far-field ($r \gg R, r \ll D$) behaviour
of the chameleon takes its critical form. We have found this to be
well approximated by: $$ \phi \sim \bar{\phi}=A\phi_{c}-
\left(\frac{\gamma(n)}{\vert n\vert}\right)^{1/\vert n+2 \vert}
\left(MR\right)^{\frac{n+4}{n+1}}\frac{1}{r}. $$ Performing the
matching to the outer approximation we find that:
\begin{eqnarray*}
A &=& 1 - \frac{1}{3}\left(1-\cosh(m_c D) + m_c D\sinh(m_c
D)\right),\\ \left(\frac{\gamma(n)}{\vert n\vert}\right)^{1/\vert n+2 \vert}
\left(MR\right)^{\frac{n+4}{n+2}} &=&
\frac{\phi_{c}}{(n+1)m_{c}}\left(\cosh(m_c D)m_c D -\sinh(m_c
D)\right) \\ &=& \frac{n
(MD)^3}{3}\left(\frac{M}{\phi_c}\right)^{n+1}\frac{3\left(\cosh(m_c
D)m_c D -\sinh(m_c D)\right)}{(m_c D)^3} \\ &\approx& \frac{n
(MD)^3}{3}\left(\frac{M}{\phi_c}\right)^{n+1}.
\end{eqnarray*}
We therefore have that: $$ m_c^{(crit)} \approx \frac{\sqrt{3\vert
n+1 \vert}}{D} \left(\frac{R}{D}\right)^{\frac{n+4}{2(n+1)}}\left(\frac{\gamma(n)}{3}\right)^{\frac{1}{2\vert n+1\vert}}. $$
When $n > 0$, a similar analysis finds that $$ \frac{D_{\ast}}{D}
>
\left[3(n+1)\left(1-\frac{R}{D}^{\frac{3}{n+1}}\right)\right]^{\frac{n+1}{3}},
$$ and that $$ \frac{n
(MD)^3}{3}\left(\frac{M}{\phi_c}\right)^{n+1} \approx MR
\left(\frac{n(n+1)M^2}{m_c^2}\right)^{1/(n+2)}. $$ We therefore find $$ m_c^{(crit)} \approx \frac{\sqrt{3\vert n+1 \vert}}{D}
\left(\frac{R}{D}\right)^{1/2}. $$

In the $n=-4$ case, critical behaviour will actually emerge whenever
$\ln(D/R)/y_0 \gtrsim 1$. Non-linear effects are still responsible
for this critical behaviour but it is not necessarily the
case that the particle have developed thin-shells.  Indeed, the thin-shell
condition requires that $m_{\phi}(\phi_p)R \gtrsim 4$, whereas
critical behaviour (for $\ln(D/R) \approx 11$ as is typical),
emerges whenever $m_{\phi}(\phi_p)R \gtrsim 1.4$.   When the
particles do have thin-shells, $y_0=1/4$ (see appendix
\ref{singnon4App}) and so critical behaviour is seen whenever $D
\gtrsim 1.3 R$.

We find the $\beta$-independent critical mass for $n=-4$ theories using eq.
(\ref{phi4match2}) and taking $y_0 \ll \ln(D/R)$.  It follows
that, the critical value of $m_c$ for $n=-4$ theories is given by:
$$ m_{c}^{(crit)} = X / D, $$ where $X$ satisfies $$ X\cosh
X-\sinh X \approx  \frac{3\sqrt{3}}{\sqrt{2\ln(r/R)}}. $$ When
$\ln(D/R) =11$, we find $m_c^{crit}D \approx 1.4$.
\subsection{Summary}
In this appendix, we have performed a very detailed analysis of
the way in which the chameleon behaves inside a body that has a
thin-shell and which is made up of many small particles.  In this way,
we have been able to calculate the average chameleon mass inside
such a body.  Despite the in depth nature of the analysis, our
results can be summarised in a very succinct fashion.

The average mass, $m_c$, of the chameleon field inside a
thin-shelled body of average density $\rho_c$ is $$ m_{c} =
\min\left(\sqrt{n(n+1)\lambda}M\left(\frac{\beta \rho_c}{\lambda
\vert n\vert M_{pl}M^3}\right)^{\frac{n+2}{2(n+1)}},
m_{c}^{(crit)}(R,D,n)\right), $$ where $R$ is the radius of the
particles that make up the body, and $2D$ is the average
inter-particle separation.  The critical mass, $m_c^{(crit)}$, is
given by:
\begin{eqnarray*}
m_c^{(crit)} &\approx& \frac{\sqrt{3\vert n+1 \vert}}{D}
\left(\frac{R}{D}\right)^{\frac{q(n)}{2}}S(n), \qquad n \neq -4, \\
m_c^{(crit)} &\approx& X /D, \qquad n = -4,
\end{eqnarray*}
where $q(n) = \min((n+4)/(n+1),1)$, $S(n) =1 $ if $n>0$, $S(n) = (\gamma(n)/3)^{1/2\vert n + 1 \vert}$ if $n <0$, and
$$
X\cosh X-\sinh X \approx  \frac{3\sqrt{3}}{\sqrt{2\ln(D/R)}}.
$$
The dependence of $m_c^{(crit)}D$ vs. $\ln(D/R)$ is shown in FIG. \ref{FIGmaD}.  We can clearly see that $D \gg R$ implies $m_c^{(crit)}D\ll 1$ when $n \neq -4$. When $n=-4$, $m_c^{(crit)}$ can be seen to be less than $\sqrt{3}$ whenever $\ln(D/R) \gtrsim 2$.      

\section{Evaluation of $\alpha_{\phi}$ for white-dwarfs \label{alphaphiApp}}
In this appendix, we evaluate $\alpha_{\phi}$ for white-dwarfs
under a more accurate approximation than that used in section
\ref{compact}. This accurate evaluation allows for the effect of a
chameleon on the general relativistic stability of white dwarfs to
be studied in more detail. The effect of the chameleon on
compact objects such as these is considered in section
\ref{compact}.

The chameleon contribution to the energy of the white-dwarf was
found, in section \ref{compact}, to be: $$ W_{\phi} =
\frac{n+1}{n} \int d^3 x \frac{\beta\phi}{M_{pl}}\rho. $$ For the
chameleon theory to be valid, we must require that the chameleon
force is weak compared to gravity inside the white dwarf.  We
shall therefore ignore the chameleon corrections in the
behaviour of $\rho$ when evaluating $W_{\phi}$, because they must
be sub-leading order. In this way, we are able to accurately find
the leading order behaviour of $W_{\phi}$.  By leading order we
mean: ignoring general relativistic and chameleon corrections. It
is important to stress that we are \emph{not} ignoring special
relativistic corrections, which are very important; we are merely
assuming that the only volume force inside the white dwarf is, to
leading order, Newtonian gravity. This approximation is similar to
the one used in \cite{bholes} to evaluate the general relativistic
corrections to the energy of the white dwarf.

We shall assume that the polytropic equation of state, $P = K
\rho^{\Gamma}$, holds everywhere.  We will also approximate the
white dwarf to be spherically symmetric. Defining $\rho = \rho_c
\theta^{p}$ and $r=a\xi$, hydrostatic equilibrium implies that, to
leading order, $\theta$ satisfies the \emph{Lane-Emden equation}
\cite{bholes}: $$ \frac{1}{\xi} \frac{d}{d\xi} \xi^2 \frac{d
\theta}{\xi} = -\theta^{p}, $$ where $$ a = \left[\frac{(p+1)K
M_{pl}^2\rho_c^{(1/p-1)}}{4\pi}\right]^{1/2}. $$ We define
$\rho_{c}$ to be the density of matter in the centre of the star.
The boundary conditions $\theta(0)=1$, $\theta'(0)=1$ follow.  The
index $p$ is related to $\Gamma$ by $\Gamma \equiv 1+1/p$.  This
equation must be solved numerically.  The surface of the star is
at $r=R=a\xi_1$, which is defined to be the point where $\rho =
0$.  We will mostly be interested in the case of relativistic
matter, $\Gamma=4/3 \rightarrow p =3$.  For $p=3$ we have $\xi_1 =
6.89685$. The chameleon potential is then given by: $$ W_{\phi} =
\frac{n+1}{n} \frac{\beta \phi(\rho_c)}{M_{pl}}a^3 \rho_c
\int^{\xi_{1}}_{0} d \xi \xi^2 \theta^{\frac{p n}{n+1}}. $$ We
evaluate this integral numerically assuming a relativistic
equation of state (i.e $p=3$) and find: $$ W_{\phi} =
\frac{n+1}{n} \alpha_{\phi} \frac{\beta \phi(\rho_c)}{M_{pl}}
M_{star} $$ where $M_{star}=m_{u}N$ is the mass of the star. The
values of $\alpha_{\phi}$ are given in table \ref{table1}.  As $n
\rightarrow \pm \infty$ we find $\alpha_{\phi} \rightarrow 1$.
%
%\TABLE[tbp]{
%\begin{table}[tbp]
%\begin{center}
%\caption{Values of $\alpha_{\phi}$ for different $n$ \label{table1}}
%\begin{tabular}{|c|c||c|c|}
%\hline n & $\alpha_{\phi}$ & n & $\alpha_{\phi}$ \\ \hline -12 &
%0.8493 & 4 & 1.523 \\ \hline -10 & 0.8207 & 5 & 1.410 \\ \hline -8 &
%0.7786 & 6 & 1.337 \\ \hline -6 & 0.7110 & 7 & 1.286 \\ \hline -4 &
%0.5853 & 8 & 1.249 \\ \hline 1 & 3.613 & 9 & 1.220 \\ \hline 2 & 2.144
%& 10 & 1.197 \\ \hline 3 & 1.720 & 11 & 1.178 \\ \hline
%\end{tabular}
%\end{center}
%\end{table}
%}
%\TABLE[Hht]{
\begin{table}[tbp]
\begin{center}
\caption{Values of $\alpha_{\phi}$ for different $n$ \label{table1}}
\begin{tabular}{|c|c|p{0.5 cm}|c|c|}
\hline n & $\alpha_{\phi}$ & &n & $\alpha_{\phi}$ \\
\cline{1-2}\cline{4-5} -12 & 0.8493 & & 4 & 1.523 \\
\cline{1-2}\cline{4-5} -10 & 0.8207 & &5 & 1.410 \\
\cline{1-2}\cline{4-5} -8 & 0.7786 & &6 & 1.337 \\
\cline{1-2}\cline{4-5} -6 & 0.7110 & &7 & 1.286 \\
\cline{1-2}\cline{4-5} -4 & 0.5853 & &8 & 1.249 \\
\cline{1-2}\cline{4-5} 1 & 3.613 & &9 & 1.220 \\
\cline{1-2}\cline{4-5} 2 & 2.144 && 10 & 1.197 \\
\cline{1-2}\cline{4-5} 3 & 1.720 & &11 & 1.178 \\ \hline
\end{tabular}
\end{center}
\end{table}
%}
%\vspace{2in}
%

\end{document}